\documentclass[oneside,12pt]{Latex/Classes/PhDthesisPSnPDF}

\ifpdf  
    \pdfcatalog { /PageMode (/UseOutlines)
                  /OpenAction (fitbh)  }
\fi

\title{Medium Access Control for Dynamic Spectrum Sharing in Cognitive Radio Networks}

\ifpdf
  \author{\href{mailto:lethanh@emt.inrs.ca}{Le Thanh Tan}}
  \collegeordept{\href{http://www.emt.inrs.ca/emt}{Centre \'{E}nergie Mat\'{e}riaux \& T\'{e}l\'{e}communications \\ Institut National de la Recherche Scientifique}}
  \university{\href{http://inrs.ca}{Universit\'{e} du Qu\'{e}bec}}

  \crest{\includegraphics[width=8cm]{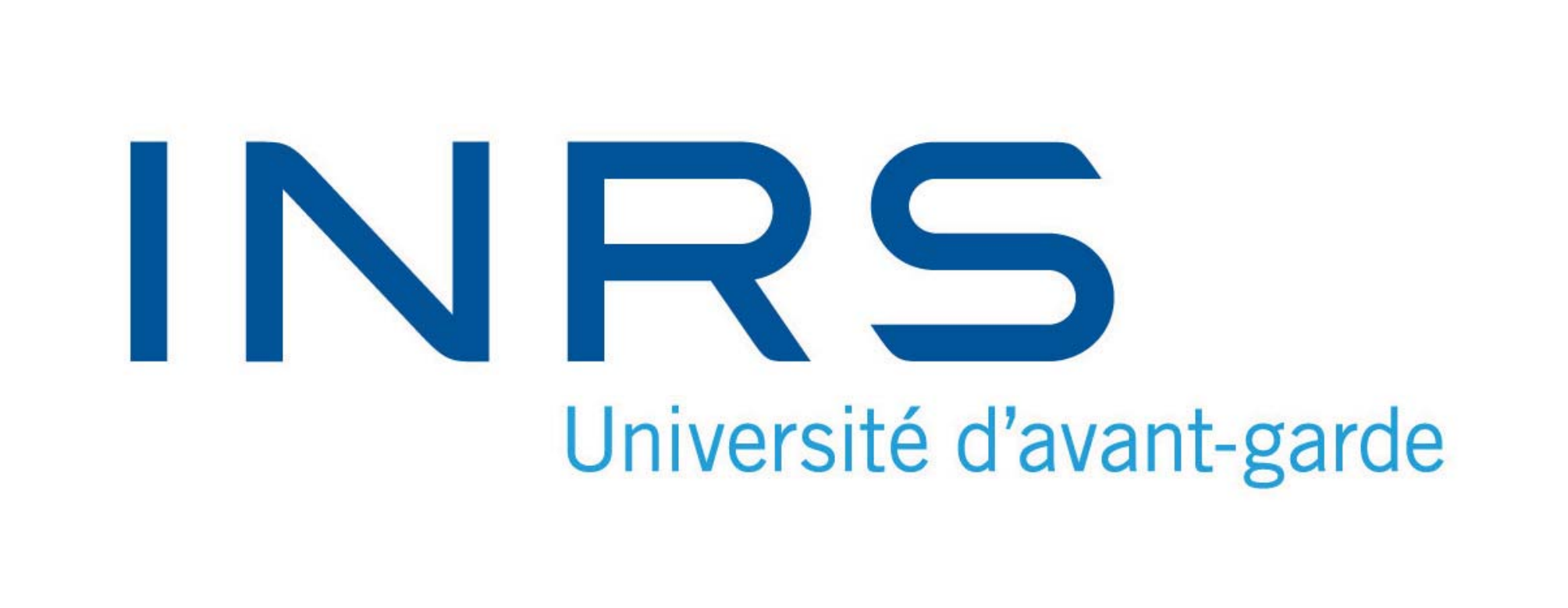}}
  
\else
  \author{Le Thanh Tan}
  \collegeordept{Centre \'{E}nergie Mat\'{e}riaux \& T\'{e}l\'{e}communications \\ Institut National de la Recherche Scientifique}
  \university{Universit\'{e} du Qu\'{e}bec}
  \crest{\includegraphics[width=8cm]{logo}}

\fi

\renewcommand{\submittedtext}{A dissertation submitted to INRS-EMT in partial fulfillment of the requirements for the degree of}
\degree{Doctor of Philosophy}
\degreedate{Montr\'{e}al, Qu\'{e}bec, Canada, August 2015\\ \copyright\ All rights reserved to Le Thanh Tan, 2015.} 

\usepackage[cmex10]{amsmath}
\usepackage{amsthm}
\usepackage {amssymb}
\usepackage{algorithmic}
\usepackage{array}
\usepackage{mdwmath}
\usepackage{mdwtab}
\usepackage{eqparbox}
\usepackage{subfigure}
\usepackage{hyperref}
\usepackage{algorithm}
\usepackage{algorithmic}
\usepackage{zref-user}
\usepackage{lipsum}
\usepackage{longtable}
\usepackage{caption}
\usepackage{lscape}
\usepackage{epstopdf}
\renewcommand{\baselinestretch}{1.86}

\newcommand{\argmin}{\operatornamewithlimits{argmin}}
\newcommand{\argmax}{\operatornamewithlimits{argmax}}
\newcommand{\beq}{\begin{equation}}
\newcommand{\eeq}{\end{equation}}
\newcommand{\beqn}{\begin{eqnarray}}
\newcommand{\eeqn}{\end{eqnarray}}
\newcommand{\beqno}{\begin{eqnarray*}}
\newcommand{\eeqno}{\end{eqnarray*}}
\newcommand{\bma}{\begin{displaymath}}
\newcommand{\ema}{\end{displaymath}}
\newcommand{\bnu}{\begin{enumerate}}
\newcommand{\enu}{\end{enumerate}}
\newcommand{\bce}{\begin{center}}
\newcommand{\ece}{\end{center}}
\newcommand{\btb}{\begin{tabular}}
\newcommand{\etb}{\end{tabular}}
\newcommand\NoIndent[1]{%
  \begingroup
  \par
  \parshape0
  #1\par
  \endgroup
}

\begin{document}

%\language{english}

% sets line spacing
\renewcommand\baselinestretch{1.2}
\baselineskip=18pt plus1pt

%: ----------------------- generate cover page ------------------------

\maketitle  % command to print the title page with above variables

%: ----------------------- cover page back side ------------------------
% Your research institution may require reviewer names, etc.
% This cover back side is required by Dresden Med Fac; uncomment if needed.

\newpage
\vspace{10mm}
1. External examiners: Prof. Wessam Ajib, Universit\'{e} du Qu\'{e}bec \`{a} Montr\'{e}al

\hspace{41.5mm} Prof. Jean-Fran\c cois Frigon, \'{E}cole Polytechnique de Montr\'{e}al

\vspace{10mm}
2. Internal examiners: Prof. Andr\'{e} Girard, INRS--EMT % Tiago Falk

\vspace{10mm}
3. Supervisor: Prof. Long Le, INRS--EMT 

\vspace{20mm}
Day of the defense:

\vspace{20mm}
\hspace{70mm}Signature from head of PhD committee:

%: ----------------------- abstract ------------------------

% Your institution may have specific regulations if you need an abstract and where it is to be placed in the document. The default here is just after title.

% The original template provides and abstractseparate environment, if your institution requires them to be separate. I think it's easier to print the abstract from the complete thesis by restricting printing to the relevant page.
 %\begin{abstractseparate}
   %\input{Abstract/abstract}
 %\end{abstractseparate}

%: ----------------------- tie in front matter ------------------------

\frontmatter
%: Declaration of originality

% Thesis statement of originality -------------------------------------

% Depending on the regulations of your faculty you may need a declaration like the one below. This specific one is from the medical faculty of the university of Dresden.

\begin{declaration}        %this creates the heading for the declaration page

%I herewith declare that I have produced this paper without the prohibited assistance of third parties and without making use of aids other than those specified; notions taken over directly or indirectly from other sources have been identified as such. This paper has not previously been presented in identical or similar form to any other German or foreign examination board.
%
%The thesis work was conducted from XXX to YYY under the supervision of PI at ZZZ.

I hereby declare that I am the sole author of this dissertation. This is a true copy of the dissertation, including any required final revisions, as accepted by my examiners.

I understand that my dissertation may be made electronically available to the public.

\vspace{10mm}

Montreal,

\end{declaration}

% ----------------------------------------------------------------------

% Thesis Abstract -----------------------------------------------------

%\begin{abstractslong}    %uncommenting this line, gives a different abstract heading
\begin{abstracts}        %this creates the heading for the abstract page

The proliferation of wireless services and applications over the past decade has led to the rapidly increasing demand in wireless spectrum.
Hence, we have been facing a critical spectrum shortage problem even though several measurements have indicated 
that most licensed radio spectrum is very underutilized. These facts have motivated the development of dynamic spectrum 
access (DSA) and cognitive radio techniques to enhance the efficiency and flexibility of spectrum utilization.

In this dissertation, we investigate design, analysis, and optimization issues for joint spectrum sensing and
cognitive medium access control (CMAC) protocol engineering for cognitive radio networks (CRNs). 
The joint spectrum sensing and CMAC design is considered under the interweave spectrum sharing
paradigm and different communications settings. Our research has resulted in four major research contributions,
which are presented in four corresponding main chapters of this dissertation.  

First, we consider the CMAC protocol design with parallel spectrum sensing for both single-channel
and multi-channel scenarios, which is presented in Chapter 5. The considered setting captures the case where each secondary user (SU) is 
equipped with multiple transceivers to perform sensing and access of spectrum holes on several channels simultaneously.
 
Second, we study the single-transceiver-based CMAC protocol engineering for hardware-constrained CRNs, which is covered
in Chapter 6. In this setting, each SU performs sequential sensing over the assigned channels and access one available
channel for communication by using random access. We also investigate the channel assignment problem for SUs to maximize
the network throughput.

Third, we design a distributed framework integrating our developed CMAC protocol and cooperative sensing for multi-channel and 
heterogeneous CRNs, which is presented in details in Chapter 7. The MAC protocol is based on the p-persistent carrier sense
multiple access (CSMA) mechanism and a general cooperative sensing adopting the a-out-of-b aggregation rule is employed.
Moreover, impacts of reporting errors in the considered cooperative sensing scheme are also investigated.

Finally, we propose an asynchronous Full--Duplex cognitive MAC (FDC-MAC) exploiting the full-duplex (FD) capability
of SUs' radios for simultaneous spectrum sensing and access. The research outcomes of this research are presented in Chapter 8.
Our design enables to timely detect the PUs' activity during transmission and adaptive reconfigure the sensing time
and SUs' transmit powers to achieve the best performance. Therefore, the proposed FDC--MAC protocol
is more general and flexible compared with existing FD CMAC protocols proposed in the literature.
 
We develop various analytical models for throughput performance analysis of our proposed CMAC protocol designs.
Based on these analytical models, we develop different efficient algorithms to configure the CMAC protocol including channel 
allocation, sensing time, transmit power, contention window to maximize the total throughput of the secondary network.
Furthermore, extensive numerical results are presented to gain further insights and to evaluate the performance of our CMAC protocol designs.
Both the numerical and simulation results confirm that our proposed CMAC protocols can achieve efficient spectrum utilization 
and significant performance gains compared to existing and unoptimized designs.

\end{abstracts}
%\end{abstractlongs}

% ---------------------------------------------------------------------- 

% Thesis Acknowledgements ------------------------------------------------

%\begin{acknowledgementslong} %uncommenting this line, gives a different acknowledgements heading
\begin{acknowledgements}      %this creates the heading for the acknowlegments

%I would like to acknowledge the thousands of individuals who have coded for the LaTeX project for free. It is due to their efforts that we can generate professionally typeset PDFs now.

First of all, I wish to express my deepest thanks and gratitude to my Ph.D. advisor Prof. Long Le for his precious advices and encouragements throughout the years. 
I would like to thank him as well for the long inspiring discussions we had together, for all the confidence he put in me. 
My distinguished thanks should also go to all the jury members who have accepted to take time from their very busy schedule in order to evaluate this dissertation. 
It is quite an honor for me to have them peer review this work.
I would like also to thank all the graduate students in INRS that have collaborated with me during the last five years.
Special thanks should also go to my wife Ta Thi Huynh Nhu for her patience during all the time she spent alone in my homeland while I was doing research. 
Last but not the least, I would like to thank all my family members for their continued support, encouragement and sacrifice throughout the years, and I will be forever indebted to them for all what they have ever done for me.

\end{acknowledgements}
%\end{acknowledgmentslong}

% ------------------------------------------------------------------------

% Thesis Dedictation ---------------------------------------------------

\begin{dedication} %this creates the heading for the dedication page

%To ...
To my parents

To my wife Ta Thi Huynh Nhu

To my cute daughters Le Thanh Van and Le Ha My

\end{dedication}

% ----------------------------------------------------------------------

%: ----------------------- contents ------------------------

\setcounter{secnumdepth}{3} % organisational level that receives a numbers
\setcounter{tocdepth}{3}    % print table of contents for level 3
\tableofcontents            % print the table of contents
% levels are: 0 - chapter, 1 - section, 2 - subsection, 3 - subsection

%: ----------------------- list of figures/tables ------------------------

\listoffigures	% print list of figures

\listoftables  % print list of tables

% this file is called up by thesis.tex
% content in this file will be fed into the main document

% Glossary entries are defined with the command \nomenclature{1}{2}
% 1 = Entry name, e.g. abbreviation; 2 = Explanation
% You can place all explanations in this separate file or declare them in the middle of the text. Either way they will be collected in the glossary.
%\addcontentsline{toc}{chapter}{Glossary}
\begin{glossary} %this creates the heading for the dedication page
%\begin{landscape}
\pagestyle{empty}
\begin{center}
\begin{longtable}{l p{130mm}}
\captionsetup{font=Large}
\caption*{\textbf{GLOSSARY}}  \\
%\hline
%\textbf{Abbreviation} & \textbf{Explanation}  \\
%\hline
%\endfirsthead
%\multicolumn{2}{c}%
%{\tablename\ \thetable\ -- \textit{Continued from previous page}} \\
%\hline
%\textbf{Abbreviation} & \textbf{Explanation} \\
%\hline
%\endhead
%\hline \multicolumn{2}{r}{\textit{Continued on next page}} \\
%\endfoot
%%\hline
%\endlastfoot
\textbf{AP} & Access point \\                              % A
\textbf{ACK} & Acknowledgment  \\ 
\textbf{AWGN} & Additive white Gaussian noise  \\ 
\textbf{CMAC} & Cognitive MAC protocol \\                           % C
\textbf{CRN}  & Cognitive radio network \\
\textbf{CSCG}  & Circularly symmetric complex Gaussian \\
\textbf{CSMA}  & Carrier sense multiple access \\
\textbf{CSMA/CA} & Carrier sense multiple access with collision avoidance \\
\textbf{CTS}  & Clear-to-send \\
\textbf{DCF} & Distributed coordination function \\        % D
\textbf{DIFS} & Distributed inter-frame space \\               
\textbf{DSA} & Dynamic spectrum access \\                  
\textbf{EGC} & Equal gain combining \\                     % E
\textbf{FCC} & Federal Communication Committee \\          % F
\textbf{FD} & Full-duplex \\                             
\textbf{FDTx} & Full-duplex transmission \\
\textbf{FDC--MAC} & Full-duplex cognitive MAC protocol \\                             
\textbf{GSC} & Generalized selection combining \\          % G
\textbf{HD} & Half-duplex \\                               % H
\textbf{HDTx}  & Half-duplex transmission \\
\textbf{MAC} & Medium Access Control \\                    % M
\textbf{MC} & Markov chain \\   
\textbf{McMAC} & Multi-channel MAC protocol \\
\textbf{MRC} & Maximal ratio combining \\                  
\textbf{NAV} & Network allocation vector  \\               % N
\textbf{NP-hard} & Non-deterministic Polynomial-time hard \\
\textbf{OFDM} & Orthogonal frequency-division multiplexing  \\           % O
\textbf{OSA} & Opportunistic spectrum access  \\           
\textbf{PD} & Propagation delay \\                         % P
\textbf{PDF} & Probability density function \\
\textbf{PMF} & Probability mass function \\
\textbf{PSK} & Phase-shift keying \\                       
\textbf{PUs} & Primary users  \\                           
\textbf{QoS} & Quality of service   \\                     % Q
\textbf{QSIC} & Quality of self-interference cancellation \\
\textbf{RR} & Round-robin      \\                          % R
\textbf{RTS} & Request-to-send       \\                   
\textbf{RV} & Random variable \\
\textbf{SC} & Selection combining  \\                      % S
\textbf{SSCH} & Slotted seeded channel hopping MAC protocol \\
\textbf{SDCSS} & Semi-distributed cooperative spectrum sensing \\
\textbf{SIC} & Self-interference cancellation \\
\textbf{SINR} & Signal-to-interference-plus-noise ratio \\
\textbf{SLC} & Square-law combining \\                   
\textbf{SIFS} & Short inter-frame space \\                   
\textbf{SLS} & Square-law selection  \\                   
\textbf{SNR} & Signal-to-noise ratio  \\                   
\textbf{SSC} & Switch and stay combining \\                         
\textbf{SUs} & Secondary users  \\                         
\textbf{WiFi}  & Wireless Fidelity  \\                     % W
\textbf{WRAN}  & Wireless regional area network  \\    
\end{longtable}
\end{center}
\newpage
\pagestyle{plain}
%\end{landscape}
\end{glossary} %this creates the heading for the dedication page
\mainmatter
%\include{0_frontmatter/glossary_manual}

%\makenomenclature
%\nomenclature{$a$}{The number of angels per unit area}%
%\nomenclature{$N$}{The number of angels per needle point}%
%\nomenclature{$A$}{The area of the needle point}%
%
%\printnomenclature

%\renewcommand{\chaptername}{} % uncomment to print only "1" not "Chapter 1"

%: ----------------------- subdocuments ------------------------

% Parts of the thesis are included below. Rename the files as required.
% But take care that the paths match. You can also change the order of appearance by moving the include commands.
%\include{9_backmatter/declaration}
%\include{1_introduction/introduction}	% background information

%\include{10_summary/Summary}
%\include{11_summary_francaise/Summary_Francaise}

% this file is called up by thesis.tex
% content in this file will be fed into the main document

%: ----------------------- introduction file header -----------------------
%\chapter{Chapter 1. Introduction}
\chapter{Introduction}

% the code below specifies where the figures are stored
\ifpdf
    \graphicspath{{1/figures/PNG/}{1/figures/PDF/}{1/figures/}{1_introduction/figures/}}
\else
    \graphicspath{{1/figures/EPS/}{1/figures/}}
\fi

% ----------------------------------------------------------------------
%: ----------------------- introduction content ----------------------- 
% ----------------------------------------------------------------------

The proliferation of wireless services and applications over the past decade has led to the rapidly increasing demand in wireless spectrum.
Hence, we have been facing a critical spectrum shortage problem. Several recent measurements have, however, reported that 
a large portion of licensed radio spectrum is very underutilized in spatial and temporal domains \cite{Datla09}, \cite{Zhao07}.
These facts have motivated the development of dynamic spectrum access (DSA) techniques to enhance the efficiency and flexibility of spectrum utilization.
DSA can be categorized into three major models, namely dynamic exclusive use, open sharing, and hierarchical access models \cite{Zhao07}. 
The third model which is also refereed to as opportunistic spectrum access (OSA), which provides fundamental ground for an extremely
active research theme, i.e., the cognitive radio research. OSA in cognitive radio networks
can be further divided into three access paradigms, namely underlay, overlay, and interweave \cite{Zhao07, Goldsmith09, III00}. 

The research in this dissertation considers the interweave paradigm where the licensed
spectrum is shared between the primary and secondary networks whose users are refereed to 
as primary and secondary users (PUs and SUs), respectively. 
In particular, SUs can opportunistically exploit spectrum holes (i.e., idle spectrum in time, frequency, and space) for their data 
transmission as long as they do not severely interfere the transmissions of PUs. 
This access principle implies that PUs have strictly higher priority than SUs in accessing the underlying spectrum;
hence, SUs can only access the licensed spectrum if PUs do not occupy them.
Toward this end, SUs can perform spectrum sensing to explore spectrum holes and adopt suitable spectrum access mechanisms to share the 
discovered available spectrum with one another \cite{Liang08}.

Although the spectrum sensing and access functions are tightly coupled, they are usually not treated jointly in the existing multi-user cognitive radio literature.
Moreover, it is desirable to employ a distributed cognitive MAC protocol for spectrum sharing in many wireless applications, which is usually more cost-efficient
compared to the centralized cognitive MAC counterpart. An efficient cognitive MAC protocol should achieve good performance in certain performance measures such as
throughput (spectrum utilization), delay, fairness, and energy consumption.
This dissertation aims to engineer the distributed cognitive MAC protocol with extensive performance 
analysis and optimization for several practically relevant cognitive network settings.

\section{Dynamic Spectrum Access}

%Fig. 1
\begin{figure}[!t] 
\centering
\includegraphics[width=150mm]{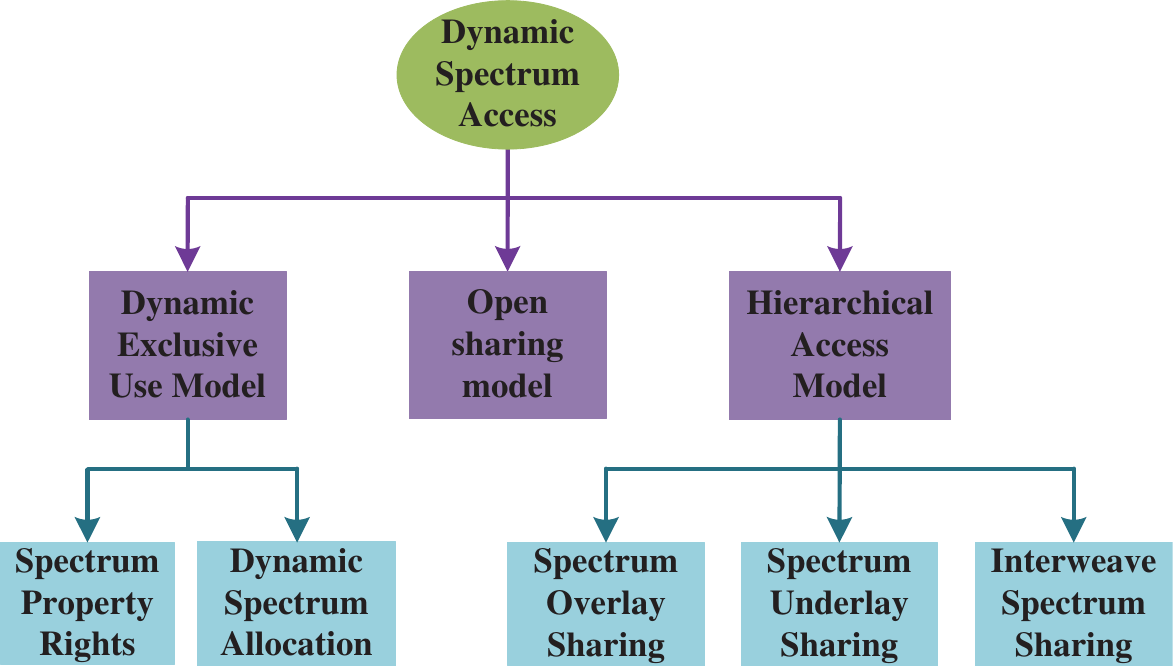}
\caption{The broad landscape of Dynamic Spectrum Access (DSA) \cite{Zhao07, Goldsmith09}.}
\label{taxoDSA_Chap3}
\end{figure}

To resolve the under-utilization of wireless spectrum and support the increasing spectrum demand of the wireless sector,
spectrum management authorities in many countries (e.g., Federal Communication Committee (FCC) in US)
have recently adopted more flexible spectrum management policies for certain parts of the
wireless spectrum such as TV bands compared to the rigid and non-dynamic policies employed in the past.
In the following, we describe three different DSA models, namely, dynamic exclusive use, open sharing, and hierarchical access models \cite{Zhao07} 
where the DSA taxonomy is illustrated in Fig.~\ref{taxoDSA_Chap3}.

%\begin{itemize}
%\item 
\textbf{Dynamic exclusive use model}: 
This model can be realized through two different approaches, namely, the spectrum property right and dynamic spectrum allocation.
In the spectrum property right approach \cite{Hatfield05}, the unused spectrum of the licensees can be leased or traded, which
can result in more flexibility in the spectrum management. Here, the market and economy play a critical role in achieving efficient
use of the spectrum. The dynamic spectrum allocation approach was proposed by the European DRiVE project \cite{Xu00} where the spatio--temporal 
traffic statistics are exploited for dynamic spectrum assignment to different services, which use the assigned spectrum exclusively.

%\item 
\textbf{Open sharing model}: 
This model is also called as \textit{Spectrum Commons} in \cite{Lehr05} where the spectrum can be shared by a number of peer users in 
a specific region under the open sharing basis. This spectrum sharing model is motivated by the very
successful deployment of wireless services in the unlicensed spectrum band (e.g., WiFi).
This spectrum sharing model can be realized in the centralized or distributed manner \cite{Raman05, Huang05}.

%\item 
\textbf{Hierarchical access model}: This model classifies the spectrum access users into PUs (i.e., licensed users) and SUs.
Specifically, it adopts a hierarchical access structure where PUs must be protected from the interference created by SUs. 
There are three main approaches for opportunistic spectrum sharing under this model, namely underlay, overlay, and interweave \cite{Zhao07, Goldsmith09, III00}. 
%\end{itemize}

This dissertation focuses on the cognitive MAC protocol design for cognitive radio networks
under the hierarchical access model, which will be described in more details in the following section. 

\section{Hierarchical Access Model and Cognitive MAC Protocol}

\subsection{Hierarchical Access Model}

Recent changes in spectrum regulation and management have motivated the development of the hierarchical spectrum access
techniques for efficient spectrum sharing between the primary users/network and secondary users/network.
In this spectrum sharing model, SUs are allowed to access the spectrum as long as
transmissions from PUs can be satisfactorily protected from the interference caused by the SUs. 
As mentioned previously, one of the three approaches can be adopted for hierarchical spectrum access, i.e.,
the underlay, overlay, and interweave access paradigms \cite{Zhao07, Goldsmith09, III00}, which are
discussed in the following.

%\begin{itemize}
%\item 
In the underlay paradigm, PUs' and SUs' transmissions can co-exist on the same frequency band at the same time; however,
 the interference created by SUs at each PU must remain below an allowable limit \cite{Le08, Kim081}.  
Advanced communications and signal processing techniques can be employed for efficient interference management and mitigation
to maintain this interference constraint. In particular, we can set the transmit power of the secondary signal
to maintain the interference constraint at the PU's receiver by using power control techniques. Furthermore, the more advanced beamforming
technique can be adopted in the multi-antenna secondary system to achieve good performance for SUs while nulling interference
toward PUs. In general, SUs need to acquire suitable information related to PUs such as channel state information,
PUs' locations to maintain the interference constraint. This is, however, difficult to achieve in practice.

%\item 
For the overlay paradigm, PUs' and SUs' signals can also co-exist at the same frequency band simultaneously;
however, the SU is assumed to have knowledge about the PUs' codebooks and messages, which can be achieved in different ways \cite{Goldsmith09}.
Knowledge about the PUs' codebooks and/or messages can be used to mitigate or cancel the interference seen at the PUs' and SUs' receivers.
Also, SUs can use the knowledge about the PUs' messages to eliminate the interference generated by the PUs' transmissions seen at the SUs' receivers.
Moreover, SUs can use part of their energy to enhance and assist the communications of PUs through cooperative communications
 \cite{Ozgur07, zuo111, zuo10} and use the remaining energy for their communications.  
With this cooperation, the interference due to the SUs' signals to the PUs' receivers can be compensated by the cooperation gain 
while the SUs can still exploit the spectrum for their transmissions \cite{Ozgur07, zuo111, zuo10}. 
Implementation of the overlay spectrum sharing paradigm requires SUs to acquire information about PUs' messages before the PUs begin their transmissions, 
which is not easy to achieve in practice. 

Finally, in the interweave paradigm, SUs opportunistically exploit the idle spectrum in time and/or frequency domains, which 
are called spectrum holes or white spaces, for data communication \cite{Kim08,Su08,Su07,Nan07,Cor07,Hsu07,
Le11,wang11,zhang113,choi11,zhang11,zhang112,jeon12,jha11,su081,mo08,ko10,Le12,Park11,tan2015distributed,tan2015distgen,tan2014joint,Tanconf2012,Tanconf2012b,Tanconf2013,Tanconf2015,Tan10,Tan101,Tan2011using,Duong10,Tan2010primary}. To identify
spectral holes, SUs must employ a suitable spectrum sensing strategy. Upon correctly
discovering spectrum holes on the licensed spectrum, SUs can transmit at high power levels
without subject to the interference constraints at SUs as in the underlay spectrum sharing paradigm. 
For the case where the SUs can synchronize their transmissions with PUs' idle time intervals
and perfect sensing can be achieved, SUs will not create any interference to active PUs. 
However, half-duplex SUs cannot sense the spectrum and transmit simultaneously;
therefore, SUs could not detect the event in which an PU changes its status from idle to active during their transmissions.
Consequently, SUs employing half-duplex radios may cause interference to active PUs.

%\end{itemize}

In comparison with the underlay and overlay paradigms, the interweave approach requires to acquire less 
information about the PUs and it can also utilize the spectrum more efficiently since there is no transmit power constraint \cite{Goldsmith09}. 
Moreover, PUs are not required to change or adapt their communication strategies or parameters to realize the spectrum sharing with SUs.
Inspired by these advantages, we focus on the MAC protocol design issues for this interweave spectrum sharing paradigm in this dissertation. 

\subsection{Cognitive MAC Protocol}

Different from conventional MAC protocols, a CMAC protocol must integrate the spectrum sensing function to identify spectrum holes
before sharing the available spectrum through a spectrum access mechanism.
In addition, a CMAC protocol must be designed appropriately considering the communication capability of SUs' radio, i.e., half--duplex (HD) 
or full--duplex (FD) radio.
Most existing research works have considered the design and analysis of HD CMAC protocols (e.g., see \cite{Cor09, Yu09} and references therein) where 
SUs are synchronized with each other to perform periodic spectrum sensing and access. 
% with the cycle $T_{\sf cycle}$. To protect the PU, we assume that SUs must stop their transmission and evacuate from the channel within the maximum delay of $T_{\sf eva}$, which is %referred to as channel evacuation time.
Due to the HD constraint, SUs typically employ a two-stage sensing/access procedure where they sense the spectrum in the first stage before accessing available channels for data transmission in the second stage \cite{Kim08,Su08,Su07,Nan07,Cor07,Hsu07,Le11,wang11,zhang113,choi11,zhang11,zhang112,jeon12,jha11,su081,mo08,ko10,Le12,Park11,tan2014joint,Tanconf2012,Tanconf2012b,Tanconf2013,Tanconf2015,Tan10,Tan101,Tan2011using,Duong10,Tan2010primary}. 
Some other works assume that the primary and secondary networks are synchronized with each other so exact idle intervals on the spectrum of interest
are known to the SUs \cite{su081,Su07,Su08}. This assumption would, however, be difficult to achieve in practice. 

In an HD CMAC protocol, if an PU changes from the idle to active status when the SUs are occupying the spectrum, then transmissions from SUs 
can cause strong interference to active PUs. With recent advances in the full-duplex technologies 
(e.g., see \cite{Duarte12,Everett14,Sabharwal14,Korpi14,Jain11,Choi10}), 
some recent works consider FD spectrum access design for cognitive radio networks (CRNs) \cite{Afifi14,Cheng15} where each SU can perform 
sensing and transmission simultaneously. This implies that SUs may be able to detect the PUs' active status while they are utilizing the licensed spectrum 
with the FD radio. However, self-interference due to simultaneous sensing and transmission of
FD radios may lead to performance degradation of the SUs' spectrum sensing.
Therefore, FD CMAC protocols must be designed appropriately to manage the FD self-interference by using suitable mechanisms such as power control. 

%\section{Research Background and Motivations}
\section{Research Challenges and Motivations}

Efficient design of CRNs imposes many new challenges that are not present in the conventional wireless  networks \cite{Haykin05,Cabric04,
Cor09,Gavrilovska14,Ahmed14}. These design challenges are originated from the variations of white spaces in  time and space as well as the 
time-varying wireless channel quality. SUs who have the lower priority than PUs in accessing the licensed spectrum
must tune their operations and transmission parameters so as not to cause harmful interference to active PUs.
We can highlight some major challenges in designing opportunistic cognitive MAC (CMAC) protocols for CRNs as follows:
\textit{1)} Efficient joint spectrum sensing and access design; \textit{2)} The trade-off between spectrum utilization and interference to PUs; 
\textit{3)} Fair spectrum access among SUs; \textit{4)} Hardware limitations; \textit{5)} Multi-channel hidden/exposed terminal problem; 
\textit{6)} Control channel configuration; \textit{7)} Spectrum heterogeneity seen by SUs; \textit{8)} QoS provisioning; 
\textit{9)} Asynchronicity between SU and PU networks.

In this dissertation, we aim to design, analyze, and optimize CMAC protocols for CRNs under different practically relevant scenarios. 
Specifically, the first three contributions are related to the design of
synchronous CMAC protocols for the HD CRNs in three different settings whilst the last contribution involves the engineering of an 
asynchronous CMAC protocol to FD CRNs. To motivate our research, we briefly discuss the existing CMAC literature and describe
its limitations for different considered scenarios in the following.

%In the first contribution, we propose a two--stage CMAC protocol with parallel sensing 
%where each SU is equipped with multiple HD transceivers, which enable SUs to sense or access different channels simultaneously.
%In this setting, the required sensing time could be very short; hence, SUs can efficiently exploit available channels for
 %data transmission in the access phase.

\subsection{CMAC with Parallel Sensing}

In many existing works, the two--stage CMAC protocol with parallel sensing is considered where SUs are equipped with
 multiple HD transceivers; hence, they are able to sense or access multiple channels simultaneously. 
% \cite{Ma05, Konda08, wang08, Xie12, Liu10, zhang11}.
In this setting, the sensing time in the first sensing phase can be quite short, hence SUs can efficiently exploit the spectrum
holes to transmit data in the second transmission phase.
Existing works considering this scenario have the following limitations: they \textit{i)} usually assume perfect spectrum sensing \cite{Ma05, Konda08}; 
\textit{ii)} only analyze the network throughput performance but very few of them consider optimizing the network and protocol
 parameters to maximize the network performance \cite{zhang11}.

Efficient CMAC design with parallel sensing presents the following challenges.
The design must integrate both spectrum sensing and access functions in
the CMAC protocol. For distributed implementation, contention-based MAC mechanism
based on the popular carrier sense multiple access (CSMA) principle can be
employed for contention resolution. Furthermore, protocol optimization,
which adapts different protocol parameters such as contention windows, back-off durations,
and access probability to maximize the network performance (e.g., throughput) while
appropriately protecting active PUs is an important
issue to investigate. Toward this end, performance analysis for the developed CMAC protocol is
needed. Our dissertation indeed makes some contributions along these lines. 

\subsection{CMAC with Sequential Sensing}

Deploying multiple transceivers for each SU as in the previous case leads to high implementation complexity and cost. 
Therefore, the scenario in which each SU has a single transceiver would be preferred in many practical
applications. % \cite{jia08, Cor07, Kim08, Su08, Su07, Nan07, Le208, Hsu07}.
A CMAC protocol in this setting, however, must employ sequential sensing where spectrum sensing for
multiple channels is sensed by each SU one by one in a sequential manner. Moreover, each SU can access at most one 
idle channel for communications. 
  
Spectrum sensing, which is integrated into the CMAC protocol, should be carefully designed to achieve efficient
tradeoff between sensing time overhead and achieved performance. Different sensing designs have been proposed
to deal with this scenario, i.e., sensing-period and optimal channel sensing order 
optimization \cite{Kim08}, random- and negotiation-based spectrum-sensing schemes \cite{Su08,Su07, Nan07, Le208, Hsu07}.
Moreover, each SU may only sense a subset of channels to reduce the spectrum sensing time.
Furthermore, the channel assignment optimization for SUs to determine the optimal subset of channels
allocated for each SU is important, which is, however, not well investigated in the literature.

Therefore, efficient MAC protocol engineering in this setting must consider the interactions
between achieved sensing/network performance and channel assignments.
Moreover, fairness among SUs could be considered in designing the CMAC and channel assignment.
Our dissertation makes some important contributions in resolving some of these issues.

\subsection{CMAC with Cooperative Sensing}

In order to enhance the spectrum sensing performance, cooperative spectrum sensing can be employed where a fusion 
center can be installed (e.g., at an access point (AP)) to collect individual sensing data
from SUs based on which it makes final sensing results and broadcasts them to the SUs \cite{Gan07, Gane07}.
Detailed design of such cooperative spectrum sensing can vary depending on the underlying
aggregation rules and hard- or soft-decision strategy \cite{Chaud12, Lee13}. 
In addition, the distributed cooperative spectrum sensing would be preferred to the centralized one in most practical deployments.
In distributed  cooperative spectrum sensing, the sensing tasks are shared among SUs (i.e., there is fusion center (AP))
and each SU performs sensing independently and then makes its own decision of sensing outcomes with some suitable 
exchanges of sensing data. 

In \cite{Sala10, Park11,zhang11,Seu10}, different multi-channel CMAC protocols were proposed considering either 
parallel or sequential spectrum sensing method. 
In these existing works, design and optimization of the cooperative spectrum sensing parameters are pursued. 
However, they do not consider spectrum access issues or they assume the availability of multiple transceivers for 
simultaneously sensing all channels. 

In addition, the MAC protocol in these works are  non-optimized standard window-based CSMA MAC protocol, which is known to 
achieved smaller throughput than the optimized $p$-persistent CSMA MAC protocol \cite{Cali00}.
Furthermore, either the single-channel setting or homogeneous network scenario (i.e., SUs experience the same channel condition 
and spectrum statistics for different channels) was assumed in these works. 
Finally, existing cooperative spectrum sensing schemes rely on a central controller
to aggregate sensing results for white space detection (i.e., centralized design). 
We have developed a CMAC protocol employing distributed cooperative spectrum
sensing in this dissertation, which overcomes several limitations of existing CMAC protocols mentioned above.

\subsection{Full--Duplex MAC Protocol for Cognitive Radio Networks}

As discussed earlier, SUs in the HD CRN must employ the two-stage sensing/access procedure due to the HD constraint.
This constraint also requires SUs be synchronized during the spectrum sensing stage, which could be difficult to achieve in practice.
In general, HD-MAC protocols may not exploit white spaces very efficiently since significant sensing time may be required,
which would otherwise be utilized for data transmission. Moreover, SUs may not timely detect the PUs' activity during 
their transmissions, which can cause severe interference to active PUs.

Recent advances in FD technologies \cite{Duarte12, Everett14, Sabharwal14} has opened
opportunities to develop FD cognitive MAC protocols that can overcome many aforementioned limitations
of HD CMAC protocols. With FD radios,  SUs can indeed perform spectrum sensing and access simultaneously.
However, the presence of self-interference, which is caused by power leakage from the transmitter
to the receiver of a FD transceiver, may indeed lead to sensing/transmission performance degradation. 

Several FD CMAC protocols have been recently proposed. The authors in \cite{Afifi14} investigate three operation modes for the FD cognitive
radio network (i.e., transmission-only, transmission-sensing, and transmission-reception modes) and the optimal parameter configurations by solving
 three corresponding optimization problems. 
In \cite{Cheng15}, another FD CMAC protocol is developed where both PUs and SUs are assumed to employ the same $p$-persistent MAC protocol for
 channel contention resolution. This design is not applicable to CRNs where PUs should have higher spectrum access priority compared to SUs.
In general, it is desirable to design a FD CMAC protocol with following characteristics: \textit{i)} a distributed
FD CMAC protocol can operate efficiently in an asynchronous manner where SUs are not required be synchronized with each other; \textit{ii)}
 SUs must timely detect the PUs' reactivation during their transmissions to protect active PUs; \textit{iii)} the
FD CMAC protocol can be easily reconfigured where its parameters can be adapted to specific channel 
state and network conditions. Our developed FD CMAC protocol satisfactorily achieve these requirements.

\section{Research Contributions and Organization of the Dissertation}

%Fig. 1
\begin{figure}[!t] 
\centering
\includegraphics[width=120mm]{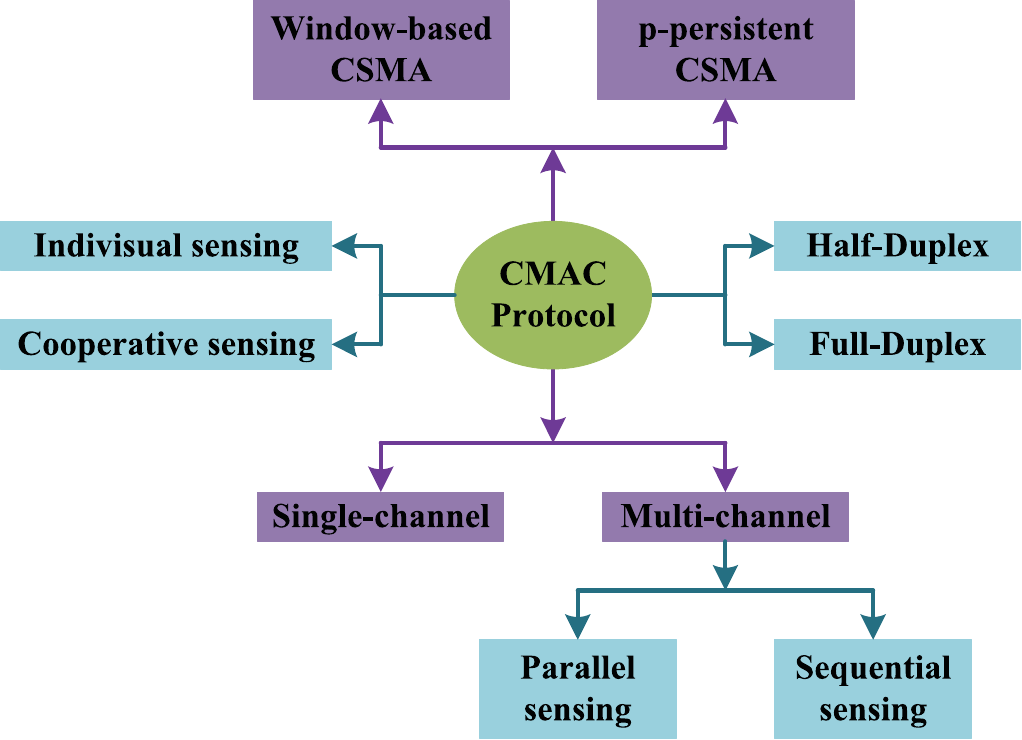}
\caption{Positioning our contributions within the broad CMAC landscape.}
\label{ContributionPositioning_Chap3}
\end{figure}

The overall objective of this dissertation is to design, analyze, and optimize the cognitive MAC protocol for efficient
dynamic spectrum sharing in CRNs. Our main contributions, which are highlighted in Fig.~\ref{ContributionPositioning_Chap3},
are described in the following.

\begin{enumerate}

\item
\textit{CMAC protocol design with parallel sensing \cite{Le11}:} We consider the setting where each SU can perform parallel 
sensing and exploit all available channels for data transmissions.
We develop a synchronous CMAC protocol integrating the parallel spectrum sensing function. We then analyze the throughput performance
 and study the optimization of its access and sensing parameters for throughput maximization. This work fundamentally
 extends the throughput-sensing optimization framework in  \cite{Liang08}, which was proposed for the single-SU setting.

%This work has been published in \cite{Le11}.

\item 
\textit{CMAC protocol and channel assignment with sequential sensing \cite{Le12, Tanconf2012b}:} This contribution covers the joint
 sensing and access design for the scenario where each SU
performs sequential sensing over multiple channels and can access at most one idle channel for communications \cite{Le12}. 
We devise and analyze the saturation throughput performance of the proposed CMAC protocol. 
Then, we develop efficient channel allocation for SUs to maximize the total throughput of the secondary network.
Furthermore, we also investigate a fair channel allocation problem where each node is allocated a subset of channels which 
are sensed and accessed periodically by using a CMAC protocol.

\item 
\textit{Distributed CMAC protocol and cooperative sensing design \cite{tan2014joint}:} 
We propose a distributed cooperative spectrum and $p$-persistent CMAC protocol for multi-channel and heterogeneous CRNs \cite{tan2014joint}. 
In particular, we develop the distributed cooperative spectrum sensing where we assume SUs directly exchange sensing results to make decisions 
on all channels' statuses by using the general a-out-of-b aggregation rule. 
We conduct the performance analysis and configuration optimization for the proposed CMAC protocol considering both perfect and imperfect exchanges of 
sensing results. We also propose a different joint cooperative spectrum and contention window-based MAC protocol for multi-channel and 
heterogeneous CRNs in \cite{Tanconf2013}.

\item 
\textit{Asynchronous full--duplex MAC protocol for cognitive radio networks \cite{tan2015distgen}:} 
We propose the FD cognitive MAC protocol (FDC--MAC) which employs the distributed $p$-persistent CSMA access mechanism and 
FD spectrum sensing \cite{tan2015distgen}.
Our design exploits the fact that FD SUs can perform spectrum sensing and access simultaneously, which enable them to detect the PUs' activity during transmission. 
Each data frame is divided into the sensing and access stages to timely detect the PUs' transmission and enable SUs' performance optimization.
Furthermore, we develop a mathematical model to analyze the throughput performance of the proposed FDC--MAC protocol. 
Then, we propose an algorithm to configure the CMAC protocol so that efficient self-interference management and sensing overhead control can be achieved.
The proposed FDC-MAC protocol design is very flexible, which can be configured to operate in the HD mode or
having simultaneous sensing and access for the whole data frame as in \cite{tan2015distributed}.

\end{enumerate}

The remaining of this dissertation is organized as follows. In chapter 4, we present the research background and literature review. 
In chapter 5, we discuss the joint MAC and sensing design under parallel sensing. We describe the proposed protocol design framework 
under sequential sensing in chapter 6. The developed distributed cooperative sensing and MAC design are presented in chapter 7.
Chapter 8 describes the distributed MAC protocol design for FD cognitive radio networks. 
Chapter 9 summarizes the contributions of the dissertation and point out some future research directions.

\chapter{Background and Literature Review}

% the code below specifies where the figures are stored
\ifpdf
    \graphicspath{{2/figures/PNG/}{2/figures/PDF/}{2/figures/}}
\else
    \graphicspath{{2/figures/EPS/}{2/figures/}}
\fi

% ----------------------------------------------------------------------
%: ----------------------- introduction content ----------------------- 
% ----------------------------------------------------------------------
%\section{Introduction}

%Design of efficient Medium Access Control (MAC) protocols is one of the most important and challenging tasks in engineering communication systems \cite{Pah94}. 
%For many applications, it is desirable to deploy the distributed MAC protocol, which is usually more cost-effective compared to the centralized MAC counterpart.
%MAC protocol enables multiple wireless users or devices to coordinate their transmissions in a common neighborhood so that the transmission from one node does not interfere with 
%those from other nodes. In general, an efficient MAC protocol should achieve good performance in terms of throughput (i.e., spectrum utilization), delay, fairness, and energy consumption.
%In emerging wireless applications, providing quality-of-service (QoS) differentiations for different types of users and networks
%as well as jointly designing MAC protocols with other network functionalities present remaining challenges. This thesis aims to resolve some of such challenges.

In this chapter, we present the research background and literature survey on different research issues studied in our dissertation. 
In particular, some background on spectrum sensing is presented where basics of spectrum sensing and more advanced spectrum sensing 
methods such as cooperative spectrum sensing are discussed. 
Then, we describe fundamentals of MAC protocols for conventional single- and multi-channel wireless networks. Finally, we provide a 
comprehensive review and taxonomy of the state-of-the-art CMAC protocols.

\section{Research Background}

\subsection{Spectrum Sensing}

Spectrum sensing and channel probing, which aim at acquiring real-time spectrum/channel information required by the 
cognitive MAC layer, are critical components of CRNs. In particular, spectrum sensing performs the
following tasks \cite{Haykin09}: \textit{i)} detection of spectrum holes; \textit{ii)} determination of spectral 
resolution of each spectrum hole; \textit{iii)} estimation of the spatial directions of incoming interfering signal; \textit{iv)} signal 
classification. Among these tasks, detection of spectrum holes, which is probably the most important one, boils down to a binary hypothesis-testing problem.
Therefore, detection of spectrum holes on a narrow frequency band is usually refereed to as spectrum sensing, which
aims at deciding the presence or absence of PUs in the underlying band. Some extensive reviews of spectrum sensing 
techniques for CRNs can be found in \cite{Yu09,Haykin09, Axell12, Akyildiz11}.

We now describe the spectrum sensing in some more details.
Let $B$ be the signal bandwidth, $f_s$ be the sampling frequency, $\tau$ be the observation time over which signal samples are 
collected, then $N = \left\lceil \tau f_s\right\rceil$ is the number of samples (we assume $N = \tau f_s$ is an integer for simplicity). 
Let $s(n)$ denote the PU's signal with zero mean and variance $\sigma_s$, $u(n)$ be the additive white Gaussian noise (AWGN) with zero mean and 
variance $N_0$, $\gamma = \frac{\sigma_s}{N_0}$ be the received signal-to-noise ratio (SNR). 
Note that $s(n)$ could capture the fading and multi-path effects of the wireless channels.

Let $\mathcal{H}_0$ and $\mathcal{H}_1$ represent two events (hypotheses) corresponding to the cases where
PUs are absent or present in the underlying spectrum, respectively.
The sampled signal received at the SU, denoted as $y(n)$, corresponding to these two hypotheses can be written as 
\beqn
\label{EQN_Hypo_Chap2}
\mathcal{H}_0 : && y(n) = u(n) \nonumber\\
\mathcal{H}_1 : && y(n) = s(n) + u(n).
\eeqn
 
Let $Y$ denote the test statistic and $\lambda$ be the decision threshold.
The objective of narrow-band spectrum sensing is to make a decision on presence or absence of the PUs' signals (i.e., choose
hypothesis $\mathcal{H}_0$ or $\mathcal{H}_1$) based on the received signals (observations). 
Such decision can be made by comparing the test statistic with the threshold as follows: 
\beqn
\label{EQN_Hypo1_Chap2}
\mathcal{H}_0 : && Y < \lambda \nonumber\\
\mathcal{H}_1 : && Y > \lambda.
\eeqn

To quantify the spectrum sensing performance, we usually employ two important performance measures, namely detection probability $\mathcal{P}_d$ and
false-alarm probability $\mathcal{P}_f$. 
In particular, $\mathcal{P}_d$ captures the probability that a spectrum sensor successfully detects a busy channel and $\mathcal{P}_f$
represents the event where a spectrum sensor returns a busy state for an idle channel (i.e., a transmission opportunity is overlooked).
Therefore, the detection and false-alarm probabilities can be expressed as
\beqn
\label{EQN_Hypo2_Chap2}
\mathcal{P}_d = \Pr\left(Y > \lambda \left|\mathcal{H}_1\right.\right) \nonumber\\
\mathcal{P}_f = \Pr\left(Y > \lambda \left|\mathcal{H}_0\right.\right).
\eeqn
A spectrum sensing algorithm is more efficient if it achieves higher $\mathcal{P}_d$  and lower $\mathcal{P}_f$. With higher $\mathcal{P}_d$,
active PUs would be better protected and  lower $\mathcal{P}_f$ means that the white space is likely not overlooked by SUs (cognitive radios). 

There are many different spectrum sensing strategies proposed in the literature, which can be categorized into
the following three main groups, namely energy detection, matched-filter detection, and feature detection where the first one is 
non-coherent detection and the others belong to the coherent detection.
In coherent detection, a spectrum sensor requires a priori knowledge of PU's signal to coherently detect the presence of the PU.
In contrast, a spectrum sensor in non-coherent detection does not require a priori knowledge of PU's signal for detection.
Furthermore, SUs can perform spectrum sensing independently or several SUs can collaborate to perform detection, which
are termed individual sensing and cooperative spectrum sensing, respectively in this dissertation.

\subsubsection{Individual Sensing}

%\vspace{0.2cm}
%\noindent
%\textit{\textbf{$\bigstar$ Energy Detection}} 

\begin{enumerate}

\item {\textbf{\textit{Energy Detection:}}

\NoIndent{

We first discuss the energy detection which is one popular spectrum sensing method since it is simple and does not require a priori knowledge of PU's signal.
As the name suggests, this sensing method detects the PUs' signal by using the energy of the received signal.
Specifically, the test statistic is based on the energy of received signal, i.e.,
\beqn
Y = \frac{1}{N}\sum_{i = 1}^{N} \left|y(n)\right|^2
\eeqn
Consider the scenario with a single antenna and a single sensor.
Under $\mathcal{H}_0$, $Y$ involves the sum of the squares of $N$ standard Gaussian variates with zero mean and variance $N_0$. 
Therefore, $Y$ follows a central chi-squared distribution with $2N$ degrees of freedom, i.e., $Y \sim \chi^2_{2B}$. 
Under $\mathcal{H}_1$, the test statistic $Y$ follows a non-central distribution $\chi^2$ with $2N$ degrees of freedom and a 
non-centrality parameter $2\gamma$ \cite{Urkowitz67}. Therefore, we can summarize the test statistic under the two hypotheses as 
\beqn
\label{EQN_Hypo_Chap2_1}
\mathcal{H}_0 : && Y \sim \chi^2_{2B} \nonumber\\
\mathcal{H}_1 : && Y \sim \chi^2_{2B} \left(2\gamma\right).
\eeqn

We now derive the detection and false alarm probabilities where we assume that transmission signals from PUs are complex-valued phase-shift keying (PSK) signals, whereas the noise is independent and identically distributed (i.i.d.) circularly symmetric complex Gaussian $\mathcal{CN}(0,N_0)$.
For large $N$, the probability density function (PDF) of $Y$ under hypothesis $\mathcal{H}_0$ and $\mathcal{H}_1$ can be approximated by Gaussian distributions
 with mean $\mu_0 = N_0$, $\mu_1 = \left(\gamma+1\right)N_0$ and variance $\sigma_0^2 = \frac{1}{N} N_0^2$, $\sigma_1^2 = \frac{1}{N} \left(\gamma+1\right) N_0^2$, respectively. Therefore, the detection and false-alarm probabilities given in (\ref{EQN_Hypo2_Chap2}) can be rewritten as \cite{Liang08}
\beqn
\label{EQN_Pd_Chap2_approx}
\mathcal{P}_d \left(\lambda, \tau \right) = \mathcal{Q}\left(\left(\frac{\lambda}{N_0} - \gamma - 1 \right) \sqrt{\frac{\tau f_s}{2 \gamma + 1}}\right), 
\eeqn
\beqn
\label{EQN_Pf_Chap2_approx}
\mathcal{P}_f \left(\lambda, \tau \right) = \mathcal{Q}\left(\left(\frac{\lambda}{N_0} - 1 \right) \sqrt{\tau f_s}\right). 
\eeqn
Recall that $\lambda$ is the detection threshold for an energy detector, $\gamma$ is the SNR of the PU's signal at the SU, $f_s$ is the sampling frequency, 
$N_0$ is the noise power, and $\tau$ is the sensing interval.
Moreover, $\mathcal{Q}\left(\circ\right)$ is defined as $\mathcal{Q}\left( x \right) = \left( {1/\sqrt {2\pi } } \right)\int_x^\infty  {\exp \left( { - {t^2}/2} \right)dt}$. 

Similarly, for other kinds of noise and primary signals, we can derive the detection and false-alarm probabilities as presented in \cite{Liang08}.
For the i.i.d. and correlated fading channels and multi-antenna setting, the probabilities of detection and false alarm can be 
found in \cite{Digham03, Digham07, Duong10, Tan2010primary}.}} 

\item {\textbf{\textit{Other Sensing Mechanisms:}}

\NoIndent{

There are other spectrum sensing methods proposed for CRNs, e.g., waveform-based sensing \cite{Tang05, Mishra05, Geirhofer06}, cyclostationarity-based sensing, radio identification based sensing, multi-taper spectral estimation \cite{Shankar05, Cabric04}, matched-filtering \cite{Cabric04}, wavelet transform based estimation, 
Hough transform, and time-frequency analysis \cite{Haykin05, Haykin09}.
In addition, the wavelet approach can be employed for detecting edges in the power spectral density of a wideband channel \cite{Tian06}.
The wavelet-based sensing method proposed in \cite{Tian06} is extended in \cite{Tian07, Fanzi11, Tan10, Tan101, Tan2011using} by using sub-Nyquist sampling 
which is termed as the compressed spectrum sensing.}
}
\end{enumerate}

%\vspace{0.2cm}
%\noindent
%\textit{\textbf{$\bigstar$ Other Sensing Mechanisms}}

\subsubsection{Cooperative Spectrum Sensing}

Cooperative spectrum sensing has been proposed to improve the sensing performance where several SUs collaborate with each other to identify spectrum holes.
The detection performances of cooperative spectrum sensing in terms of detection and false-alarm probabilities can be significantly 
better than those due to individual spectrum sensing thanks to the spatial diversity gain.
However, it is required to collect, share, and combine individual sensing information to make final sensing decisions. 
In addition, spectrum sensors can make soft or hard decisions based on their measurements and then share their decisions to others \cite{Chaud12}.

When the number of spectrum sensors that collaborate to sense one particular channel increases, 
the sensing overhead is increased because more sensing information must be exchanged, which would consume more system resources.
Therefore, optimized design of cooperative sensing is important, e.g., we can optimize the number of sensors assigned to
sense each channel to achieve good balance between sensing performance and overhead.
In the following, we describe cooperative spectrum sensing with hard and soft decisions.

%\vspace{0.2cm}
%\noindent
%\textbf{\textit{$\bigstar$ Cooperative spectrum sensing with hard decisions}} 

\begin{enumerate}

\item \textbf{\textit{Cooperative spectrum sensing with hard decisions}} 

\NoIndent{
Cooperative spectrum sensing can be realized via centralized and distributed implementations.
In the centralized approach, a central unit (e.g., an AP) collects sensing information from SUs, makes sensing decisions 
then broadcasts them to all SUs. In the distributed sensing method, all SUs perform sensing on their assigned
channels then exchange the sensing results with others. Finally, SUs make their sensing decisions independently by themselves.

\begin{figure}[!t]
\centering
\mbox{\subfigure[]{\includegraphics[width=2.5in]{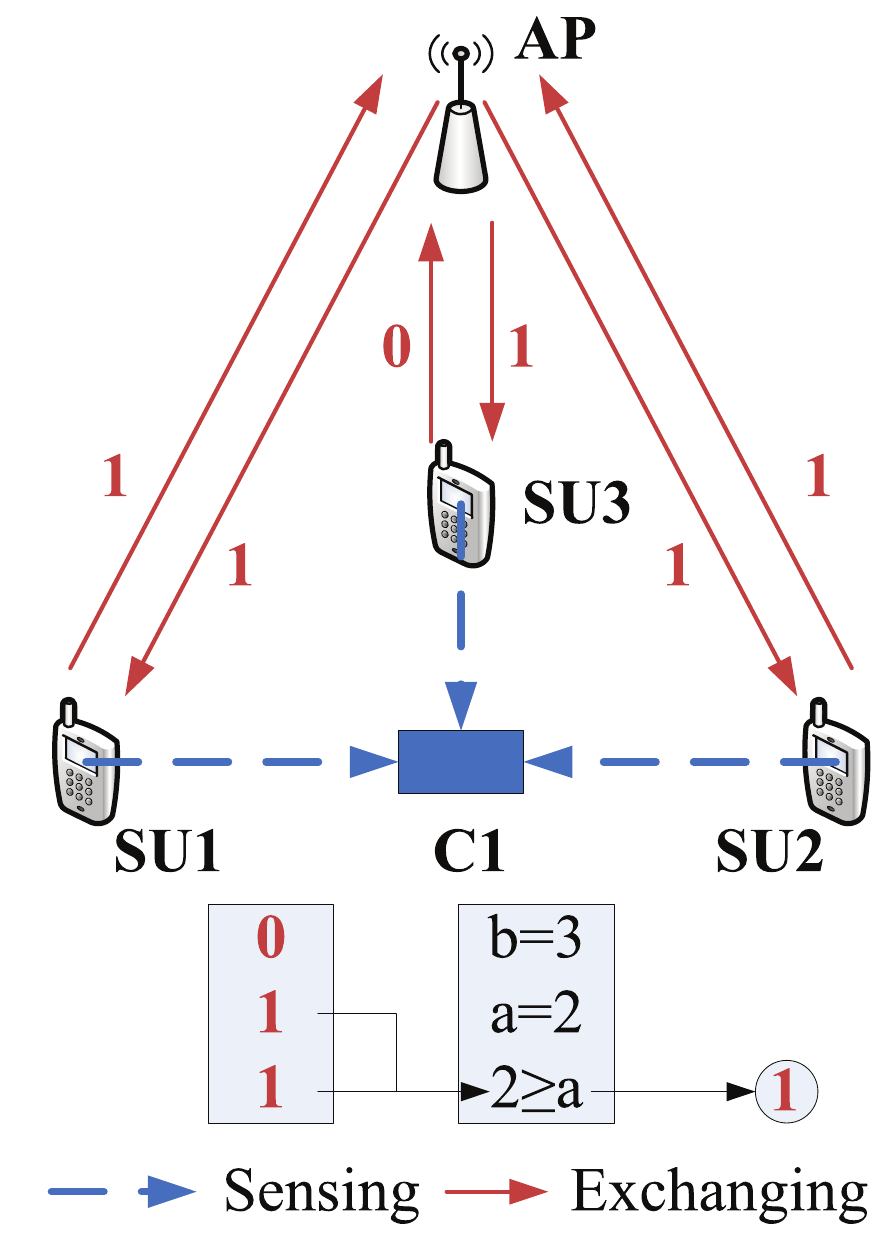} \label{CSS_eg_Chap2}}  
\subfigure[]{\includegraphics[width=2.5in]{DCSS_eg} \label{DCSS_eg_Chap2}} }
\caption{Cooperative sensing examples with (a) centralized processing,  (b) distributed processing.}
\end{figure}

We now study the centralized spectrum sensing algorithm.
We assume that each SU $i$ is assigned in advance a set of channels $\mathcal{S}_i$ for sensing
at the beginning of each sensing cycle. 
Upon completing the channel sensing, each SU $i$ sends the idle/busy states of all channels in $\mathcal{S}_i$ to the central unit
for further processing. Suppose that the channel status of each channel can be represented by one bit (e.g., 1 for idle and 0 for busy status).
Upon collecting sensing results from all SUs, the central unit decides the idle/busy status for all channels,
then it broadcasts the list of available channels to all SUs for exploitation. 

Consider a general cooperative sensing rule, namely $a$-out-of-$b$ rule, which is employed by the central unit to determine the idle/busy 
status of each channel based on reported sensing results from all SUs. 
In this scheme, the central unit declares that a channel is idle if $a$ or more SUs out of $b$ SUs report that the underlying channel is idle. 
The $a$-out-of-$b$ rule covers different other rules including OR, AND and Majority rules as special cases. 
In particular, when $a = 1$, it is the OR rule; when $a = b$, it is the AND rule; and when $a = \left\lfloor b/2\right\rfloor$, it is the Majority rule. 

We now analyze the cooperative sensing performance for a particular channel $j$. 
Let $\mathcal{S}^U_j$ denote the set of SUs that sense channel $j$ and $b_j = \left|\mathcal{S}^U_j\right|$ be the number of SUs sensing channel $j$. 
Then, the detection and false alarm probabilities for this channel can be calculated respectively as \cite{Wei11}
\beqn
\mathcal{P}_u^j\left( {\vec \varepsilon ^j}, {\vec \tau^j} , a_j  \right) = \sum_{l=a_j}^{b_j} \sum_{k=1}^{C_{b_j}^l} \prod_{i_1 \in \Phi^k_l} \mathcal{P}_u^{i_1j} \prod_{i_2 \in \mathcal{S}_j^{U} \backslash \Phi^k_l} \mathcal{\bar P}_u^{i_2j} \label{eq1_css_1_Chap2},
\eeqn
where $u$ represents $d$ or $f$ as we calculate the probability of detection $\mathcal{P}_d^j$ or false alarm $\mathcal{P}_f^j$, respectively; 
$\mathcal{P}_d^{ij}$ and $\mathcal{P}_f^{ij}$ are the probabilities of detection and false alarm at SU $i$ for channel $j$, respectively; 
$\mathcal{\bar P}$ is defined as $\mathcal{\bar P} = 1-\mathcal{P}$; 
$\Phi^k_l$ in (\ref{eq1_css_1_Chap2}) denotes a particular set with $l$ SUs whose sensing outcomes suggest that channel $j$ is busy given that 
this channel is indeed busy and idle as $u$ represents $d$ and $f$, respectively. In this calculation, we generate all possible sets
$\Phi^k_l$ where there are indeed $C_{b_j}^l $ combinations. Also, ${\vec \varepsilon ^j} = \left\{\varepsilon ^{ij} \right\}$, 
${\vec \tau^j} = \left\{\tau^{ij}\right\}$, $i \in \mathcal{S}_j^U$ represent the set of detection thresholds and sensing times, respectively. 

To illustrate the operations of the $a$-out-of-$b$ rule, let us consider a simple example for the centralized cooperative spectrum sensing 
implementation shown in Fig.~\ref{CSS_eg_Chap2}. Here, we assume that 3 SUs collaborate to sense channel one (C1) with $a = 2 $ and $b = 3$.
After sensing the channel, all SUs report their sensing outcomes to an AP. 
Here, the AP receives the reporting results comprising two ``1s'' and one ``0'' where ``1'' means that the channel is busy and ``0''  means 
channel is idle. Because the total number of ``1s'' is two which is larger than or equal to $a=2$, the AP outputs the ``1'' in the final 
sensing result, i.e., the channel is busy. Then, the AP broadcast this final sensing result to all SUs.

%\begin{figure*}%[!t]
%\centering
%\includegraphics[width=50mm]{DCSS_eg}
%\caption{Example for SDCSS on 1 channel.}
%\label{DCSS_eg_Chap2}
%\end{figure*}

Now we investigate the distributed cooperative spectrum sensing. 
Upon completing the channel sensing, each SU $i$ exchanges the sensing results (i.e., idle/busy status of all channels in $\mathcal{S}_i$) with other SUs for 
further processing. After collecting the sensing results, each SU will decide the idle/busy status for each channel. 
Fig.~\ref{DCSS_eg_Chap2} illustrates the distributed cooperative spectrum sensing using the $a$-out-of-$b$ rule where 3 SUs collaborate to sense 
channel one with $a = 2$ and $b = 3$. After sensing the channel, all SUs exchange their sensing outcomes. 
SU3 receives the reporting results comprising two ``1s'' and one ``0''. Because the total number of ``1s'' is two which is 
larger than or equal to $a=2$, SU3 outputs the ``1'' in the final sensing result, i.e., the channel is busy.
}

%\vspace{0.2cm}
%\noindent
%\textbf{\textit{$\bigstar$ Cooperative spectrum sensing with soft decisions}} 

\item \textbf{\textit{Cooperative spectrum sensing with soft decisions}} 

\NoIndent{
In this sensing method, the SUs must send their measurements (not the decisions) to the central unit and then the central unit will make final
sensing decisions and broadcast them to all SUs. Performance analysis of this method for the i.i.d., correlated fading channels, and multi-antenna 
settings is conducted in \cite{Digham03, Digham07, Duong10, Tan2010primary}.
Here, different combining techniques such as maximal ratio combining (MRC), selection combining (SC), equal gain combining (EGC), switch and stay 
combining (SSC), square-law selection (SLS), square-law combining (SLC), and generalized
selection combining (GSC) schemes can be employed to exploit the spatial diversity for sensing performance enhancement.}

\end{enumerate}

%\subsection{MAC Protocol}
%\subsubsection{MAC Framework}
%\vspace{0.2cm}
%\noindent
%\textbf{\textit{A window-based CSMA MAC protocol:}} 

\subsection{MAC Protocol in Traditional Wireless Networks}

\subsubsection{Single-channel MAC Protocols}

There are many random-access-based MAC protocols, which have been developed over the past decades and employed in different wireless
systems and standards. Popular MAC protocols include 
 ALOHA, Slotted ALOHA, $p$-persistent carrier sense multiple access (CSMA) \cite{Cali00} (including non-persistent, $p$-persistent,
and $1$-persistent) and CSMA/CA (CSMA with collision avoidance) \cite{bian00} (or window-based CSMA) MAC protocols.
In these MAC protocols, active stations (users) perform contention to capture the channel for data transmission 
in the distributed manner. Moreover, suitable mechanisms are adopted to mitigate the potential
collisions among users (e.g., the well-known backoff mechanism for contention resolution in the CSMA/CA protocol).
We provide more detailed discussions and performance analysis for some of these popular MAC protocols in the following where
 a single channel is shared by multiple users.

%\vspace{0.2cm}
%\noindent
%\textbf{\textit{$\clubsuit$ A Window-based CSMA MAC Protocol}}

\begin{enumerate}

\item \textbf{\textit{Window-based CSMA MAC Protocol}}

\NoIndent{

\begin{figure}[!t]
\centering
\includegraphics[width=130mm]{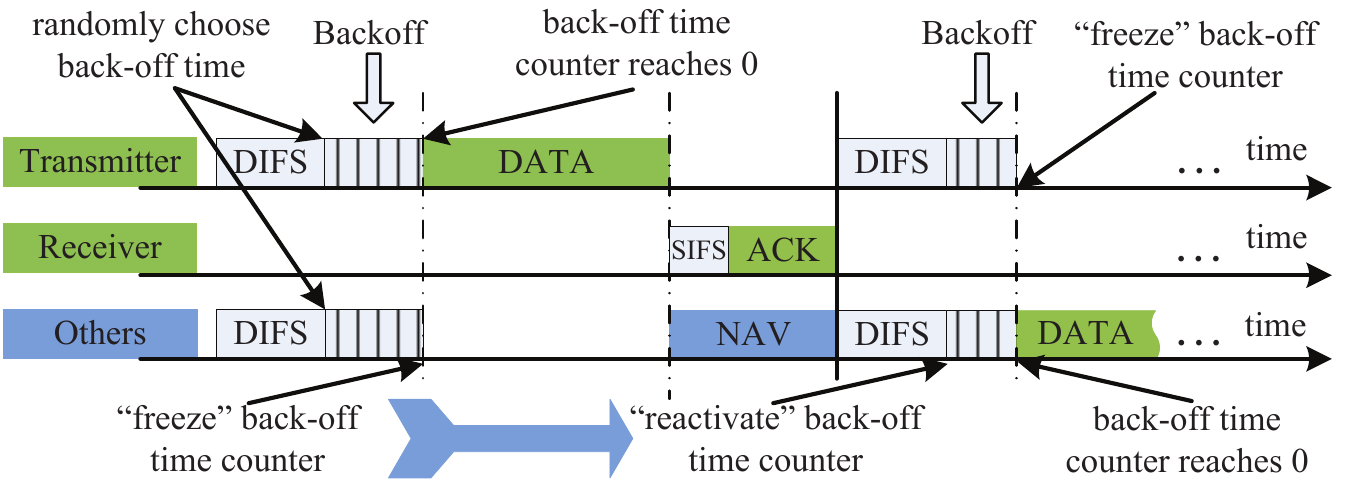}
\caption{Example of basic access mechanism for window--based CSMA MAC protocol \cite{bian00}.}
\label{CSMABasic_Chap2}
\end{figure}

\begin{figure}[!t]
\centering
\includegraphics[width=150mm]{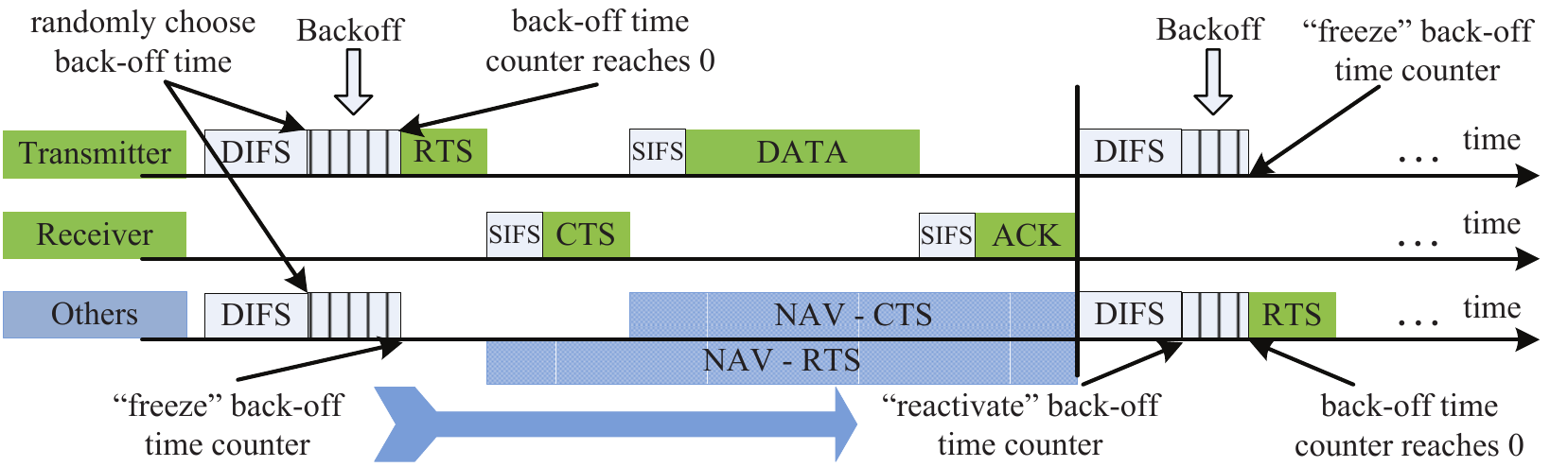}
\caption{Example of RTS/CTS access mechanism for window--based CSMA MAC protocol \cite{bian00}.}
\label{CSMARTSCTS_Chap2}
\end{figure}

\begin{enumerate}

\item \textbf{\textit{MAC protocol}}

\NoIndent{

%The description of the  window-based CSMA MAC protocol is presented as follows. We assume that the length of 
%each cycle is sufficiently large so that users can transmit several packets during the data transmission phase. 

To capture the channel for data transmission, all active users employ the same contention resolution mechanism, 
which is described in the following \cite{bian00}.
To avoid collisions among contending users, each user takes a random waiting time before access, which is chosen
based on the so-called contention window $W$. Moreover, the value of this contention window $W$ is doubled
after each collision until the contention window reaches $2^m W_{0}$ where $W_{0}$ is the minimum value of
contention window and $m$ is called maximum back-off stage.

%Exponential back-off with minimum contention window $W$ and maximum back-off stage $m$  is employed in the contention phase.

Suppose that the current back-off stage of a particular user is $i$ then it starts the contention by choosing a random 
back-off time uniformly distributed in the range $[0,2^i W_0-1]$, $0 \leq i \leq m$. 
This user then starts decrementing its back-off time counter while carrier sensing transmissions from other users. 
Let $\sigma$ denote a mini-slot interval, each of which corresponds one unit of the back-off time counter. 
Upon hearing a transmission from any other users, the underlying user will ``freeze'' its back-off time counter and reactivate when the channel is sensed idle again. 
Otherwise, if the back-off time counter reaches zero, the underlying user wins the contention and transmits its data. 

Either two-way or four-way handshake with Request-to-send/Clear-to-send (RTS/CTS) exchange can be employed. 
In the four-way handshake, the transmitter sends RTS to the receiver and waits until it successfully receives CTS from the receiver before sending a data packet. 
Note that the RTS and CTS contain the information of the packet length, hence other users can obtain these information by listening to the channel.
With these information, users can update the so-called network allocation vector (NAV) which indicates the period of a busy channel.
In both handshake mechanisms, after sending each data packet the transmitter expects an acknowledgment (ACK) from the receiver to 
indicate a successful reception of the packet. 
Standard small intervals, namely short inter-frame space (SIFS) and distributed inter-frame space (DIFS), are used before the back-off countdown 
process and ACK packet transmission as described in \cite{bian00}. 
We refer to the CSMA MAC protocol using the two-way handshaking technique as a basic access scheme in the following. 
Timing diagram for basic and RTS/CTS based CSMA MAC protocols are illustrated in Figs.~\ref{CSMABasic_Chap2} and \ref{CSMARTSCTS_Chap2}, respectively.}

%\vspace{0.2cm}
%\noindent
%\textbf{\textit{Markov Chain Model for CSMA Protocol and Throughput Analysis}}

\item \textbf{\textit{Markov Chain Model for Throughput Analysis of CSMA Protocol}}

\NoIndent{

%In the analysis of CSMA/CA in CRNs, the throughput, delay and other QoS parameters of a node are usually developed and then based on those, the parameter optimization is 
%performed to configure the network operation.

%\begin{figure*}%[!t]
\begin{figure}[!t]
\centering
\includegraphics[width=120mm]{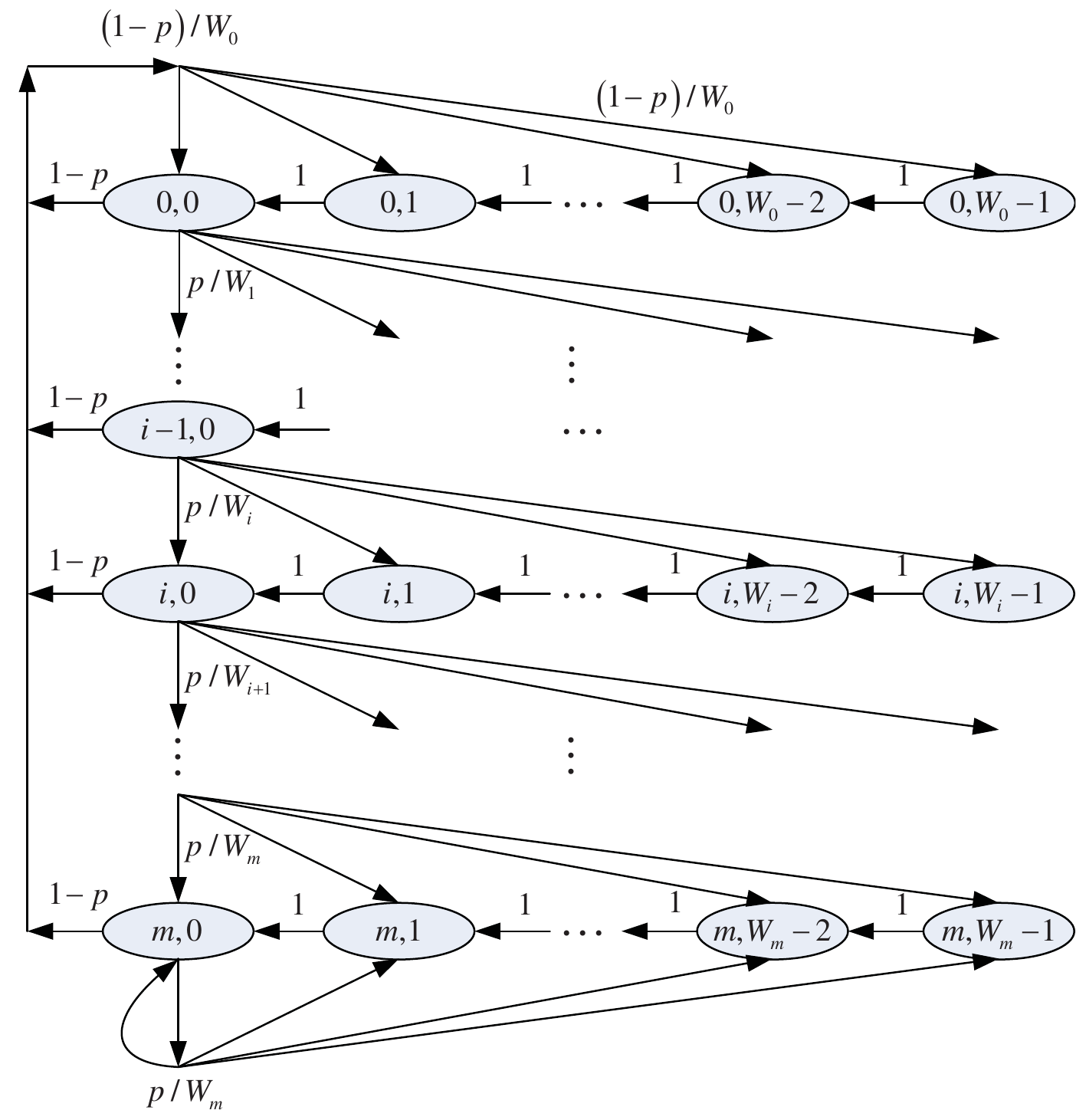}
\caption{Markov chain model for window--based CSMA MAC protocol \cite{bian00}.}
\label{Markovchain_Chap2}
\end{figure}
%\end{figure*}

In the following, we present the throughput analysis of the CSMA/CA protocol for a network with $n_0$ users \cite{bian00}.
We consider the 2D Markov chain (MC) for CSMA/CA MAC protocol $\left(s(t), c(t) \right)$ where 
$s(t)$ represents the backoff stage of the station at time $t$ where $s(t)=[0,m]$. 
Moreover, we have $c(t)=[0,W_i-1]$ where $W_i = 2^i W_0$ describes the states of the backoff counter. 
Fig.~\ref{Markovchain_Chap2} shows the state transition diagram of this MC. 

Let $b_{i,k} = {\sf lim}_{t \rightarrow \infty} \Pr\left(s(t) = i, c(t) = k\right)$ ($i \in [0,m]$, $k \in [0,W_i-1]$) denote the stationary 
probability of the Markov chain.  
For convenience, we define $W = W_0$ in the following derivations.
Using the analysis as in \cite{bian00}, we can arrive at the following the relationship for the steady-state probabilities: 
$b_{i,0} = p^i b_{0,0}$ (for $i \in \left[0, m\right)$), $b_{m-1,0} p = (1-p) b_{m,0}$ or $b_{m,0} =  \frac{p}{1-p} b_{0,0}$, 
and $b_{i,k} = \frac{W_i-k}{W_i} b_{i,0} $ (for $i \in \left[0, m\right],  k \in \left[0, W_i-1\right] $).
Since we have $\sum_{i,k} b_{i,k} = 1 $, substitute the above results for all $b_{i,k}$ and perform some manipulations, we can obtain
\beqn
\label{Stead_EQN_Chap2}
1=\frac{b_{0,0}}{2} \left\{W \left[\sum_{i=0}^{m-1} (2p)^i + \frac{(2p)^m}{1-p}\right] + \frac{1}{1-p}\right\}. 
\eeqn
From these results, we can find the relationship among $b_{0,0} $, $p$, $W$ as follows \cite{bian00}:
\beqn
b_{0,0} = \frac{2(1-2p)(1-p)}{(1-2p)(W+1)+pW\left[1-(2p)^m\right]}.
\eeqn

The throughput can be calculated by using the technique developed by Bianchi in \cite{bian00} where we approximately assume a 
fixed transmission probability $\phi$ in a generic slot time. 
Specifically, Bianchi shows that this transmission probability can be calculated from the following two equations \cite{bian00}
\beqn \label{phi_Chap3}
\phi  = \frac{{2\left( {1 - 2p} \right)}}{{\left( {1 - 2p} \right)\left( {W + 1} \right) + Wp\left( {1 - {{\left( {2p} \right)}^m}} \right)}},
\eeqn
\beqn \label{p_Chap3}
p = 1-\left(1-\phi\right)^{n_0-1},
\eeqn
where $m$ is the maximum back-off stage, $p$ is the conditional collision probability (i.e., the probability that a collision occurs
given that there is one user transmitting its data).

The probability that at least one user transmits its data packet can be written as
\begin{equation}
\label{eq_Chap3_8a} %eq8
{{\mathcal{P}}_t} = 1 - {\left( {1 - \phi } \right)^{{n_0}}}.
\end{equation}
However, the probability that a transmission on the channel is successful given there is at least one user transmitting can be written as
\begin{equation}
\label{eq_Chap3_9a}% eq9
{{\mathcal{P}}_s} = \frac{{{n_0}\phi {{\left( {1 - \phi } \right)}^{{n_0} - 1}}}}{{{{\mathcal{P}}_t}}}.
\end{equation}
The average duration of a generic slot time can be calculated as 
\begin{equation}
\label{eq_Chap3_10a}% eq10
{{\bar T}_{sd}} = \left( {1 - {{\mathcal{P}}_t}} \right){T_e} + {{\mathcal{P}}_t}{{\mathcal{P}}_s}{T_s} + {{\mathcal{P}}_t}\left( {1 - {{\mathcal{P}}_s}} \right){T_c},
\end{equation}
where $T_e = \sigma$, $T_s$ and $T_c$ represent the duration of an empty slot, the average time the channel is sensed busy due to a successful transmission, 
and the average time the channel is sensed busy due to a collision, respectively. 
These quantities can be calculated as \cite{bian00}
\\
\textit{For basic mechanism:}
\begin{equation}
\zlabel{eq_Chap3_11a}% eq11
\left\{ \!\!\!{\begin{array}{*{20}{c}}
   {{T_s} = T_s^{1} = H + PS + SIFS + 2PD \!+ \!ACK \!+\! DIFS} \hfill  \\
   {{T_c} = T_c^{1} = H + PS + DIFS + PD} \hfill  \\
   {H = {H_{PHY}} + {H_{MAC}}} \hfill  \\
\end{array}} \right.\!\!\!\!,
\end{equation}
where $H_{PHY}$ and $H_{MAC}$ are the packet headers for physical and MAC layers, $PS$ is the average packet size, $PD$ is the propagation delay, $SIFS$ is the length of a short inter-frame space, $DIFS$ is the length of a distributed inter-frame space, $ACK$ is the length of an acknowledgment. 
\\
\textit{For RTS/CTS mechanism:}
\beqn \label{TsTc_Chap3}
\left\{ {\begin{array}{*{20}{c}}
   \begin{array}{l}
 {T_s} = T_s^2 = H + PS + 3SIFS + 2PD + RTS + CTS + ACK + DIFS \\ 
 \end{array} \hfill  \\
   {{T_c} = T_c^2 = H + DIFS + RTS + PD} \hfill  \\
\end{array}} \right.,
\eeqn
where $RTS$ and $CTS$ represent the length of RTS and CTS control packets, respectively.

Based on these quantities, we can express the normalized throughput as follows:
\begin{equation}
\label{eq_Chap3_13a}%eq13
\mathcal{T} = \frac{\mathcal{P}_s\mathcal{P}_t PS}{{\bar T}_{sd}}.
\end{equation}
}
\end{enumerate}
}
 
%\vspace{0.2cm}
%\noindent
%\textbf{\textit{A $p$-persistent CSMA MAC protocol:}} 

%\vspace{0.2cm}
%\noindent
%\textbf{\textit{$\clubsuit$ A $p$-persistent CSMA MAC Protocol}}

\item \textbf{\textit{The $p$-persistent CSMA MAC Protocol}}

\NoIndent{

\begin{figure}[!t]
\centering
\includegraphics[width=120mm]{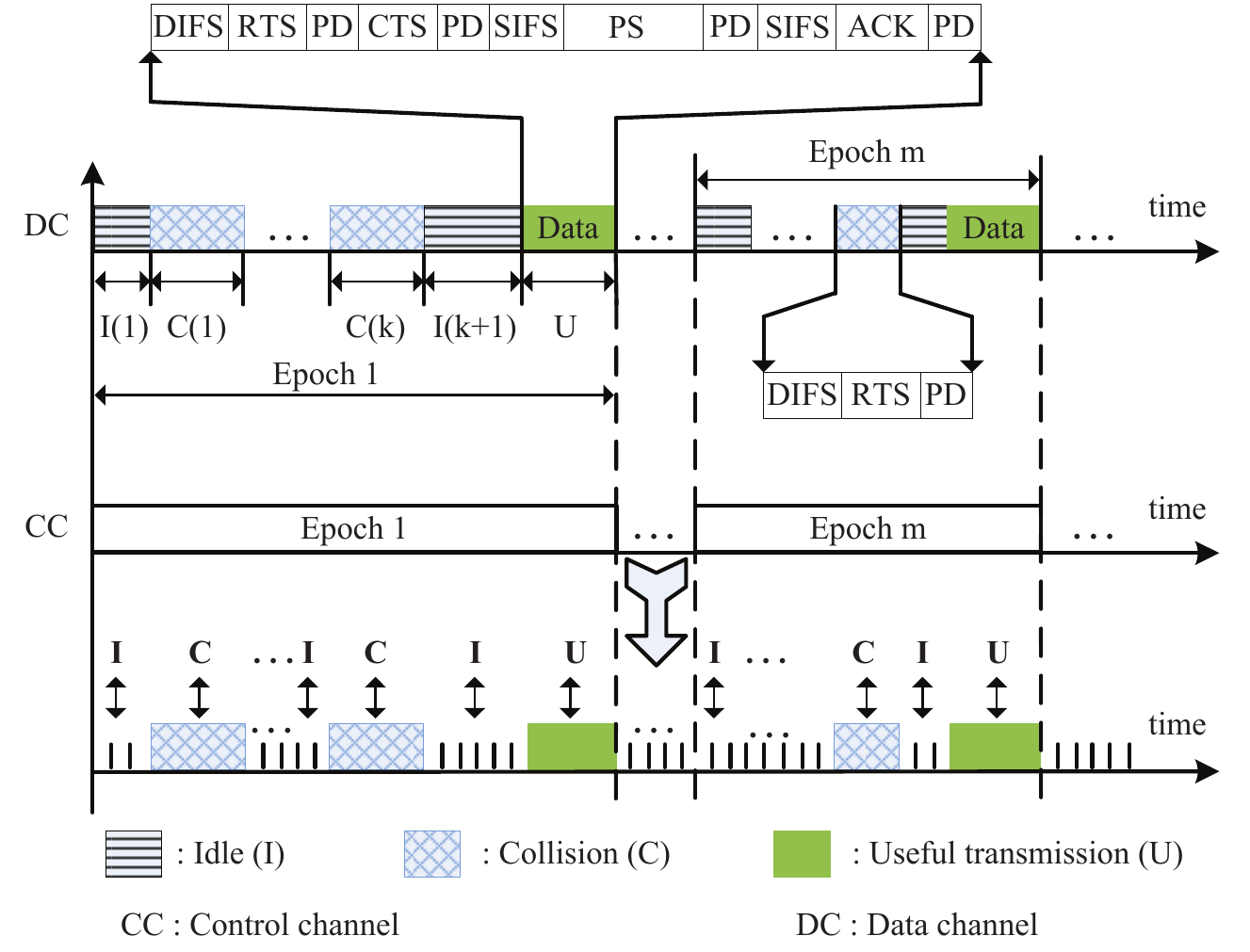}
\caption{Time diagram for $p$-persistent CSMA MAC protocol \cite{Cali00}.}
\label{persistentcsma_Chap2}
\end{figure}

\begin{enumerate}

\item \textbf{\textit{MAC protocol}}

\NoIndent{
We briefly describe the $p$-persistent CSMA MAC protocol and then present the saturation throughput analysis for this protocol.
In this protocol, each user attempts to transmit on the chosen channel with a probability of $p$ if it senses an available channel 
(i.e., no other users transmit data) \cite{Cali00}. 
In case the user decides not to transmit (with probability of $1-p$), it will carrier sense the channel and attempt to transmit again 
in the next slot with probability $p$.
If there is a collision, the user will wait until the channel is available and attempt to transmit with probability $p$ as before.

The basic 2-way or 4-way handshake with RTS/CTS \cite{bian00} can be employed to reserve a channel for data transmission. 
An ACK from the receiver is transmitted to the transmitter to indicate the successful reception of a packet.
The timing diagram of this MAC protocol is presented in Fig.~\ref{persistentcsma_Chap2} which will be further clarified later. 
In this MAC protocol, transmission time is divided into time slot (we call time slot or slot size interchangeably). 
The average packet size is assumed to be $PS$ time slots.  }

%\vspace{0.2cm}
%\noindent
%\textbf{\textit{Saturation Throughput Analysis}}

\item \textbf{\textit{Saturation Throughput Analysis}}

\NoIndent{

In the following, we will derive the saturation throughput $\mathcal{T}$
where each user always has data packets ready to transmit. 
As shown in Fig.~\ref{persistentcsma_Chap2}, each contention and access cycle (called
epoch in this figure) between two consecutive successful packet transmissions comprises several idle and busy periods 
denoted as $I$ and $B$, respectively. In particular, an epoch starts with an idle period $I(1)$ and then followed by several collisions ($C(i)$) and idle 
periods ($I(i + 1)$) and finally ends with a successful transmission ($U$). Note that an idle period is the time interval between two consecutive 
packet transmissions (a collision or a successful transmission). 

%For simplicity, we also use $I$ to denote the idle duration and $C$ to represent the collision duration during which there are at least two concurrent data transmissions.

Let us define ${\bar T}_{\sf cont}$ as the average time due to contention, collisions, and RTS/CTS exchanges before a successful packet transmission; 
$T_S$ is the total time due to data packet transmission, ACK control packet, and overhead between these data and ACK packets. 
Then, the saturation throughput $\mathcal{T}$ can be written as 
\beqn
\label{sat_T_Chap2}
\mathcal{T} = \frac{T_S}{{\bar T}_{\sf cont} + T_S}.
\eeqn

To calculate ${\bar T}_{\sf cont} $, we define some further parameters as follows.
Let denote $T_C$ as the duration of a collision; $T_S$ as the duration required for a successful data transmission (including the overhead);
${\bar T}_S$ is the required time for successful RTS/CTS transmission. 
These quantities can be calculated under the 4-way handshake as 
\beqn
\label{TCTSTI}
\left\{ {\begin{array}{*{20}{c}}
   T_S = PS + 2SIFS + 2PD + ACK   \hfill\\
   {\bar T}_S = DIFS + RTS + CTS + 2PD \hfill  \\
   T_C = RTS + DIFS + PD \hfill  \\
\end{array}} \right.,
\eeqn
where $PS$ is the packet size, $ACK$ is the length of an ACK packet, $SIFS$ is the length of a short interframe space, $DIFS$ is the length of a distributed interframe space, $PD$ is the propagation delay where $PD$ is usually relatively small compared to the slot time $\sigma$. 

Let $T_I^i$ be the $i$-th idle duration between two consecutive RTS/CTS transmissions
 (they can be collisions or successes). Then, $T_I^{i,j_2}$ can be calculated based on its probability mass function (pmf),
 which is derived as follows. Recall that all quantities are defined in terms of number of time slots. Now, suppose there are  
$n_0$ users contending to capture the channel, let $\mathcal{P}_S$, $\mathcal{P}_C$ and $\mathcal{P}_I$ denote the probabilities that a generic slot corresponds
 to a successful transmission, a collision, and an idle slot, respectively. These quantities can be calculated as follows:
\beqn
\mathcal{P}_S = n_0 p\left(1-p\right)^{n_0-1} \\
\mathcal{P}_I = \left(1-p\right)^{n_0} \\
\mathcal{P}_C = 1-\mathcal{P}_S-\mathcal{P}_C,
\eeqn
where $p$ is the transmission probability of any user in a generic slot. 
Note that ${\bar T}_{\sf cont}$ is a random variable (RV) consisting of several intervals corresponding to idle periods, collisions, and one 
successful RTS/CTS transmission. Hence, this quantity can be calculated as 
\beqn
\label{T_cont_Chap2}
{\bar T}_{\sf cont} = \sum_{i=1}^{N_c} \left(T_C+ T_I^i\right) + T_I^{N_c+1} + {\bar T}_S,
\eeqn
where $N_c$ is the number of collisions before the first successful RTS/CTS exchange, 
which is a geometric RV with parameter $1-\mathcal{P}_C/\mathcal{\bar P}_I$ (where $\mathcal{\bar P}_I = 1 - \mathcal{P}_I$). 
Its pmf can be expressed as
\beqn
\label{N_c_cal_Chap2}
 f_{X}^{N_c} \left(x\right) = \left(\frac{\mathcal{P}_C}{\mathcal{\bar P}_I}\right)^{x} \left(1-\frac{\mathcal{P}_C}{\mathcal{\bar P}_I}\right), \: x = 0, 1, 2, \ldots
\eeqn
Also, $T_I^i$ represents the number of consecutive idle slots, which is also a geometric RV with parameter $1-\mathcal{P}_I$ with the following pmf
\beqn
\label{T_I_cal_Chap2}
f_{X}^{I} \left(x\right) = \left(\mathcal{P}_I\right)^{x} \left(1-\mathcal{P}_I\right), \: x = 0, 1, 2, \ldots
\eeqn
Therefore, ${\bar T}_{\sf cont}$  can be written as follows \cite{Cali00}:
\beqn
{\bar T}_{\sf cont}  = {\bar N}_c T_C + {\bar T}_I \left({\bar N}_c + 1\right) + {\bar T}_S \label{T_contgeo_Chap2},
\eeqn
where ${\bar T}_I$ and ${\bar N}_c$ can be calculated as
\beqn
{\bar T}_I &=& \frac{\left(1-p\right)^{n_0}}{1-\left(1-p\right)^{n_0}} \\
{\bar N}_c &=& \frac{1-\left(1-p\right)^{n_0}}{n_0 p\left(1-p\right)^{n_0-1}}-1. 
\eeqn
These expressions are obtained by using the pmfs of the corresponding RVs given in (\ref{N_c_cal_Chap2}) and (\ref{T_I_cal_Chap2}), respectively \cite{Cali00}.}
\end{enumerate}

}

%\vspace{0.2cm}
%\noindent
%\textbf{\textit{$\clubsuit$ Other MAC Protocols}}

\item \textbf{\textit{Other MAC Protocols}}

\NoIndent{

\begin{enumerate}

%\vspace{0.2cm}
%\noindent
%\textbf{\textit{The ALOHA protocol}} 

\item \textbf{\textit{ALOHA protocol}}

\NoIndent{

The ALOHA protocol was initially developed for satellite communication \cite{Abramson70}, which has been then employed for other wireless networks such as 
wireless sensor networks \cite{Baccelli06,Kahn78}. The original design objective of the ALOHA protocol is to provide a random access mechanism to multiplex 
a large number of users that communicate with a satellite using a single communication channel \cite{Abramson70}. 
Whenever the satellite correctly receives a frame, it will broadcast an ACK including the addresses of the source user to all users. 
Hence, the source user can recognize a possible collision and re-transmit the frame in the case that it does not receive the ACK of the transmitted frame.

There is also another method without ACK where the satellite  will rebroadcast the received data frame from a source user. 
Therefore, the source user can listen to the channel and decode the received broadcast frame from the satellite to know the outcome of its transmitted data
frame. In this method, the satellite ignores all corrupted frames due to collisions or interference, and hence the source users must re-send their frames.
Different from the CSMA/CA MAC protocol, a user with data backlogs in the ALOHA protocol will transmit its data with a certain
 probability without carrier sensing the channel. There are two basic types of ALOHA protocols, namely pure ALOHA and slotted ALOHA. 
In the pure ALOHA, a user can start transmission at any time whereas in the slotted ALOHA, all users have to be synchronized and perform data transmissions 
in fixed-size time slots. }

%\vspace{0.2cm}
%\noindent
%\textbf{The other persistent CSMA protocols}

\item \textbf{\textit{Other persistent CSMA protocols}}

\NoIndent{
There are different kinds of persistent CSMA protocols, namely
1-persistent CSMA, non-persistent CSMA beside the $p$-persistent CSMA \cite{Demirkol06,Tay04,Cali00}. 
The 1-persistent CSMA protocol operates as follows. 
Whenever a user senses the idle channel, it will transmit the data packet immediately (i.e., it
transmits packet data with a probability of 1). 
If the channel is sensed to be busy, a user keeps listening the channel and transmits immediately when the channel becomes idle. 
In the case of collisions, each user will wait and start over again.

We now discuss the main difference between non-persistent and 1-persistent CSMA protocols. 
In the non-persistent CSMA protocol, if a user senses the busy channel, it waits
for a period of time, and senses the channels again. 
Hence, the non-persistent CSMA protocol can reduce the collision probability
and the efficiency as well while the 1-persistent CSMA protocol
increases the efficiency and also the collision probability. 
The $p$-persistent CSMA protocol can result in more efficient trade-off between 
transmission efficiency and the collision probability, which
can, therefore, achieve better performance than the other two counterparts.}
\end{enumerate}
}

\end{enumerate}

%\vspace{0.2cm}
%\noindent
%\textbf{\textit{$\bigstar$ Multi-channel MAC Protocols}}

\subsubsection{Multi-channel MAC Protocols}

When there are multiple channels for data transmissions, the overall network throughput and communication delay 
can be improved because of the increasing spectrum resources.
The critical design issues here are how to efficiently arrange simultaneous transmissions on multiple
channels by using distributed contention resolution.
In this section, we present some important multi-channel MAC protocols which are categorized 
accordingly four different approaches employed to perform data channel arrangement for the users \cite{mo08}.
The first three approaches are the dedicated control channel, common hopping, and split phase approaches
 which result in the so-called single rendezvous protocols, while the last approach
leads to the multiple rendezvous MAC protocol.

\begin{figure}[!t]
\centering
\includegraphics[width=120mm]{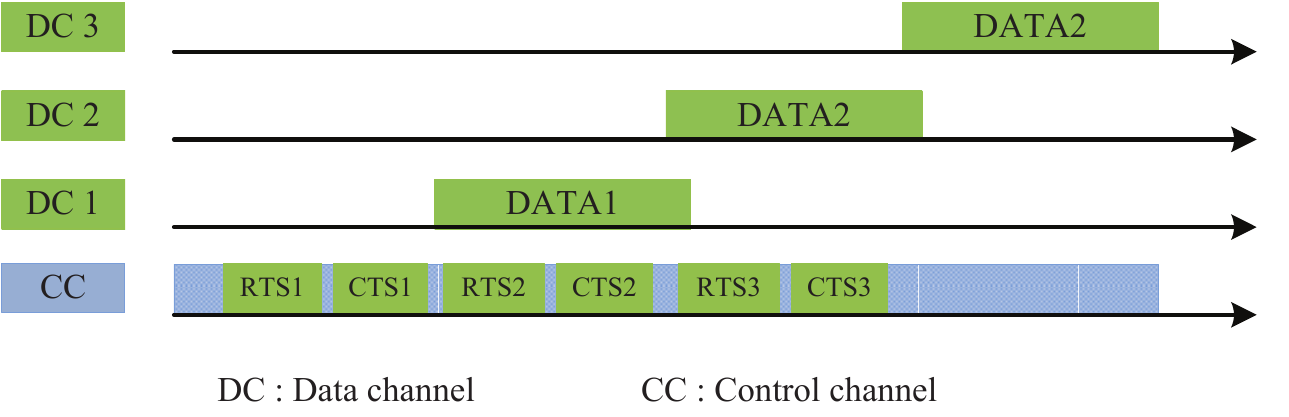}
\caption{Dedicated control channel mechanism for multi-channel MAC protocols \cite{mo08}.}
\label{MMAC_Chap2}
\end{figure}

\begin{figure}[!t]
\centering
\includegraphics[width=120mm]{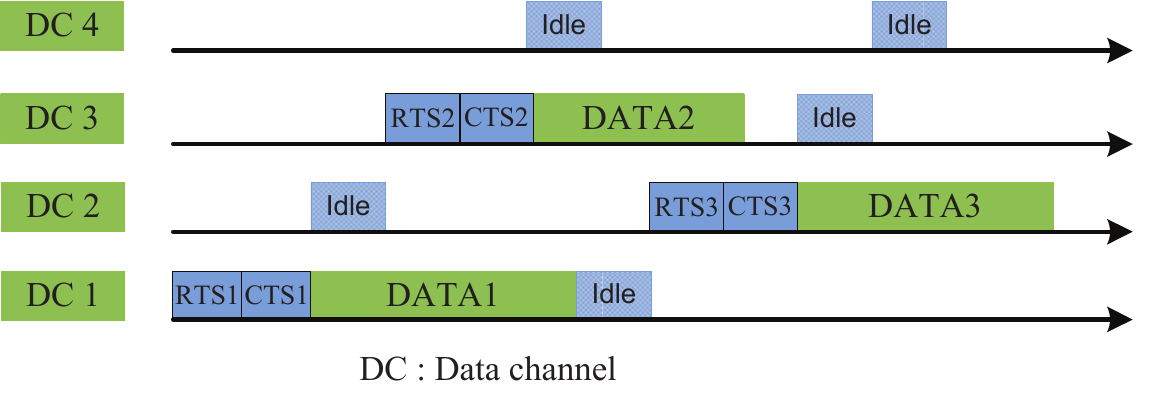}
\caption{Common hopping mechanism for multi-channel MAC protocols \cite{mo08}.}
\label{MMAC_CHM_Chap2}
\end{figure}

\begin{figure}[!t]
\centering
\includegraphics[width=120mm]{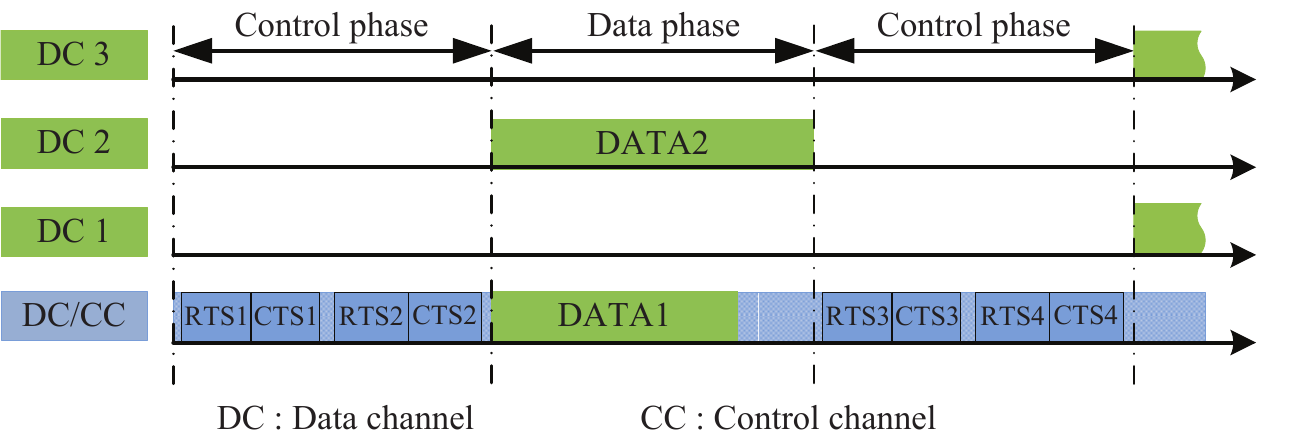}
\caption{Split phase mechanism for multi-channel MAC protocols \cite{mo08}.}
\label{MMAC_SPM_Chap2}
\end{figure}

%\vspace{0.2cm}
%\noindent
%\textbf{\textit{$\clubsuit$ A Dedicated Control Channel Mechanism}}

\begin{enumerate}

\item \textbf{\textit{Dedicated Control Channel Approach}}

\NoIndent{
In this approach, each user is equipped with two transceivers where the first transceiver is used for channel agreement operating 
on the control channel while the second  transceiver is used for data transmission on a data channel.
MAC protocol design in this case is quite simple since every user can always have the knowledge of other users' channel agreements
and network state by listening to the control channel. 
Moreover, we can efficiently reduce loads for busy channels with large number of sharing users during the channel selection process.

Operations of a multi-channel MAC protocol using the dedicated control channel approach
and RTS/CTS handshake is illustrated in Fig.~\ref{MMAC_Chap2}. Here,
RTS and CTS exchanges are performed on the control channel and the RTS and CTS packets 
can include information of channel agreements obtained by using certain channel selection criteria.
Moreover, the RTS and CTS packets can also carry a NAV to inform the duration of busy time on the selected channel.
After the successful RTS/CTS exchange, the user transmits data on the selected channel.}

\item \textbf{\textit{Common Hopping Approach}}

\NoIndent{

Users are required to have only one transceiver in the common hopping approach; 
hence, they perform a channel agreement on one common channel \cite{mo08, Tzamaloukas00}.
To achieve this goal, all users follow the same hopping pattern and they can negotiate
to choose one data channel through exchanging the RTS/CTS control messages. After the
channel selection, the involved pair of users stops the hopping and jumps to the
chosen channel for data transmission. After the transmission is completed, 
the users return to the hopping pattern with other users. 
Detailed operations of this approach is illustrated in Fig.~\ref{MMAC_CHM_Chap2}.}

\item \textbf{\textit{Split Phase Mechanism}}

\NoIndent{

In this approach, the user is equipped with only one transceiver, which alternatively performs channel selection
and data transmission in the control and data phases. A control channel is needed for channel agreement in the control phase;
however, the control channel can be used for data transmission in the data phase.
Active users again exchange RTS/CTS control packets on the control
channel containing the chosen data channel (e.g., the idle channel with the lowest channel index).
In the second data phase, all users start their data transmissions on the chosen channels. 
Note that one channel may be chosen by multiple users. 
If it is the case, further scheduling or contention would be performed in the data phase. 
Detailed operation of the split phase mechanism is illustrated in Fig.~\ref{MMAC_SPM_Chap2}.}

%\vspace{0.2cm}
%\noindent
%\textbf{\textit{$\clubsuit$ A Parallel Rendezvous Mechanism}}

\item \textbf{\textit{Parallel Rendezvous Approach}}

\NoIndent{
Different from the previous approaches, the parallel rendezvous approach allows multiple pairs of users to make 
simultaneously channel agreements on different channels. Therefore, we can resolve the congestion problem due to the single control channel. 
However, a sophisticated coordination is required to make successful channel agreements in this design.
Toward this end, each transmitter can be assigned a hopping sequence and it then finds the intended receiver based on its hopping sequence. 
When both transmitter and receiver hops to the same channel, they can negotiate to choose the data channel for data communications.
Further information for this approach can be found in the descriptions of the SSCH protocol \cite{Bahl04} and  McMAC protocol \cite{So07}.}

\end{enumerate}

\subsection{Cognitive MAC Protocol for CRNs}

%\vspace{0.2cm}
%\noindent
%\textbf{\textit{Spectrum Sensing Integration to Cognitive MAC Protocol Design}}

%\subsubsection{CMAC Protocol Design Challenges}

Different from the conventional wireless networks, a CMAC protocol must perform both exploration and exploitation of the spectrum holes in a dynamic manner 
because the spectrum holes are opportunistic resources \cite{Liang11, Gavrilovska14,Cor09}. 
In addition, various practical constraints must be carefully considered in the CMAC design including
the availability of a control channel for channel agreements and the number of available transceivers to perform contention, sensing, and transmission.
In general, a good CMAC protocol should well balance between spectrum sensing and access times to achieve high spectrum utilization and network performance. 

Specifically, spectrum sensing should be designed so that minimum sensing time is required while maintaining
the target sensing performance (in terms of detection and/or false-alarm probabilities). 
Furthermore, spectrum sensing operations should be scheduled repeated to timely detect PUs' active status and avoid 
creating intolerable interference to PUs' transmissions. 

%\vspace{0.2cm}
%\noindent
%\textbf{\textit{Cognitive MAC Protocols}}

%\subsubsection{Cognitive MAC Protocols}

\begin{figure}[!t]
\centering
\includegraphics[width=100mm]{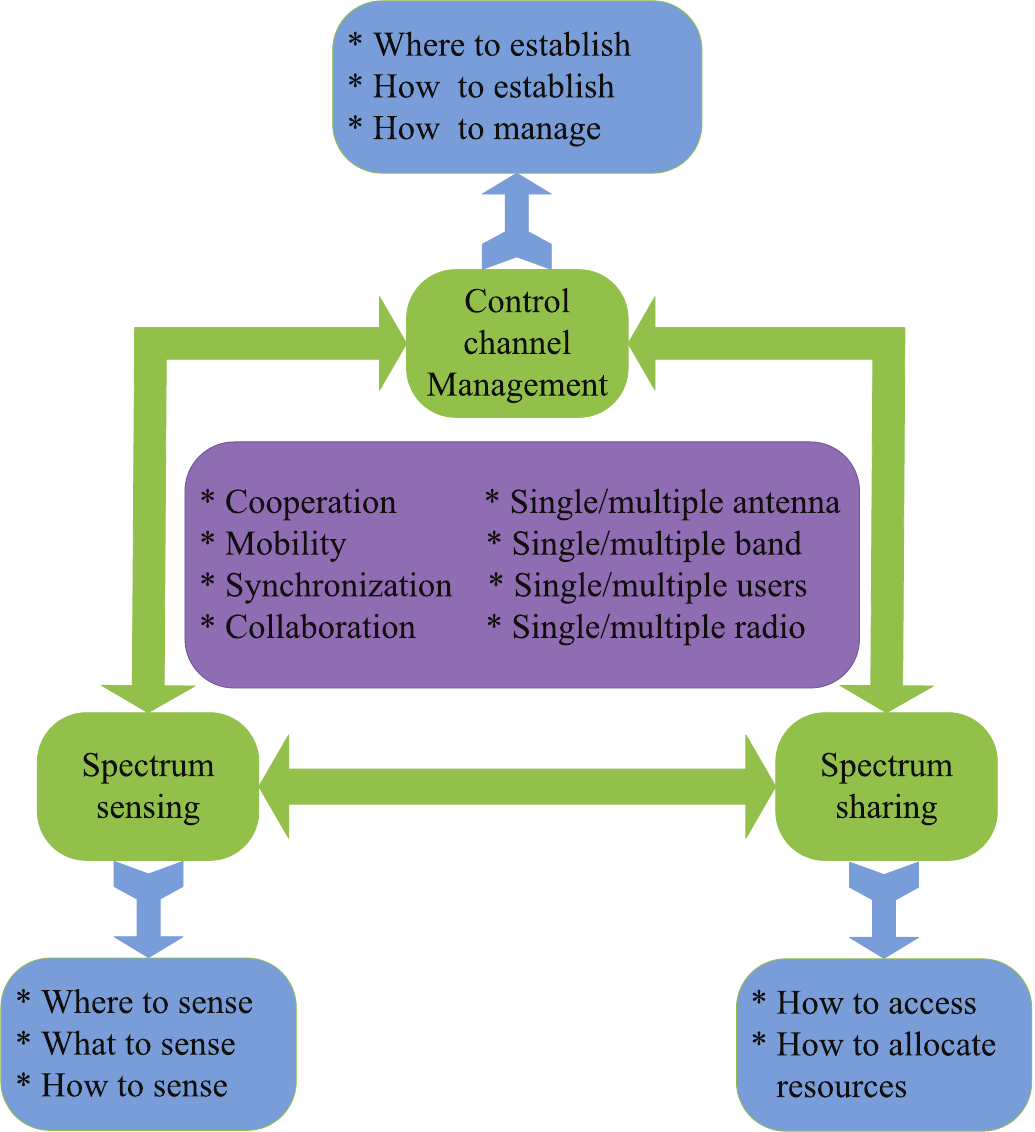}
\caption{The generic functionalities of the CMAC protocol \cite{Gavrilovska14}.}
\label{CMAC_Chap2}
\end{figure}

Basic functions of the CMAC protocol include spectrum sensing, spectrum sharing, and control channel management \cite{Gavrilovska14}, which
are tightly coupled as illustrated in Fig.~\ref{CMAC_Chap2}.
While spectrum sensing is a physical layer functionality providing information about spectrum holes for the MAC layer,
spectrum access aims at improving the spectrum utilization efficiency whilst assuring transparency and protection for PUs. 
Moreover, the spectrum sharing function aims at coordinating the medium access, allocation, and sharing of spectrum holes for SUs,
which can be implemented in either centralized or distributed manner (see \cite{Liang11, Gavrilovska14,Cor09}
Finally, the control channel management provides mechanisms for coordination and collaboration between the SUs and the spectrum sensing and sharing processes. 
In addition, the control channel management is responsible for many tasks in CRNs, e.g., allocation, establishment, and monitoring of 
available channels via broadcasting relevant control information \cite{Gavrilovska14}.

%\subsubsection{Spectrum Access Scheme}

\section{Literature Review}

Hierarchical spectrum sharing between the primary and secondary networks is one of the most important research topics
in cognitive radio literature. For this spectrum sharing paradigm, PUs have strictly higher priority 
than SUs in accessing the underlying spectrum. Here, primary and secondary networks can transmit simultaneously on the same 
spectrum with appropriate interference control to protect the primary network \cite{Kim081}, \cite{Le08}. In particular, it is 
typically required that a certain interference constraint due to SUs' transmissions must be maintained at each primary receiver.

Instead of imposing interference constraints, spectrum sensing can be employed by SUs to seek
and exploit spectrum holes over space and time \cite{Liang08}. There are several challenging issues 
related to this spectrum exploration and exploitation problem. On one hand, SUs should spend sufficient time 
for spectrum sensing so that they can correctly identify spectrum holes, which avoid creating undesirable interference for PUs. 
On the other hand, SUs wish to spend more time for data transmission for better utilization of spectrum holes. 
In what follows, we will provide the comprehensive survey of the state-of-the-art spectrum sensing and CMAC 
protocol designs.

\subsection{Spectrum Sensing}

There is a rich literature on spectrum sensing for cognitive radio networks (e.g., see \cite{Yu09} and references therein).
Spectrum sensing literature is very diverse ranging from the popular energy detection scheme to advanced cooperative sensing strategies \cite{Un08} where 
multiple SUs collaborate to achieve more reliable sensing performance \cite{Gan07,Quan08, Peh09, Wei11, Wei09, Seu10}. 
In a typical cooperative sensing strategy, each SU performs sensing independently and then sends its sensing results to 
a control center, which then makes sensing decisions on the idle/busy
status of each channel using certain aggregation rule. 

%In \cite{Peh09, Wei11, Wei09}, the optimization of cooperative sensing under an a-out-of-b rule is performed.
 %In \cite{Wei11}, the game-theoretic based method is employed to develop a cooperative spectrum sensing strategy.
%However, these works only focus on design and optimization of cooperative sensing without considering spectrum access issues.
%Furthermore, sensing optimization is mostly performed for a single channel and
 %homogeneous scenario where channel parameters such as the signal to noise power ratios (SNRs) and availability probabilities of different channels 
 %are the same. In \cite{Seu10}, the authors investigate a multi-channel scenario where each SU simultaneously senses all channels using one receiver per channel
 %and calculates the log-likelihood ratio of observed measurement, which is then employed by the control center to decide when to terminate the process. 

Cooperative spectrum sensing has been proposed to improve the sensing performance via collaborations among SUs
 \cite{Gan07, Gane07, Quan08, Peh09, Cui11, Wei11, Wei09, Seu10, Chaud12}. 
To combine individual sensing results from different SUs, a central controller (e.g., an AP). 
can employ various aggregation rules to decide whether or not a particular frequency band is available for secondary access. 
In \cite{Chaud12}, the authors studied the performance of hard decisions and soft decisions at a central controller (fusion center). 
They also investigated the impact of reporting errors on the cooperative sensing performance.
Recently, the authors of \cite{Lee13} proposed a novel cooperative spectrum sensing scheme using hard decision
combining considering feedback errors. In \cite{Quan08}, weighted data based fusion is proposed to improve 
sensing performance. 

In \cite{Peh09, Cui11, Wei11, Wei09}, optimization of cooperative sensing under the a-out-of-b aggregation rule was studied. 
In \cite{Wei11}, the game-theoretic based method was proposed for cooperative spectrum sensing. 
In \cite{Seu10}, the authors investigated the multi-channel scenario where the central controller collects 
statistics from SUs to decide whether it should stop at the current time slot. 
In \cite{Male11, Male13}, two different optimization problems for cooperative sensing were studied. 
The first one focuses on throughput maximization where throughput depends on the false alarm probability and
the second one attempts to perform interference management where the objective function is related to the detection probability.

\subsection{Cognitive MAC Protocol Design}

There is the rich literature on CMAC protocol design and analysis for CRNs \cite{Cor09,Kim08,Su08,Su07,Nan07,Le208,Cor07,Hsu07,Konda08,Do05,Shi09,bian00,
So04,Le11,jia08,Sala10, taosu10,wang11,zhang113,choi11,zhang11,zhang112,jeon12,jha11,su081,mo08,ko10,tass05,Le12,Park11} (see \cite{Cor09,Gavrilovska14,Liang11,Ahmed14} for a survey of recent works). 
In the following, we provide survey of existing works in this topic according to four main scenarios considered
in our dissertation, i.e., CMAC with parallel sensing, CMAC with sequential sensing, CMAC with cooperative sensing, and FD CMAC protocols.

%\vspace{0.2cm}
%\noindent
%\textbf{\textit{CMAC with Parallel Sensing}}

\subsubsection{CMAC with Parallel Sensing}

In \cite{Kim08}, sensing-time optimization and optimal channel sequencing algorithms were proposed to efficiently 
discover spectrum holes and to minimize the exploration delay.
Another work along this line was conducted in \cite{Su08}, where a control-channel-based MAC protocol was proposed
for SUs to exploit white spaces in the cognitive ad hoc network. In particular, the authors of this paper developed
both random- and negotiation-based spectrum-sensing schemes and performed throughput analysis for both saturation and
non-saturation scenarios.
There are several other proposed synchronous CMAC protocols that rely on a control channel for
spectrum negotiation and access \cite{Su07, Nan07, Le208, Hsu07}. 
Since in these existing works, the spectrum sensing and access aspects are addressed separately;
development of a concrete CMAC framework considering both aspects and optimized configuration
for the sensing and access parameters would be important research issues to tackle.

%\vspace{0.2cm}
%\noindent
%\textbf{\textit{CMAC with Sequential Sensing}}

\subsubsection{CMAC with Sequential Sensing}

The above CMAC protocols use parallel sensing with the requirement that each SU is equipped by multiple transceivers.
However, having multiple transceivers at the SUs will increase the complexity and deployment cost; therefore, the MAC protocol with a single
transceiver would be preferred in many CRN deployments \cite{jia08, Cor07, Kim08, Su08, Su07, Nan07, Le208, Hsu07}.
There have been many existing works that propose different CMAC protocols with sequential sensing where each SU 
 must perform sequential sensing over multiple channels and can access at most one idle channel for communications.  
Here, the channel assignment is an important design task since it would reduce the sensing time
while efficiently exploring spectrum holes if each SU is assigned a ``best'' subset of channels for sensing.
Reduction of sensing time through effective channel assignment for sensing can then result in improving cognitive
 network throughput.

%\vspace{0.2cm}
%\noindent
%\textbf{\textit{CMAC with Cooperative Sensing}}

\subsubsection{CMAC with Cooperative Sensing}

In \cite{Sala10}, a multi-channel MAC protocol was proposed considering location-dependent detection performance
of SUs on different channels so that white spaces can be efficiently exploited while satisfactorily protecting PUs.
In \cite{Park11,zhang11}, the authors conducted design and analysis for a CMAC protocol using cooperative 
spectrum sensing where parallel spectrum sensing on different channels was assumed at each SU. 

Most existing works focused on designing and optimizing parameters for the cooperative
spectrum sensing algorithm; however, they did not consider spectrum access issues. 
Furthermore, either the single channel setting or multi-channel scenario with parallel sensing was assumed.
 Moreover, existing cooperative spectrum sensing schemes rely on a central controller
to aggregate sensing results for white space detection (i.e., centralized design). 
Finally, homogeneous environments (i.e., SUs experience the same channel condition and spectrum
statistics for different channels) have been commonly assumed in the literature, which would not 
be very realistic.

%\vspace{0.2cm}
%\noindent
%\textbf{\textit{Full--Duplex CMAC Protocol for Cognitive Radio Networks}}

\subsubsection{Full--Duplex CMAC Protocol for CRNs}

Despite recent advances on self-interference cancellation (SIC) techniques for FD radios \cite{Duarte12, Everett14, Sabharwal14} (e.g., propagation SIC, 
analog-circuit SIC, and digital baseband SIC), self-interference still exists due to various reasons 
such as the limitation of hardware and channel estimation errors.
The FD technology has been employed for more efficient spectrum access design in CRNs \cite{Afifi14, Cheng15,tan2015distributed, Kim12,
Zheng13,Kim15} where SUs can perform sensing and transmission simultaneously. 

In \cite{Afifi14}, the authors considered the cognitive FD-MAC design assuming that SUs perform sensing in multiple small time slots 
to detect the PU's activity during their transmissions, which may not be very efficient.
Furthermore, they proposed three operation modes for the SU network, i.e., transmission-only, transmission-sensing, and transmission-reception modes. 
Then, they study the optimal parameter configurations for these modes by solving three corresponding optimization problems. 
In practice, it would be desirable to design a single adaptable MAC protocol, which can be configured to operate in an 
optimal fashion depending on specific channel and network conditions. 

In \cite{Cheng15}, a FD-MAC protocol which allows simultaneous spectrum access of the SU and PU networks was developed. 
In addition, both PUs and SUs are assumed to employ the same $p$-persistent MAC protocol for channel contention resolution.
This design is, however, not applicable to the hierarchical spectrum access in the CRNs where PUs should have higher spectrum access 
priority compared to SUs. Moreover, engineering of a cognitive FD relaying network was considered in \cite{Kim12, Zheng13,Kim15} where various 
resource allocation algorithms to improve the outage probability are proposed.
In addition, the authors in \cite{Ramirez13} developed the joint routing and distributed resource allocation for FD wireless networks.
In \cite{Choi15}, Choi et al. studied the distributed power allocation for a hybrid FD/HD system where all network nodes operate in the
HD mode but the AP communicates using the FD mode.

\chapter{Distributed MAC Protocol for Cognitive Radio Networks: Design, Analysis, and Optimization} % top level followed by section, subsection
\zlabel{Chapter3}

%: ----------------------- paths to graphics ------------------------

% change according to folder and file names
\ifpdf
    \graphicspath{{3/figures/PNG/}{3/figures/PDF/}{3/figures/}}
\else
    \graphicspath{{3/figures/EPS/}{3/figures/}}
\fi
%\usepackage[cmex10]{amsmath}
%\usepackage {amssymb}
%\usepackage{algorithmic}
%\usepackage{array}
%\usepackage{mdwmath}
%\usepackage{mdwtab}
%\usepackage{eqparbox}
%%\usepackage{url}
%\usepackage{hyperref}
%\usepackage{algorithm}
%\usepackage{algorithmic}

%\renewcommand{\baselinestretch}{1.86}

%\newcommand{\argmin}{\operatornamewithlimits{argmin}}
%\newcommand{\argmax}{\operatornamewithlimits{argmax}}
%\newcommand{\beq}{\begin{equation}}
%\newcommand{\eeq}{\end{equation}}
%\newcommand{\beqn}{\begin{eqnarray}}
%\newcommand{\eeqn}{\end{eqnarray}}
%\newcommand{\beqno}{\begin{eqnarray*}}
%\newcommand{\eeqno}{\end{eqnarray*}}
%\newcommand{\bma}{\begin{displaymath}}
%\newcommand{\ema}{\end{displaymath}}
%\newcommand{\bnu}{\begin{enumerate}}
%\newcommand{\enu}{\end{enumerate}}
%\newcommand{\bce}{\begin{center}}
%\newcommand{\ece}{\end{center}}
%\newcommand{\btb}{\begin{tabular}}
%\newcommand{\etb}{\end{tabular}}
%: ----------------------- contents from here ------------------------

The content of this chapter was published in IEEE Transactions on Vehicular Technology in the following paper:

L.~T.~ Tan, and L.~B.~ Le, ``Distributed MAC Protocol for Cognitive Radio Networks: Design, Analysis,and Optimization,'' {\em IEEE Trans. Veh. Tech.}, vol. 60, no. 8, pp. 3990--4003, 2011.

\section{Abstract}

In this paper, we investigate the joint optimal sensing and distributed MAC protocol design problem for cognitive radio networks.
We consider both scenarios with single and multiple channels.
For each scenario, we design a synchronized MAC protocol for dynamic spectrum sharing among multiple secondary users (SUs), which
incorporates spectrum sensing for protecting active primary users. 
We perform saturation throughput analysis for the corresponding proposed MAC protocols that 
explicitly capture spectrum sensing performance. Then, we find their optimal
configuration by formulating throughput maximization problems subject to detection probability constraints 
for primary users. In particular, the optimal solution of the optimization problem returns the required sensing time for
primary users' protection and optimal contention window for maximizing total throughput of the secondary network.
Finally, numerical results are presented to illustrate developed theoretical findings in the paper and significant performance
gains of the optimal sensing and protocol configuration.

\section{Introduction}

Emerging broadband wireless applications have been demanding unprecedented increase in radio spectrum resources.
As a result, we have been facing a serious spectrum shortage problem. However, several recent
measurements reveal very low spectrum utilization in most useful frequency bands \cite{Zhao07}.
To resolve this spectrum shortage problem, the Federal Communications Commission (FCC) has opened licensed bands for 
unlicensed users' access. This important change in spectrum regulation has resulted in growing research interests on dynamic 
spectrum sharing and cognitive radio in both industry and academia. In particular, IEEE has established an IEEE 802.22 workgroup to 
build the standard for WRAN based on CR techniques \cite{Stevenson09}.

Hierarchical spectrum sharing between primary networks and secondary networks is one of the most widely studied dynamic
spectrum sharing paradigms. For this spectrum sharing paradigm, primary users (PUs) typically have strictly higher priority 
than SUs (SUs) in accessing the underlying spectrum. One potential approach for dynamic spectrum sharing
is to allow both primary and secondary networks to transmit simultaneously on the same frequency with appropriate interference control
to protect the primary network \cite{Kim081,Le08}. In particular, it is typically required that a certain
interference temperature limit due to SUs' transmissions must be maintained at each primary receiver.
Therefore, power allocation for SUs should be carefully performed to meet stringent interference requirements in this 
spectrum sharing model.

Instead of imposing interference constraints for PUs,  spectrum sensing can be adopted by SUs to search for
and exploit  spectrum holes (i.e., available frequency bands) \cite{Liang08, Peh09}. Several
challenging technical issues are related to this spectrum discovery and exploitation problem. On one hand, SUs should spend sufficient time 
for spectrum sensing so that they do not interfere with active PUs. On the other hand, SUs should efficiently
exploit spectrum holes to transmit their data by using an appropriate spectrum sharing mechanism. Even though these aspects are tightly
coupled with each other, they have not been treated thoroughly in the existing literature.

In this paper, we make a further bold step in designing, analyzing, and optimizing Medium Access Control (MAC) protocols for cognitive radio networks,
considering sensing performance captured in detection and false alarm probabilities. 
In particular, the contributions of this paper can be summarized as follows.
\begin{enumerate} 
\item We design distributed synchronized MAC protocols for
cognitive radio networks incorporating spectrum sensing operation for both single and multiple channel scenarios.
\item We analyze 
saturation throughput of the proposed MAC protocols. 
\item We perform throughput maximization of the proposed MAC protocols against their key parameters, namely sensing time and minimum
contention  window.
\item We present numerical results to illustrate performance of the proposed
MAC protocols and the throughput gains due to optimal protocol configuration.
\end{enumerate}  

The remaining of this paper is organized as follows. In Section~\ref{RelatedWorks}, we discuss some important related works in the literature. Section ~\ref{SystemModel_Chap3} describes system and sensing models. MAC protocol design,
throughput analysis, and optimization for the single channel case are performed in Section ~\ref{SingleChan_Chap3}.
The multiple channel case is considered in Section~\ref{MultipleChan_Chap3}. Section ~\ref{Results_Chap3} presents 
numerical results followed by concluding remarks in Section ~\ref{conclusion_Chap3}.

\section{Related Works}
\label{RelatedWorks}

Various research problems and solution approaches have been considered for a dynamic spectrum sharing problem in
 the literature.  In \cite{Kim081}, \cite{Le08}, a dynamic power allocation problem for cognitive radio
networks was investigated considering fairness among SUs and interference constraints for primary users. 
When only mean channel gains averaged over short term fading can be estimated, the authors proposed more relaxed
protection constraints in terms of interference violation probabilities for the underlying fair power allocation problem.
In \cite{Dev06}, information theory limits of cognitive radio channels were derived. Game theoretic approach for 
dynamic spectrum sharing was considered in \cite{wang08}, \cite{niyato08}. 

There is a rich literature on spectrum sensing for cognitive radio networks (e.g., see \cite{Yu09} and references therein).
Classical sensing schemes based on, for example, energy detection techniques or advanced cooperative sensing strategies \cite{Un08} where multiple
SUs collaborate with one another to improve the sensing performance have been investigated in the literature.
 There are a large number of papers
considering MAC protocol design and analysis for cognitive radio networks \cite{Cor09, Kim08, Su07, Nan07, Cor07, Konda08, Hsu07, Su08} (see \cite{Cor09} for a survey 
of recent works in this topic). However, these existing works either assumed perfect spectrum sensing or did not explicitly model the sensing imperfection in their design and analysis. In \cite{Liang08}, optimization of sensing and throughput tradeoff under a detection probability constraint was investigated. It was shown that the detection constraint is met with equality at optimality. However, this optimization tradeoff was only investigated
for a simple scenario with one pair of SUs. The extension of this sensing and throughput tradeoff to wireless fading channels 
was considered in \cite{zuo10}.

There are also some recent works that propose to exploit cooperative relays to improve sensing and throughput performance
of cognitive radio networks. In particular, a novel selective fusion spectrum sensing and best relay data transmission
scheme was proposed in \cite{zuo111}. A closed-form expression for the spectrum hole utilization efficiency of the
proposed scheme was derived and significant performance improvement compared with other sensing and transmission schemes
was demonstrated through extensive numerical studies. In \cite{zuo112}, a selective relay based cooperative spectrum sensing scheme
was proposed that does not require a separate channel for reporting sensing results. In addition, the proposed scheme can achieve
excellent sensing performance with controllable interference to primary users. These existing works, however, only consider
a simple setting with one pair of SUs.

\section{System and Spectrum Sensing Models}
\label{SystemModel_Chap3}

In this section, we describe the system and spectrum sensing models. Specifically, sensing performance in terms of
detection and false alarm probabilities are explicitly described.

\subsection{System Model}
\label{System_Chap3}

%Fig. 1
\begin{figure}[!t] 
\centering
\includegraphics[width=90mm]{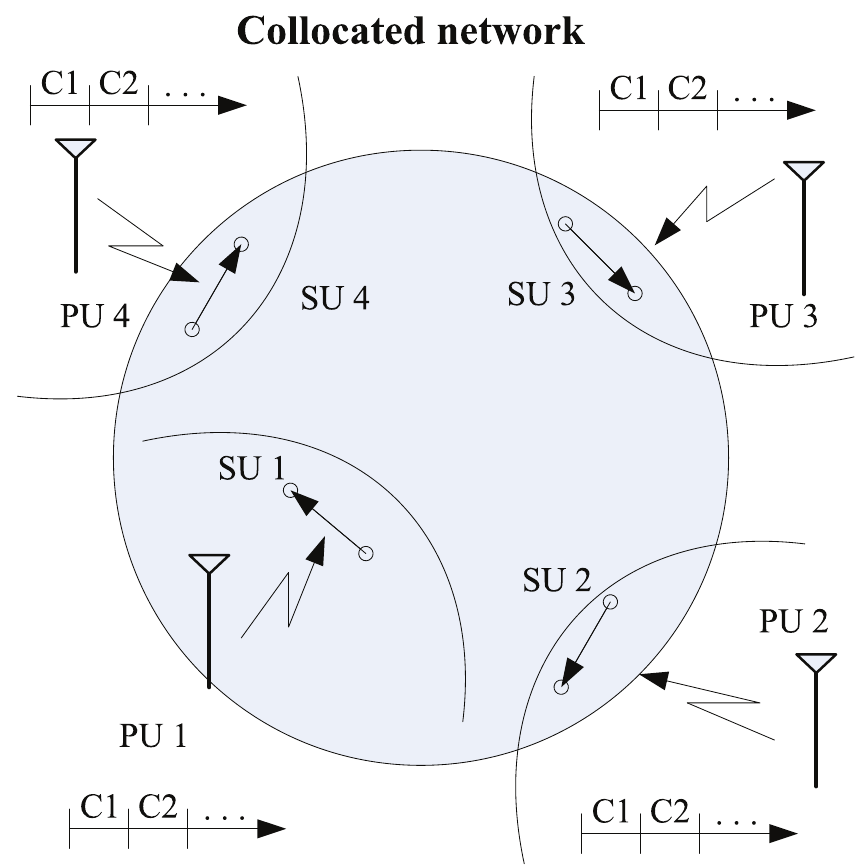}
\caption{Considered network and spectrum sharing model (PU: primary user, SU: secondary user).}
\label{Fig1_Chap3}
\end{figure}

We consider a network setting where $N$ pairs of SUs opportunistically exploit available frequency bands, which belong a primary network, for their data transmission. Note that the optimization model in \cite{Liang08} is a special case of our model with only one pair of SUs. 
In particular, we will consider both scenarios in which one or multiple radio channels are exploited by these SUs. 
We will design synchronized MAC protocols for both scenarios assuming that each channel can be in idle or busy state for a predetermined periodic interval, which is referred to as a cycle in this paper.

We further assume that each pair of SUs can overhear transmissions from other pairs of SUs (i.e., collocated networks). 
In addition, it is assumed that transmission from each individual pair of SUs affects one different primary receiver. 
It is straightforward to relax this assumption to the scenario where each pair of SUs affects more than one primary receiver and/or each primary receiver is affected by more than one  pair of SUs.
The network setting under investigation is shown in Fig.~\ref{Fig1_Chap3}. 
In the following, we will refer to pair $i$ of SUs as secondary link $i$ or flow $i$ interchangeably.

\vspace{0.2cm}
\noindent
\textit{Remark 1:} In practice, SUs can change their idle/busy status any time (i.e., status changes can occur
in the middle of any cycle). Our assumption on synchronous channel status changes is only needed to estimate the
system throughput. In general, imposing this assumption would not sacrifice the accuracy of our network throughput calculation 
 if primary users maintain their idle/busy status  for 
sufficiently long time on average. This is actually the case for many practical scenarios
such as in TV bands as reported by several recent studies (see \cite{Stevenson09} and references therein). In addition, our 
MAC protocols developed under this assumption would result in very few collisions with primary users because
the cycle time is quite small compared to typical active/idle periods of PUs.

\subsection{Spectrum Sensing}
\label{Ss_Chap3}

We assume that secondary links rely on a distributed synchronized MAC protocol to share available frequency channels. 
Specifically, time is divided into fixed-size cycles and it is assumed that secondary links can perfectly synchronize with each other (i.e., there is no synchronization error) \cite{Konda08}, 
\cite{Shi09}. 
It is assumed that each secondary link performs spectrum sensing at the beginning of each cycle and only proceeds to contention with other links to transmit on available channels if its sensing outcomes indicate at least one available channel (i.e., channels not being used by nearby PUs). 
For the multiple channel case, we assume that there are $M$ channels and each secondary transmitter is equipped with $M$ sensors to  sense all channels simultaneously. 
Detailed MAC protocol design will be elaborated in the following sections.

Let $\mathcal{H}_0$ and $\mathcal{H}_1$ denote the events that a particular PU is idle and active, respectively (i.e., the underlying channel is available and busy, respectively) in any cycle. 
In addition, let ${{\mathcal{P}}^{ij}}\left( {{\mathcal{H}_0}} \right)$ and  ${{\mathcal{P}}^{ij}}\left( {{\mathcal{H}_1}} \right) = 1 - {{\mathcal{P}}^{ij}}\left( {{\mathcal{H}_0}} \right)$ be the probabilities that channel $j$ is available and not available at secondary link $i$, respectively.
We assume that SUs employ an energy detection scheme and let $f_s$ be the sampling frequency used in the sensing period whose length is $\tau$ for all secondary links. 
There are two important performance measures, which are used to quantify the sensing performance, namely detection and false alarm probabilities. 
In particular, detection event occurs when a secondary link successfully senses a busy channel and false alarm represents the situation when a spectrum sensor returns a busy state for an idle channel (i.e., a transmission opportunity is overlooked).

Assume that transmission signals from PUs are complex-valued PSK signals while the noise at the secondary links is independent and identically distributed circularly symmetric complex Gaussian $\mathcal{CN}\left( {0,{N_0}} \right)$ \cite{Liang08}. 
Then, the detection and false alarm probability for the channel $j$ at secondary link $i$ can be calculated as \cite{Liang08}
\beqn
\label{eq_Chap3_1}
{{\mathcal{P}}_d^{ij}}\left( {{\varepsilon ^{ij}} ,\tau } \right) = \mathcal{Q}\left( {\left( {\frac{{\varepsilon ^{ij}} }{{{N_0}}} - {\gamma ^{ij}}  - 1} \right)\sqrt {\frac{{\tau {f_s}}}{{2{\gamma ^{ij}}  + 1}}} } \right), 
\eeqn
\beqn
 {{\mathcal{P}}_f^{ij}}\left( {{\varepsilon ^{ij}} ,\tau } \right) = \mathcal{Q}\left( {\left( {\frac{{\varepsilon ^{ij}} }{{{N_0}}} - 1} \right)\sqrt {\tau {f_s}} } \right) 
 = \mathcal{Q}\left( {\sqrt {2{\gamma ^{ij}}  + 1} {\mathcal{Q}^{ - 1}}\left( {{{\mathcal{P}}_d^{ij}}\left(  {{\varepsilon ^{ij}} ,\tau }  \right)} \right)+\sqrt {\tau {f_s}} {\gamma ^{ij}} } \right),  \label{eq_Chap3_2}
\eeqn
where $i \in \left[ {1,N} \right]$ is the index of a SU link, $j \in \left[ {1,M} \right]$ is the index of a channel, ${\varepsilon ^{ij}} $ is the detection threshold for an energy detector, ${\gamma ^{ij}} $ is the signal-to-noise ratio (SNR) of the PU's signal at the secondary link, $f_s$ is the sampling frequency, $N_0$ is the noise power, $\tau$ is the sensing interval, and $\mathcal{Q}\left( . \right)$ is defined as $\mathcal{Q}\left( x \right) = \left( {1/\sqrt {2\pi } } \right)\int_x^\infty  {\exp \left( { - {t^2}/2} \right)dt}$. 
In the analysis performed in the following sections, we assume a homogeneous scenario where sensing performance on different channels is the same for each SU.
In this case, we denote these probabilities for SU $i$ as  ${\mathcal{P}}_f^{i}$ and ${\mathcal{P}}_d^{i}$ for brevity.

\vspace{0.2cm}
\noindent
\textit{Remark 2:} For simplicity, we do not consider the impact of wireless channel fading in modeling 
the sensing performance in (\ref{eq_Chap3_1}), (\ref{eq_Chap3_2}). This enables us to gain insight into the investigated spectrum sensing and access problem
while keeping the problem sufficiently tractable. Extension of the model to capture wireless fading 
will be considered in our future works. Relevant results published in some recent 
works such as those in \cite{zuo10} would be useful for these further studies.

\vspace{0.2cm}
\noindent
\textit{Remark 3:} The analysis performed in the following sections can be easily extended to the case
where each secondary transmitter is equipped with only one spectrum sensor or each secondary transmitter only senses a subset
of all channels in each cycle. Specifically, we will need to adjust the sensing time for some 
spectrum sensing performance requirements. In particular, if only one spectrum sensor is available at each secondary transmitter,
then the required sensing time should be $M$ times larger than the case in which each transmitter has $M$ spectrum sensors.

\section{MAC Design, Analysis and Optimization: Single Channel Case}
\label{SingleChan_Chap3}

We consider the MAC protocol design, its throughput analysis and optimization for the single channel case in this section.

\subsection{MAC Protocol Design}
\label{MACDesigna_Chap3}

% Fig. 2
\begin{figure}[!t]
\centering
\includegraphics[width=120mm]{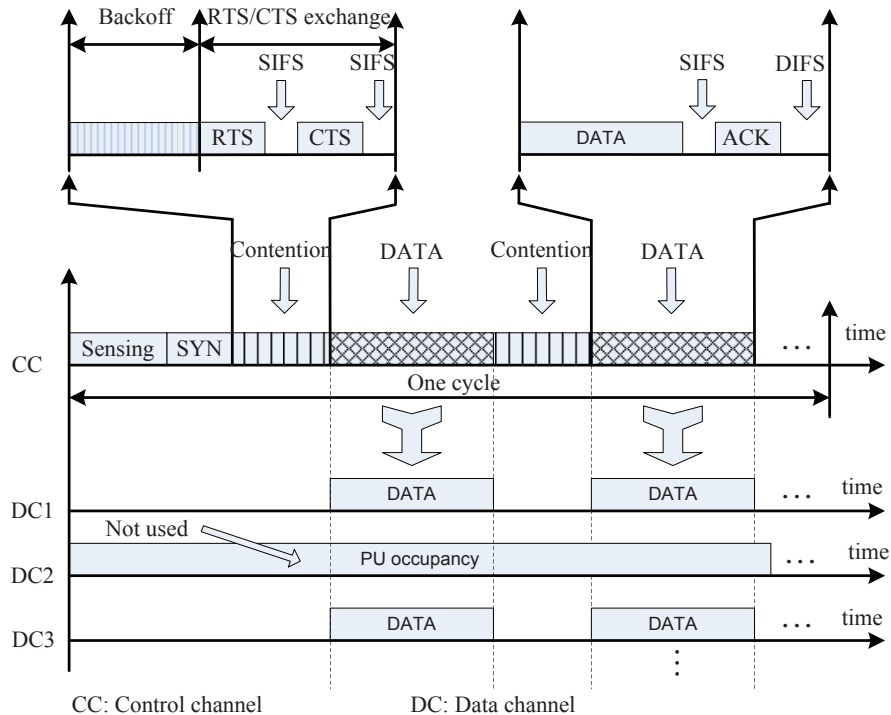}
\caption{Timing diagram of the proposed multi-channel MAC protocol.}
\label{MACoperation_Chap3}
\end{figure}

%% Fig. 2
%\begin{figure}[!t]
%\centering
%\includegraphics[width=70mm]{Figure2}
%\caption{Timing diagram of the proposed multi-channel MAC protocol.}
%\label{Fig2}
%\end{figure}

We now describe our proposed synchronized MAC for dynamic spectrum sharing among secondary flows.
We assume that each fixed-size cycle of length $T$ is divided into 3 phases, namely sensing phase, synchronization phase, and data transmission phase. 
During the sensing phase of length $\tau$, all SUs perform spectrum sensing on the underlying channel.
Then, only secondary links whose sensing outcomes indicate an available channel proceed to the next phase (they will be called active SUs/links in the following). 
In the synchronization phase, active SUs broadcast beacon signals for synchronization purposes. 
Finally, active SUs perform contention and transmit data in the data transmission phase. 
The timing diagram of one particular cycle is illustrated in Fig.~\ref{MACoperation_Chap3}.
For this single channel scenario, synchronization, contention, and data transmission occur on the same channel.

We assume that the length of each cycle is sufficiently large so that SUs can transmit several packets during the data transmission phase. 
Indeed, the current 802.22 standard specifies the spectrum evacuation time upon the return of PUs is 2 seconds, which is a relatively large interval. 
Therefore, our assumption would be valid for most practical cognitive systems. 
During the data transmission phase, we assume that active SUs employ a standard contention technique to capture the channel similar to that in the CSMA/CA protocol.
Exponential back-off with minimum contention window $W$ and maximum back-off stage $m$ \cite{bian00} is employed in the contention phase.
For brevity, we refer to $W$ simply as contention window in the following.
Specifically, suppose that the current back-off stage of a particular SU is $i$ then it starts the contention by choosing a random back-off time uniformly distributed in the range $[0,2^i W-1]$, $0 \leq i \leq m$. 
This user then starts decrementing its back-off time counter while carrier sensing transmissions from other secondary links.

Let $\sigma$ denote a mini-slot interval, each of which corresponds one unit of the back-off time counter. 
Upon hearing a transmission from any secondary link, each secondary link will ``freeze'' its back-off time counter and reactivate when the channel is sensed idle again. 
Otherwise, if the back-off time counter reaches zero, the underlying secondary link wins the contention. 
Here, either two-way or four-way handshake with RTS/CTS will be employed to transmit one data packet on the available channel. 
In the four-way handshake, the transmitter sends RTS to the receiver and waits until it successfully receives CTS before sending a data packet. 
In both handshake schemes, after sending the data packet the transmitter expects an acknowledgment (ACK) from the receiver to indicate a successful reception of the packet. 
Standard small intervals, namely DIFS and SIFS, are used before back-off time decrements and ACK packet transmission as described in \cite{bian00}. 
We refer to this two-way handshaking technique as a basic access scheme in the following analysis.

\subsection{Throughput Maximization}
\label{TputOpt_Chap3}

Given the sensing model and proposed MAC protocol, we are interested in finding its optimal configuration to achieve the maximum throughput subject to protection constraints for primary receivers. 
Specifically, let $\mathcal{NT}(\tau, W)$ be the normalized total throughput, which is a function of sensing time $\tau$ and contention window $W$. 
Suppose that each primary receiver requires that detection probability achieved by its conflicting primary link $i$ be at least $\overline{P}_d^i$. 
Then, the throughput maximization problem can be stated as follows:

\vspace{0.2cm}
\noindent
\textbf{Problem 1:} 
%\vspace{0.1cm}
\begin{equation}
\label{eq_Chap3_3a}% eq3
\begin{array}{l}
 {\mathop {\max }\limits_{\tau ,W}} \quad {\mathcal{NT}} \left( {\tau ,W} \right)  \\ 
 \mbox{s.t.}\,\,\,\, {\mathcal{P}}_d^i\left( {{\varepsilon ^i},\tau } \right) \geq \mathcal{\bar {P}}_d^i, \quad i=1, 2,\cdots, N \\
 \quad \quad 0 < \tau  \le {T},  \quad 0< W \leq W_{\sf max}, \\
 \end{array}\!\!
\end{equation}
where $W_{\sf max}$ is the maximum contention window and recall that $T$ is the cycle interval.
In fact, optimal sensing $\tau$ would allocate sufficient time to protect primary receivers and
optimal contention window would balance between reducing collisions among active secondary links
and limiting protocol overhead. 

\subsection{Throughput Analysis and Optimization}
\label{ThroughputAO_Chap3}

We perform saturation throughput analysis and solve the optimization problem (\ref{eq_Chap3_3a}) in this subsection.
Throughput analysis for the cognitive radio setting under investigation is more involved compared to standard MAC protocol throughput analysis (e.g., see \cite{Shi09}, \cite{bian00}) because the number of active secondary links participating in the contention in each cycle varies depending on the sensing outcomes. 
Suppose that all secondary links have same packet length. 
Let  $\Pr \left( {n = {n_0}} \right)$ and $\mathcal{T}\left( {\tau ,\phi \left| {n = {n_0}} \right.} \right)$ be the probability that $n_0$ secondary links participating in the contention and the conditional normalized throughput when $n_0$ secondary links join the channel contention, respectively. 
Then, the normalized throughput  can be calculated as 
\begin{equation}
\label{ntputa_Chap3} % ntput
\mathcal{NT} = \sum\limits_{{n_0} = 1}^N {\mathcal{T}\left( {\tau , W\left| {n = {n_0}} \right.} \right) \Pr \left( {n = {n_0}} \right)},
\end{equation}
where recall that $N$ is the number of secondary links, $\tau $ is the sensing time, $W$ is the contention window.
In the following, we show how to calculate $\Pr \left( {n = {n_0}} \right)$ and $\mathcal{T}\left( {\tau ,\phi \left| {n = {n_0}} \right.} \right)$.

\subsubsection{Calculation of $\Pr \left( {n = {n_0}} \right)$} 
\label{CalPrn0_Chap3}

Note that only secondary links whose sensing outcomes in the sensing phase indicate an available channel
proceed to contention in the data transmission phase. This case can happen for a particular secondary link $i$ in the following two scenarios:

\begin{itemize}

\item 
The PU is not active, and no false alarm is generated by the underlying secondary link.

\item
The PU is active, and secondary link $i$ mis-detects its presence.

\end{itemize}
Therefore, secondary link $i$ joins contention in the data transmission phase with probability
\begin{equation}
\label{eq_Chap3_4}
{{\mathcal{P}}_{\text{idle}}^i} = \left[ {1 - {{\mathcal{P}}_f^i}\left( {{\varepsilon ^i},\tau } \right)} \right]{{\mathcal{P}}^i}\left( {{\mathcal{H}_0}} \right) + {{\mathcal{P}}_m^i}\left( {{\varepsilon ^i} ,\tau} \right){{\mathcal{P}}^i} \left( {{\mathcal{H}_1}} \right),
\end{equation}
where ${{\mathcal{P}}_m^i}\left( {{\varepsilon ^i} ,\tau} \right)  = 1- {{\mathcal{P}}_d^i}\left( {{\varepsilon ^i} ,\tau} \right)$ is the mis-detection probability.
Otherwise, it will be silent for the whole cycle and waits until the next cycle. This occurs with probability
\beqn \label{Pbusy1ch_Chap3}
\begin{array}{l}
 \mathcal{P}_{\text{busy}}^i = 1 - \mathcal{P}_{\text{idle}}^i =  \mathcal{P}_f^i\left( {{\varepsilon ^i},\tau } \right){\mathcal{P}^i}\left( {{\mathcal{H}_0}} \right) + \mathcal{P}_d^i\left( {{\varepsilon ^i},\tau } \right){\mathcal{P}^i}\left( {{\mathcal{H}_1}} \right) \\ 
 \end{array}.
\eeqn
We assume that interference of active PUs to the SU is negligible; therefore, a transmission from any secondary link only fails when it collides with transmissions from other secondary links.
Now, let $\mathcal{S}_k$ denote one particular subset of all secondary links having exactly $n_0$ secondary links. 
There are $C_N^{{n_0}} = \frac{{N!}}{{{n_0}!(N - {n_0})!}}$ such sets $\mathcal{S}_k$. 
The probability of the event that $n_0$ secondary links join contention in the data transmission phase can be calculated as
\begin{equation}
\label{eq_Chap3_5}
\Pr \left( {n = {n_0}} \right) = \sum\limits_{k = 1}^{C_N^{{n_0}}} {\prod\limits_{i \in {\mathcal{S}_k}} {{\mathcal{P}}_{\text{idle}}^i} \prod\limits_{j \in \mathcal{S}\backslash {\mathcal{S}_k}} {{\mathcal{P}}_{\text{busy}}^j} },
\end{equation}
where  $\mathcal{S}$ denotes the set of all $N$ secondary links, and ${\mathcal{S}\backslash {\mathcal{S}_k}}$ is the complement of $\mathcal{S}_k$ with  $N-n_0$ secondary links. 
If all secondary links have the same $SNR_p$ and  the same probabilities ${{\mathcal{P}}^i}\left( {{\mathcal{H}_0}} \right)$ and ${{\mathcal{P}}^i}\left( {{\mathcal{H}_1}} \right)$, then we have
${\mathcal{P}}_{\text{idle}}^i = {\mathcal{P}}_{\text{idle}}$ and ${\mathcal{P}}_{\text{busy}}^i = {\mathcal{P}}_{\text{busy}} = 1-{\mathcal{P}}_{\text{\text{idle}}}$ for all $i$. 
In this case, (\ref{eq_Chap3_5}) becomes
\begin{equation}
\label{eq_Chap3_6}
\Pr \left( {n = {n_0}} \right) = C_N^{{n_0}}{\left( {1 - {{\mathcal{P}}_{\text{busy}}}}\right)^{{n_0}}}{\left( {{{\mathcal{P}}_{\text{busy}}}}  \right)^{N - {n_0}}},
\end{equation}
where all terms in the sum of (\ref{eq_Chap3_5}) become the same.

\vspace{0.2cm}
\noindent
\textit{Remark 4:} In general, interference from active PUs will impact transmissions of
SUs. However, strong interference from PUs would imply high SNR of sensing signals collected at PUs.
In this high SNR regime, we typically require small sensing time while still satisfactorily protecting PUs. Therefore, for the 
case in which interference from active PUs to SUs is small, sensing time will have the most significant impact
on the investigated sensing-throughput tradeoff. Therefore, consideration of this setting enables us to gain better 
insight into the underlying problem. Extension to the more general case is possible by explicitly calculating transmission rates
achieved by SUs as a function of SINR. Due to the space constraint, we will not explore this issue further in this paper.

\subsubsection{Calculation of Conditional Throughput}
\label{Con_thoughput_Chap3}

The conditional throughput can be calculated by using the technique developed by Bianchi in \cite{bian00} where we approximately assume a fixed transmission probability $\phi$ in a generic slot time. 
Specifically, Bianchi shows that this transmission probability can be calculated from the following two equations \cite{bian00}
\beqn \label{phi_Chap3}
\phi  = \frac{{2\left( {1 - 2p} \right)}}{{\left( {1 - 2p} \right)\left( {W + 1} \right) + Wp\left( {1 - {{\left( {2p} \right)}^m}} \right)}},
\eeqn
\beqn \label{p_Chap3}
p = 1-\left(1-\phi\right)^{n-1},
\eeqn
where $m$ is the maximum back-off stage, $p$ is the conditional collision probability (i.e., the probability that a collision is observed when a data packet is transmitted on the channel).

Suppose there are $n_0$ secondary links participating in contention in the third phase, the probability of the event that at least one secondary link transmits its data packet can be written as
\begin{equation}
\label{eq_Chap3_8a} %eq8
{{\mathcal{P}}_t} = 1 - {\left( {1 - \phi } \right)^{{n_0}}}.
\end{equation}
However, the probability that a transmission occurring on the channel is successful given there is at least one secondary link transmitting can be written as
\begin{equation}
\label{eq_Chap3_9a}% eq9
{{\mathcal{P}}_s} = \frac{{{n_0}\phi {{\left( {1 - \phi } \right)}^{{n_0} - 1}}}}{{{{\mathcal{P}}_t}}}.
\end{equation}
The average duration of a generic slot time can be calculated as 
\begin{equation}
\label{eq_Chap3_10a}% eq10
{{\bar T}_{sd}} = \left( {1 - {{\mathcal{P}}_t}} \right){T_e} + {{\mathcal{P}}_t}{{\mathcal{P}}_s}{T_s} + {{\mathcal{P}}_t}\left( {1 - {{\mathcal{P}}_s}} \right){T_c},
\end{equation}
where $T_e = \sigma$, $T_s$ and $T_c$ represent the duration of an empty slot, the average time the channel is sensed busy due to a successful transmission, and the average time the channel is sensed busy due to a collision, respectively. These quantities can be calculated as \cite{bian00}
\\
\textit{For basic mechanism:}
\begin{equation}
\zlabel{eq_Chap3_11a}% eq11
\left\{ \!\!\!{\begin{array}{*{20}{c}}
   {{T_s} = T_s^{1} = H + PS + SIFS + 2PD \!+ \!ACK \!+\! DIFS} \hfill  \\
   {{T_c} = T_c^{1} = H + PS + DIFS + PD} \hfill  \\
   {H = {H_{PHY}} + {H_{MAC}}} \hfill  \\
\end{array}} \right.\!\!\!\!,
\end{equation}
where $H_{PHY}$ and $H_{MAC}$ are the packet headers for physical and MAC layers, $PS$ is the packet size, which is assumed to be fixed in this chapter, $PD$ is the propagation delay, $SIFS$ is the length of a short inter-frame space, $DIFS$ is the length of a distributed inter-frame space, $ACK$ is the length of an acknowledgment. 
\\
\textit{For RTS/CTS mechanism:}
\beqn \label{TsTc_Chap3}
\left\{ {\begin{array}{*{20}{c}}
   \begin{array}{l}
 {T_s} = T_s^2 = H + PS + 3SIFS + 2PD + RTS + CTS + ACK + DIFS \\ 
 \end{array} \hfill  \\
   {{T_c} = T_c^2 = H + DIFS + RTS + PD} \hfill  \\
\end{array}} \right.,
\eeqn
where we abuse notations by letting $RTS$ and $CTS$ represent the length of $RTS$ and $CTS$ control packets, respectively.

Based on these quantities, we can express the conditional normalized throughput as follows:
\begin{equation}
\label{eq_Chap3_13a}%eq13
\mathcal{T} \left( \tau ,\phi \left| {n = {n_0}} \right. \right) = \left\lfloor {\frac{T - \tau }{{\bar T}_{sd}}} \right\rfloor \frac{\mathcal{P}_s\mathcal{P}_t PS}{T},
\end{equation}
where $\left\lfloor  .  \right\rfloor $ denotes the floor function and recall that $T$ is the duration of a cycle. Note
that $\left\lfloor {\frac{T - \tau }{{\bar T}_{sd}}} \right\rfloor$ denotes the average number of generic slot times
in one particular cycle excluding the sensing phase. Here, we omit the length of the synchronization phase, which is assumed to be negligible.

\subsubsection{Optimal Sensing and MAC Protocol Design}
\label{Optimal_SSa_Chap3}

Now, we turn to solve the throughput maximization problem formulated in (\ref{eq_Chap3_3a}). 
Note that we can calculate the normalized throughput given by (\ref{ntputa_Chap3}) by using $\Pr \left( {n = {n_0}} \right)$ calculated from (\ref{eq_Chap3_5}) and the conditional throughput calculated from (\ref{eq_Chap3_13a}). 
It can be observed that the detection probability ${\mathcal{P}}_d^i\left( {{\varepsilon ^i},\tau } \right)$ in the primary protection constraints ${\mathcal{P}}_d^i\left( {{\varepsilon ^i},\tau } \right) \ge \,\mathcal{\bar {P}}_d^i$ depends on both detection threshold ${\varepsilon ^i}$ and the optimization variable $\tau$. 

We can show that by optimizing the normalized throughput over $\tau$ and $W$ while fixing detection thresholds ${\varepsilon ^i} = {\varepsilon ^i_0}$ where ${\mathcal{P}}_d^i\left( {{\varepsilon ^i_0},\tau } \right) = \,\mathcal{\bar {P}}_d^i, \: i=1, 2,\cdots , N$, we can achieve almost the maximum throughput gain. 
The intuition behind this observation can be interpreted as follows.
If we choose ${\varepsilon ^i} < {\varepsilon ^i_0}$ for a given $\tau$, then both ${\mathcal{P}}_d^i\left( {{\varepsilon ^i},\tau }\right)$ and ${\mathcal{P}}_f^i\left( {{\varepsilon ^i},\tau }\right)$ increase compared to the case ${\varepsilon ^i} = {\varepsilon ^i_0}$.
As a result, ${{\mathcal{P}}_{\text{busy}}^i}$ given in (\ref{Pbusy1ch_Chap3}) increases. 
Moreover, it can be verified that the increase in ${{\mathcal{P}}_{\text{busy}}^i}$ will lead to the shift of the probability distribution $\Pr \left( {n = {n_0}} \right)$ to the left. 
Specifically, $\Pr \left( {n = {n_0}} \right)$ given in (\ref{eq_Chap3_5}) increases for small $n_0$ and decreases for large $n_0$ as ${{\mathcal{P}}_{\text{busy}}^i}$ increases. 
Fortunately, with appropriate choice of contention window $W$ the conditional throughput ${\mathcal{T}} \left( { \tau ,W\left| {n = n_0} \right.} \right)$ given in (\ref{eq_Chap3_13a}) is quite flat for different $n_0$ (i.e., it only decreases slightly when $n_0$ increases). 
Therefore, the normalized throughput given by (\ref{ntputa_Chap3}) is almost a constant when we choose ${\varepsilon ^i} < {\varepsilon ^i_0}$.

In the following, we will optimize the normalized throughput over $\tau$ and $W$ while choosing detection thresholds such that ${\mathcal{P}}_d^i\left( {{\varepsilon ^i_0},\tau } \right) = \,\mathcal{\bar {P}}_d^i, \: i=1, 2,\cdots , N$. 
From these equality constraints and (\ref{eq_Chap3_2}) we have
\beqn \label{Pfoa_Chap3}%Pfo
 {{\mathcal{P}}_f^i} = \mathcal{Q} \left( \alpha^i + \sqrt {\tau {f_s}} {{\gamma ^i} } \right)
\eeqn
where $\alpha^i =\sqrt {2{\gamma ^i}  + 1} {\mathcal{Q}^{ - 1}} \left( {\bar{\mathcal{P}}_d^i} \right)$.
Hence, the optimization problem (\ref{eq_Chap3_3a}) becomes independent of all detection thresholds ${\varepsilon ^i}, \: i=1,2,\cdots , N$.
Unfortunately, this optimization problem is still a mixed integer program (note that $W$ takes integer values), which is difficult to solve. 
In fact, it can be verified even if we allow $W$ to be a real number, the resulting optimization problem is still not convex because the objective function is not concave \cite{R6}. 
Therefore, standard convex optimization techniques cannot be employed to find the optimal solution for the optimization problem under investigation. 
Therefore, we have to rely on numerical optimization \cite{R7} to find the optimal configuration for the proposed MAC protocol.
Specifically, for a given contention window $W$ we can find the corresponding optimal sensing time $\tau$ as follows:

\vspace{0.2cm}
\noindent
\textbf{Problem 2:} \\
\begin{equation}
\label{eq_Chap3_14}
{\mathop {\max }_{0<\tau \leq T} } \quad {{\mathcal{NT}} \!\!\left( { \tau, W} \right)}  
= \sum\limits_{{n_0} = 1}^N \!\!\! {{\mathcal{T}} \left( { \tau ,W\left| {n = n_0} \right.} \right) \!\Pr \left( {n = {n_0}} \right)}.
\end{equation}

This optimization problem is not convex because its objective function is not concave in general. 
However, we will prove that ${{\mathcal{NT}} \!\!\left( { \tau } \right)}$ is an unimodal function in the range of $\left[0,T\right]$. 
Specifically, ${{\mathcal{NT}} \!\!\left( { \tau } \right)}$ is monotonically increasing in $\left[0,{\overline{\tau}}\right)$ while it is monotonically decreasing in $\left(\overline{\tau},T\right]$ for some $0 < \overline{\tau} \leq T$. 
Hence, ${{\mathcal{NT}} \!\!\left( { \overline{\tau} } \right)}$ is the only global maximum in the entire range of $\left[0,T\right]$.
This property is formally stated in the following proposition.

\vspace{0.2cm}
\noindent
\textbf{Proposition 1:} The objective function ${{\mathcal{NT}} \!\!\left( { \tau } \right)}$ of (\ref{eq_Chap3_14}) satisfies
the following properties
\begin{enumerate}
\item %{\beqn \label{eq15}
$\mathop {\lim }\limits_{\tau  \to {T}}   \frac{{\partial {\mathcal{NT}}}}{{\partial \tau }}  < 0$,

\item
$\mathop {\lim }\limits_{\tau  \to 0} \frac{{\partial {\mathcal{NT}}}}{{\partial \tau }} =  + \infty$, %\label{eq17}
%\eeqn}

\item {there is an unique ${\overline{\tau}}$ where ${\overline{\tau}}$ is in the range of $\left[0,T\right]$ such that $\frac{{\partial \mathcal{{NT}}\left( {{\overline{\tau} }} \right)}}{{\partial \tau }} = 0$,}
	\item the objective function ${{\mathcal{NT}} \!\!\left( { \tau } \right)}$ is bounded from above.
\end{enumerate}

\begin{proof} The proof is provided in Appendix~\ref{Proof_Chap3}. \end{proof}

We would like to discuss the properties stated in Proposition 1. 
Properties 1, 2, and 4 imply that there must be at least one $\tau$ in  $\left[0,T\right]$ that maximizes ${\mathcal{{NT}}\left( \tau  \right)}$. 
The third property implies that indeed such an optimal solution is unique. 
Therefore, one can find the globally optimal $(W^*,\tau^*)$ by finding optimal $\tau$ for each $W$ in its feasible range $[1,W_{\sf max}]$. 
The procedure to find $(W^*,\tau^*)$ can be described in Alg.~ \ref{mainalg_Chap3}. 
Numerical studies reveal that this algorithm has quite low computation time for practical values of $W_{\sf max}$ and $T$.

\begin{algorithm}[h]%\leesize
\caption{\textsc{Optimization of Cognitive MAC Protocol}}
\label{mainalg_Chap3}
%\algsetup{indent=1.5em}
\begin{algorithmic}[1]

\STATE For each integer value of $W \in [1,W_{\sf max}]$, find the optimal $\tau$ according to (\ref{eq_Chap3_14}), i.e.,
\beqn
\overline{\tau}(W) = {\mathop {\argmax  }_{0<\tau \leq T} } \quad {{\mathcal{NT}} \!\!\left( { \tau, W} \right)}  
\eeqn
\STATE The globally optimal $(W^*,\tau^*)$ can then be found as
\beqn
(W^*,\tau^*) = \argmax_{W, \overline{\tau}(W)}  \quad {{\mathcal{NT}} \!\!\left( { \overline{\tau}(W), W} \right)}.
\eeqn
\end{algorithmic}
\end{algorithm}

\subsection{Some Practical Implementation Issues}

Implementation for the proposed optimal MAC protocol configuration can be done as follows.
Each SU will need to spend some time to estimate the channel availability probabilities,
channel SNRs, and the number of SUs sharing the underlying spectrum. When these
system parameters have been estimated, each SU can independently calculate the optimal
sensing time and minimum contention window and implement them. Therefore, implementation for optimal MAC protocol can
be performed in a completely distributed manner, which would be very desirable.

\section{MAC Design, Analysis, and Optimization: Multiple Channel Case}
\label{MultipleChan_Chap3}

We consider the MAC protocol design, analysis and optimization for the multi-channel case in this section.

\subsection{MAC Protocol Design}
\label{MACDesign_Chap3}

We propose a synchronized multi-channel MAC protocol for dynamic spectrum sharing in this subsection.
To exploit spectrum holes in this case, we assume that there is one control channel which belongs to the secondary network (i.e., it is always available) and $M$ data channels which can be exploited by SUs. 
We further assume that each transmitting SU employ a reconfigurable transceiver which can be tuned to the control channel or
vacant channels for data transmission easily. In addition, we assume that this transceiver can turn on and off
the carriers on the available or busy channels, respectively (e.g., this can be achieved by the OFDM technology). 

There are still three phases for each cycle as in the single-channel case.
However, in the first phase, namely the sensing phase of length $\tau$, all SUs simultaneously perform spectrum sensing on all $M$ underlying channels. 
Because the control channel is always available, all SUs exchange beacon signals to achieve synchronization in the second phase. 
Moreover, only active secondary links whose sensing outcomes indicate at least one vacant channel participate in the third phase (i.e., data transmission phase).  
As a result, the transmitter of the winning link in the contention phase will need to inform its receiver about the available channels. 
Finally, the winning secondary link will transmit data on all vacant channels in the data transmission phase. 
The timing diagram of one particular cycle is illustrated in Fig.~\ref{MACoperation_Chap3}.

Again, we also assume that the length of each cycle is sufficiently large such that secondary links can transmit several packets on each available channel during the data transmission phase. 
In the data transmission phase, we assume that active secondary links adopt the standard contention technique to capture the channels similar to that employed by the CSMA/CA protocol using exponential back-off and either two-way or four-way handshake as described in Section ~\ref{SingleChan_Chap3}.
For the case with two-way handshake, both secondary transmitters and receivers need to perform spectrum sensing.
With four-way handshake, only secondary transmitters need to perform spectrum sensing and the RTS message will contain additional information about the available channels on which the receiver will receive data packets.
Also, multiple packets (i.e., one on each available channel) are transmitted by the winning secondary transmitter. 
Finally, the ACK message will be sent by the receiver to indicate successfully received packets on the vacant channels.

\subsection{Throughput Maximization}
\label{TputOpt_1_Chap3}

In this subsection, we discuss how to find the optimal configuration to maximize the normalized throughput under sensing constraints for PUs.
Suppose that each primary receiver requires that detection probability achieved by its conflicting primary link $i$ on channel $j$ be at least $\overline{P}_d^{ij}$. 
Then, the throughput maximization problem can be stated as follows:

\vspace{0.2cm}
\noindent
\textbf{Problem 3:} 
\begin{equation}
\label{eq_Chap3_3}
\begin{array}{l}
 {\mathop {\max }\limits_{\tau ,W}} \quad {\mathcal{NT}} \left( {\tau ,W} \right)  \\ 
 \mbox{s.t.}\,\,\,\, {\mathcal{P}}_d^{ij} \left( {{\varepsilon ^{ij}},\tau } \right) \geq \mathcal{\bar {P}}_d ^{ij}, i\in\left[1, N\right], j\in\left[1, M\right]\\
 \quad \quad 0 < \tau  \le {T},  \quad 0< W \leq W_{\sf max}, \\
 \end{array}\!\!
\end{equation}
where $\mathcal{P}_d^{ij}$ is the detection probability for SU $i$ on channel $j$, $W_{\sf max}$ is the maximum contention window and recall that $T$ is the cycle interval.
We will assume that for each SU $i$, ${\mathcal{P}}_d\left( {{\varepsilon ^{ij}},\tau } \right)$ and $\mathcal{\bar {P}}_d ^{ij}$ are the same for all channel $j$, respectively.
This would be valid because sensing performance (i.e., captured in ${\mathcal{P}}_d\left( {{\varepsilon ^{ij}},\tau } \right)$ and ${\mathcal{P}}_f\left( {{\varepsilon ^{ij}},\tau } \right)$) depends on detection thresholds $\epsilon^{ij}$ and the SNR $\gamma^{ij}$, which would be the same for different channels $j$. 
In this case, the optimization problem reduces to that of the same form as (\ref{eq_Chap3_3a}) although the normalized throughput ${\mathcal{NT}} \left( {\tau ,W} \right)$ will need to be derived for this multi-channel case. 
For brevity, we will drop all channel index $j$ in these quantities whenever possible.

%\vspace{10pt}
\subsection{Throughput Analysis and Optimization}
\label{ThroughputAO_1_Chap3}

We analyze the saturation throughput and show how to obtain an optimal solution for \textbf{Problem 3 }.
Again we assume that all secondary links transmit data packets of the same length. 
Let  $\Pr \left( {n = {n_0}} \right)$, $\mathbf{E}\left[l\right]$ and $\mathcal{T}\left( {\tau ,\phi \left| {n = {n_0}} \right.} \right)$ denote the probability that $n_0$ secondary links participating in the contention phase, the average number of vacant channels at the winning SU link, and the conditional normalized throughput when $n_0$ secondary links join the contention,  respectively. 
Then, the normalized throughput  can be calculated as 
\begin{equation}
\label{ntput_Chap3}
\mathcal{NT} = \sum\limits_{{n_0} = 1}^N {\mathcal{T}\left( {\tau , W\left| {n = {n_0}} \right.} \right) \Pr \left( {n = {n_0}} \right) \frac{{\mathbf{E}\left[ {{l}} \right]}}{M}},
\end{equation}
where recall that $N$ is the number of secondary links, $M$ is the number of channels, $\tau $ is the sensing time, $W$ is the contention window.
Note that this is the average system throughput per channel.
We will calculate $\mathcal{T}\left( {\tau ,\phi \left| {n = {n_0}} \right.} \right)$ using (\ref{eq_Chap3_13a}) for the proposed MAC protocol with four-way handshake and exponential random back-off. In addition, we also show how to calculate $\Pr \left( {n = {n_0}} \right)$ .

\subsubsection{Calculation of $\Pr \left( {n = {n_0}} \right)$ and $\mathbf{E} \left[ l\right]$} 

Recall that only secondary links whose sensing outcomes indicate at least one available channel participate in contention in the data transmission phase. 
Again,  as in the single channel case derived in Section \ref{CalPrn0_Chap3} the sensing outcome at SU $i$ indicates that channel $j$ is available or busy with probabilities ${\mathcal{P}^{i}_{\text{idle}}}$ and ${\mathcal{P}^{i}_{\text{busy}}}$, which are in the same forms with (\ref{Pbusy1ch_Chap3}) and (\ref{eq_Chap3_4}), respectively (recall that we have dropped the channel index $j$ in these quantities). 
Now, $\Pr \left( {n = {n_0}} \right)$ can be calculated from these probabilities. 
Recall that secondary link $i$ only joins the contention if its sensing outcomes indicate at least one vacant channel. 
Otherwise, it will be silent for the whole cycle and waits until the next cycle. 
This occurs if its sensing outcomes indicate that all channels are busy.

To gain insight into the optimal structure of the optimal solution while keeping mathematical details sufficiently tractable, we will consider the homogeneous case in the following where  ${\mathcal{P}^{i}_f}$,  ${\mathcal{P}^{i}_d}$ (therefore, ${\mathcal{P}^{i}_{\text{idle}}}$ and ${\mathcal{P}^{i}_{\text{busy}}}$) are the same for all SUs $i$. 
The obtained results, however, can extended to the general case even though the corresponding expressions will be more lengthy and tedious.
For the homogeneous system, we will simplify $\mathcal{P}^{i}_{\text{SUidle}}$ and $\mathcal{P}^{i}_{\text{SUbusy}}$ to $\mathcal{P}_{\text{SUidle}}$ and $\mathcal{P}_{\text{SUbusy}}$, respectively for brevity. 
Therefore, the probability that a particular channel is indicated as busy or idle by the corresponding spectrum sensor can be written as
\beq \label{Pbus1_Chap3}
{\mathcal{P}_{\text{busy}}} = {\mathcal{P}_f}\mathcal{P}\left( {{\mathcal{H}_0}} \right) + {\mathcal{P}_d}\mathcal{P}\left( {{\mathcal{H}_1}} \right),
\eeq
\beqn \label{Pidl_Chap3}
{\mathcal{P}_{\text{idle}}} = 1 - {\mathcal{P}_{\text{busy}}}.
\eeqn
Let $Pr\left(l = l_0\right)$ denote the probability that $l_0$ out of $M$ channels are indicated as available by the spectrum sensors.
Then, this probability can be calculated as
\beqn \label{Pl0_Chap3}
\Pr \left( {l = {l_0}} \right) = C_{M}^{l_0} \mathcal{P}_{\text{idle}}^{{l_0}}\mathcal{P}_{\text{busy}}^{M - {l_0}}.
\eeqn
Now, let $\mathcal{P}_{\text{SUidle}}$ be the probability that a particular secondary link $i$ participates in the contention (i.e., its spectrum sensors indicate at least one available channel) and $\mathcal{P}_{\text{SUbusy}}$ be the probability that secondary link $i$ is silent (i.e., its spectrum sensors indicate that all channels are busy). 
Then, these probabilities can be calculated as
\beqn \label{PSUbus_Chap3} 
{\mathcal{P}_{\text{SUbusy}}} = \Pr \left( {l = 0} \right) = \mathcal{P}_{\text{busy}}^M, 
\eeqn
\beqn \label{PSUidl_Chap3}
{\mathcal{P}_{\text{SUidle}}} = \sum\limits_{{l_0} = 1}^M {\Pr \left( {l = {l_0}} \right) = 1 - \mathcal{P}_{\text{SUbusy}}}.
\eeqn
 
Again we assume that a transmission from a particular secondary link only fails if it collides with transmissions from other secondary links.
The probability that $n_0$ secondary links join the contention can be calculated by using (\ref{PSUidl_Chap3}) and (\ref{PSUbus_Chap3}) as follows: 
\beqn \label{Pn0_Chap3}
 \begin{array}{l}
 \Pr \left( {n = {n_0}} \right) = C_N^{n_0} \mathcal{P}_{\text{SUidle}}^{{n_0}}\mathcal{P}_{\text{SUbusy}}^{N - {n_0}} = C_N^{n_0} \left( 1 - \mathcal{P}_{\text{busy}}^M \right)^{n_0}\mathcal{P}_{\text{busy}}^{M\left( {N - {n_0}} \right)} \\ 
 \end{array}.
\eeqn
From (\ref{Pl0_Chap3}), we can calculate the average number of available channels, denoted by the expectation $\mathbf{E}\left[l\right]$, at one particular secondary link as
\beqn \label{El_Chap3}
\begin{array}{l}
 \mathbf{E}[l] = \sum\limits_{{l_0} = 0}^M {{l_0}\Pr \left( {l = {l_0}} \right)}  = \sum\limits_{{l_0} = 0}^M l_0 C_M^{l_0} \mathcal{P}_{\text{idle}}^{l_0} \mathcal{P}_{\text{busy}}^{M - l_0} = M{\mathcal{P}_{\text{idle}}} = M\left( {1 - {\mathcal{P}_{\text{busy}}}} \right) \\ 
 \end{array}.
\eeqn

\subsubsection{Optimal Sensing and MAC Protocol Design}
\label{Optimal_SS_Chap3}

We now tackle the throughput maximization problem formulated in (\ref{eq_Chap3_3}). In this case, the normalized throughput given by (\ref{ntput_Chap3}) can be calculated by using $\Pr \left( {n = {n_0}} \right)$ in (\ref{Pn0_Chap3}), the conditional throughput in (\ref{eq_Chap3_13a}), and the average number of available channels in (\ref{El_Chap3}). 
Similar to the single-channel case, we will optimize the normalized throughput over $\tau$ and $W$ while choosing a detection threshold such that ${\mathcal{P}}_d\left( {{\varepsilon _0},\tau } \right) = \,\mathcal{\bar {P}}_d$. 
Under these equality constraints, the false alarm probability can be written as
\beqn \label{Pfo_Chap3}
 {{\mathcal{P}}_f} = \mathcal{Q} \left( \alpha + \sqrt {\tau {f_s}} {{\gamma } } \right)
\eeqn
where $\alpha =\sqrt {2{\gamma }  + 1} {\mathcal{Q}^{ - 1}} \left( {\bar{\mathcal{P}}_d} \right)$. 
Hence, \textbf{Problem 3} is independent of detection thresholds.
Again, for a given contention window $W$ we can find the corresponding optimal sensing time $\tau$ in the following optimization problem

\vspace{0.1cm}
\noindent
\textbf{Problem 4:} 
\beqn \label{Prob2_Chap3}
\begin{array}{l}
 \begin{array}{*{20}{c}}
   {\mathop {\max }\limits_\tau  } & {\mathcal{\widetilde{NT}}\left( \tau  \right) \buildrel \Delta \over = NT\left( {\tau ,W} \right)\left| {_{W = \bar W}} \right.}  \\
\end{array} \\ 
 \begin{array}{*{20}{c}}
   {s.t.} & {0 \le \tau  \le }  \\
\end{array}T \\ 
 \end{array}.
\eeqn 

Similar to the single-channel case, we will prove that ${\mathcal{\widetilde{NT}}\left( \tau  \right)}$ is a unimodal function in the range of $\left[0,T\right]$. 
Therefore, there is a unique global maximum in the entire range of $\left[0,T\right]$. 
This is indeed the result of several properties stated in the following proposition. 

\vspace{0.2cm}
\noindent
\textbf{Proposition 2:} The function ${\mathcal{\widetilde{NT}}\left( \tau  \right)}$ satisfies the following properties
\begin{enumerate}
	\item {$\mathop {\lim \,}\limits_{\tau  \to 0} \frac{{\partial \mathcal{\widetilde{NT}}\left( \tau  \right)}}{{\partial \tau }} > 0$,}
	\item {$\mathop {\lim \,}\limits_{\tau  \to T} \frac{{\partial \mathcal{\widetilde{NT}}\left( \tau  \right)}}{{\partial \tau }} < 0$,}	
	\item {there is an unique $\overline{\tau}$ where $\overline{\tau}$ is in the range of $\left[0,T\right]$ such that $\frac{{\partial \mathcal{\widetilde{NT}}\left( {\overline{\tau}} \right)}}{{\partial \tau }} = 0$,}
	\item {and the objective function ${\mathcal{\widetilde{NT}}\left( \tau  \right)}$ is bounded from above.}
\end{enumerate}
Therefore, it is a unimodal function in the range of $\left[0,T\right]$.

\begin{proof} The proof is provided in Appendix~\ref{Proof_Chap3_P2}. \end{proof}

Therefore, given one particular value of $W$ we can find a unique optimal $\overline{\tau}(W)$ for the optimization problem (\ref{Prob2_Chap3}). 
Then, we can find the globally optimal $(W^*,\tau^*)$ by finding optimal $\tau$ for each $W$ in its feasible range $[1,W_{\sf max}]$.
The procedure to find $(W^*,\tau^*)$ is the same as that described in Alg.~\ref{mainalg_Chap3}.

%\vspace{10pt}
\section{Numerical Results}
\label{Results_Chap3}

We present numerical results to illustrate throughput performance of the proposed cognitive MAC protocols.
We take key parameters for the MAC protocols from Table II in \cite{bian00}. Other parameters are chosen as follows: cycle time is $T = 100 ms$; mini-slot (i.e., generic empty slot time) is $\sigma = 20 {\mu} s$; sampling frequency for spectrum sensing is $f_s = 6 MHz$;  bandwidth of PUs' QPSK signals is $6 MHz$. 
In addition, the exponential back-off mechanism with the maximum back-off stage $m$ is employed to reduce collisions.

\subsection{Performance of Single Channel MAC Protocol}
\label{Results_1_Chap3}

For the results in this section, we choose other parameters of the cognitive network as follows. 
The signal-to-noise ratio of PU signals at secondary links $SNR_p^i$ are chosen randomly in the range $[-15, -20] dB$. 
The target detection probability for secondary links and the probabilities  ${{\mathcal{P}}^i}\left( {{\mathcal{H}_0}} \right)$ are chosen randomly in the intervals $[0.7, 0.9]$ and $[0.7, 0.8]$, respectively. The basic scheme is used as a handshaking mechanism for the MAC protocol.

% Fig. 3

\begin{figure}[!t]
\centering
\includegraphics[width=90mm]{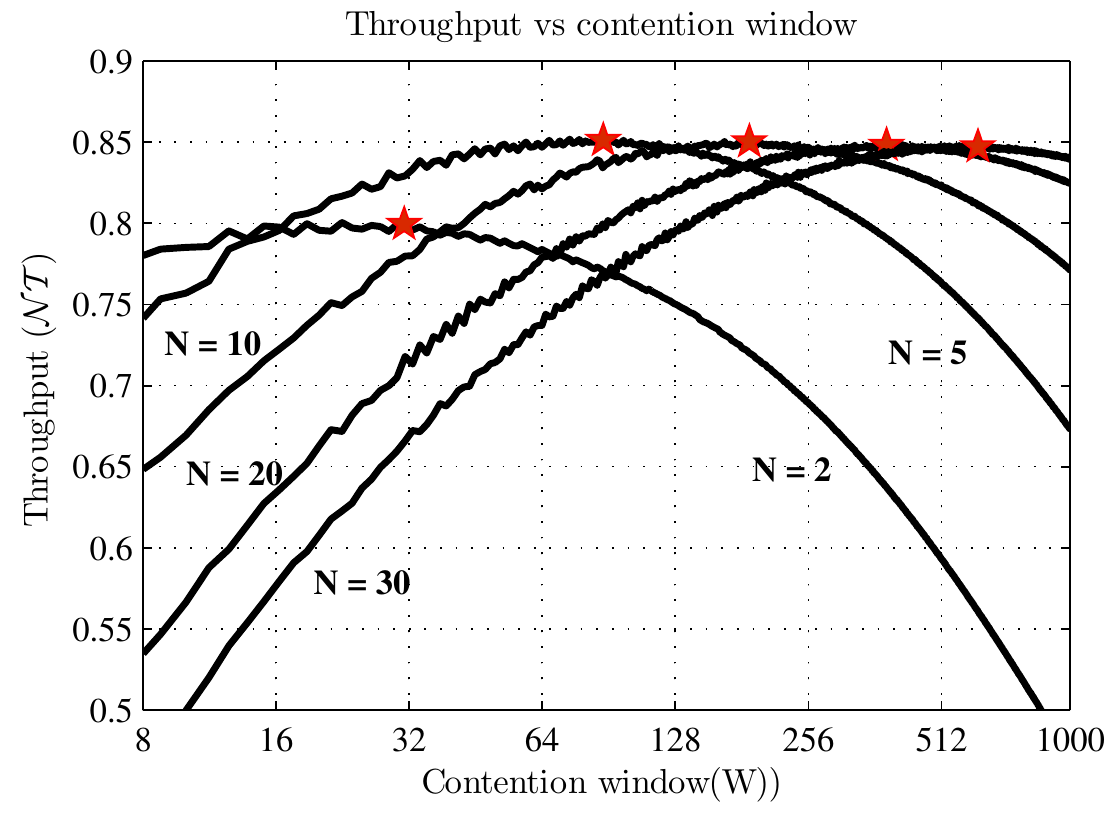}
\caption{Normalized throughput versus contention window $W$ for $\tau = 1 ms$, $m = 3$, different $N$ and basic access mechanism.}
\label{Fig3_Chap3}
\end{figure}

In Fig.~\ref{Fig3_Chap3}, we show the normalized throughput $\mathcal{NT}$ versus contention window $W$ for different values of $N$ 
when the sensing time is fixed at $\tau = 1 ms$  and the maximum backoff stage is chosen at $m = 3$ for one particular realization of system parameters. 
The maximum throughput on each curve is indicated by a star symbol. 
This figure indicates that the maximum throughput is achieved at larger $W$ for larger $N$. 
This is expected because larger contention window can alleviate collisions among active secondary for larger number of secondary links.
It is interesting to observe that the maximum throughput can be larger than 0.8 although  ${{\mathcal{P}}^i}\left( {{\mathcal{H}_0}} \right)$ are chosen in the range $[0.7, 0.8]$. 
This is due to a multiuser gain because secondary links are in conflict with different primary receivers.

% Fig. 4

\begin{figure}[!t]
\centering
\includegraphics[width=90mm]{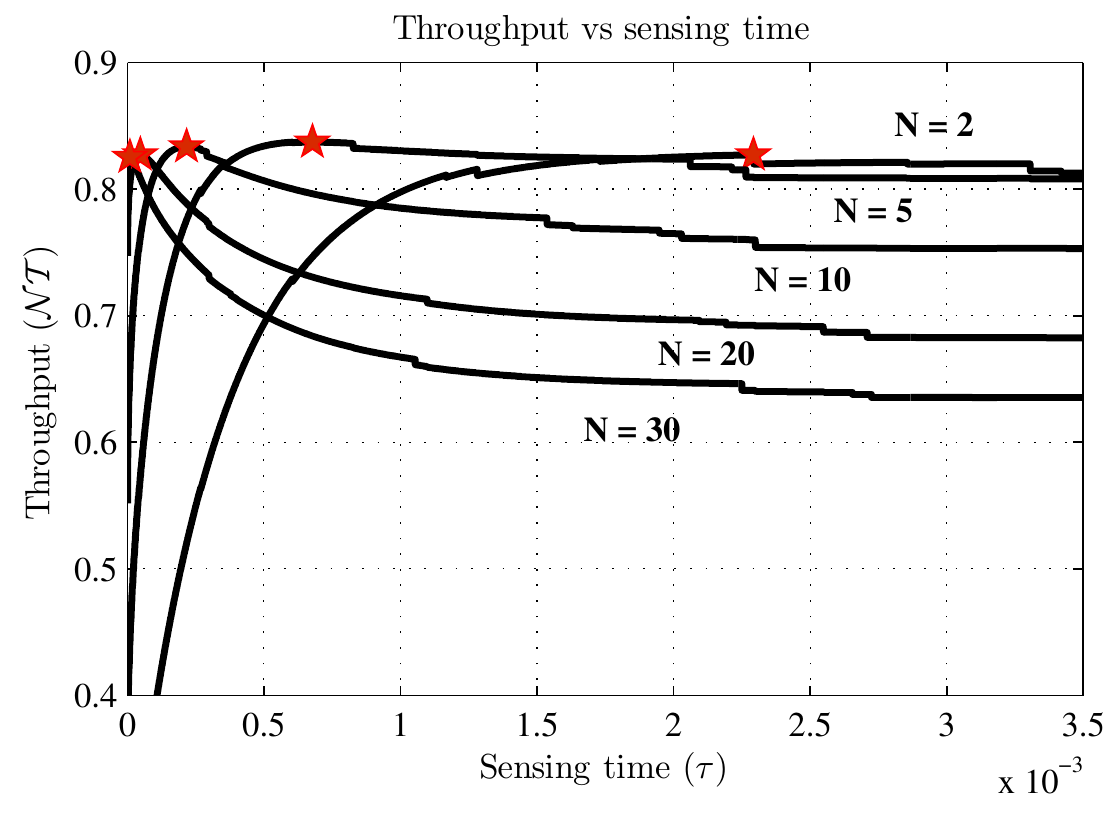}
\caption{Normalized throughput versus the sensing time $\tau$ for $W = 32$, $m = 3$ , different $N$ and basic access mechanism.}
\label{Fig4_Chap3}
\end{figure}

In Fig.~\ref{Fig4_Chap3} we present the normalized throughput $\mathcal{NT}$ versus sensing time $\tau$ for a fixed contention window $W=32$, maximum backoff stage $m = 3$, and different number of secondary links $N$. 
The maximum throughput is indicated by a star symbol on each curve. 
This figure confirms that the normalized throughput $\mathcal{NT}$ increases when $\tau$ is small and decreases with large $\tau$ as being proved in Proposition 1. 
Moreover, for a fixed contention window the optimal sensing time indeed decreases with the number of secondary links $N$. 
Finally, the multi-user diversity gain can also be observed in this figure.

% Fig. 5

\begin{figure}[!t]
\centering
\includegraphics[width=90mm]{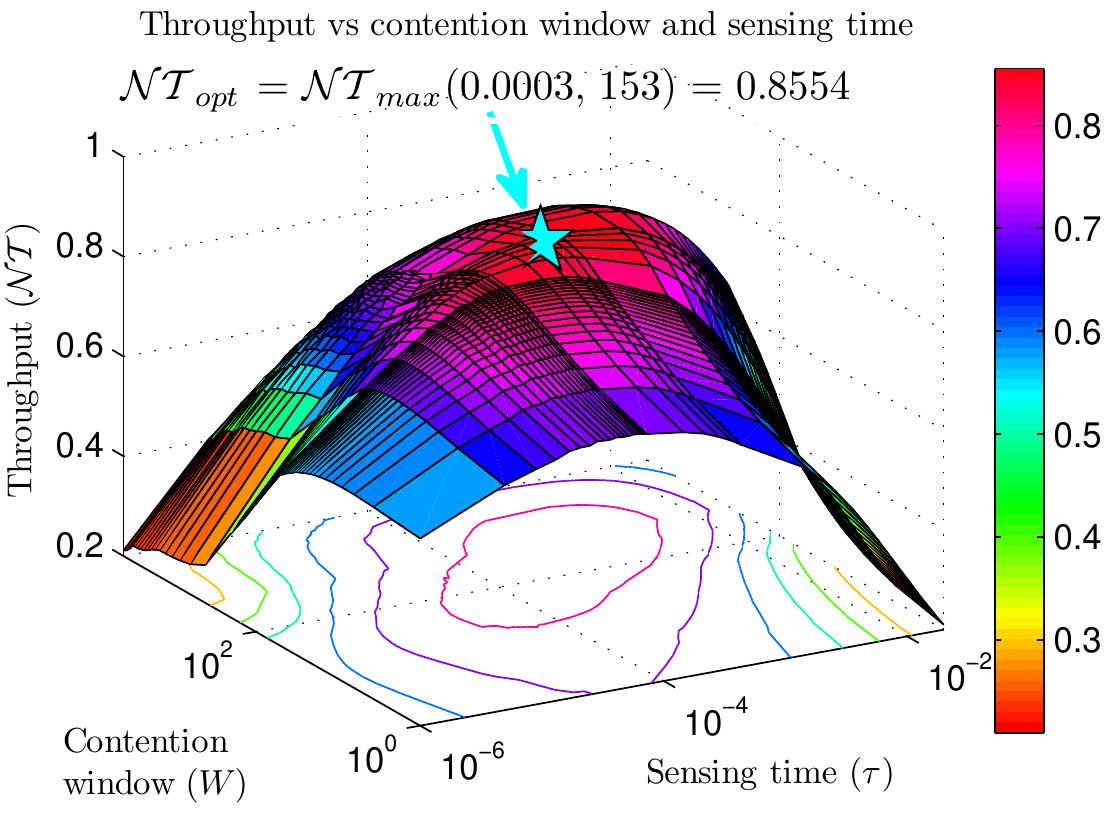}
\caption{Normalized throughput versus sensing time $\tau$ and contention window $W$ for $N = 15$, $m = 4$ and basic access mechanism.}
\label{Fig5_Chap3}
\end{figure}

To illustrate the joint effects of contention window $W$ and sensing time $\tau$, we show the normalized throughput $\mathcal{NT}$ 
versus $\tau$ and contention window $W$ for $N = 15$ and $m = 4$ in Fig.~\ref{Fig5_Chap3}.
We show the globally optimal parameters $(\phi^*,\tau^*)$ which maximize the normalized throughput $\mathcal{NT}$  of the proposed cognitive MAC protocol by a star symbol in this figure. 
This figure reveals that the performance gain due to optimal configuration of the proposed MAC protocol is very significant. 
Specifically, while the normalized throughput $\mathcal{NT}$ tends to be less sensitive to the contention window $W$, it decreases 
significantly when the sensing time $\tau$ deviates from the optimal value $\tau^*$.
Therefore, the proposed optimization approach would be very useful in achieving the largest throughput performance for the secondary
network.

\subsection{Performance of Multi-Channel MAC Protocol}
\label{Results_2_Chap3}

In this section, we present numerical results for the proposed multi-channel MAC protocol. Although, we analyze the homogeneous scenario in Section~\ref{MultipleChan_Chap3} for brevity,  we present simulation results for the heterogeneous settings in this subsection. 
The same parameters for the MAC protocol as in Section~\ref{Results_Chap3} are used. However, this model covers for the case in which each secondary link has multiple channels. In addition, some key parameters are chosen as follows. The SNRs of the signals from the primary user to secondary link $i$ (i.e., $SNR_p^{ij}$) are randomly chosen in the range of $\left[-15, -20 \right] dB$. The target detection probabilities $\mathcal{\bar P}_d^{ij} $ and the probabilities $\mathcal{P}^{ij} \left(\mathcal{H}_0\right)$ for channel $j$ at secondary link $i$ are randomly chosen in the intervals $\left[0.7, 0.9 \right] $ and $\left[0.7, 0.8 \right] $, respectively. Again the exponential backoff mechanism with the maximum backoff stage $m$ is employed to reduce collisions.

\begin{figure}[!t]
\centering
\includegraphics[width=90mm]{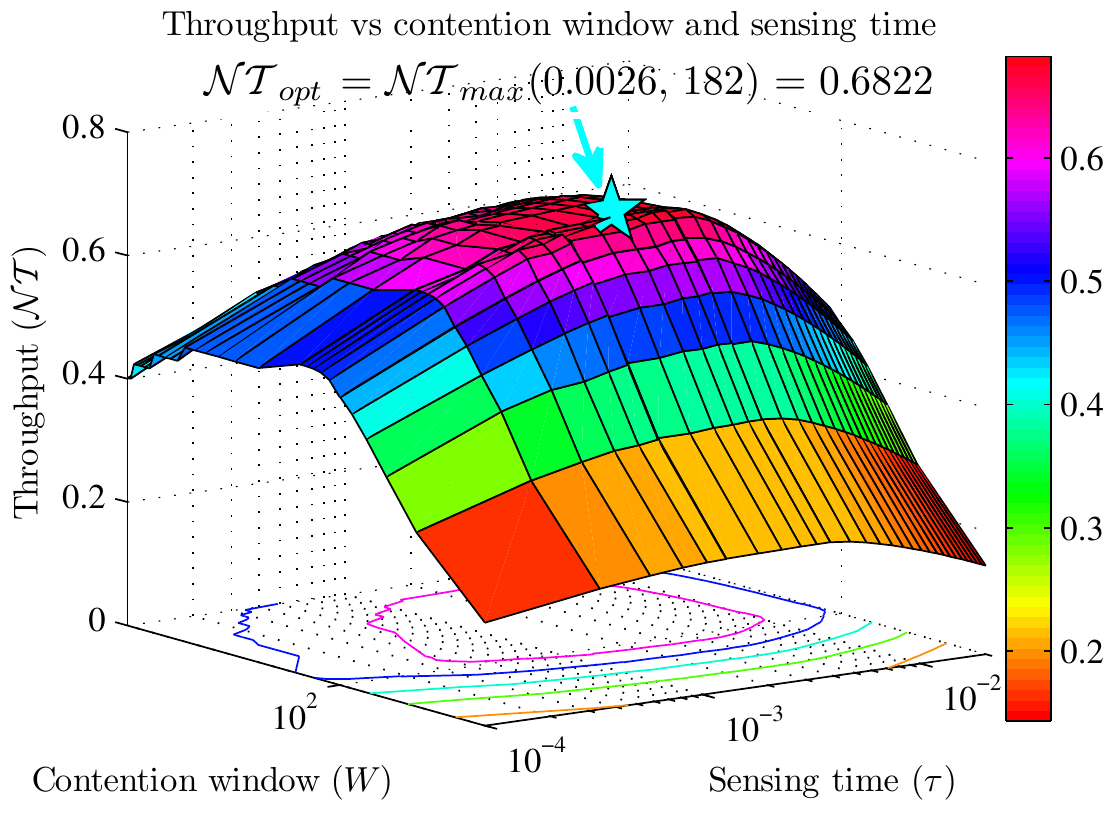}
\caption{Normalized throughput versus sensing time $\tau$ and contention window $W$ for $N = 10$, $m = 4$ , $M = 5$ and basic access mechanism.}
\label{Fig6_Chap3}
\end{figure}

In Fig.~\ref{Fig6_Chap3}, we illustrate the normalized throughput $\mathcal{NT}$ versus sensing times $\tau$ and contention windows $W$ for $N = 10$, $M = 5$ and $m = 4$ and the basic access mechanism. 
We show the optimal configuration $\left(\tau^*, W^*\right)$, which maximizes the normalized throughput $\mathcal{NT}$ of the proposed multichannel MAC protocol. 
Again it can be observed that the normalized throughput $\mathcal{NT}$ tends to be less sensitive to the contention window $W$ while it significantly decreases when the sensing time $\tau$ deviates from the optimal sensing time $\tau^*$.

\begin{table}
%\small
\footnotesize
\centering
\caption{Comparison between the normalized throughputs of basic access and RTS/CTS access}
\label{table_Chap3}
\begin{tabular}{|c|c|c|c|c|c|}
\hline 
\multicolumn{6}{|c|}{BASIC ACCESS ($N$,$M$,$m$) = (10,5,4)}\tabularnewline
\hline
\hline 
\multicolumn{2}{|c|}{} & \multicolumn{4}{c|}{$\tau(ms)$}\tabularnewline
\cline{3-6} 
\multicolumn{2}{|c|}{$\mathcal{NT}$} & 1 & 2.6 & 10 & 20\tabularnewline
\hline 
 & 16 & 0.4865 & 0.5545 & 0.5123 &  0.4677\tabularnewline
\cline{2-6} 
 & 64 & 0.5803  & 0.6488  & 0.6004 & 0.5366\tabularnewline
\cline{2-6} 
$W$ & 182 & 0.6053  & \textbf{\emph{\Large 0.6822}} & 0.6323 & 0.5594\tabularnewline
\cline{2-6} 
 & 512 & 0.6014  & 0.6736  & 0.6251 & 0.5567\tabularnewline
\cline{2-6} 
 & 1024 & 0.5744  & 0.6449  & 0.5982 & 0.5312\tabularnewline
\hline
\end{tabular}\begin{tabular}{|c|c|c|c|c|c|}
\hline 
\multicolumn{6}{|c|}{RTS/CTS ACCESS ($N$,$M$,$m$) = (10,5,4)}\tabularnewline
\hline
\hline 
\multicolumn{2}{|c|}{} & \multicolumn{4}{c|}{$\tau(ms)$}\tabularnewline
\cline{3-6} 
\multicolumn{2}{|c|}{$\mathcal{NT}$} & 1 & 2.6 & 10 & 20\tabularnewline
\hline 
 & 16 & 0.6029 & 0.6654 & 0.6236 & 0.5568\tabularnewline
\cline{2-6} 
 & 60 & 0.6022 & \textbf{\emph{\Large 0.6733}} & 0.6231 & 0.5568\tabularnewline
\cline{2-6} 
$W$ & 128 & 0.5954 & 0.6707 & 0.6175 & 0.5533\tabularnewline
\cline{2-6} 
 & 512 & 0.5737  & 0.6444 & 0.5982 & 0.5323\tabularnewline
\cline{2-6} 
 & 1024 & 0.5468 & 0.6134 & 0.5692 & 0.5059\tabularnewline
\hline
\end{tabular}

\begin{tabular}{|c|c|c|c|c|c|}
\hline 
\multicolumn{6}{|c|}{BASIC ACCESS ($N$,$M$,$m$) = (5,3,4)}\tabularnewline
\hline
\hline 
\multicolumn{2}{|c|}{} & \multicolumn{4}{c|}{$\tau(ms)$}\tabularnewline
\cline{3-6} 
\multicolumn{2}{|c|}{$\mathcal{NT}$} & 1 & 2.3 & 10 & 20\tabularnewline
\hline 
 & 16 & 0.5442  & 0.6172  & 0.5647 & 0.5079 \tabularnewline
\cline{2-6} 
 & 64 & 0.6015  & 0.6757  & 0.6302 & 0.5565\tabularnewline
\cline{2-6} 
$W$ & 100 & 0.6094  & \textbf{\emph{\Large 0.6841}}  & 0.6345 & 0.5665\tabularnewline
\cline{2-6} 
 & 512 & 0.5735  & 0.6443  & 0.5983 & 0.5324\tabularnewline
\cline{2-6} 
 & 1024 & 0.5210  & 0.5866  & 0.5447 & 0.4842\tabularnewline
\hline
\end{tabular}\begin{tabular}{|c|c|c|c|c|c|}
\hline 
\multicolumn{6}{|c|}{RTS/CTS ACCESS ($N$,$M$,$m$) = (5,3,4)}\tabularnewline
\hline
\hline 
\multicolumn{2}{|c|}{} & \multicolumn{4}{c|}{$\tau(ms)$}\tabularnewline
\cline{3-6} 
\multicolumn{2}{|c|}{$\mathcal{NT}$} & 1 & 2.5 & 10 & 20\tabularnewline
\hline 
 & 22 & 0.5972 & \textbf{\emph{\Large 0.6789}} & 0.6177 & 0.5529\tabularnewline
\cline{2-6} 
 & 64 & 0.5931 & 0.6674 & 0.6217 & 0.5483\tabularnewline
\cline{2-6} 
$W$ & 128 & 0.5876 & 0.6604 & 0.6131 & 0.5441 \tabularnewline
\cline{2-6} 
 & 512 & 0.5458 & 0.6128 & 0.5691 & 0.5057\tabularnewline
\cline{2-6} 
 & 1024 & 0.4965 & 0.5591 & 0.5189 & 0.4610\tabularnewline
\hline
\end{tabular}
\end{table} 

In order to study the joint effect of contention window $W$ and sensing time $\tau$ in greater details, we show the normalized throughput $\mathcal{NT}$ versus $W$ and $\tau$ in Table \ref{table_Chap3}. 
In this table, we consider both handshaking mechanisms, namely basic access and RTS/CTS access schemes.
Each set of results applies to a particular setting with certain number of secondary links $N$, number of channels $M$ and 
maximum backoff stage $m$. In particular, we will consider two settings, namely $\left(N,M,m\right) = \left(10, 5, 4\right)$ 
and $\left(N,M,m\right) = \left(5, 3, 4\right)$. 
Optimal normalized throughput is indicated by a bold number. It can be confirmed from this table that as
 $\left(\tau, W\right)$ deviate from the optimal $\left(\tau^*, W^*\right)$, the normalized throughput decreases significantly. 

This table also demonstrates potential effects of the number of secondary links $N$ on the network throughput and optimal configuration for the MAC protocols.
In particular, for secondary networks with the small number of secondary links, the probability of collision is lower than that for networks with the large number of secondary links. 
We consider the two scenarios corresponding to different combinations $\left(N, M, m\right)$. 
The first one which has a smaller number of secondary links $N$ indeed requires smaller contention window $W$ and maximum back-off stage $m$ to achieve the maximum throughput. 
Finally, it can be observed that for the same configuration of $\left(N, M, m\right) $, the basic access mechanism slightly outperforms the RTS/CTS access mechanism, while the RTS/CTS access mechanism can achieve the optimal normalized throughput at lower $W$ compared to the basic access mechanism.

%\vspace{10pt}
\section{Conclusion}
\label{conclusion_Chap3} 
In this paper, we have proposed MAC protocols for cognitive radio networks that explicitly take into account spectrum sensing performance.
Specifically, we have derived normalized throughput of the proposed MAC protocols and determined their optimal configuration for throughput 
maximization. These studies have been performed for both single and multiple channel scenarios subject to protection
constraints for primary receivers. Finally, we have presented numerical results to confirm important theoretical findings in the paper and to
show significant performance gains achieved by the optimal configuration for proposed MAC protocols.

%\appendices
\section{Appendices}

\subsection{Proof of Proposition 1}
\label{Proof_Chap3}

We start the proof by defining the following quantities: 
$\varphi^j := -\frac{ \left( \alpha^j + \sqrt{\tau f_s}\gamma^j \right)^2} {2}$ 
and ${c_{{n_0}}} := \frac{{{{\mathcal{P}}_s}{{\mathcal{P}}_t}PS}}{{{T}}}$. 
Taking the derivative of $\mathcal{NT}$ versus $\tau$, we have
\begin{equation}
\label{eq16}
\begin{array}{l}
 \!\!\!\!\frac{\partial \mathcal{NT}}{\partial \tau } = \sum\limits_{n_0=1}^N {c_{n_0}\sum\limits_{k = 1}^{C_N^{n_0}} {} }  \\ 
 \!\!\!\!\left\{\! \!\!\!\begin{array}{l}
 \left( \frac{ - 1}{\bar T_{sd}} \right)\prod\limits_{i \in \mathcal{S}_k} \!\!{\mathcal{P}_{idle}^i}\!\! \prod\limits_{j \in \mathcal{S} \backslash \mathcal{S}_k} \!\!{\mathcal{P}_{busy}^j}\!\!  + \left\lfloor {\frac{T - \tau }{\bar T_{sd}}} \right\rfloor \sqrt {\frac{f_s}{8\pi\tau }}  \times  \\ 
 \left[ \!\!\!\begin{array}{l}
 \sum\limits_{i \in \mathcal{S}_k} {\gamma ^i \exp \left( \varphi ^i \right) \mathcal{P}^i\left(\mathcal{H}_0 \right) \!\!\prod\limits_{l \in \mathcal{S}_k \backslash i} \!\!{\mathcal{P}_{idle}^l} \!\!\prod\limits_{j \in \mathcal{S} \backslash \mathcal{S}_k} \!\!{\mathcal{P}_{busy}^j} }  \\ 
  - \!\! \!\!\sum\limits_{j \in \mathcal{S} \backslash \mathcal{S}_k} \!\!\!\! {\gamma ^j \exp \left( \varphi ^j \right) \mathcal{P}^j\left(\mathcal{H}_0 \right) \!\!\!\!\!\prod\limits_{l \in \mathcal{S} \backslash \mathcal{S}_k \backslash j} \!\!\!\!{\mathcal{P}_{busy}^l}\!\! \prod\limits_{i \in  \mathcal{S}_k} \!\!{\mathcal{P}_{idle}^i} }  \\ 
 \end{array} \!\!\!\right] \\ 
 \end{array} \!\!\!\!\right\} \\ 
 \end{array}\!\!\!.
\end{equation}
From this we have
\begin{equation}
\label{eq17}
\mathop {\lim }\limits_{\tau  \to {T}}\!\!\!\! \frac{{\partial {\mathcal{NT}}}}{{\partial \tau }} \!\!=\!\! \sum\limits_{{n_0}=1}^N \!\!{{c_{{n_0}}}\sum\limits_{k = 1}^{C_N^{n_0}} \!\!{\left(\! {\frac{{ - 1}}{{{\bar T_{sd}}}}} \!\right)\!\!\!\prod\limits_{i \in {{\mathcal{S}}_k}} \!\!\!{{\mathcal{P}}_{idle}^i}\!\!\! \!\!\prod\limits_{j \in {\mathcal{S}}\backslash {{\mathcal{S}}_k}} \!\!\!\!\!{{\mathcal{P}}_{busy}^j} }  < 0}.
\end{equation}
Now, let us define the following quantity 
\begin{equation}
\label{eq18}
{\mathcal{K}_\tau } \!\! \buildrel \Delta \over = \!\! \sum\limits_{{n_0}=1}^N \!\!\!{c_{n_0}\!\!\sum\limits_{k = 1}^{C_N^{n_0}}\!\! {\left[\!\!\! \begin{array}{l}
 \sum\limits_{i \in {{\mathcal{S}}_k}} \!\!{{\gamma ^i}\exp \left( {{\varphi ^i}} \right)\!\! \mathcal{P}^i\!\left(\mathcal{H}_0 \right) \!\!\!\!\!\prod\limits_{l \in {{\mathcal{S}}_k}\backslash i} \!\!\!\!{{\mathcal{P}}_{idle}^l} \!\!\!\!\prod\limits_{j \in {\mathcal{S}}\backslash {{\mathcal{S}}_k}} \!\!\!\! {\mathcal{P}_{busy}^j}   }  \\ 
 -\!\!\!\!\!\!\sum\limits_{j \in {\mathcal{S}}\backslash {{\mathcal{S}}_k}}\!\! \!\!\!\!{{\gamma ^j}\exp \left( {{\varphi ^j}} \right) \!\!\mathcal{P}^j\!\left(\mathcal{H}_0 \right) \!\!\!\!\!\!\!\!\prod\limits_{l \in {\mathcal{S}}\backslash {{\mathcal{S}}_k}\backslash j} \!\!\!\!\!\!{{\mathcal{P}}_{busy}^l} \!\!\!\prod\limits_{i \in  {{\mathcal{S}}_k}} \!\!\!{{\mathcal{P}}_{idle}^i} }  \\ 
 \end{array} \!\!\!\!\right]} }.
\end{equation}

Then, it can be shown that ${\mathcal{K}_\tau }>0$ as being explained in the following. 
First, it can be verified that the term $c_{n_0} $ is almost a constant for different $n_0$. Therefore, to highlight intuition behind the 
underlying property (i.e., ${\mathcal{K}_\tau }>0$), we substitute $\mathcal{K} = c_{n_0}$ into the above equation. Then, ${\mathcal{K}_\tau }$ in (\ref{eq18}) reduces to
\beqn \label{Ktau1ch1}
{\mathcal{K}_\tau } \!\!=\!\! {\mathcal{K}_a}\!\!\sum\limits_{{n_0} = 1}^N \!\!{C_N^{n_0}\left(\! {{n_0}\mathcal{P}_{idle}^{{n_0} - 1}\mathcal{P}_{busy}^{N - {n_0}}\!-\!\left( {N \!-\! {n_0}} \right)\mathcal{P}_{busy}^{N - {n_0} - 1}} \!\right)} ,
\eeqn
where ${\mathcal{K}_a} = \mathcal{K}\gamma \exp \left( \varphi  \right)\mathcal{P}\left( {{\mathcal{H}_0}} \right)$. Let define the following quantities $x =  \mathcal{P}_{busy}$, $x \in {\mathbb{R}_x} \buildrel \Delta \over = \left[ {{\mathcal{P}_d}\mathcal{P}\left( {{\mathcal{H}_1}} \right),\mathcal{P}\left( {{\mathcal{H}_0}} \right) + {\mathcal{P}_d}\mathcal{P}\left( {{\mathcal{H}_1}} \right)} \right]$. After some manipulations, we have
\beqn \label{Ktau1ch}
{\mathcal{K}_\tau } = {\mathcal{K}_a}\sum\limits_{{n_0} = 1}^N {f\left( x \right)\left( {\frac{{{n_0}}}{{x\left( {1 - x} \right)}} - \frac{N}{x}} \right)} ,
\eeqn
where $f\left( x \right) = C_N^{n_0} {\left( {1 - x} \right)^{{n_0}}}{x^{N - {n_0}}}$ is the binomial mass function \cite{Kris06} with $p = 1 - x$ and $q = x$. Because the total probabilities and the mean of this binomial distribution are 1 and $Np = N\left(1 - x\right)$, respectively, we have 
\beqn \label{bioCDF}
\sum\limits_{{n_0} = 0}^N {f\left( x \right)}  = 1,
\eeqn
\beqn \label{biomean}
\sum\limits_{{n_0} = 0}^N {{n_0}f\left( x \right)}  = N\left( {1 - x} \right).
\eeqn
It can be observed that in (\ref{Ktau1ch}), the element corresponding to $n_0 =0$ is missing.  Apply the results in
 (\ref{bioCDF}) and (\ref{biomean}) to (\ref{Ktau1ch}) we have
\beqn \label{Ktau1ch2}
{\mathcal{K}_\tau } = {\mathcal{K}_a} N x^{N-1} > 0,  \: \forall x.
\eeqn
Therefore, we have
\begin{equation}
\label{eq19}
\mathop {\lim }\limits_{\tau  \to 0} \frac{{\partial {\mathcal{NT}}}}{{\partial \tau }} =  + \infty.
\end{equation}
Hence, we have completed the proof for first two properties of Proposition 1.

In order to prove the third property, let us find the solution of $\frac{{\partial {\mathcal{NT}}}}{{\partial \tau }} = 0$. 
After some simple manipulations and using the properties of the binomial distribution, this equation reduces to
\beqn \label{hg1ch}
h\left(\tau \right) = g\left(\tau \right),
\eeqn
where 
\beq
g\left( \tau  \right) = {\left( {\alpha  + \gamma \sqrt {{f_s}\tau } } \right)^2},
\eeq
and
\beq
 h\left( \tau  \right) = 2\log \left( {\mathcal{P}\left( {{\mathcal{H}_0}} \right)\gamma \sqrt {\frac{{{f_s}}}{{8\pi }}} \frac{{T - \tau }}{{\sqrt \tau  }}} \right) + {h_1}\left( x \right)
 \eeq 
where 
${h_1}\left( x \right) = 2\log \frac{{{\mathcal{K}_\tau }/{\mathcal{K}_a}}}{{\sum\limits_{{n_0} = 1}^N {C_N^{n_0} f\left( x \right)} }} = 2\log \frac{{N{x^{N - 1}}}}{{1 - {x^N}}}$. 

To prove the third property, we will show that $h\left(\tau \right)$ intersects $g\left(\tau \right)$ only once.
We first state one important property of $h\left(\tau \right)$ in the following lemma.

\vspace{0.2cm}
\noindent
\textbf{Lemma 1:} $h\left(\tau \right)$ is an decreasing function.

\begin{proof}
Taking the first derivative of $h(.)$, we have 
\beqn \label{partialh1ch} 
\frac{{\partial h}}{{\partial \tau }} = \frac{{ - 1}}{\tau } - \frac{2}{{T - \tau }} + \frac{{\partial {h_1}}}{{\partial x}}\frac{{\partial x}}{{\partial \tau }}. 
\eeqn
We now derive $\frac{{\partial x}}{{\partial \tau }}$ and $\frac{{\partial {h_1}}}{{\partial x}}$  as follows:
\beqn \label{partialx1ch} 
\frac{{\partial x}}{{\partial \tau }} =  - \mathcal{P}\left( {{\mathcal{H}_0}} \right)\gamma \sqrt {\frac{{{f_s}}}{{8\pi \tau }}} \exp \left( { - \frac{{{{\left( {\alpha  + \gamma \sqrt {{f_s}\tau } } \right)}^2}}}{2}} \right) < 0,
\eeqn
\beqn \label{partialh11ch} 
\frac{{\partial {h_1}}}{{\partial x}} = 2 \frac{{N - 1 + {x^N}}}{{x\left( {1 - {x^N}} \right)}}>0.
\eeqn
Hence, $\frac{{\partial {h_1}}}{{\partial x}}\frac{{\partial x}}{{\partial \tau }}<0$. Using this result in (\ref{partialh1ch}), we have $\frac{{\partial h}}{{\partial \tau }} < 0$. Therefore, we can conclude that $h\left(\tau \right)$ is monotonically decreasing.
\end{proof}

We now consider function $g\left(\tau\right)$. Take the derivative of $g\left(\tau\right)$, we have
\beq \label{dgfun}
\frac{{\partial g}}{{\partial \tau }} = \left( \alpha + \gamma\sqrt{f_s \tau} \right) \frac{ \gamma\sqrt{f_s} }{\sqrt{\tau}}.
\eeq
Therefore, the monotonicity property of $g\left(\tau\right)$ only depends on $y = \alpha + \gamma\sqrt{f_s \tau}$.
Properties 1 and 2 imply that there must be at least one intersection between $h\left(\tau \right)$ and $g\left(\tau\right)$.
We now prove that there is indeed a unique intersection. To proceed, we consider two different regions for $\tau$ as follows:

${\mathbf{\Omega}_1} = \left\{ {\tau \left| {\alpha  + \gamma \sqrt {{f_s}\tau }  < 0,\,  \tau  \le T} \right.} \right\}
= \left\{  0< \tau < \frac{\alpha^2 }{\gamma^2 f_s} \right\}$

and 

${\mathbf{\Omega}_2} = \left\{ {\tau \left| {\alpha  + \gamma \sqrt {{f_s}\tau }  \geq 0,\,  \tau  \le T} \right.} \right\}
= \left\{  \frac{\alpha^2 }{\gamma^2 f_s} \leq \tau \leq T \right\}$.

From the definitions of these two regions, we have $g\left(\tau\right)$ decreases in ${\mathbf{\Omega}_1}$ and increases in ${\mathbf{\Omega}_2}$.
To show that there is a unique intersection between $h\left(\tau \right)$ and $g\left(\tau\right)$, we prove the following.

\vspace{0.2cm}
\noindent
\textbf{Lemma 2:} The following statements are correct:
\bnu

\item
If there is an intersection between $h\left(\tau \right)$ and $g\left(\tau\right)$ in ${\mathbf{\Omega}_2}$ then it is the only intersection in this region and there is no
intersection in ${\mathbf{\Omega}_1}$.

\item If there is an intersection between $h\left(\tau \right)$ and $g\left(\tau\right)$ in ${\mathbf{\Omega}_1}$ then it is the only intersection in this region and there is no
intersection in ${\mathbf{\Omega}_2}$.

\enu

\begin{proof}
We now prove the first statement. Recall that $g\left(\tau\right)$ monotonically increases in ${\mathbf{\Omega}_2}$; therefore, 
$g\left(\tau\right)$ and $h\left(\tau\right)$ can intersect at most once in this region (because $h\left(\tau\right)$ decreases). 
In addition, $g\left(\tau\right)$ and $h\left(\tau\right)$ cannot intersection in ${\mathbf{\Omega}_1}$ for this case if
we can prove that $\frac{{\partial h}}{{\partial \tau }} < \frac{{\partial g}}{{\partial \tau }}$. This is because both
functions decrease in ${\mathbf{\Omega}_1}$. We will prove that $\frac{{\partial h}}{{\partial \tau }} < \frac{{\partial g}}{{\partial \tau }}$
in lemma 3 after this proof.

We now prove the second statement of lemma 2. Recall that we have $\frac{{\partial h}}{{\partial \tau }} < \frac{{\partial g}}{{\partial \tau }}$.
Therefore, there is at most one intersection between $g\left(\tau\right)$ and $h\left(\tau\right)$ in ${\mathbf{\Omega}_1}$. In addition,
it is clear that there cannot be any intersection between these two functions in ${\mathbf{\Omega}_2}$ for this case.
\end{proof}

\vspace{0.2cm}
\noindent
\textbf{Lemma 3:} We have $\frac{{\partial h}}{{\partial \tau }} < \frac{{\partial g}}{{\partial \tau }}$.

\begin{proof} 
From (\ref{partialh1ch}), we can see that lemma 3 holds if we can prove the following stronger result
\beqn \label{partialh1g1ch}
\frac{-1}{\tau}+\frac{{\partial h_1}}{{\partial \tau }} < \frac{{\partial g}}{{\partial \tau }},
\eeqn
where $\frac{{\partial {h_1}}}{{\partial \tau }} = \frac{{\partial {h_1}}}{{\partial x}}\frac{{\partial x}}{{\partial \tau }}$, $\frac{{\partial x}}{{\partial \tau }}$ is derived in (\ref{partialx1ch}), $\frac{{\partial {h_1}}}{{\partial x}}$ is derived in (\ref{partialh11ch})
and  $\frac{{\partial g}}{{\partial \tau }}$ is given in (\ref{dgfun}).

To prove (\ref{partialh1g1ch}), we will prove the following
\beqn \label{partialh1g11chalter}
 \!\!- \!\frac{1}{\tau} \!+ \!
 \frac{y\mathcal{P}\left( {\mathcal{H}_0} \right)\gamma \sqrt {\frac{{{f_s}}}{\tau }} }{{\mathcal{P}\left( {{\mathcal{H}_0}} \right) \!+\! \sqrt {2\pi } \mathcal{P}\left( {{\mathcal{H}_1}} \right)\!\left( {1 - {\mathcal{\bar P}_d}} \right)\! \left(  - y \right)\!\exp\! \left( \frac{y^2}{2} \right)}} 
 \!< \!\frac{\partial g} {\partial \tau },
\eeqn
where $y = \left(\alpha + \gamma \sqrt{f_s \tau} < 0\right)$.
Then, we show that
\beq \label{eq50}
\!\!\!\!\!\!\frac{{\partial h_1}}{{\partial \tau }} < \frac{{y\mathcal{P}\left( {{\mathcal{H}_0}} \right)\gamma \sqrt {\frac{{{f_s}}}{\tau }} }}{{\mathcal{P}\left( {{\mathcal{H}_0}} \right) + \sqrt {2\pi } \mathcal{P}\left( {{\mathcal{H}_1}} \right)\left( {1 - {\mathcal{\bar P}_d}} \right)\left( { - y} \right)\exp \left( {\frac{{{y^2}}}{2}} \right)}}.
\eeq
Therefore, the result in (\ref{partialh1g1ch}) will hold. Let us prove (\ref{eq50}) first.
First, let us prove the following 
\beqn \label{h2lower1cha}
\frac{{\partial {h_1}}}{{\partial x}} > \frac{2}{{1 - x}}.
\eeqn
Using the result in $\frac{{\partial {h_1}}}{{\partial x}}$ from (\ref{partialh11ch}), (\ref{h2lower1cha}) is equivalent to
\beqn \label{h2lower1ch}
2\frac{{N - 1 + {x^N}}}{{x\left( {1 - {x^N}} \right)}} > \frac{2}{{1 - x}},
\eeqn
After some manipulations, we get
\beqn \label{h2lower1cha1}
\left( {1 - x} \right)\left( {N - 1 - \left( {x + {x^2} +  \cdots  + {x^{N - 1}}} \right)} \right) > 0.
\eeqn
It can be observed that $0<x<1$ and  $0<x^i<1$, $i \in \left[1,N-1\right]$. So $N-1-\left(x+x^2+\cdots+x^{\left(N-1\right)}\right) >0$; hence (\ref{h2lower1cha1}) holds. Therefore, we have completed the proof for (\ref{h2lower1cha}).

We now show that the following inequality holds
\beqn \label{onex1ch}
\frac{2}{{1 - x}} > \frac{{2\sqrt {2\pi } \left( { - y} \right)\exp \left( {\frac{{{y^2}}}{2}} \right)}}{{\mathcal{P}\left( {{\mathcal{H}_0}} \right) + \sqrt {2\pi } \mathcal{P}\left( {{\mathcal{H}_1}} \right)\left( {1 - {\mathcal{\bar P}_d}} \right)\left( { - y} \right)\exp \left( {\frac{{{y^2}}}{2}} \right)}}.
\eeqn
This can be proved as follows.
In \cite{Geor07}, it has been shown that $\mathcal{Q}\left(t\right)$ with $t > 0$ satisfies  
\beqn \label{Qfunction}
\frac{1}{{\mathcal{Q}\left( t \right)}} > \sqrt {2\pi } t\exp \left( {\frac{{{t^2}}}{2}} \right).
\eeqn
Apply this result to $\mathcal{P}_f = \mathcal{Q}\left(y\right) = 1 - \mathcal{Q}\left(-y\right)$ with $y = \left( {\alpha  + \gamma \sqrt {{f_s}\tau } } \right) < 0$ 
we have
\beqn \label{Qfunction1ch}
\frac{1}{{1 - {P_f}}} > \sqrt {2\pi } \left( { - y} \right)\exp \left( {\frac{{{y^2}}}{2}} \right).
\eeqn
After some manipulations, we obtain
\beqn \label{Pf1ch}
{\mathcal{P}_f} > 1 - \frac{1}{{\sqrt {2\pi } \left( { - y} \right)\exp \left( {\frac{{{y^2}}}{2}} \right)}}.
\eeqn
Recall that we have defined $x = {\mathcal{P}_f}\mathcal{P}\left( {{\mathcal{H}_0}} \right) + {\mathcal{\bar P}_d}\mathcal{P}\left( {{\mathcal{H}_1}} \right)$. Using the result in (\ref{Pf1ch}), we can obtain the  lower bound of $\frac{2}{1-x}$ given in (\ref{onex1ch}).
Using the results in (\ref{h2lower1cha}) and (\ref{onex1ch}), and the fact that $\frac{{\partial x}}{{\partial \tau }} < 0$, we 
finally complete the proof for (\ref{eq50}).

To complete the proof of the lemma, we need to prove that (\ref{partialh1g11chalter}) holds.
Substitute $\frac{{\partial g}}{{\partial \tau }}$ from (\ref{dgfun}) to (\ref{partialh1g11chalter}) and make some further manipulations, we have
\beqn \label{finalproof1ch}
\!\!\frac{- 1}{ y\left( y \!-\! \alpha  \right)} \!> \!1\! -\! \frac{y\mathcal{P}\left( \mathcal{H}_0 \right)\gamma \sqrt {\frac{f_s}{\tau }} }{\!\mathcal{P}\left(\! \mathcal{H}_0 \!\right) \!+\! \sqrt {2\pi } \mathcal{P}\!\left(\! \mathcal{H}_1 \!\right)\left( 1 \!-\! \mathcal{\bar P}_d \right)\!\left( - y \right)\exp \!\left( \!\frac{y^2}{2} \!\right)}.
\eeqn
Let us consider the LHS of (\ref{finalproof1ch}).  We have $0<y-\alpha={\gamma \sqrt {{f_s}\tau } }<-\alpha$; therefore, we have
 $0<-y<-\alpha$. Apply the Cauchy-Schwarz inequality to $-y$ and $y - \alpha$, we have the following
\beqn \label{leftside1ch} 
0 <- y\left( {y - \alpha } \right) \leq {\left( {\frac{{ - y + y - \alpha }}{2}} \right)^2} = \frac{{{\alpha ^2}}}{4}.
\eeqn
Hence
\beqn \label{leftside1ch1} 
\frac{1}{{ - y\left( {y - \alpha } \right)}} \ge \frac{4}{{{\alpha ^2}}} = \frac{4}{{\left( {2\gamma  + 1} \right){{\left( {{\mathcal{Q}^{ - 1}}\left( {{\mathcal{\bar P}_d}} \right)} \right)}^2}}} > 1.
\eeqn
It can be observed that the RHS of (\ref{finalproof1ch}) is less than 1. Therefore, (\ref{finalproof1ch}) holds, which
implies that (\ref{partialh1g11chalter}) and (\ref{partialh1g1ch}) also hold. 
\end{proof}

Finally, the last property holds because because $\Pr \left( {n = {n_0}} \right)<1$ and conditional throughput are all bounded from above. 
Therefore, we have completed the proof of Proposition 1.

\subsection{Proof of Proposition 2}
\label{Proof_Chap3_P2}

To prove the properties stated in  Proposition 2, we first find the derivative of ${\mathcal{\widetilde{NT}}\left( \tau  \right)}$. 
Again, it can be verified that $\frac{{{\mathcal{P}_t}{\mathcal{P}_s}PS}}{{{T}}}$ is almost a constant for different $n_0$. To demonstrate 
the proof for the proposition, we substitute this term as a constant value, denoted as $\mathcal{K}$, in the throughput formula. In addition, for large $T$, $\left\lfloor {\frac{{{T} - \tau }}{{{{\bar T}_{sd}}}}} \right\rfloor$ is very close to ${\frac{{{T} - \tau }}{{{{\bar T}_{sd}}}}}$. Therefore,  $\mathcal{\widetilde{NT}}$
can be accurately approximated as
\beqn \label{NT_tau}
\mathcal{\widetilde{NT}} \!\!\left( \tau  \right)\!\! =\!\! \sum\limits_{n_0 = 1}^N {\!\!\mathcal{K} C_N^{n_0} \left( T - \tau  \right)\!\! \left( 1 - x^M \right)^{n_0}\!\!x^{M\left( N - n_0 \right)}\left( 1 - x \right)},
\eeqn
where
$\mathcal{K} = \frac{{{P_t}{P_s}PS}}{{{T}}}$, and $x = P_{busy}$. Now, let us define the following function
\beqn \label{fx}
f'\left( x \right) = {\left( {1 - {x^M}} \right)^{{n_0}}}{x^{M\left( {N - {n_0}} \right)}}\left( {1 - x} \right).
\eeqn
Then, we have
\beqn \label{partialfx}
\frac{{\partial f'}}{{\partial x}} = f'\left( x \right)\left[ {\frac{{ - 1}}{{1 - x}} - \frac{{M{n_0}}}{{1 - {x^M}}}{x^{M - 1}} + \frac{{M\left( {N - {n_0}} \right)}}{x}} \right],
\eeqn
and $\frac{{\partial x}}{{\partial \tau }} $ is the same as (\ref{partialx1ch}).
Hence, the first derivation of ${\mathcal{\widetilde{NT}}\left( \tau  \right)}$ can be written as
\beqn \label{deriNT_tau} 
\begin{array}{l}
\!\!\!\! \frac{\partial \mathcal{\widetilde{NT}}\left( \tau  \right)}{\partial \tau } = \sum\limits_{n_0 = 1}^N {\mathcal{K}C_N^{n_0}\left[  - f'\left( x \right) + \left( T - \tau  \right)\frac{\partial f'}{\partial x}\frac{\partial x}{\partial \tau } \right]}  \\ 
 \,\,\,\,\,\,\,\,\,\,\,\,\,\,\,\,\,\,\,\,\, = \sum\limits_{n_0 = 1}^N {\mathcal{K}C_N^{n_0}f'\left( x \right)}  \times  \\ 
 \,\,\,\,\,\,\,\,\,\,\,\,\,\,\,\,\,\,\,\left[ \!\!\begin{array}{l}
 \left( T - \tau  \right)\left[ \frac{1}{1 - x} + \frac{M n_0}{1 - x^M}{x^{M - 1}} - \frac{M\left( N - n_0 \right)}{x} \right] \\ 
  \times \mathcal{P}\left( \mathcal{H}_0 \right)\gamma \sqrt {\frac{f_s}{8\pi \tau }} \exp \left( { - \frac{\left( \alpha  + \gamma \sqrt {f_s \tau }  \right)^2}{2}} \right) - 1 \\ 
 \end{array} \!\!\right] \\ 
 \end{array}.
\eeqn 
From (\ref{Pbus1_Chap3}), the range of $x$, namely $\mathbb{R}_x$ can be expressed as $\left[ {{\mathcal{P}_d}\mathcal{P}\left( {{\mathcal{H}_1}} \right),\,\mathcal{P}\left( {{\mathcal{H}_0}} \right) + {\mathcal{P}_d}\mathcal{P}\left( {{\mathcal{H}_1}} \right)} \right]$.
Now, it can be observed that
\beq
\mathop {\lim \,}\limits_{\tau  \to T} \frac{\partial \mathcal{\widetilde{NT}}\left( \tau  \right)}{\partial \tau } =  - \sum\limits_{n_0 = 1}^N {\mathcal{K}C_N^{n_0}f'\left( x \right)}  < 0.
\eeq
Therefore, the second property of Proposition 2 holds.

Now, let us define the following quantity
\beqn \label{Kcons}
\mathcal{K'}_\tau  \!=\! \sum\limits_{n_0 = 1}^N {\!\!C_N^{n_0} \! f'\left( x \right)\!\!\left[ \frac{1}{1 - x} \!+\! \frac{M n_0 x^{M - 1}}{1 - x^M}  \!- \!\frac{M\!\left( {N \!- {n_0}} \right)}{x} \right]} .
\eeqn
Then, it can be seen that $\mathop {\lim \,}\limits_{\tau  \to 0} \frac{{\partial \mathcal{\widetilde{NT}}\left( \tau  \right)}}{{\partial \tau }} =  + \infty  > 0$ if ${\mathcal{K'}_\tau } > 0,\,\forall M,\,N,\,x \in {\mathbb{R}_x}$. This last property is stated and proved in the following lemma.

\vspace{0.2cm}
\noindent
\textbf{Lemma 4:}
\label{Prop3}
${\mathcal{K'}_\tau } > 0,\,\forall M,\,N,\,x \in {\mathbb{R}_x}$.
\begin{proof} 
Making some manipulations to (\ref{Kcons}), we have 
\beqn \label{Kcons1}
\begin{array}{l}
 \mathcal{K'}_\tau  = \left( 1 - \frac{\left( 1 - x \right)M}{x} \right)\!\! \sum\limits_{{n_0} = 1}^N \!\!{C_N^{n_0} \left( 1 - x^M \right)^{n_0}  x^{M\left( N - n_0 \right)}}  \\ 
 \,\,\,\,\,\,\,\,\,\,\,\,  + \frac{{M\left( {1 - x} \right)}}{{x\left( {1 - {x^M}} \right)}}\sum\limits_{{n_0} = 1}^N {C_N^{n_0} n_0 \left( 1 - x^M \right)^{n_0} x^{M\left( N - n_0 \right)}}.  \\ 
 \end{array}
\eeqn
It can be observed that $\sum\limits_{n_0 = 1}^N {C_N^{n_0} \left( 1 - x^M \right)^{n_0} x^{M\left( N - n_0 \right)}} $ and $\sum\limits_{n_0 = 1}^N {C_N^{n_0}n_0 \left( 1 - x^M \right)^{n_0} x^{M\left( N - n_0 \right)}} $ represent a cumulative distribution
 function (CDF) and the mean of a binomial distribution \cite{Kris06} with parameter $p$, respectively  missing the term corresponding
  to $n_0 =0$ where $p = 1 - x^M$. Note that the CDF and mean of such a distribution are 1 and 
  $Np = N\left(1-x^M\right)$, respectively. Hence, (\ref{Kcons1}) can be rewritten as
\beqn \label{Kcons2}
\mathcal{K'}_\tau \! = \!\left(\! 1 \!- \!\frac{\left( 1\! - \!x \right)\!M}{x} \right)\!\left( 1 \!-\! x^{MN} \!\right) \!+\! \frac{M\left( 1 \!-\! x \right)}{x\left( 1 \!-\! x^M \right)}N\left( 1\! -\! x^M \right).
\eeqn
After some manipulations, we have
\beqn \label{Kcons3}
{\mathcal{K'}_\tau } = 1 - {x^{MN}} + MN{x^{MN - 1}}\left( {1 - x} \right) > 0, \: {\forall x}.
\eeqn
Therefore, we have completed the proof.
\end{proof} 

Hence, the first property of Proposition 1 also holds.

To prove the third property, let us consider the following equation $\frac{{\partial \mathcal{\widetilde{NT}}\left( \tau  \right)}}{{\partial \tau }} = 0$. After some manipulations, we have the following equivalent equation
\beqn \label{gh}
g\left(\tau \right) = h' \left( \tau \right),
\eeqn
where
\beqn \label{gtau}
g\left( \tau  \right) = {\left( {\alpha  + \gamma \sqrt {{f_s}\tau } } \right)^2},
\eeqn 
\beqn \label{hx}
h'\left( \tau  \right) = 2\log \left( {\mathcal{P}\left( {\mathcal{H}_0} \right)\gamma \sqrt {\frac{f_s}{8\pi }} \frac{T - \tau }{\sqrt \tau  }} \right) + {h'_1}\left( x \right),
\eeqn 
\beqn \label{h1x}
h'_1 \left( x \right) = 2\log \frac{\mathcal{K'}_\tau }{\sum\limits_{n_0 = 1}^N {C_N^{n_0} f'\left( x \right)} },
\eeqn
$K'_{\tau}$ is given in (\ref{Kcons}). We have the following result for $h'\left( \tau  \right)$.

\vspace{0.2cm}
\textbf{Lemma 5:} $h'\left( \tau  \right)$ monotonically decreases in $\tau$.

\begin{proof}
The derivative of  $h'\left( \tau  \right)$ can be written as
\beqn \label{heq} 
\frac{{\partial h'}}{{\partial \tau }} = \frac{{ - 1}}{\tau } - \frac{2}{{T - \tau }} + \frac{{\partial {h'_1}}}{{\partial \tau }}.
\eeqn

In the following, we will show that $\frac{{\partial {h'_1}}}{{\partial x}} > 0$ for all $x \in {\mathbb{R}_x}$, all $M$ and $N$, and $\frac{{\partial x}}{{\partial \tau }} < 0$. Hence $\,\frac{{\partial {h'_1}}}{{\partial \tau }} = \frac{{\partial {h'_1}}}{{\partial x}}\frac{{\partial x}}{{\partial \tau }} < 0$. From this, we have $\frac{{\partial h'}}{{\partial \tau }}<0$; therefore, the property stated in lemma 5 holds.

We now show that $\frac{{\partial {h'_1}}}{{\partial x}} > 0$ for all $x \in {\mathbb{R}_x}$, all $M$ and $N$.
Substitute $\mathcal{K'}_\tau$ in (\ref{Kcons3}) to (\ref{h1x}) and exploit the property of the CDF of the binomial distribution function, we have
\beqn \label{h1x1}
\begin{array}{l}
 {h'_1}\left( x \right) = 2\log \frac{1 - x^{MN} + MN x^{MN - 1}\left( 1 - x \right)}{\left( 1 - x \right)\sum\limits_{{n_0} = 1}^N {C_N^{n_0} \left( 1 - x^M \right)^{n_0} x^{M\left( N - n_0 \right)}} } \\ 
 \,\,\,\,\,\,\,\,\,\,\,\,\, = 2\log \frac{1 - x^{MN} + MNx^{MN - 1} \left( 1 - x \right)}{\left( 1 - x \right)\left( 1 - x^{MN} \right)} \\ 
 \end{array}.
\eeqn

Taking the first derivative of  ${h'_1}\left( x \right)$ and performing some manipulations, we obtain
\beqn \label{h1x2} 
\frac{{\partial {h'_1}}}{{\partial x}} = 2\frac{{\left( \begin{array}{l}
 r\left( {r - 1} \right){x^{r - 2}}{\left( {1 - x} \right)^2}\left( {1 - {x^r}} \right) \\ 
  + {\left( {1 - {x^r}} \right)^2} + {r^2}{x^{2\left( {r - 1} \right)}}{\left( {1 - x} \right)^2} \\ 
 \end{array} \right)}}{{\left( {1 - {x^r} + r{x^{\left( {r - 1} \right)}}\left( {1 - x} \right)} \right)\left( {1 - x} \right)\left( {1 - {x^r}} \right)}},
\eeqn
where $r = MN$.
It can be observed that there is no negative term in (\ref{h1x2}); hence, $\frac{{\partial {h'_1}}}{{\partial x}} > 0$ for all $x \in {\mathbb{R}_x}$, all $M$ and $N$. Therefore, we have proved the lemma.
\end{proof}

To prove the third property, we show that 
$g\left( \tau \right)$ and $h'\left( \tau \right)$ intersect only once in the range of $\left[ 0, T\right]$. 
This will be done using the same approach as that in Appendix A. Specifically, we will consider two regions
${\mathbf{\Omega}_1}$ and  ${\mathbf{\Omega}_2}$ and prove two properties stated in Lemma 2 for this case.
As in Appendix A, the third property holds if we can prove
$ - \frac{1}{\tau }+\frac{{\partial {h'_1}}}{{\partial \tau }} < \frac{{\partial g}}{{\partial \tau }}$. 
It can be observed that all steps used to prove this inequality are the same as those in the proof of (\ref{partialh1g1ch}) 
for Proposition 1. Hence, we need to prove
\beqn \label{h2lower}
\frac{{\partial {h'_1}}}{{\partial x}} > \frac{2}{{1 - x}}.
\eeqn
Substitute $\frac{{\partial {h'_1}}}{{\partial x}}$ from (\ref{h1x2}) to (\ref{h2lower}), this inequality reduces to
\beqn \label{h2lower1}
2\frac{{\left( \begin{array}{l}
 r\left( {r - 1} \right){x^{r - 2}}{\left( {1 - x} \right)^2}\left( {1 - {x^r}} \right) \\ 
  + {\left( {1 - {x^r}} \right)^2} + {r^2}{x^{2\left( {r - 1} \right)}}{\left( {1 - x} \right)^2} \\ 
 \end{array} \right)}}{{\left( {1 - {x^r} + r{x^{\left( {r - 1} \right)}}\left( {1 - x} \right)} \right)\left( {1 - x} \right)\left( {1 - {x^r}} \right)}}> \frac{2}{1-x}.
\eeqn
After some manipulations, this inequality becomes equivalent to 
\beqn \label{h2lower2}
rx^{\left(r-2\right)}\left(1-x\right)^2\left[r-\left(1+x+x^2+\cdots+x^{\left(r-1\right)}\right)\right]>0.
\eeqn
It can be observed that $0<x<1$ and  $0<x^i<1$, $i \in \left[0,r-1\right]$. Hence, we have $r-\left(1+x+x^2+\cdots+x^{\left(r-1\right)}\right) >0$ which shows that (\ref{h2lower2}) indeed holds. Therefore, (\ref{h2lower}) holds and we have completed the proof of the third property. Finally,
the last property of the Proposition is obviously correct. Hence, we have completed the proof of Proposition 2.

%\end{appendices}			
% this file is called up by thesis.tex
% content in this file will be fed into the main document

%: ----------------------- name of chapter  -------------------------
%\chapter{Chapter 4. Channel Assignment with Access Contention Resolution for Cognitive Radio Networks} % top level followed by section, subsection
\chapter{Channel Assignment With Access Contention Resolution for Cognitive Radio Networks} % top level followed by section, subsection
\zlabel{Chapter4}

%: ----------------------- paths to graphics ------------------------

% change according to folder and file names
\ifpdf
    \graphicspath{{4/figures/PNG/}{4/figures/PDF/}{4/figures/}}
\else
    \graphicspath{{4/figures/EPS/}{4/figures/}}
\fi

%: ----------------------- contents from here ------------------------

The content of this chapter was published in IEEE Transactions on Vehicular Technology in the following paper:

L.~T.~ Tan, and L.~B.~ Le, ``Channel Assignment With Access Contention Resolution for Cognitive Radio Networks,'' {\em IEEE Trans. Veh. Tech.}, vol. 61, no. 6, pp.  2808--2823, 2012.

\section{Abstract}
%\begin{abstract}

In this paper, we consider the channel allocation problem for throughput maximization in
cognitive radio networks with hardware-constrained secondary users. In particular,
we assume that secondary users (SUs) exploit spectrum holes on a set of channels
where each SU can use at most one available channel for communication. 
We present the optimal brute-force search algorithm and its complexity for this non-linear integer optimization problem.
Because the optimal solution has exponential complexity with the numbers of channels
and SUs, we develop two low-complexity channel assignment algorithms that can efficiently utilize spectrum
opportunities on these channels. In the first algorithm, SUs are 
assigned distinct sets of channels. We show that this algorithm achieves the maximum throughput limit
if the number of channels is sufficiently large. In addition, we propose an overlapping
channel assignment algorithm, that can improve the throughput performance compared to
the non-overlapping channel assignment counterpart. Moreover, we design a distributed
MAC protocol for access contention resolution and  integrate it into
the overlapping channel assignment algorithm. We then analyze
the saturation throughput and the complexity of the proposed channel assignment algorithms.
We also present several potential extensions, including the development of greedy channel assignment 
algorithms under max-min fairness criterion and throughput analysis, considering sensing errors.  
Finally, numerical results are presented
to validate the developed theoretical results and illustrate the performance gains due to the proposed
channel assignment algorithms.
%\end{abstract}

%\begin{IEEEkeywords}
%Channel assignment, MAC protocol, spectrum sensing, throughput maximization, cognitive radio.
%\end{IEEEkeywords}
%\IEEEpeerreviewmaketitle

\section{Introduction}

Emerging broadband wireless applications have been demanding unprecedented increase in radio spectrum resources.
As a result, we have been facing a serious spectrum shortage problem. However, several recent
measurements reveal very low spectrum utilization in most useful frequency bands \cite{Zhao07}.
Cognitive radio technology is a promising technology that can fundamentally improve the
spectrum utilization of licensed frequency bands through secondary spectrum access.
However, transmissions from primary users (PUs) should be satisfactorily protected from 
secondary spectrum access due to their strictly higher access priority.

Protection of primary communications can be achieved through interference avoidance or interference
control approach (i.e., spectrum overlay or spectrum underlay) \cite{Zhao07}.
For the interference control approach, transmission powers of SUs
should be carefully controlled so that the aggregated interference they create
at primary receivers does not severely affect ongoing primary communications \cite{Le08}.
In most practical scenarios where direct coordination between PUs and SUs
is not possible and/or if distributed communications strategies are desired, it would be very
difficult to maintain these interference constraints. The interference avoidance approach
instead protects primary transmissions by requiring SUs to perform spectrum
sensing to discover spectrum holes over which they can transmit data \cite{Liang08}, \cite{Yu09}. 
This paper focuses on developing efficient channel assignment  %\cite{Cor09}, \cite{Liang08}
algorithms for a cognitive radio network with hardware-constrained secondary nodes
using the interference avoidance spectrum sharing approach. 

In particular, we consider the scenario where each SU can exploit at most one
available channel for communications. This can be the case if SUs
are equipped with only one radio employing a narrow-band RF front end \cite{So04}. In addition, it is 
assumed that white spaces are so dynamic that it is not affordable for each SU to sense all channels to discover
available ones and/or to exchange sensing results with one another. Under this setting, we
are interested in determining a set of channels allocated to each SU in advance
so that  maximum network throughput can be achieved in a distributed manner. To the best of
our knowledge, this important problem has not been considered before. The contributions of
this paper can be summarized as follows.

\begin{itemize}

\item 
We formulate the channel assignment problem for throughput maximization  as an integer optimization problem.
We then derive user and total network throughput for the case SUs are assigned distinct
sets of channels. We present the optimal brute-force search algorithm and analyze its complexity.

\item
We develop two greedy non-overlapping and overlapping channel assignment algorithms to solve the underlying NP-hard problem.
We prove that the proposed non-overlapping channel assignment algorithm achieves the maximum throughput as the number of
channels is sufficiently large. For the overlapping channel assignment algorithm, we design a medium access control (MAC) protocol for 
access contention resolution and we integrate the MAC protocol overhead analysis into the channel assignment algorithm.

\item
We analyze the saturation throughput and complexity of the proposed channel assignment algorithms. Moreover, we investigate
the impact of contention collisions on the developed throughput analytical framework.

\item
We show how to extend the proposed channel assignment algorithms when max-min fairness is considered.
We also extend the throughput analytical model to consider sensing errors and propose an alternative MAC
protocol that can relieve congestion on the control channel.

\item 
We demonstrate through numerical studies the interactions among various MAC protocol parameters and suggest its configuration.
We show that the overlapping channel assignment algorithm can achieve noticeable network throughput improvement
 compared to the non-overlapping counterpart. In addition, we present the throughput gains due to both proposed 
channel assignment algorithms compared to the round-robin algorithms, which do not exploit the heterogeneity in the
channel availability probabilities.

\end{itemize}

The remaining of this paper is organized as follows. In Section~\ref{Relworks_Chap4}, we discuss important related works on spectrum sharing
algorithms and MAC protocols. Section~\ref{SystemModel_Chap4} describes the system model and problem formulation. 
We present the non-overlapping channel assignment algorithm and describe its performance in Section~\ref{nonover_Chap4}.
The overlapping channel assignment and the corresponding MAC protocol are developed in Section~\ref{over_Chap4}. Performance analysis of
the overlapping channel assignment algorithm and the MAC protocol is presented in Section~\ref{Tputderive_Chapter4}. Several potential extensions
are discussed in Section~\ref{Poten_Chapter4}. Section~\ref{Results_Chap4} 
demonstrates numerical results followed by concluding remarks in Section~\ref{conclusion_Chap4}.

\section{Related Works}
\label{Relworks_Chap4}

Developing efficient spectrum sensing and access mechanisms for cognitive radio networks has been a very active 
research topic in the last several years \cite{Liang08, Cor09, Le11, Kim08, Su08, Su07, Nan07, Le208, Cor07, Hsu07, Konda08, Do05, Sala09, shu06, chen07}. %\cite{Cor09}-\cite{chen07}. 
A great survey of recent works on MAC protocol design and analysis is given in \cite{Cor09}.
In \cite{Liang08}, it was shown that by optimizing the sensing time, a significant throughput gain
can be achieved for a SU. In \cite{Le11}, we extended the result in \cite{Liang08} to the multi-user setting where we design,
analyze, and optimize a MAC protocol to achieve optimal tradeoff between sensing time and contention overhead. In fact, in \cite{Le11}, we 
assumed that each SU can use all available channels simultaneously. Therefore, the channel assignment problem
and the exploitation of multi-user diversity  do not exist in this setting, which is the topic of our current paper. Another related
effort along this line was conducted in \cite{Kim08} where sensing-period optimization and optimal channel-sequencing algorithms were
proposed to efficiently discover spectrum holes and to minimize the exploration delay.

In \cite{Su08}, a control-channel-based MAC protocol was proposed for secondary users to exploit white spaces in the
cognitive ad hoc network setting. In particular, the authors of this paper developed both random and negotiation-based
spectrum sensing schemes and performed throughput analysis for both saturation and non-saturation scenarios.
There exists several other synchronous cognitive MAC protocols, which rely on a control channel for spectrum negotiation and access
including those in \cite{Su07, Nan07, Le208, Cor07, Hsu07}.
A synchronous MAC protocols without using a control channel was proposed and studied in \cite{Konda08}.
In \cite{Do05}, a MAC layer framework was developed to dynamically reconfigure MAC and physical layer protocols.
Here, by monitoring current network metrics the proposed framework can achieve great performance by selecting the
best MAC protocol and its corresponding configuration. 

In \cite{Sala09}, a power-controlled MAC protocol was developed to efficiently exploit spectrum access opportunities
while satisfactorily protecting PUs by respecting interference constraints. Another
power control framework was described in \cite{shu06}, which aims to meet the rate requirements of SUs
and interference constraints of PUs. A novel clustering algorithm was devised in \cite{chen07} for
network formation, topology control, and exploitation of spectrum holes in a cognitive mesh network. It was shown that the proposed
clustering mechanism can efficiently adapt to the changes in the network and radio transmission environment.

Optimal sensing and access design for cognitive radio networks were designed by using optimal stopping
theory in \cite{jia08}. In \cite{Sala10}, a multichannel medium access control (McMAC)
protocol was proposed taking into account the distance
among users so that the white spaces can be efficiently exploited while satisfactorily protecting PUs.
Different power and spectrum allocation algorithms were devised to maximize the secondary network throughput in
\cite{taosu10, wang11, zhang113}. Optimization of spectrum sensing and access in which either cellular or TV bands can be
employed was performed in \cite{choi11}.

In \cite{zhang11}, cooperative sequential spectrum sensing and packet scheduling
were designed for cognitive radios which are equipped with multiple spectrum sensors.
An energy-efficient MAC protocol was proposed for cognitive radio networks in \cite{zhang112}.
Spectrum sensing, access, and power control algorithms were developed  considering QoS protection 
for PUs and QoS provisioning for SUs in \cite{jeon12, jha11}. Finally,
a channel hopping based MAC protocol was proposed in \cite{su081} for cognitive radio networks to alleviate the congestion
problem in the fixed control channel design. All these existing works, however, did not consider the scenario where
cognitive radios have hardware constraints which allows them to access at most one channel at any time. Moreover,
exploiting the multichannel diversity through efficient channel assignment is very critical to optimize the throughput
performance of the secondary network for this problem. We will investigate this problem considering its unique design issues in this paper.

%\vspace{10pt}
\section{System Model and Problem Formulation}
\label{SystemModel_Chap4}

%In this section, we describe the system model and define the channel allocation problem.

\subsection{System Model}
\label{System}

We consider a collocated cognitive radio network in which $M$ SUs exploit spectrum opportunities in  $N$ channels. We assume that 
any SU can hear the transmissions of other SUs. In addition, each SU can use at most one channel for 
its data transmission.
In addition, time is divided fixed-size cycle where SUs perform sensing on assigned channels
at the beginning of each cycle to explore available channels for communications.
 We assume that perfect sensing can be achieved with no sensing error. Extension to the imperfect
 spectrum sensing will be discussed in Section~\ref{Imper_sens_Chapter4}.
It is assumed that SUs transmit at a constant rate with the normalized value of one.

\subsection{Problem Formulation}
\label{ProbForm}

We are interested in performing channel assignment to maximize the system throughput. Let $T_i$
denote the throughput achieved by SU $i$. 
Let $x_{ij}$ describe the channel assignment decision where $x_{ij}=1$ if channel $j$ is assigned to SU
 $i$ and $x_{ij}=0$, otherwise. The throughput maximization problem can be formally written as follows:
\vspace{0.0cm}
\beqn
%\label{Tput}
\max\limits_\textbf{x} \sum\limits_{i = 1}^M {{T_i}}.  \label{obj1_Chapter4} 
\eeqn
For non-overlapping channel assignments, we have following constraints
\beqn
 \sum\limits_{i = 1}^M {{x_{ij}} = 1}, \quad  \mbox{for\:all}\: j  \label{con1_Chapter4}.
\eeqn
We can derive the throughput achieved by SU $i$ for non-overlapping channel assignment as follows.
Let $S_i$ be the set of channels solely assigned to SU $i$. Let $p_{ij}$ be the probability that 
channel $j$ is available at SU $i$. For simplicity, we assume that $p_{ij}$ are independent from one another.
This assumption holds when each SU impacts different set of PUs on each channel. This can indeed be the case 
because spectrum holes depend on space. Note, however, that 
this assumption can be relaxed if the dependence structure of these probabilities is available. Under this assumption, $T_i$ can be calculated as
\beq \label{tput1_Chapter4}
T_i = 1 - \prod_{j \in \mathcal{S}_i} \overline{p}_{ij} = 1 - \prod\limits_{j = 1}^N {{{\left( {{{\bar p}_{ij}}} \right)}^{{x_{ij}}}}}
\eeq
where $\overline{p}_{ij} = 1 - {p}_{ij}$ is the probability that channel $j$ is not available for 
SU $i$. In fact,  $1 - \prod_{j \in \mathcal{S}_i} \overline{p}_{ij}$ is the probability that
there is at least one channel available for SU $i$. Because each SU can use
at most one available channel, its maximum throughput is 1. 
In the overlapping channel assignment scheme, constraints in (\ref{con1_Chapter4}) are not needed. 
From this calculation, it can be observed that the optimization problem (\ref{obj1_Chapter4})-(\ref{con1_Chapter4})
is a non-linear integer program, which is a NP-hard problem (interest readers can refer to Part VIII of reference
\cite{Lee12} for detailed treatment of this hardness result).

\subsection{Optimal Algorithm and Its Complexity}

Due to the non-linear and combinatorial structure of the formulated channel assignment problem, it would be impossible
to explicitly determine its optimal closed form solution. However, we can employ the brute-force search (i.e., exhaustive search) to determine the best channel assignment that results in the maximum total throughput. In particular, we can enumerate all possible channel assignment solutions then determine
the best one by comparing their achieved throughput. This solution method requires a throughput analytical model
 that calculates the throughput for any particular channel assignment solution. We will develop such a model in Section \ref{tputana_Chapter4} of this paper.

We now quantify the complexity of the optimal brute-force search algorithm.
Let us consider SU $i$ (i.e., $i \in \left\{1, \ldots, M\right\}$). Suppose we assign $k$ channels to this SU $i$ where $k \in \left\{1, \ldots, N\right\}$). Then, there are $C_N^k$ ways to do so. Since $k$ can take any values in $k \in \left\{1, \ldots, N\right\}$, the total number of ways
to assign channels to SU $i$ is $\sum \limits_{k = 1}^N C_N^k \approx 2^N$. Hence, the total number of ways to assign channels to all SUs is asymptotically equal to $\left(2^N\right)^M = 2^{NM}$. Recall that 
we need to calculate the throughputs achieved by $M$ SUs for each potential assignment to determine the best one.  Therefore, the complexity of the optimal brute-force search algorithm is $\mathcal{O}(2^{NM})$.
Given the exponentially large complexity required to find the optimal channel assignment solution, we will develop 
sub-optimal and low-complexity channel assignment algorithms in the following sections. In particular, we consider two different 
channel assignment schemes: 1) non-overlapping channel assignment and 2) overlapping channel assignment.

\section{Non-overlapping Channel Assignment Algorithm}
\label{nonover_Chap4}

We develop a low-complexity  algorithm for non-overlapping channel assignment in this section. 
Recall that $\mathcal{S}_i$ is the set of channels solely assigned for SU
 $i$ (i.e., $\mathcal{S}_i \cap \mathcal{S}_j = \emptyset, \: i \neq j$). The greedy
channel assignment algorithm iteratively allocates channels to SUs that achieves the maximum increase
in the throughput. A detailed description of the proposed algorithm is presented in Alg.~\ref{mainalg_non_Chapter4}.
In each channel allocation iteration, each SU $i$ calculates its increase in throughput
if the best available channel (i.e., channel  $j_i^* = \mathop {\arg \max }\limits_{j \in {\mathcal{S}_a}} \: {p_{ij}}$)
is allocated. This increase in throughput can be calculated as follows:
\beqn
\label{Tlem1_Chapter4}
 \Delta {T_i} = T_i^a - T_i^b = \left[ {1 - \left( {1 - {p_{ij_i^*}}} \right)\prod\limits_{j \in \mathcal{S}_i } {(1 - {p_{ij}})} } \right] 
 %\nonumber \\ 
 - \left[ {1 - \prod\limits_{j \in \mathcal{S}_i } {(1 - {p_{ij}})} } \right] %\nonumber\\ 
= {p_{ij_i^*}} \prod\limits_{j \in \mathcal{S}_i } {(1 - {p_{ij}})}.  
\eeqn
Based on (\ref{Tlem1_Chapter4}), it can be observed that $\Delta {T_i}$ will quickly decrease over allocation iterations
because $\prod\limits_{j \in \mathcal{S}_i } {(1 - {p_{ij}})}$ tends to zero as the set $\mathcal{S}_i$ is expanded. We have the
following property for the resulting channel assignment due to Alg.~\ref{mainalg_non_Chapter4}.

\begin{algorithm}[h]%\leesize
\algsetup{linenosize=\scriptsize}
  \scriptsize %%\small \footnotesize \scriptsize  %\tiny
\caption{\textsc{Non-Overlapping Channel Assignment}}
\label{mainalg_non_Chapter4}
%\algsetup{indent=1.5em}
\begin{algorithmic}[1]

\STATE Initialize the set of available channels  ${\mathcal{S}_a} := \left\{ {1,2, \ldots ,N} \right\}$ and $\mathcal{S}_i := \emptyset$ for $i=1, 2,\ldots , M$

\FOR{$i = 1$ to $M$}

\STATE $j_i^* = \mathop {\argmax }\limits_{j \in {\mathcal{S}_a}} \: {p_{ij}}$

\IF {$\mathcal{S}_{i} \neq 0$}

\STATE Find $\Delta {T_i} = T_i^a - T_i^b$,
where $T_i^a$ and $T_i^b$ is the throughputs after and before assigning channel $j_i^*$.

\ELSE

\STATE Find $\Delta {T_i} = p_{ij_i^*}$,

\ENDIF

\ENDFOR

%\STATE Find ${i^*} = \mathop {\arg \max }\limits_i {p_{ij_i^*}}$

\STATE ${i^*} = \argmax_i \Delta {T_i} $.

\STATE Assign channel $j_{i^*}^*$ to user $i^*$.

\STATE Update $\mathcal{S}_a = \mathcal{S}_a\backslash j_{i^*}^*$.

\STATE If $\mathcal{S}_a $ is empty, terminate the algorithm. Otherwise, return to step 2. 
\end{algorithmic}
\end{algorithm}

\vspace{0.1cm}
\noindent
\textbf{Proposition 1:} If we have $N >> M$, then the throughput achieved by any SU $i$ due to Alg.~\ref{mainalg_non_Chapter4} is very close
to the maximum value of 1.

\begin{proof}
This proposition can be proved by showing that if the number of channels is much larger than the number of
SUs (i.e., $N >> M$) then each SU will be assigned a large number of channels.
Recall that Alg.~\ref{mainalg_non_Chapter4} assigns channels to a particular SU $i$ based on the increase-in-throughput metric $\Delta {T_i}$.
This property can be proved by observing that if a particular SU $i$ has been assigned a large
number of channels, its $\Delta {T_i}$ is very close to zero. Therefore, other SUs who have been assigned a small
number of channels will have a good chance to receive more channels. As a result, all SUs are assigned a large
number of channels if $N >> M$. According to (\ref{tput1_Chapter4}), throughput achieved by SU $i$ will reach its maximum
value of 1 if its number of assigned channels is sufficiently large. Hence, we have proved the proposition.
\end{proof}

In practice, we do not need a very large number of channels to achieve the close-to-maximum throughput. In particular,
if each channel is available for secondary spectrum access with probability at least 0.8 then the throughput achieved
by a SU assigned three channels is not smaller than $1-(1-0.8)^3=0.992$, which is less than $1\%$ below the maximum
throughput. Note that after running Alg.~\ref{mainalg_non_Chapter4}, we can establish the set of channels allocated to each SU, 
from which we calculate its throughput by using (\ref{tput1_Chapter4}). Then,
the total throughput of the secondary network can be calculated by summing the throughputs of all SUs. 
 When the number of channel is not sufficiently large, we can potentially improve the system throughput by allowing
overlapping channel assignment. We develop such an overlapping channel assignment algorithm in the next section. 

\section{Overlapping Channel Assignment}
\label{over_Chap4}

Overlapping channel assignment can improve the network throughput by exploiting the multiuser diversity gain.
However, a MAC protocol is needed to resolve the access contention under the overlapping channel assignments.
The MAC protocol incurs overhead that offsets the throughput gain due to the multiuser diversity. Hence, a sophisticated
channel assignment algorithm is needed to balance the protocol overhead and throughput gain.

\begin{figure}[!t]
\centering
\includegraphics[width=120mm]{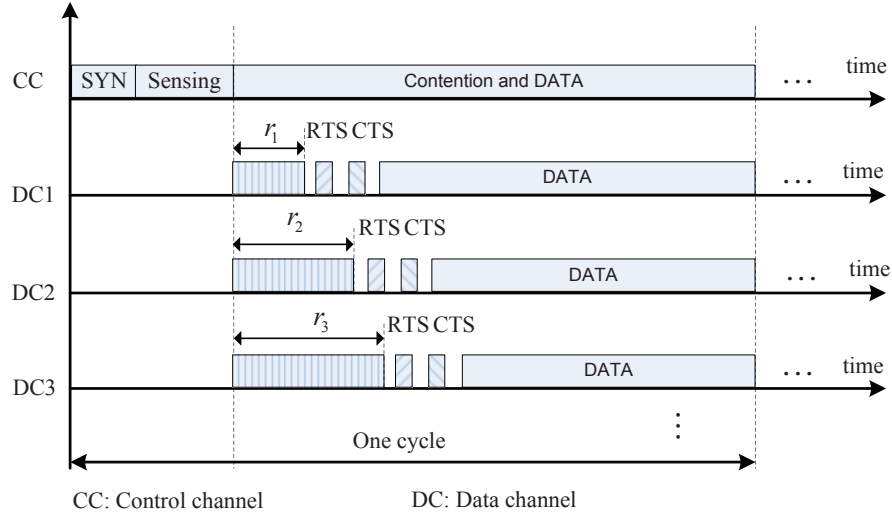}
\caption{Timing diagram for the proposed multi-channel MAC protocol.}
\label{MACoperation}
\end{figure}

\subsection{MAC Protocol}
\label{over_MACPro_Chapter4}

Let $\mathcal{S}_i$ be the set of
channels  solely assigned  for SU $i$ and $\mathcal{S}_{i}^{\sf com}$ be the set of  channels assigned for SU $i$ and some other SUs. Let denote $\mathcal{S}_i^{\sf tot} = \mathcal{S}_i \cup \mathcal{S}_i^{\sf com}$, which is the set of all channels assigned to SU $i$. 
Assume that there is one control channel, which is always available and used for access contention resolution.
We consider the following MAC protocol run by any particular SU $i$, which belongs the class of synchronized MAC
protocol \cite{Shi09}. The MAC protocol is illustrated in Fig.~\ref{MACoperation} where
synchronization and sensing phases are employed before the channel contention and transmission phase in each cycle. 
A synchronization message is exchanged among SUs during the synchronization phase to establish
the same starting epoch of each cycle. After sensing the assigned channels in the sensing phase,
each SU $i$ proceeds as follows. If there is at least one channel in $\mathcal{S}_i$ available, then SU $i$ chooses
one of these available channels randomly for communication. If this is not the case, SU $i$ will choose one available channel
 in $\mathcal{S}_i^{\sf com}$ randomly (if there is any channel in this set available). For brevity, we simply call \textit{users} instead of
 \textit{SUs}  when there is no confusion. Then, it chooses
a random backoff value which is uniformly distributed in the interval $[0, W-1]$ (i.e.,  $W$ is the contention window) and 
starts decreasing its backoff counter while listening on the control channel.

\begin{table}[ht]
\centering
\caption{Channel Assignment Example (M=3, N =6) }
\label{table_J2_Chapter4}
\begin{tabular}{|c|c|c|c|c|c|c|}
 \hline 
 & $\mathcal{S}_{1}$ & $\mathcal{S}_{2}$ & $\mathcal{S}_{3}$ & $\mathcal{S}_{1}^{\text{com}}$ & $\mathcal{S}_{2}^{\text{com}}$ & $\mathcal{S}_{3}^{\text{com}}$  \\
\hline \hline
% &  &  & \tabularnewline
%\hline 
C1 & x &  &   &  &  &  \\
\hline 
C2 &  & x &   &  &  &   \\
\hline 
C3 &  &  & x  &  &  &   \\
\hline 
C4 &  &  &   & x &  x &  \\
\hline 
C5 &  &  &  &  &  x  & x \\
\hline 
C6 &  &  &  &  x &  x  & x \\
\hline 
\end{tabular}
\end{table}

If it overhears transmissions of RTS/CTS from any other users, it will freeze from decreasing its backoff counter until the control
channel is free again. As soon as a user's backoff counter reaches zero, its transmitter transmits an RTS message
containing a chosen channel to its receiver. If the receiver successfully receives the RTS, it will reply with
CTS and user $i$ starts its communication on the chosen channel for the remaining of the cycle.
If the RTS/CTS message exchange fails due to collisions, the corresponding user will quit the contention and wait until the next cycle.
In addition, by overhearing RTS/CTS messages of neighboring users, which convey information about the channels chosen for communications, other users compared these channels with their chosen ones. 

Any user who has its chosen channel 
coincides with the overheard channels quits the contention and waits until the next cycle. Otherwise, it will continue to decrease its 
backoff counter before exchanging RTS/CTS messages. 
Note that the fundamental aspect that makes this MAC protocol different from
that proposed in \cite{Le11} is that in \cite{Le11} we assumed each winning user can use all available channels for
communications while at most one available channel can be exploited by hardware-constrained secondary users in the current paper.
Therefore, the channel assignment problem does not exist for the setting considered in  \cite{Le11}.
One example of overlapping channel assignment for three users and six channels is illustrated in Table~\ref{table_J2_Chapter4}. 
where channel assignments are indicated by an ``x''.

\begin{algorithm}[h]%\leesize
\algsetup{linenosize=\scriptsize}
  \scriptsize %%\small \footnotesize \scriptsize  %\tiny
%\algsetup{linenosize=\tiny}
  %\small %\footnotesize %\small %\tiny
  %\scriptsize
\caption{\textsc{Overlapping Channel Assignment}}
\label{mainalg_ove_Chapter4}
%\algsetup{indent=1.5em}
\begin{algorithmic}[1]

\STATE Initialize the sets of allocated channels for all users  $\mathcal{S}_i := \emptyset$ for $i=1, 2,\ldots , M$ and $\delta_0$

\STATE Run Alg.~\ref{mainalg_non_Chapter4} to obtain non-overlapping channel assignment solution.

\STATE Let the group of channels shared by $l$ users be $\mathcal{G}_l$
and $\mathcal{U}_j$ be the set of users sharing channel $j$ and set $\mathcal{U}_j^\text{temp} := \mathcal{U}_j, \: \forall j = 1, 2, \ldots, N$.

\STATE continue := 1; $h$ = 1; updoverhead := 0

\WHILE {$\text{continue}  = 1$}

\STATE Find the group of channels shared by $h$ users, $\mathcal{G}_{h}$

\FOR {$j=1$ to $\left|\mathcal{G}_{h}\right|$}

\FOR {$l=1$ to $M$}

\IF{$l \in \mathcal{U}_j $}

\STATE $\Delta T_{l}^{h, \text{est}} (j) = 0$

\ELSE

\STATE User $l$ calculates $\Delta T_l^{h, \text{est}} (j)$ assuming channel $j$ is allocated to user $l$

\ENDIF

\ENDFOR

\STATE ${l^*_j} = \argmax_{l}  \Delta T_l^{h, \text{est}} (j) $.

\ENDFOR

\STATE ${j_{l^*}^*} = \argmax_j  \Delta T_{l^*_j}^{h, \text{est}} (j) $.

\IF{ $\Delta T_{l^*}^{h, \text{est}} (j_{l^*}^*) \leq \epsilon$ and \text{updoverhead} = 1}

\STATE Set: \text{continue} := 0

\STATE Go to step 35

\ENDIF

\IF{ $\Delta T_{l^*}^{h, \text{est}} (j_{l^*}^*) > \epsilon$}

\STATE Temporarily assign channel $j_{l^*}^*$ to user $l^*$, i.e., 
 update $\mathcal{U}_{j_{l^*}^*}^{\text{temp}} = \mathcal{U}_{j_{l^*}^*}  \cup \left\{ l^* \right\} $;

\STATE Calculate $W$ and $\delta$ with  $\mathcal{U}_{j_{l^*}^*}^{\text{temp}}$ by using methods in Sections \ref{ConWinCal_Chapter4} and \ref{Overcal_Chapter4}, respectively. 

\IF{ $\left| \delta - \delta_0 \right| > \epsilon_{\delta}$}

\STATE Set: \text{updoverhead} := 1

\STATE Return Step 7 using the updated $\delta_0 =\delta $

\ELSE 

\STATE Update  $\mathcal{U}_{j_{l^*}^*} := \mathcal{U}_{j_{l^*}^*}^{\text{temp}}$ 
 (i.e., assign channel $j_{l^*}^*$ to user $l^*$), calculate $W$ and $\delta_0$ with $\mathcal{U}_{j_{l^*}^*}$, and
 update $\mathcal{G}_{h}$
 
 \STATE Update: \text{updoverhead} := 0

\ENDIF

\ENDIF

\STATE Return Step 7 

\STATE $h=h+1$

\ENDWHILE

\end{algorithmic}
\end{algorithm}

\noindent
\textit{Remark 1:} We focus on the saturation-buffer scenario in this paper. In practice, cognitive radios may 
have low backlog or, sometimes, even empty buffers. In addition, because 
the data transmission phase is quite large compared to a typical packet size, we should allow users to transmit several packets to
 completely fill the transmission phase in our MAC design. This condition can be realized by allowing only sufficiently backlogged 
  users to participate in the sensing and access contention processes.

\subsection{Overlapping Channel Assignment Algorithm}

We develop an overlapping channel assignment algorithm that possesses two phases as follows.
We run Alg.~\ref{mainalg_non_Chapter4} to obtain the non-overlapping channel
assignment solution in the first phase. Then, we perform overlapping channel assignment
by allocating channels that have been assigned to some users
to other users in the second phase. 
We devise a greedy overlapping channel assignment algorithm using the increase-of-throughput metric similar to that
employed in Alg.~\ref{mainalg_non_Chapter4}. 
However, calculation of this metric exactly turns out to be a complicated
task. Hence, we employ an estimate of the increase-of-throughput, which is derived as follows
to perform channel assignment assuming that the MAC protocol overhead  is $\delta<1$. 
In fact, $\delta$ depends on the outcome of the channel assignment algorithm (i.e., sets of
channels assigned to different users). We will show
how to calculate $\delta$ and integrate it into this channel assignment algorithm later. 

Consider a case where channel $j$ is the common channel of users $i_1, i_2, \ldots, i_{\mathcal{MS}}$. Here,
$\mathcal{MS}$ is the number of users sharing this channel. We are interested in estimating
 the increase in throughput for a particular user $i$ if channel $j$ is assigned to this user.
 Indeed, this increase of throughput can be achieved because user $i$ may be able to exploit
 channel $j$ if this channel is not available or not used by other users $i_1, i_2, \ldots, i_{\mathcal{MS}}$.
 To estimate the increase of throughput, in the remaining of this paper we are only interested in
 a practical scenario where all $p_{ij}$ are close to 1 (e.g., at least 0.8). This assumption would be reasonable, given several recent measurements reveal that spectrum utilization of useful frequency bands
 is very low (e.g., less that $15\%$). Under this assumption,  
 we will show that the increase-of-throughput for user $i$ can be estimated as
\beqn \label{upith_Chapter4}
 \Delta T_{i}^{\mathcal{MS}, \text{est}} (j) =  
(1-1/\mathcal{MS})(1-\delta) p_{ij} \left( \prod_{h \in \mathcal{S}_i} \overline{p}_{ih} \right) \label{Del1_Chapter4} %\hspace{0.8cm} %\nonumber\\ \times
  \left(1-\prod_{h \in \mathcal{S}_i^{\text{com}}}\overline{p}_{ih} \right)  \sum_{k=1}^{\mathcal{MS}} \left[ \overline{p}_{i_k j} \left(\prod_{q=1, q \neq k}^{\mathcal{MS}}p_{i_q j}  \right) \right]  \\
+(1-\delta) p_{ij}  \prod_{h \in \mathcal{S}_i} \overline{p}_{ih}    \prod_{h \in \mathcal{S}_i^{\text{com}}}\overline{p}_{ih}  \label{Del2_Chapter4} %\hspace{3.2cm} \nonumber\\  \times 
 \prod_{q=1}^{\mathcal{MS}}p_{i_q j}   \prod_{q=1}^{\mathcal{MS}}\left(1-\prod_{h \in \mathcal{S}_{i_q}} \overline{p}_{i_q h}\right)   \hspace{0cm}\\
\hspace{0cm} + (1-1/\mathcal{MS})(1-\delta) p_{ij} \prod_{h \in \mathcal{S}_i} \overline{p}_{ih}  \left(1-\prod_{h \in \mathcal{S}_i^{\text{com}}}\overline{p}_{ih} \right)   % \nonumber \\ \times
  \prod_{q=1}^{\mathcal{MS}}p_{i_q j}  \prod_{q=1}^{\mathcal{MS}}\left(1-\prod_{h \in \mathcal{S}_{i_q}} \overline{p}_{i_q h}\right). \label{Del3_Chapter4}
\eeqn

This estimation is obtained by listing all possible scenarios/events in which user $i$ can exploit channel $j$ to
increase its throughput. Because the user throughput is bounded by 1, we only count events that occur with
non-negligible probabilities. In particular, under the assumption that $p_{ij}$ are high (or $\overline{p}_{ij}$ are small)
we only count events whose probabilities have at most two such elements $\overline{p}_{ij}$ in the product. 
In addition, we can determine the increase of throughput for user $i$ by comparing its achievable throughput
before and after channel $j$ is assigned to it. It can be verified we have the following 
events for which the average increases of throughput are significant.
 
\begin{itemize}

\item Channel $j$ is available for all users $i$ and $i_q$, $q = 1, 2, \ldots, \mathcal{MS}$  except $i_k$ where $k = 1, 2, \ldots, \mathcal{MS}$. 
In addition, all channels in $S_i$ are not available and there is at least one channel in $\mathcal{S}_i^{\text{com}}$ available for user $i$. User $i$ can achieve a
maximum average throughput of $1-\delta$  by exploiting channel $j$, while its minimum average throughput before being assigned channel $i$
is at least $(1-\delta)/\mathcal{MS}$ (when user $i$ needs to share the available channel in $\mathcal{S}_i^{\text{com}}$ with $\mathcal{MS}$ 
other users). The increase of throughput for this case is at most $(1-1/\mathcal{MS})(1-\delta)$ and the upper-bound for the increase of
throughput of user $i$ is written in (\ref{Del1_Chapter4}).

\item Channel $j$ is available for user $i$ and all users $i_q$, $q = 1, 2, \ldots, \mathcal{MS}$ but each user $i_q$ uses other available channel in $\mathcal{S}_{i_q}$ for his/her transmission. Moreover, there is no channel in $\mathcal{S}_i^{\text{tot}}$ available. In this case, the increase of throughput for user $i$ is $1-\delta$ and the average increase of throughput of user $i$ is written in (\ref{Del2_Chapter4}).

\item Channel $j$ is available for user $i$ and all users $i_q$, $q = 1, 2, \ldots, \mathcal{MS}$ but each user $i_q$ uses other available channel in $\mathcal{S}_{i_q}$ for transmission. Moreover, there is at least one channel in $\mathcal{S}_i^{\text{com}}$ available. In this case, the increase of throughput for user $i$ is upper-bounded by $(1-1/\mathcal{MS})(1-\delta)$ and the average increase of
throughput of user $i$ is written in (\ref{Del3_Chapter4}).

\end{itemize}

The detailed description of the algorithm is given in Alg.~\ref{mainalg_ove_Chapter4}. This algorithm has outer and inter loops where the outer loop
increases the parameter $h$, which represents the maximum of users allowed to share any particular channel (i.e., $\mathcal{MS}$ in
the above estimation of the increase of throughput) and the inner loop performs channel allocation for one particular value of $h=\mathcal{MS}$. 
In each assignment iteration of the inner loop, we assign one ``best'' channel $j$ to user $i$ that achieves maximum $\Delta T_{i}^{h,\text{est}} (j)$.
This assignment continues until the maximum $\Delta T_{i}^{h,\text{est}} (j)$ is less than a pre-determined number $\epsilon>0$.
As will be clear in the throughput analysis developed later, it is beneficial to maintain at least one channel in each set $S_i$.
This case is because the throughput contributed by channels in $S_i$ constitutes a significant fraction of the total throughput.
Therefore, we will maintain this constraint when running Alg.~\ref{mainalg_ove_Chapter4}.

\subsection{Calculation of Contention Window }
\label{ConWinCal_Chapter4}

We show how to calculate contention window $W$ so that collision probabilities among contending secondary users are sufficiently small.
In fact, there is a trade-off between collision probabilities and the average overhead of the MAC protocol, which depends on $W$. In particular, larger values of $W$ reduce collision probabilities at the cost of higher protocol overhead and vice versa. Because there can be several collisions
during the contention phase each of which occurs if two or more users randomly choose the same value of backoff time.
In addition, the probability of the first collision is largest because the number of contending users decreases
for successive potential collisions.

Let $\mathcal{P}_c$ be the probability of the first collision. In the following, we determine contention window $W$ by imposing a
 constraint $\mathcal{P}_c \leq \epsilon_P $ where $\epsilon_P $ controls the collision probability and overhead tradeoff. Let us calculate $\mathcal{P}_c$ as a function of $W$ assuming that there are $m$ secondary 
 users in the contention phase. Without loss of generality, assume that the random backoff times of $m$  users are ordered as $r_1 \leq r_2 \leq \ldots \leq r_m$. The conditional probability of the first collision if there are $m$ users in the contention stage can be written as
\beqn
\label{Pfirstc_Chapter4}
\mathcal{P}_c^{(m)} &=& \sum _{j=2}^{m} \Pr \left( j \: \text{users collide} \right) \nonumber\\
&=& \sum_{j=2}^m \sum_{i=0}^{W-2} C_m^j \left( \frac{1}{W}\right)^j \left( \frac{W-i-1}{W}\right)^{m-j}
\eeqn
where each term in the double-sum represents the probability that $j$ users collide when they
choose the same backoff value equal to $i$. 
Hence, the probability of the first collision can be calculated as 
\beqn \label{pc_Chapter4}
\mathcal{P}_c = \sum_{m=2}^M \mathcal{P}_c^{(m)} \times \Pr\left\{m \: \text{ users contend}\right\},
\eeqn
where $\mathcal{P}_c^{(m)}$ is given in (\ref{Pfirstc_Chapter4}) and $\Pr\left\{m  \: \text{users contend}\right\}$
is the probability that $m$ users join the contention phase. To compute $\mathcal{P}_c$, we now
derive $\Pr\left\{m  \: \text{users contend}\right\}$. It can be verified that 
user $i$ joins contention if all channels in $\mathcal{S}_i$ are busy and there is at least one channel in $\mathcal{S}_i^{\sf com}$ available. The probability of this event can be written as
\beqn
\mathcal{P}_{\sf con}^{(i)} &=& \Pr \left\{ \text{all channels in} \: \mathcal{S}_i \: \text{ are busy}, %\right. \nonumber\\ && \left.
 \exists ! \: \text{some channels in} \:\mathcal{S}_i^{\sf com} \: \text{are available }\right\} \nonumber\\
&=&\left( \prod_{j \in \mathcal{S}_i} \overline{p}_{ij} \right) \left( 1- \prod_{j \in \mathcal{S}_i^{\sf com}} \overline{p}_{ij} \right).
\eeqn
The probability of the event that $m $ users join the contention phase is 
\beqn \label{Pmusercon_Chapter4}
\Pr \left\{ m \: \text{users contend} \right\} = \sum_{n=1}^{C_M^m} \left( \prod_{i \in {\Lambda}_n} \mathcal{P}_{\sf con}^{(i)}\right) 
%\hspace{1.5cm} \nonumber \\ \times
 \left( \prod_{j \in {\Lambda}_M \backslash {\Lambda}_n} \mathcal{\overline{P}}_{\sf con}^{(j)}\right) 
\eeqn
where ${\Lambda}_n$ is one particular set of $m$ users, ${\Lambda}_M$ is the set of all $M$ users ($\left\{ 1,2, \ldots , M \right\}$).
Substitute the result in (\ref{Pmusercon_Chapter4}) into (\ref{pc_Chapter4}), we can calculate $\mathcal{P}_c$. Finally,
we can determine $W$ as 
\beqn \label{Window_Chapter4}
W=  \min \left\{{W} \: \text{such that} \:  \mathcal{P}_c(W) \leq \epsilon_P \right\}
\eeqn
where for clarity we denote $\mathcal{P}_c(W)$, which is given in (\ref{pc_Chapter4}) as a function of $W$.

\subsection{Calculation of MAC Protocol Overhead}
\label{Overcal_Chapter4}

Using the contention window calculated in (\ref{Window_Chapter4}), we can quantify the average overhead of the proposed MAC protocol.
Toward this end, let $r$ be the average value of the backoff value chosen by any SU.
Then, we have $r = (W-1)/2$ because the backoff counter value is uniformly chosen
in the interval $[0,W-1]$. As a result, average overhead can be calculated as follows:
\beqn \label{overhead_Chapter4}
\delta\left(W\right) = \frac { \left[ W-1 \right]\theta/2 + t_{\sf RTS} + t_{\sf CTS} + 3 t_{\sf SIFS} + t_{\sf SEN}+ t_{\sf SYN}} {\sf T_{\sf cycle}}
\eeqn
where $\theta$ is the time corresponding to one backoff unit;  $t_{\sf RTS}$,  $t_{\sf CTS}$, $t_{\sf SIFS}$
are the corresponding time of RTS, CTS and SIFS messages; $t_{\sf SEN}$ is the
sensing time; $t_{\sf SYN}$ is the length of the synchronization message; 
and $\sf T_{\sf cycle}$ is the cycle time.

\subsection{Update $\delta$ inside Alg.~\ref{mainalg_ove_Chapter4}}

The overhead $\delta$ depends on the channel assignment outcome, which is not known when we are 
running Alg.~\ref{mainalg_ove_Chapter4}. Therefore, in each allocation step we update $\delta$ based on the current channel assignment outcome.
Because $\delta$ does not change much in two consecutive allocation decisions, Alg.~\ref{mainalg_ove_Chapter4} runs smoothly in practice.

\subsection{Practical Implementation Issues}

To perform channel assignment, we need to know $p_{ij}$ for all users and channels. Fortunately, we only need to perform estimation 
of $p_{ij}$ once these values change,
which would be infrequent in practice. These estimation and channel assignment tasks can be performed by one secondary node or collaboratively
performed by several of them. For example, for the secondary network supporting communications between $M$ secondary nodes and a single secondary
base station (BS), the BS can take the responsibility of estimating $p_{ij}$ and performing channel assignment. Once the channel assignment solution
has been determined and forwarded to all SUs, each SU will perform spectrum sensing and run the underlying MAC
protocol to access the spectrum in each cycle. 

It is emphasized again that although sensing and MAC protocol are performed and run in every cycle, 
estimating $p_{ij}$ and performing channel assignment (given these $p_{ij}$) are only performed if the values of $p_{ij}$ change,
which should be infrequent. Therefore, it would be affordable to estimate $p_{ij}$ accurately by employing sufficiently long
sensing time. This is because for most spectrum sensing schemes, including an energy detection scheme, mis-detection and false alarm probabilities
tend to zero when the sensing time increases for a given sampling frequency \cite{Liang08, Yu09}.

\section{Performance Analysis}
\label{Tputderive_Chapter4}

Suppose we have run Alg.~\ref{mainalg_ove_Chapter4} and obtained the set of users $U_j$ associated with each allocated channel $j$.
From this, we have the corresponding sets $\mathcal{S}_i$ and $\mathcal{S}_i^{\text{com}}$ for each user $i$.
Given this channel assignment outcome, we derive the throughput in the following assuming that there is no
collision due to MAC protocol access contention. We will show that by appropriately choosing
contention parameters for the MAC protocol, the throughput analysis under this assumption achieves accurate 
results. 

\subsection{Throughput Analysis}
\label{tputana_Chapter4}

Because the total throughput is the sum of throughput of all users, it is sufficient to analyze the throughput of one
particular user $i$. We will perform the throughput analysis by considering all possible sensing outcomes performed by
the considered user $i$ for its assigned channels. 
We will have the following cases, which correspond to different achievable throughput for the considered user.
\begin{itemize}
\item{ \NoIndent{Case 1: If there is at least one channel in $\mathcal{S}_i$ available, then user $i$ will exploit this available
channel and achieve the throughput of one. Here, we have
\beqn
T_i \left\{\text{Case 1} \right\}  = \Pr\left\{\text{Case 1} \right\} = 1-\prod\limits_{j\in \mathcal{S}_i} \bar{p}_{ij}.
\eeqn }}

\item{\NoIndent{$\text{Case 2}$: In this case, we consider scenarios where all channels in $\mathcal{S}_i$ are not 
available, at least one channel in $\mathcal{S}_i^{\text{ com}}$ is available, and user $i$ chooses the available channel $j$ for transmission.
Suppose that channel $j$ is shared  by $\mathcal{MS}_j$ secondary users including user $i$ (i.e., $\mathcal{MS}_j = |\mathcal{U}_j|$). 
The following four possible groups of users $i_k$, $k=1, \ldots, \mathcal{MS}_j$ share channel $j$.
\begin{itemize}
\item{ \textbf{Group I}: channel $j$ is available for user $i_k$ and user $i_k$ has at least 1 channel in $\mathcal{S}_{i_k}$ available.}
\item{ \textbf{Group II}: channel $j$ is not available for user $i_k$. }
\item{ \textbf{Group III}: channel $j$ is available for user $i_k$, all channels in $\mathcal{S}_{i_k}$ are not available and another channel
 $j'$ in $\mathcal{S}_{i_k}^{\text{com}}$ is available for user $i_k$. In addition, user $i_k$ chooses channel $j'$ for transmission in the contention stage.}
\item{ \textbf{Group IV}: channel $j$ is available for user $i_k$, all channels in $\mathcal{S}_{i_k}$ are not available. In addition, user $i_k$  chooses channel $j$ for transmission in the contention stage. Hence, user $i_k$ competes with user $i$ for channel $j$.}

\end{itemize}

The throughput achieved by user $i$ in this case can be written as
\beqn
\label{Tputa1delta_Chapter4}
T_i\left( \text{ Case 3} \right) = %\hspace{3.4cm} \nonumber \\
 (1-\delta) \Theta_i \sum \limits_{A_1 = 0}^{\mathcal{MS}_j} \sum \limits_{A_2 = 0}^{\mathcal{MS}_j-A_1} \sum \limits_{A_3=0}^{\mathcal{MS}_j-A_1-A_2} %\nonumber \\
 \Phi_1(A_1) \Phi_2(A_2)  \Phi_3(A_3) \Phi_4(A_4)
 %\hspace{4cm} 
\eeqn
where the following conditions hold.
\begin{itemize}
\item $\Theta_i$ is the probability that all channels in $\mathcal{S}_i$ are not available and user $i$ chooses 
some available channel $j$ in $\mathcal{S}_i^{\text{com}}$ for transmission.
\item $\Phi_1(A_1)$ denotes the probability that there are $A_1$ users belonging to Group I described above
among $\mathcal{MS}_j$ users sharing channel $j$.
\item $\Phi_2(A_2)$ represents the probability that there are $A_2$ users belonging to Group II 
among $\mathcal{MS}_j$ users sharing channel $j$.
\item $\Phi_3(A_3)$ describes the probability that there are $A_3$ users belonging to Group III
among $\mathcal{MS}_j$ users sharing channel $j$.
\item $\Phi_4(A_4)$ denotes the probability that there are $A_4 = \mathcal{MS}_j-A_1-A_2-A_3$ 
remaining users belonging to Group IV  scaled by $1/(1+A_4)$ where $A_4$ is the number of users 
excluding user $i$ competing with user $i$ for channel $j$.
\end{itemize}
We now proceed to calculate these quantities. We have
\beqn \label{group0_Chapter4}
\Theta_i =  \prod_{k \in \mathcal{S}_i} \overline{p}_{ik} \sum \limits _{B_i =1}^{H_i} \sum \limits_{h=1}^{C_{H_i}^{B_i}} \sum \limits_{j \in {\Psi}^h_i}  \frac{1}{B_i} \prod_{j_1 \in \Psi^h_i } p_{ij_1} \!\!\!\!\!\!\!\!\!\prod_{j_2 \in \mathcal{S}_i^{\text{com}} \backslash \Psi^h_i} \!\!\!\!\!\!\!\!\overline{p}_{ij_2} 
\eeqn
where $H_i$ denotes the number of channels in $\mathcal{S}_i^{\text{com}}$. The first product term in (\ref{group0_Chapter4}) represents
the probability that all channels in $\mathcal{S}_i$ are not available for user $i$. The second term in (\ref{group0_Chapter4}) describes the
probability that user $i$ chooses an available channel $j$ among $B_i$ available channels in $\mathcal{S}_i^{\text{com}}$ for transmission.
Here, we consider all possible subsets of $B_i$ available channels and for one such particular case $\Psi^h_i$ describes the corresponding
set of $B_i$ available channels, i.e.,
\beqn
\Phi_1(A_1) = \sum \limits_{c_1=1}^{C_{\mathcal{MS}_j}^{A_1}} \prod \limits_{m_1 \in  \Omega_{c_1}^{(1)}  } \left(p_{m_1j} \left(1-\prod \limits_{l \in \mathcal{S}_{m_1}}\overline{p}_{m_1 l}\right)\right).   \label{groupI_Chapter4}
\eeqn
In (\ref{groupI_Chapter4}), we consider all possible subsets of size $A_1$ belonging to Group I (there are $C_{\mathcal{MS}_j}^{A_1}$ such subsets).
Each term inside the sum represents the probability for the corresponding event whose set of $A_1$ users is denoted by $\Omega_{c_1}^{(1)}$, i.e.,
\beqn
\Phi_2(A_2) = \sum \limits_{c_2=1}^{C_{\mathcal{MS}_j-A_1}^{A_2}} \prod \limits_{m_2 \in \Omega_{c_2}^{(2)}  }  \overline{p}_{m_2j}. \label{groupII_Chapter4} 
\eeqn
In (\ref{groupII_Chapter4}), we capture the probability that channel $j$ is not available for $A_2$ users in group II whose possible sets are denoted by $\Omega_{c_2}^{(2)}$, i.e., 
\beqn
\Phi_3(A_3) = \sum \limits_{c_3=1}^{C_{\mathcal{MS}_j-A_1-A_2}^{A_3}} \prod \limits_{m_3 \in \Omega_{c_3}^{(3)}  } \left(p_{m_3j} \prod_{l_3 \in \mathcal{S}_{m_3}} \overline{p}_{m_3l_3}\right)  \label{1groupIII_Chapter4} \hspace{3.0cm} \\ 
\hspace{0cm} \times   \left[  \sum \limits_{n=0}^{\beta} \sum_{q=1}^{C_{\beta}^n} 
\prod_{h_1 \in \mathcal{S}^{\text{com},q}_{j,m_3} }  p_{m_3 h_1} 
\prod_{h_2 \in \overline{\mathcal{S}}^{\text{com},q}_{j,m_3} }  \overline{p}_{m_3 h_2}  %\right. \nonumber \\   \left.  \times
  \left( 1- \frac{1}{ n+1} \right) \right]. \label{2groupIII_Chapter4} 
\eeqn
For each term in (\ref{1groupIII_Chapter4}) we consider different possible subsets of $A_3$ users, which are denoted by $\Omega_{c_3}^{(3)}$.
Then, each term in (\ref{1groupIII_Chapter4})  represents the probability that channel $j$ is available for each user $m_3 \in \Omega_{c_3}^{(3)}$ 
while all channels in  $\mathcal{S}_{m_3}$ for the user $m_3$ are not available. 
In (\ref{2groupIII_Chapter4}), we consider all possible sensing outcomes for channels in $\mathcal{S}_{m_3}^{\text{com}}$ performed by user
 $m_3 \in \Omega_{c_3}^{(3)}$.
In addition, let $\mathcal{S}^{\text{com}}_{j,m_3} = \mathcal{S}_{m_3}^{\text{com}} \backslash \left\{ j \right\}$  
and $\beta = |\mathcal{S}^{\text{com}}_{j,m_3}|$.
Then, in (\ref{2groupIII_Chapter4}) we consider all possible scenarios in which there are $n$ channels in $\mathcal{S}^{\text{com}}_{j,m_3}$ available; and user $m_3$ chooses a channel different from channel $j$ for transmission (with probability $\left( 1- \frac{1}{ n+1} \right)$) where
$\mathcal{S}^{\text{com}}_{j,m_3} = \mathcal{S}^{\text{com},q}_{j,m_3} \cup \overline{\mathcal{S}}^{\text{com},q}_{j,m_3} $ and
$\mathcal{S}^{\text{com},q}_{j,m_3} \cap \overline{\mathcal{S}}^{\text{com},q}_{j,m_3} = \emptyset$. We have
\beqn
\Phi_4(A_4) =  \left(\frac{1}{1+ A_4}\right) \prod \limits_{m_4 \in \Omega^{(4)} } \left(p_{m_4j} \prod_{l_4 \in \mathcal{S}_{m_4}} \overline{p}_{m_4l_4}\right)  \label{1groupIV_Chapter4}   \hspace{3cm}\\
\times \left[ \sum \limits_{m=0}^{\gamma} \sum_{q=1}^{C_{\gamma}^{m}}   %\right.
\prod_{h_1 \in \mathcal{S}^{\text{com},q}_{j,m_4} }  p_{m_4 h_1} 
\prod_{h_2 \in \overline{\mathcal{S}}^{\text{com},q}_{j,m_4} }  \overline{p}_{m_4 h_2}  % \right. \nonumber \\  \left.  
\left( \frac{1}{ m+1} \right)  \right].
 \label{2groupIV_Chapter4} 
\eeqn
The sensing outcomes captured in (\ref{1groupIV_Chapter4}) and (\ref{2groupIV_Chapter4}) are similar to those in (\ref{1groupIII_Chapter4}) and (\ref{2groupIII_Chapter4}).
However, given three sets of $A_1$, $A_2$, and $A_3$ users, the set $\Omega^{(4)}$ can be determined whose size is $|\Omega^{(4)}| = A_4$. 
Here, $\gamma$ denotes cardinality of the set $\mathcal{S}_{j,m_4}^{\text{com}} = \mathcal{S}_{m_4}^{\text{com}} \backslash \left\{ j \right\}$.
Other sets are similar to those in (\ref{1groupIII_Chapter4}) and (\ref{2groupIII_Chapter4}). However, all users in $\Omega^{(4)}$ choose channel $j$ for transmission
in this case. Therefore, user $i$ wins the contention with probability $1/(1+A_4)$ and its achievable throughput is $(1-\delta)/(1+A_4)$.

}}
\end{itemize}

Summarizing all considered cases, the throughput achieved by user $i$ is given as
\beqn
T_i = T_i \left\{\text{Case 1} \right\} + T_i \left\{ \text{Case 3} \right\} .
\eeqn
In addition, the total throughput of the secondary network $\mathcal{T}$ is the sum of throughputs achieved by all SUs.

\subsection{Impacts of Contention Collision}
\label{conovh_Chapter4}

We have presented the saturation throughput analysis assuming that there is no contention collision.
Intuitively, if the MAC protocol is designed such that collision probability is sufficiently small
then the impact of collision on the throughput performance would be negligible. For our MAC protocol,
users  perform contention resolution in case 2 considered in the previous throughput analysis,
which occurs with a small probability. Therefore, if the contention window in (\ref{Window_Chapter4}) is chosen
for a sufficiently small $\epsilon_P$, then contention collisions would have negligible impacts on the network throughput.
We formally state this intuitive result in the following proposition.

\vspace{0.1cm}
\noindent
\textbf{Proposition 2:} The throughput $\mathcal{T}$ derived in the previous sub-section has an error, which can be upper-bounded as 
\beq \label{error_Chapter4}
E_t \leq  \epsilon_P \sum_{i=1}^M \prod\limits_{j\in \mathcal{S}_i} \bar{p}_{ij} \left( 1 - \prod\limits_{j\in 
\mathcal{S}_i^{\text{com}}} \bar{p}_{ij}  \right)
\eeq
where $\epsilon_P$ is the target collision probability that is used to determine the contention window in (\ref{Window_Chapter4}).

\begin{proof}
As aforementioned, contention collision can only occur in case 2 of the previous throughput analysis. The probability covering all
possible events for user $i$ in this case is $\prod\limits_{j\in \mathcal{S}_i} \bar{p}_{ij} \left( 1 - \prod\limits_{j\in 
\mathcal{S}_i^{\text{com}}} \bar{p}_{ij}  \right)$. In addition, the maximum average throughput that a particular user $i$ 
can achieve is $1-\delta<1$ (because no other users contend with user $i$ to exploit a chosen channel). In addition, if contention collision
happens then user $i$ will quit the contention and may experience a maximum average throughput loss of $1-\delta$ compared to the 
ideal case with no contention collision. In addition, the collision probabilities of all potential collisions is bounded above by $\epsilon_P$.
Therefore, the average error due to the proposed throughput analysis
can be upper-bounded as in (\ref{error_Chapter4}).
\end{proof}

To illustrate the throughput error bound presented in this proposition, let us consider an example where
$\bar{p}_{ij} \leq 0.2$ and $\epsilon_P \leq 0.03$. Because the sets $S_i$ returned by Alg.~\ref{mainalg_ove_Chapter4} contain at least
one channel, the throughput error can be bounded by $M \times 0.2 \times 0.03 = 0.006M$. In addition, the total
throughput will be at least $\sum_{i=1}^M \left(1 - \prod\limits_{j\in \mathcal{S}_i} \bar{p}_{ij}\right) \geq 0.8M$ if
we only consider throughput contribution from case 1.
Therefore, the relative throughput error can be upper-bounded by $0.006M/0.8M \approx 0.75 \%$, which is quite negligible. 
This example shows that the proposed throughput analytical model is very accurate in most practical settings.

\subsection{Complexity Analysis}
\label{Compx_Chapter4}

We analyze the complexity of Alg.~\ref{mainalg_non_Chapter4} and Alg.~\ref{mainalg_ove_Chapter4} in this subsection.
Let us proceed by analyzing the steps taken in each iteration in Alg.~\ref{mainalg_non_Chapter4}.
To determine the best assignment for the first channel, we have to search over $M$ SUs and $N$ channels, which involves $MN$ cases. Similarly, to assign the second channel, we need to perform searching over secondary users and $N-1$ channels (one channel is already assigned in
the first iteration). Hence, the second assignment involves $M\left(N-1\right)$ cases. Similar analysis can be applied for other assignments in later iterations. In summary, the total number of cases involved in assigning all channels to $M$ SUs is $M\left(N+\ldots+2+1\right) = MN\left(N+1\right)/2$,
which is $O(MN^2)$. In Alg.~\ref{mainalg_non_Chapter4}, the increase of throughput used in the search is calculated by using (\ref{Tlem1_Chapter4}).

In Alg.~\ref{mainalg_ove_Chapter4}, we run Alg.~\ref{mainalg_non_Chapter4} in the first phase then perform further overlapping channel assignments using Alg.~\ref{mainalg_ove_Chapter4} in the second phase.
Hence, we need to analyze the complexity involved in the second phase (i.e., Alg.~\ref{mainalg_ove_Chapter4}).
In Alg.~\ref{mainalg_ove_Chapter4}, we increase the parameter $h$ from 1 to $M-1$ over iterations of the \textit{while loop} (to increase the number of
users who share one channel). For a particular value of $h$, we search over the channels that have been shared by $h$ users and over
all $M$ users. Therefore, we have $NM$ cases to consider for each value of $h$ each of which requires to calculate the corresponding increase
of throughput using (\ref{upith_Chapter4}). Therefore, the worst case complexity of the second phase is $NM(M-1)$, which is $O(NM^2)$.
Considering the complexity of both phases, the complexity of Alg.~\ref{mainalg_ove_Chapter4} is $O(MN^2+NM^2)=O(MN(M+N))$,
which is much lower than that of the optimal brute-force search algorithm ($O(2^{NM})$).

%In conclusion, the brute-force search algorithm has an approximate number of cases in our proposed algorithm power to $N$ ($\left[2^{NM}\right] / \left[\left(M+2^M\right) N\left(N+1\right)/2\right] \approx 2^N$). In the large scale cognitive radio networks, the number of secondary users, $M$ is high, and hence the number of channels, $N$ is also high; our proposed algorithm is more robust and efficient. 

\section{Further Extensions and Design Issues}
\label{Poten_Chapter4}

\subsection{Fair Channel Assignment}
\label{Fairness_Chapter4}

We extend the channel assignment problem to consider the max-min fairness objective, which maximizes the minimum throughput achieved by all
SUs \cite{tass05}. In particular, the max-min channel assignment problem can be stated as follows:
\vspace{0.0cm}
\beqn
\label{Tput_fair_Chapter4}
&& \mathop {\max_\textbf{x}  } \mathop {\min }_{i} \: {T_i}.  %\label{obj1} 
\eeqn
Intuitively, the max-min fairness criterion tends to allocate more radio resources for ``weak'' users to balance the throughput
performance among all users. Thanks to the exact throughput analytical model developed in Section~\ref{tputana_Chapter4}, the optimal solution
of the optimization problem (\ref{Tput_fair_Chapter4}) can be found by the exhaustive search, which, however, has extremely high computational complexity.

To resolve this complexity issue, we devise greedy fair non-overlapping and overlapping channel assignment algorithms, which are 
described in Alg.~\ref{mainalgFN_Chapter4} and Alg.~\ref{mainalgFO1_Chapter4}, respectively. In Section~\ref{Results_Chap4}, we compare the performance
of these algorithms with that of the optimal exhaustive search algorithm. These algorithms are different from
Alg.~\ref{mainalg_non_Chapter4} and Alg.~\ref{mainalg_ove_Chapter4} mainly in the way we choose the user to allocate one ``best'' channel in each iteration.
In Alg.~\ref{mainalgFN_Chapter4}, we find the set of 
users who achieve a minimum throughput in each iteration. For each user in this set, we find one available channel
that results in the highest increase of throughput. Then, we assign the ``best'' channel that achieves the maximum
increase of throughput considering all throughput-minimum users. Therefore, this assignment attempts to increase the throughput
of a weak user while exploiting the multiuser diversity.

\begin{algorithm}[h]%\leesize
\caption{\textsc{Fair Non-Overlapping Channel Assignment}}
\label{mainalgFN_Chapter4}
%\algsetup{indent=1.5em}
\begin{algorithmic}[1]

\STATE Initialize SU $i$'s set of available channels, ${\mathcal{S}_i^a} := \left\{ {1,2, \ldots ,N} \right\}$ and  $\mathcal{S}_i := \emptyset$ for $i=1, 2,\ldots , M$ where $\mathcal{S}_i$ denotes the set of channels assigned for SU $i$.

\STATE $\text{continue} := 1$

\WHILE {$\text{continue} = 1$}

\STATE Find the set of users who currently have minimum throughput $\mathcal{S}^{\text{min}} = \mathop {\argmin} \limits_i  T_i^b $

where $\mathcal{S}^{\text{min}} = \left\{i_1, \ldots, i_m\right\} \subset \left\{1,\ldots,M\right\}$ is the set of minimum-throughput SUs.

\IF {$\mathop {\mathcal{OR}} \limits_{i_l \in \mathcal{S}^{\text{min}}} \left(\mathcal{S}_{i_l}^a \neq \emptyset\right)$}

\STATE For each SU $i_l \in \mathcal{S}^{\text{min}}$ and channel $j_{i_l} \in \mathcal{S}_{i_l}^a$, find $\Delta T_{i_l}(j_{i_l}) = T_{i_l}^a - T_{i_l}^b$

where $T_{i_l}^a$ and $T_{i_l}^b$ are the throughputs after and before assigning channel $j_{i_l}$; and we set $\Delta T_{i_l} = 0$ if $\mathcal{S}_{i_l}^a = \emptyset$.

\STATE $\left\{i_{l}^{*}, j_{i_{l}^*}^* \right\}= \mathop {\argmax }\limits_{i_l \in \mathcal{S}^{\text{min}}, j_{i_l} \in {\mathcal{S}_{i_l}^a}} \: \Delta {T_{i_l}}(j_{i_l})$

\STATE Assign channel $j_{i_{l}^*}^*$ to SU $i_{l}^*$.

\STATE Update $\mathcal{S}_{i_{l}^*} = \mathcal{S}_{i_{l}^*} \cup j_{i_{l}^*}^*$ and $\mathcal{S}_k^a = \mathcal{S}_k^a \backslash j_{i_{l}^*}^*$ for all $k \in \left\{1,\ldots,M\right\}$.

\ELSE

\STATE Set $\text{continue} := 0$

\ENDIF

\ENDWHILE

\end{algorithmic}
\end{algorithm}

\begin{algorithm}[h]%\leesize
\caption{\textsc{Fair Overlapping Channel Assignment}}
\label{mainalgFO1_Chapter4}
%\algsetup{indent=1.5em}
\begin{algorithmic}[1]

%\STATE Initialize the sets of allocated channels for all users  $\mathcal{S}_i := \emptyset$ for $i=1, 2,\ldots , M$ and $\delta_0$

\STATE Run Alg.~\ref{mainalgFN_Chapter4} and obtain the sets $\mathcal{S}_i$ for all SU $i$. Initialize $\mathcal{S}_i^{\text{com}}=\emptyset$ for $i$.

\STATE continue := 1.

\WHILE {$\text{continue}  = 1$}

\STATE Find  $i^* = \mathop {\argmin} \limits_{i \in \left\{1,\ldots,M \right\}} T_i^b$ and $T_{\text{min}} = T_{i^*}^b$ where ties
are broken randomly.

\STATE $\mathcal{S}_{i^*}^{\text{Sep}} = \mathop \cup \limits_{i, i \neq i^*}  \mathcal{S}_{i} $, $\mathcal{S}_{i^*}^{\text{Uni}} = \mathop \cup \limits_i  \mathcal{S}_{i}^{\text{com}} \backslash \mathcal{S}_{i^*}^{\text{com}}$.

\STATE Run Alg.~\ref{mainalgFO2_Chapter4}.

%\STATE Call Algorithm 6.

\IF {$\mathop \mathcal{OR} \limits_i  \mathcal{S}_i^{\text{com},\text{temp}} \neq \emptyset$}

\STATE Assign $\mathcal{S}_i^{\text{com}} = \mathcal{S}_i^{\text{com},\text{temp}}$ and $\mathcal{S}_i = \mathcal{S}_i^{\text{temp}}$.

\ELSE

\STATE Set $\text{continue} := 0$.

\ENDIF

\ENDWHILE

\end{algorithmic}
\end{algorithm}

\begin{algorithm}[h]%\leesize
\algsetup{linenosize=\scriptsize}
  \scriptsize %%\small \footnotesize \scriptsize  %\tiny
\caption{\textsc{Searching Potential Channel Assignment}}
\label{mainalgFO2_Chapter4}
%\algsetup{indent=1.5em}
\begin{algorithmic}[1]

%\PRINT \texttt{?Hello, World!?}

\STATE --- \textit{Search potential channel assignment from separate sets} ---
\vspace{0.2cm}
%\COMMENT{this is a comment}

\FOR {$j \in \mathcal{S}_{i*}^{\text{Sep}}$}

\STATE Find SU $i'$ where $j \in S_{i'}$. Let $n_c = M - 2$. 

\FOR {$l = 0$ to $n_c$}

\FOR {$k = 1$ to $C_{n_c}^l$}

\STATE Find $T_{i^*}^a $, $T_{i'}^a $, and $T_m^a\left|_{m \in \mathcal{U}^l_j}\right.$, where $\mathcal{U}^l_j$ is the set of $l$ new SUs sharing channel $j$.

\IF { $\min \left(T_{i^*}^a , T_m^a\left|_{m \in \mathcal{U}^l_j}\right., T_{i'}^a \right) > T_{\text{min}}$ }

\STATE - Temporarily assign channel $j$ to SUs $i^*$, $i'$ and all SUs $m$: $\mathcal{S}_{i^*}^{\text{com},\text{temp}} = \mathcal{S}_{i^*}^{\text{com}} \cup j$, $\mathcal{S}_{i'}^{\text{com},\text{temp}} = \mathcal{S}_{i'}^{\text{com}} \cup j$, $\mathcal{S}_{i'}^{\text{temp}} = \mathcal{S}_{i'} \backslash j$ and $\mathcal{S}_m^{\text{com},\text{temp}} = \mathcal{S}_m^{\text{com}} \cup j$ . 

\STATE - Update $T_{\text{min}} = \min \left(T_{i^*}^a , T_m^a\left|_{m \in \mathcal{U}^l_j}\right. , T_{i'}^a \right)$. 

\STATE - Reset all temporary sets of other SUs to be empty.

\ENDIF

\ENDFOR

\ENDFOR

\ENDFOR

\vspace{0.4cm}

\STATE --- \textit{Search potential channel assignment from common sets} ---
\vspace{0.2cm}

\FOR {$j \in \mathcal{S}_{i^*}^{\text{Uni}}$}

\STATE Find the subset of SUs except SU $i^*$, $\mathcal{S}^{\text{Use}}$ who use channel $j$ as an overlapping channel.

\FOR {$l = 0$ to $M-1-\left|\mathcal{S}^{\text{Use}}\right|$}

\FOR {$k = 1$ to $C_{M-1-\left|\mathcal{S}^{\text{Use}}\right|}^l$}

\STATE Find $T_{i^*}^a $, $T_{i'}^a\left|_{i' \in \mathcal{S}^{\text{Use}}} \right.$, $T_m^a\left|_{m \in \mathcal{U}^l_j} \right.$, where $\mathcal{U}^l_j$ is the set of $l$ new SUs sharing channel $j$.

\IF { $\min \left(T_{i^*}^a , T_{i'}^a\left|_{i' \in \mathcal{S}^{\text{Use}}} \right., T_m^a\left|_{m \in \mathcal{U}^l_j} \right. \right) > T_{\text{min}}$}

\STATE - Temporarily assign channel $j$ to SU $i^*$, all SUs $i'$ and all SUs $m$: $\mathcal{S}_{i^*}^{\text{com},\text{temp}} = \mathcal{S}_{i^*}^{\text{com}} \cup j $, $\mathcal{S}_m^{\text{com},\text{temp}} = \mathcal{S}_m^{\text{com}} \cup j$. 

\STATE - Update $T_{\text{min}} = \min \left(T_{i^*}^a , T_{i'}^a\left|_{i' \in \mathcal{S}^{\text{Use}}} \right., T_m^a\left|_{m \in \mathcal{U}^l_j} \right. \right)$. 

\STATE - Reset all temporary sets of other SUs to be empty.

\ENDIF

\ENDFOR

\ENDFOR

\ENDFOR

\end{algorithmic}
\end{algorithm}

In Alg.~\ref{mainalgFO1_Chapter4}, we first run Alg.~\ref{mainalgFN_Chapter4} to obtain non-overlapping sets of channels for all users. Then, we seek to improve
the minimum throughput by performing overlapping channel assignments. 
In particular, we find the minimum-throughput user and an overlapping channel assignment that results in the largest increase in its throughput.
The algorithm terminates when there is no such overlapping channel assignment. The search of an overlapping channel assignment
in each iteration of Alg.~\ref{mainalgFO1_Chapter4} is performed in Alg.~\ref{mainalgFO2_Chapter4}. Specifically, we sequentially search over channels which have already been
allocated for a single user or shared by several users (i.e., channels in separate and common sets, respectively). Then, we update the
current temporary assignment with a better one (if any) during the search. This search requires throughput calculations for which
we use the analytical model developed in Section \ref{tputana_Chapter4} with the MAC protocol overhead, $\delta<1$ derived in Section~\ref{Overcal_Chapter4}. 
It can be observed that the proposed throughput analysis is very useful since it can be used to evaluate the performance of any
channel assignment solution and to perform channel assignments in greedy algorithms.

\subsection{Throughput Analysis under Imperfect Sensing}
\label{Imper_sens_Chapter4}

We extend the throughput analysis considering imperfect sensing.
The following two important performance measures are used to quantify the sensing performance: 1) detection probabilities and 2) false-alarm probabilities.
Let $\mathcal{{P}}_d^{ij}$ and $\mathcal{{P}}_f^{ij}$ be detection and false alarm probabilities, respectively, of SU $i$ on channel $j$. 
In particular, detection event occurs when a secondary link successfully senses a busy channel and false alarm
represents the situation when a spectrum sensor returns a busy state for an idle channel (i.e., a transmission opportunity
is overlooked). Also, let us define $\mathcal{{P}}_d^{ij} = 1-\mathcal{\overline{P}}_d^{ij}$ and $\mathcal{\overline{P}}_f^{ij} =  1- \mathcal{{P}}_f^{ij}$. Under imperfect sensing,  the following four scenarios are possible for channel $j$ and SU $i$.

\begin{itemize}
\item {Scenario I}: A spectrum sensor indicates that channel $j$ is available and the nearby PU is not using channel $j$ (i.e., correct
sensing). This scenario occurs with the probability $\mathcal{\overline{P}}_f^{ij}p_{ij}$.

\item {Scenario  II}: A spectrum sensor indicates that channel $j$ is available and the nearby PU is using channel $j$ (i.e., mis-detection). This scenario occurs with the probability $\mathcal{\overline{P}}_d^{ij} \overline{p}_{ij}$. In this case, potential transmission of secondary user $i$ will collide with that of the nearby primary user. We assume that both transmissions from SU $i$ and the nearby PU fail.

\item {Scenario  III}: A spectrum sensor indicates that channel $j$ is not available and the nearby PU is using channel $j$ (i.e., correct detection). 
This scenario occurs with the probability $\mathcal{P}_d^{ij} \overline{p}_{ij}$.

\item {Scenario IV}: A spectrum sensor indicates that channel $j$ is not available and the nearby PU is not using channel $j$ (i.e., false alarm). 
 This scenario occurs with the probability $\mathcal{P}_f^{ij} p_{ij}$ and the channel opportunity is overlooked.

\end{itemize}

Because SUs make channel access decisions based on their sensing outcomes, the first two scenarios can result in spectrum access on channel 
$j$ by SU $i$. Moreover, spectrum access in scenario one actually lead to successful data transmission. 
Let us define $\mathcal{P}_{\text{idle}}^{ij}=\mathcal{\overline{P}}_f^{ij}p_{ij}+\mathcal{\overline{P}}_d^{ij}\overline{p}_{ij}$ and $\mathcal{P}_{\text{busy}}^{ij}=1-\mathcal{P}_{\text{idle}}^{ij}$ as the probabilities under which SU $i$ may and may not access channel 
$j$, respectively. 
The same synchronized MAC protocol described in Section~\ref{over_MACPro_Chapter4} is assumed here.
In addition, the MAC protocol overhead can be calculated as presented in Section~ \ref{Overcal_Chapter4} where the contention window $W$ 
is determined as described in Section \ref{ConWinCal_Chapter4}. However, $p_{ij}$ and $\overline{p}_{ij}$ are substituted by 
$\mathcal{P}_{\text{idle}}^{ij}$ and $\mathcal{P}_{\text{busy}}^{ij}$, respectively in the calculation of contention window
in Section \ref{ConWinCal_Chapter4} for this case. This is because $\mathcal{P}_{\text{idle}}^{ij}$ and $\mathcal{P}_{\text{busy}}^{ij}$
capture the probabilities that channel $j$ is available and busy for user $i$ as indicated by sensing, respectively considering potential
sensing errors. Because the total throughput is the sum of throughput of all users, it is sufficient to analyze the throughput of one
particular user $i$. To analyze the throughput of user $i$, we consider the following cases.
\begin{itemize}
\item{\NoIndent{ Case 1: At least one channel in $\mathcal{S}_i$ is available and user $i$ chooses one of these available channels for its transmission. User $i$ can achieve throughput of one in such a successful access, which occurs with the following probability:
\beqn
T_i \left\{\text{Case 1} \right\}  = \Pr\left\{\text{Case 1} \right\} = \sum \limits_{k_1=1}^{\left| \mathcal{S}_i\right|} \sum \limits_{l_1=1}^{C_{\left| \mathcal{S}_i\right|}^{k_1}} 
\prod \limits_{j_1 \in \mathcal{S}_i^{l_1}} p_{ij_1} \prod \limits_{j_2 \in \mathcal{S}_i \backslash \mathcal{S}_i^{l_1}} \overline{p}_{ij_2} \label{Case1_1_Chapter4} \hspace{4.0cm} \\
\sum \limits_{k_2=1}^{k_1} \sum \limits_{l_2=1}^{C_{k_1}^{k_2}} 
\prod \limits_{j_3 \in \mathcal{S}_i^{l_2}} \mathcal{\overline{P}}_f^{ij_3} \prod \limits_{j_4 \in \mathcal{S}_i^{l_1} \backslash \mathcal{S}_i^{l_2}} \mathcal{P}_f^{ij_4} \label{Case1_2_Chapter4} \hspace{3.0cm}\\
\sum \limits_{k_3 = 0}^{\left|\mathcal{S}_i\right|-k_1} \sum \limits_{l_3=1}^{C_{\left|\mathcal{S}_i\right|-k_1}^{k_3}} \frac{k_2}{k_2+k_3} \prod \limits_{j_5 \in \mathcal{S}_i^{l_3}} \mathcal{\overline{P}}_d^{ij_5} \prod \limits_{j_6 \in \mathcal{S}_i \backslash \mathcal{S}_i^{l_1} \backslash \mathcal{S}_i^{l_3}} \mathcal{P}_d^{ij_6}. \label{Case1_3_Chapter4} \hspace{0.0cm}
\eeqn 
where we have $T_i \left\{\text{Case 1} \right\}  =  \Pr\left\{\text{Case 1} \right\}$. The quantity (\ref{Case1_1_Chapter4}) represents the
probability that there are $k_1$ actually available channels in $\mathcal{S}_i$ (which may or may not be correctly sensed by SU $i$). Here, $\mathcal{S}_i^{l_1}$ denotes a particular set of $k_1$ actually available channels whose index is $l_1$.
In addition, the quantity (\ref{Case1_2_Chapter4}) describes the
probability that there are $k_2$ available channels as indicated by sensing (the remaining available channels are overlooked due to
sensing errors) where $\mathcal{S}_i^{l_2}$ denotes the $l_2$-th set with $k_2$ available channels. For the quantity in (\ref{Case1_3_Chapter4}),
$k_3$ denotes the number of channels that are not actually available but the sensing outcomes indicate they are available (i.e., due to
mis-detection). Moreover, $k_2/(k_2+k_3)$ represents the probability that SU $i$ chooses the actually available channel for transmission
given its sensing outcomes indicate $k_2+k_3$ available channels. The remaining quantity in (\ref{Case1_3_Chapter4}) describes the probability
that the sensing outcomes due to SU $i$ incorrectly indicates $k_3$ available channels. }}

\item{\NoIndent{$\text{Case 2}$:  All channels in $\mathcal{S}_i$ are indicated as not available by sensing; there is at least one channel in $\mathcal{S}_i^{\text{ com}}$ indicated as available by sensing, and user $i$ chooses an actually available channel $j$ for transmission. 
Suppose that channel $j$ is shared  by $\mathcal{MS}_j$ secondary users including user $i$ (i.e., $\mathcal{MS}_j = |\mathcal{U}_j|$). 
There are four possible groups of users $i_k$, $k=1, \ldots, \mathcal{MS}_j$ sharing channel $j$, which are described in
the following
\begin{itemize}
\item{ \textbf{Group I}: channel $j$ is available for user $i_k$ and user $i_k$ has at least 1 channel in $\mathcal{S}_{i_k}$ available as indicated by sensing.}
\item{ \textbf{Group II}: channel $j$ is indicated as not available for user $i_k$ by sensing. }
\item{ \textbf{Group III}: channel $j$ is available for user $i_k$, all channels in $\mathcal{S}_{i_k}$ are not available and there is another channel
 $j'$ in $\mathcal{S}_{i_k}^{\text{com}}$ available for user $i_k$ as indicated by sensing. In addition, user $i_k$ chooses channel $j'$ for transmission in the contention stage.}
\item{ \textbf{Group IV}: channel $j$ is available for user $i_k$, all channels in $\mathcal{S}_{i_k}$ are not available as indicated by sensing. In addition, user $i_k$  chooses channel $j$ for transmission in the contention stage. Hence, user $i_k$ competes with user $i$ for channel $j$.}

\end{itemize}

The throughput that was achieved by user $i$ in this case can be written as
\beqn
\label{Tputa1delta1_Chapter4}
T_i\left( \text{ Case 3} \right) = %\hspace{3.4cm} \nonumber \\
 (1-\delta) \Theta_i \sum \limits_{A_1 = 0}^{\mathcal{MS}_j} \sum \limits_{A_2 = 0}^{\mathcal{MS}_j-A_1} \sum \limits_{A_3=0}^{\mathcal{MS}_j-A_1-A_2} %\nonumber \\
 \Phi_1(A_1) \Phi_2(A_2)  \Phi_3(A_3) \Phi_4(A_4).
 %\hspace{4cm} 
\eeqn
Here, we use the same notations as in the perfect sensing scenario investigated in Section~\ref{tputana_Chapter4} where the following conditions hold.
\begin{itemize}
\item $\Theta_i$ is the probability that all channels in $\mathcal{S}_i$ are indicated as not available by sensing and user $i$ chooses 
some available channel $j$ in $\mathcal{S}_i^{\text{com}}$ as indicated by sensing for transmission.
\item $\Phi_1(A_1)$ denotes the probability that there are $A_1$ users belonging to Group I described above
among $\mathcal{MS}_j$ users sharing channel $j$.
\item $\Phi_2(A_2)$ represents the probability that there are $A_2$ users belonging to Group II 
among $\mathcal{MS}_j$ users sharing channel $j$.
\item $\Phi_3(A_3)$ describes the probability that there are $A_3$ users belonging to Group III
among $\mathcal{MS}_j$ users sharing channel $j$.
\item $\Phi_4(A_4)$ denotes the probability that there are $A_4 = \mathcal{MS}_j-A_1-A_2-A_3$ 
remaining users belonging to Group IV  scaled by $1/(1+A_4)$ where $A_4$ is the number of users 
excluding user $i$ competing with user $i$ for channel $j$.
\end{itemize}
We now proceed to calculate these quantities. We have
\beqn \label{group1_0_Chapter4}
\Theta_i =  \sum \limits_{k_1=0}^{\left| \mathcal{S}_i\right|} \sum \limits_{l_1=1}^{C_{\left| \mathcal{S}_i\right|}^{k_1}} 
\prod \limits_{j_1 \in \mathcal{S}_i^{l_1}} \mathcal{P}_f^{ij_1} p_{ij_1} \prod \limits_{j_2 \in \mathcal{S}_i \backslash \mathcal{S}_i^{l_1}} \overline{p}_{ij_2} \label{group1_00_Chapter4} \hspace{6.0cm} \\
\sum \limits _{k_2 =1}^{H_i} \sum \limits_{l_2=1}^{C_{H_i}^{k_2}} 
\prod_{j_3 \in \Psi^{l_2}_i } p_{ij_3} \prod_{j_4 \in \mathcal{S}_i^{\text{com}} \backslash \Psi^{l_2}_i} \overline{p}_{ij_4} \label{group1_01_Chapter4} \hspace{5.0cm} \\
\sum \limits_{k_3=1}^{k_2} \sum \limits_{l_3=1}^{C_{k_2}^{k_3}}  \sum \limits_{j \in \Gamma_1^k} 
\prod \limits_{j_5 \in \Gamma_1^{l_3}} \mathcal{\overline{P}}_f^{ij_5} \prod \limits_{j_6 \in \Psi^{l_2}_i \backslash \Gamma_i^{l_3}} \mathcal{P}_f^{ij_6} \label{group1_02_Chapter4} \hspace{4.0cm} \\
\sum \limits_{k_4 = 0}^{H_i-k_2} \sum \limits_{l_4=1}^{C_{H_i-k_2}^{k_4}} \frac{1}{k_3+k_4} \prod \limits_{j_7 \in \Gamma_2^{l_4}} \mathcal{\overline{P}}_d^{ij_7} \prod \limits_{j_8 \in \mathcal{S}_i^{\text{com}} \backslash \Psi_i^{l_2} \backslash \Gamma_2^{l_4}} \mathcal{P}_d^{ij_8} \label{group1_03_Chapter4} \hspace{0.5cm} 
\eeqn
where $H_i$ denotes the number of channels in $\mathcal{S}_i^{\text{com}}$. The quantity in (\ref{group1_00_Chapter4}) is the probability that
all available channels in $\mathcal{S}_i$ (if any) are overlooked by user $i$ due to false alarms. Therefore, user $i$ does not access
any channels in $\mathcal{S}_i$. The quantity in (\ref{group1_01_Chapter4}) describes the
probability that there are $k_2$ actually available channels in $\mathcal{S}_i^{\text{com}}$ and
 $\Psi_i^{l_2}$ denotes such a typical set with $k_2$ available channels.
The quantity in (\ref{group1_02_Chapter4}) describes the
probability that user $i$ correctly detects $k_3$ channels out of $k_2$ available channels.
The last quantity in (\ref{group1_03_Chapter4}) excluding the factor ${1}/{(k_3+k_4)}$
denotes the probability that user $i$ mis-detects $k_4$ channels among the remaining $H_i-k_2$ busy channels in $\mathcal{S}_i^{\text{com}}$.
Finally, the factor  ${1}/{(k_3+k_4)}$ is the probability that user $i$ correctly chooses one available channels in $\mathcal{S}_i^{\text{com}}$
for transmission out of $k_3+k_4$ channels which are indicated as being available by sensing, i.e.,
\beqn
\Phi_1(A_1) = \sum \limits_{c_1=1}^{C_{\mathcal{MS}_j}^{A_1}} \prod \limits_{m_1 \in  \Omega_{c_1}^{(1)}  } \left(\mathcal{P}_{\text{idle}}^{m_1j} \left(1-\prod \limits_{l \in \mathcal{S}_{m_1}}\mathcal{P}_{\text{busy}}^{m_1 l}\right)\right).   \label{group1_I_Chapter4}
\eeqn
In (\ref{group1_I_Chapter4}), we consider all possible subsets of users of size $A_1$ that belongs to Group I (there are $C_{\mathcal{MS}_j}^{A_1}$ such subsets).
Each term inside the sum represents the probability of the corresponding event whose set of $A_1$ users is denoted by $\Omega_{c_1}^{(1)}$, i.e.,
\beqn
\Phi_2(A_2) = \sum \limits_{c_2=1}^{C_{\mathcal{MS}_j-A_1}^{A_2}} \prod \limits_{m_2 \in \Omega_{c_2}^{(2)}  }  \mathcal{P}_{\text{busy}}^{m_2j}. \label{group1_II_Chapter4} 
\eeqn
In (\ref{group1_II_Chapter4}), we capture the probability that channel $j$ is indicated as not being available by sensing for $A_2$ users in group II. Possible sets of
these users are denoted by $\Omega_{c_2}^{(2)}$, i.e., 
\beqn
\Phi_3(A_3) = \sum \limits_{c_3=1}^{C_{\mathcal{MS}_j-A_1-A_2}^{A_3}} \prod \limits_{m_3 \in \Omega_{c_3}^{(3)}  } \left(\mathcal{P}_{\text{idle}}^{m_3j} \prod_{l_3 \in \mathcal{S}_{m_3}} \mathcal{P}_{\text{busy}}^{m_3l_3}\right)  \label{1group1_III_Chapter4} \hspace{3.0cm} \\ 
\hspace{0cm} \times   \left[  \sum \limits_{n=0}^{\beta} \sum_{q=1}^{C_{\beta}^n} 
\prod_{h_1 \in \mathcal{S}^{\text{com},q}_{j,m_3} }  \mathcal{P}_{\text{idle}}^{m_3 h_1} 
\prod_{h_2 \in \overline{\mathcal{S}}^{\text{com},q}_{j,m_3} }  \mathcal{P}_{\text{busy}}^{m_3 h_2}  %\right. \nonumber \\   \left.  \times
  \left( 1- \frac{1}{ n+1} \right) \right]. \label{2group1_III_Chapter4} 
\eeqn
For each term in (\ref{1group1_III_Chapter4}) we consider different possible subsets of $A_3$ users, which are denoted by $\Omega_{c_3}^{(3)}$.
Then, each term in (\ref{1group1_III_Chapter4})  represents the probability that channel $j$ is indicated as available by sensing for each user $m_3 \in \Omega_{c_3}^{(3)}$ 
while all channels in  $\mathcal{S}_{m_3}$ are indicated as not available by sensing. 
In (\ref{2group1_III_Chapter4}), we consider all possible sensing outcomes for channels in $\mathcal{S}_{m_3}^{\text{com}}$ performed by user
 $m_3 \in \Omega_{c_3}^{(3)}$.
In addition, let $\mathcal{S}^{\text{com}}_{j,m_3} = \mathcal{S}_{m_3}^{\text{com}} \backslash \left\{ j \right\}$  
and $\beta = |\mathcal{S}^{\text{com}}_{j,m_3}|$.
Then, in (\ref{2group1_III_Chapter4}) we consider all possible scenarios in which $n$ channels in $\mathcal{S}^{\text{com}}_{j,m_3}$ are indicated as
available by sensing; and user $m_3$ chooses a channel different from channel $j$ for transmission (with probability $\left( 1- \frac{1}{ n+1} \right)$) where
$\mathcal{S}^{\text{com}}_{j,m_3} = \mathcal{S}^{\text{com},q}_{j,m_3} \cup \overline{\mathcal{S}}^{\text{com},q}_{j,m_3} $ and
$\mathcal{S}^{\text{com},q}_{j,m_3} \cap \overline{\mathcal{S}}^{\text{com},q}_{j,m_3} = \emptyset$. We have
\beqn
\Phi_4(A_4) =  \left(\frac{1}{1+ A_4}\right) \prod \limits_{m_4 \in \Omega^{(4)} } \left(\mathcal{P}_{\text{idle}}^{m_4j} \prod_{l_4 \in \mathcal{S}_{m_4}} \mathcal{P}_{\text{busy}}^{m_4l_4}\right)  \label{1group1_IV_Chapter4}   \hspace{3cm}\\
\times \left[ \sum \limits_{m=0}^{\gamma} \sum_{q=1}^{C_{\gamma}^{m}}   %\right.
\prod_{h_1 \in \mathcal{S}^{\text{com},q}_{j,m_4} }  \mathcal{P}_{\text{idle}}^{m_4 h_1} 
\prod_{h_2 \in \overline{\mathcal{S}}^{\text{com},q}_{j,m_4} }  \mathcal{P}_{\text{busy}}^{m_4 h_2}  % \right. \nonumber \\  \left.  
\left( \frac{1}{ m+1} \right)  \right].
 \label{2group1_IV_Chapter4} 
\eeqn
The sensing outcomes captured in (\ref{1group1_IV_Chapter4}) and (\ref{2group1_IV_Chapter4}) are similar to those in (\ref{1group1_III_Chapter4}) and (\ref{2group1_III_Chapter4}).
However, given three sets of $A_1$, $A_2$, and $A_3$ users, the set $\Omega^{(4)}$ can be determined whose size is $|\Omega^{(4)}| = A_4$. 
Here, $\gamma$ denotes cardinality of the set $\mathcal{S}_{j,m_4}^{\text{com}} = \mathcal{S}_{m_4}^{\text{com}} \backslash \left\{ j \right\}$.
Other sets are similar to those in (\ref{1group1_III_Chapter4}) and (\ref{2group1_III_Chapter4}). However, all users in $\Omega^{(4)}$ choose channel $j$ for transmission
in this case. Therefore, user $i$ wins the contention with probability $1/(1+A_4)$ and its achievable throughput is $(1-\delta)/(1+A_4)$.

}}
\end{itemize}

Summarize all considered cases, the throughput achieved by user $i$ is written as
\beqn
T_i = T_i \left\{\text{Case 1} \right\} + T_i \left\{ \text{Case 3} \right\} .
\eeqn
In addition, the total throughput $\mathcal{T}$ can be calculated by summing the throughputs of all SUs.

\subsection{Congestion of Control Channel}

Under our design, contention on the control channel is mild if the number of channels $N$ is relatively large compared to the number of SUs $M$.
In particular, there is no need to employ a MAC protocol if we have $N>>M$ since distinct sets of channels can be allocated for SUs
by using Alg.~\ref{mainalg_non_Chapter4}. In contrast, if the number of channels $N$ is small compared to the
number of SUs $M$ then the control channel may experience congestion due to excessive control message exchanges. The congestion
of the control channel in such scenarios can be alleviated if we allow RTS/CTS messages to be exchanged in parallel on several channels 
(i.e., multiple rendezvous \cite{mo08}).

We describe potential design of a multiple-rendezvous
MAC protocol in the following using similar ideas of a multi-channel MAC protocol (McMAC) 
in \cite{mo08, su081}. We assume that each SU hops through all channels by following a particular hopping pattern, which corresponds to 
a unique seed \cite{mo08}. In addition, each SU puts its seed in every packets so that neighboring SUs can learn its hopping pattern.
The same cycle structure as being described in Section~\ref{over_MACPro_Chapter4} is employed here. Suppose SU A wishes to transmit data SU B in a particular
cycle. Then, SU A turns to the current channel of B and senses this channel as well as its assigned channels in $\mathcal{S}_{\sf AB}^{\sf tot}$,
which is the set of allocated channels for link $AB$. If SU A's sensing outcomes indicate that the current channel of SU B is available then SU A sends RTS/CTS messages with SU B containing a chosen available communication channel. Otherwise, SU A waits until the next cycle to perform sensing and contention again. If the handshake is successful, SU A transmits data to SU B on the chosen channel in the data phase. Upon completing
data transmission, both SUs A and B return to their home hopping patterns.
In general, collisions 
among SUs are less frequent under a multiple-rendezvous MAC protocol since contentions can occur in parallel on different channels. 

%It is worth noting that the throughput analysis performed in Section VI.A and Section VII.B is still valid here except that 
%we have to derive the protocol overhead and choose an appropriate contention window under this new design. In general, collisions 
%among SUs are less frequent under
%a multiple-rendezvous MAC protocol since contentions can occur in parallel on different channels. As suggested by  \cite{mo08},
%it would not be possible to design a multi-channel MAC protocol that can work efficiently in all different scenarios.  
%Discussions on different potential designs of a multi-channel MAC protocol and their corresponding pros/cons can be found in \cite{mo08}
%and the references therein. We would like to emphasize that the focus of this
%paper is on the channel assignment issue; therefore, consideration of alternative designs of a MAC protocol is beyond its scope. 

%\vspace{10pt}
\section{Numerical Results}
\label{Results_Chap4}

\begin{figure}[!t]
\centering
\includegraphics[width=90mm]{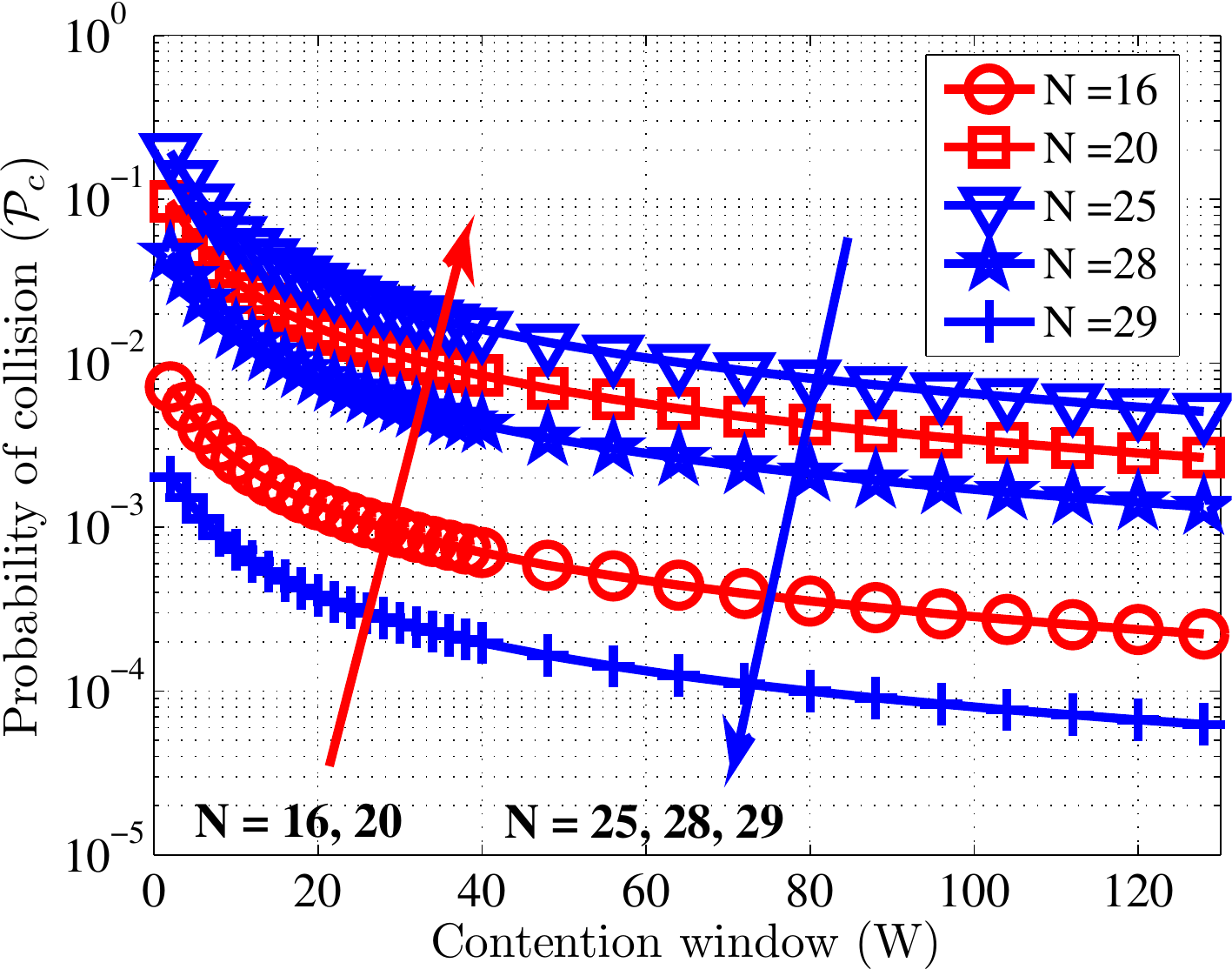}
\caption{Collision probability versus the contention window (for $M$ = 15).}
\label{het_P_M15_Aware_Blind}
\end{figure}

\begin{figure}[!t]
\centering
\includegraphics[width=90mm]{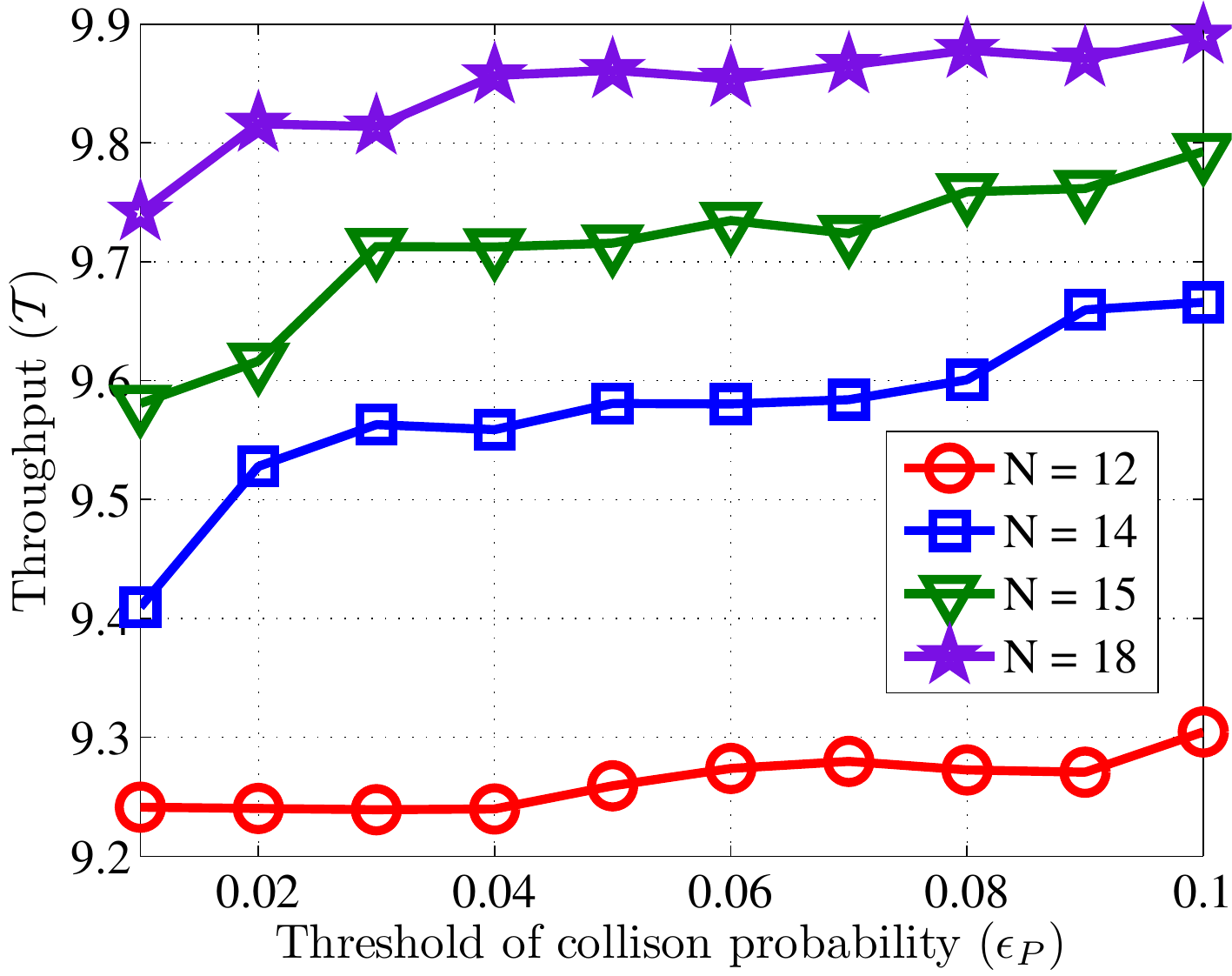}
\caption{Total throughput versus target collision probability under throughput maximization design (for $M$ = 10)}
\label{T_Pc_M10_N121518}
\end{figure}

\begin{figure}[!t]
\centering
\includegraphics[width=90mm]{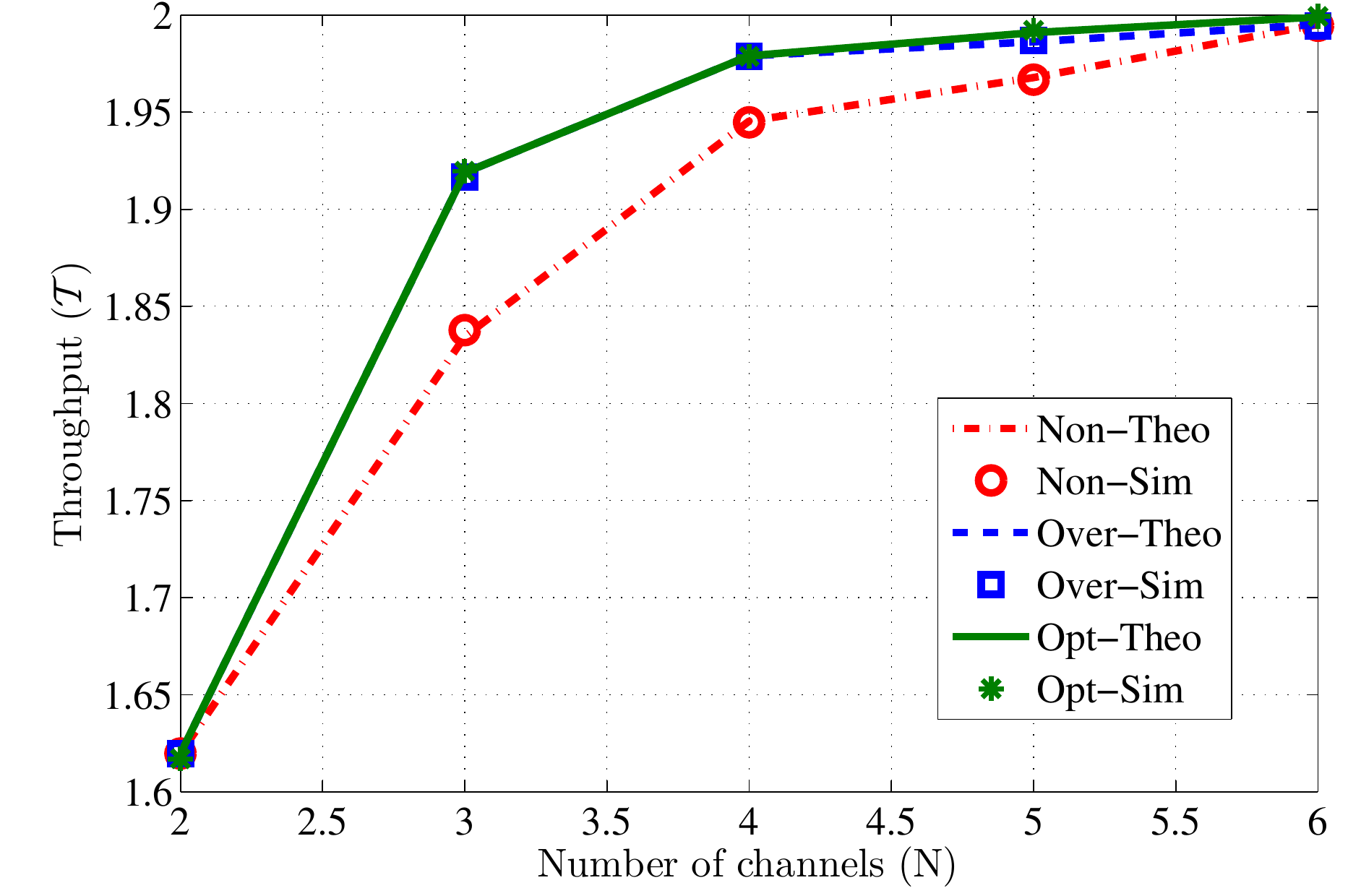}
\caption{Total throughput versus the number of channels under throughput maximization design (for $M $ = 2, Theo: Theory, Sim: Simulation, Over: Overlapping, Non: Non-overlapping, Opt: Optimal assignment).}
\label{M_2_Opt_Comparison}
\end{figure}

\begin{figure}[!t]
\centering
\includegraphics[width=90mm]{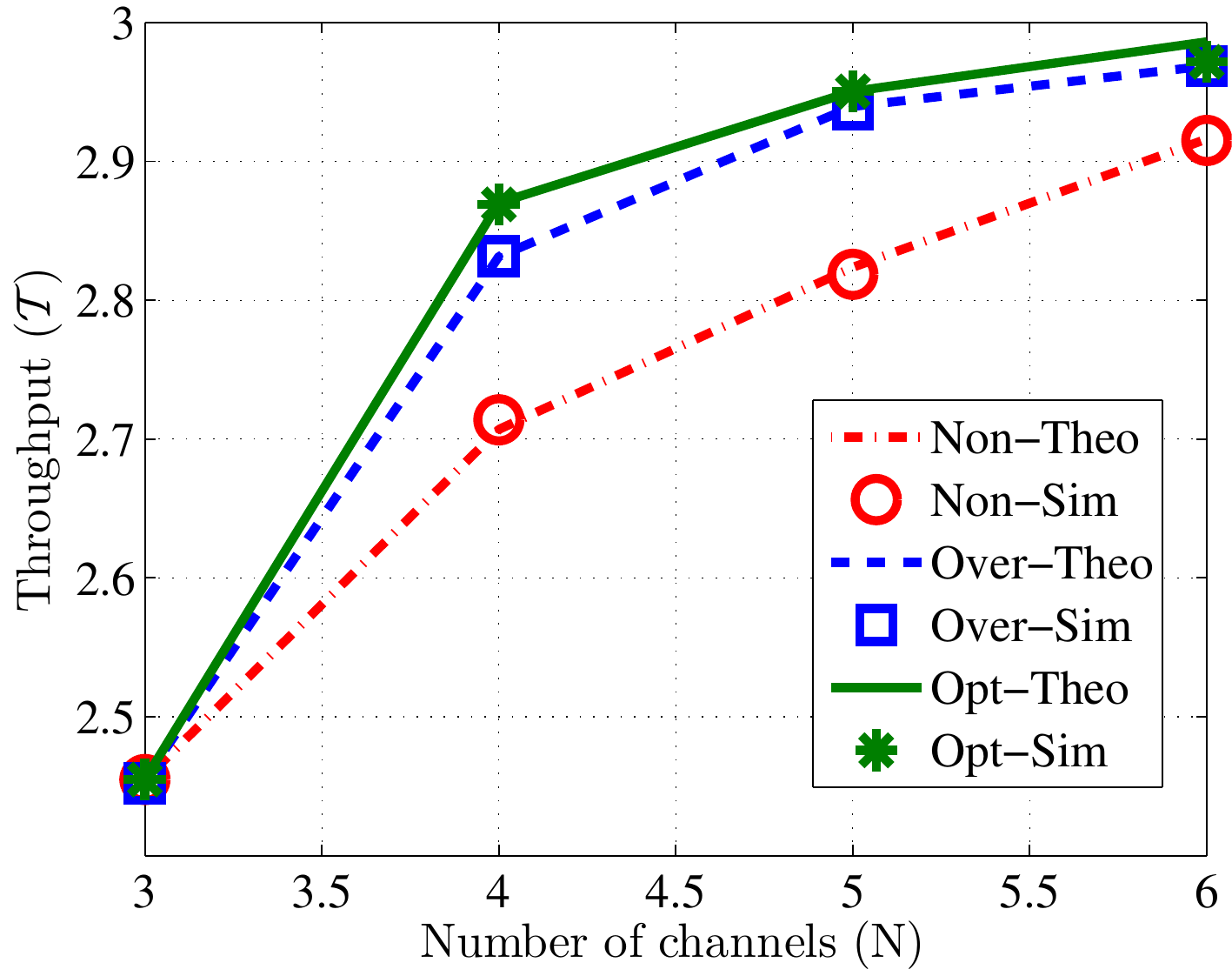}
\caption{Total throughput versus the number of channels under throughput maximization design (for $M $ = 3, Theo: Theory, Sim: Simulation, Over: Overlapping, Non: Non-overlapping, Opt: Optimal assignment).}
\label{M_3_Opt_Comparison}
\end{figure}

\begin{figure}[!t]
\centering
\includegraphics[width=90mm]{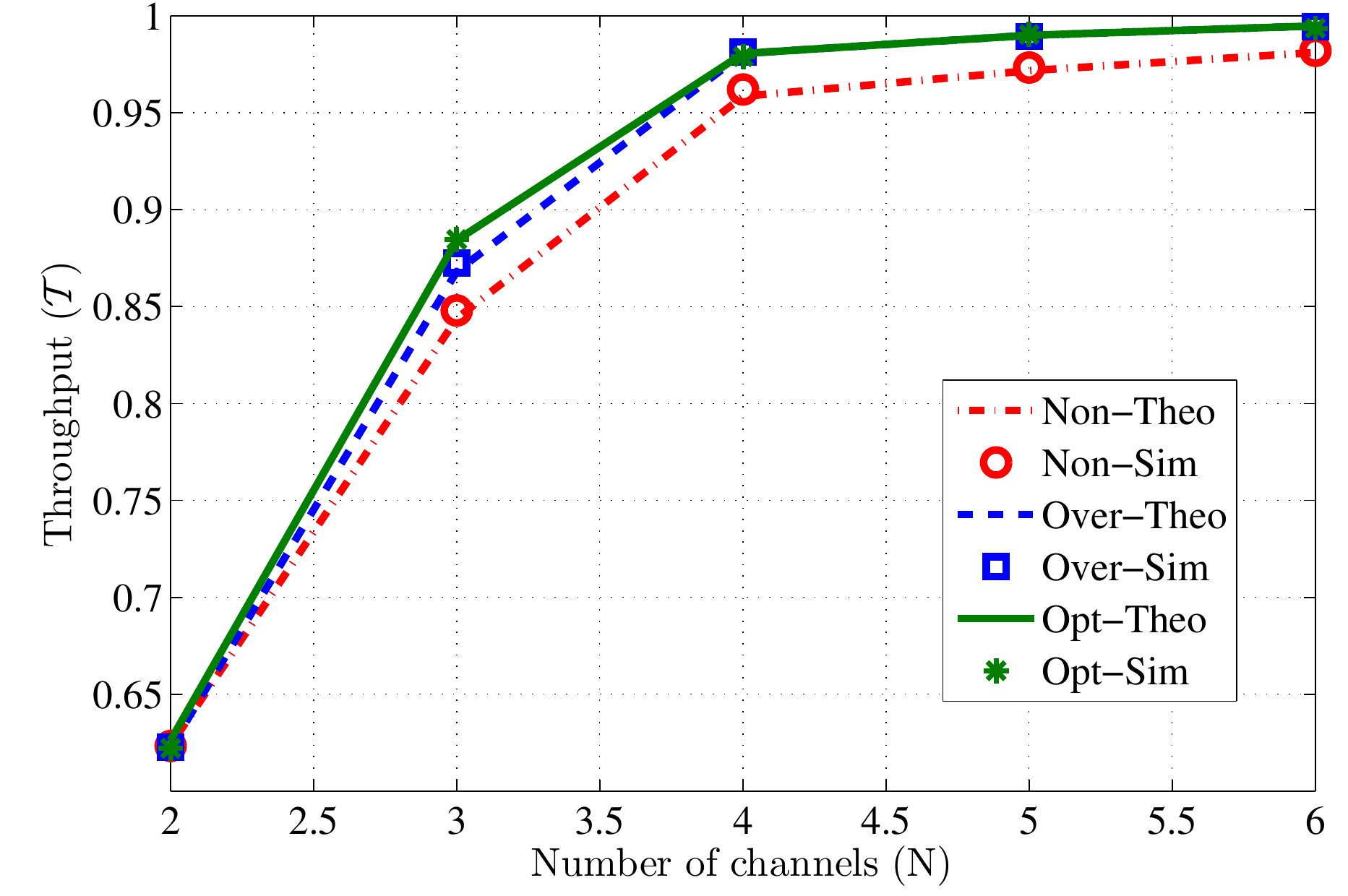}
\caption{Minimum throughput versus the number of channels under max-min fairness  (for $ M$ = 2, Theo: Theory, Sim: Simulation, Over: Overlapping, Non: Non-overlapping, Opt: Optimal assignment).}
\label{M_2_fairness_optimal_compare}
\end{figure}

\begin{figure}[!t]
\centering
\includegraphics[width=90mm]{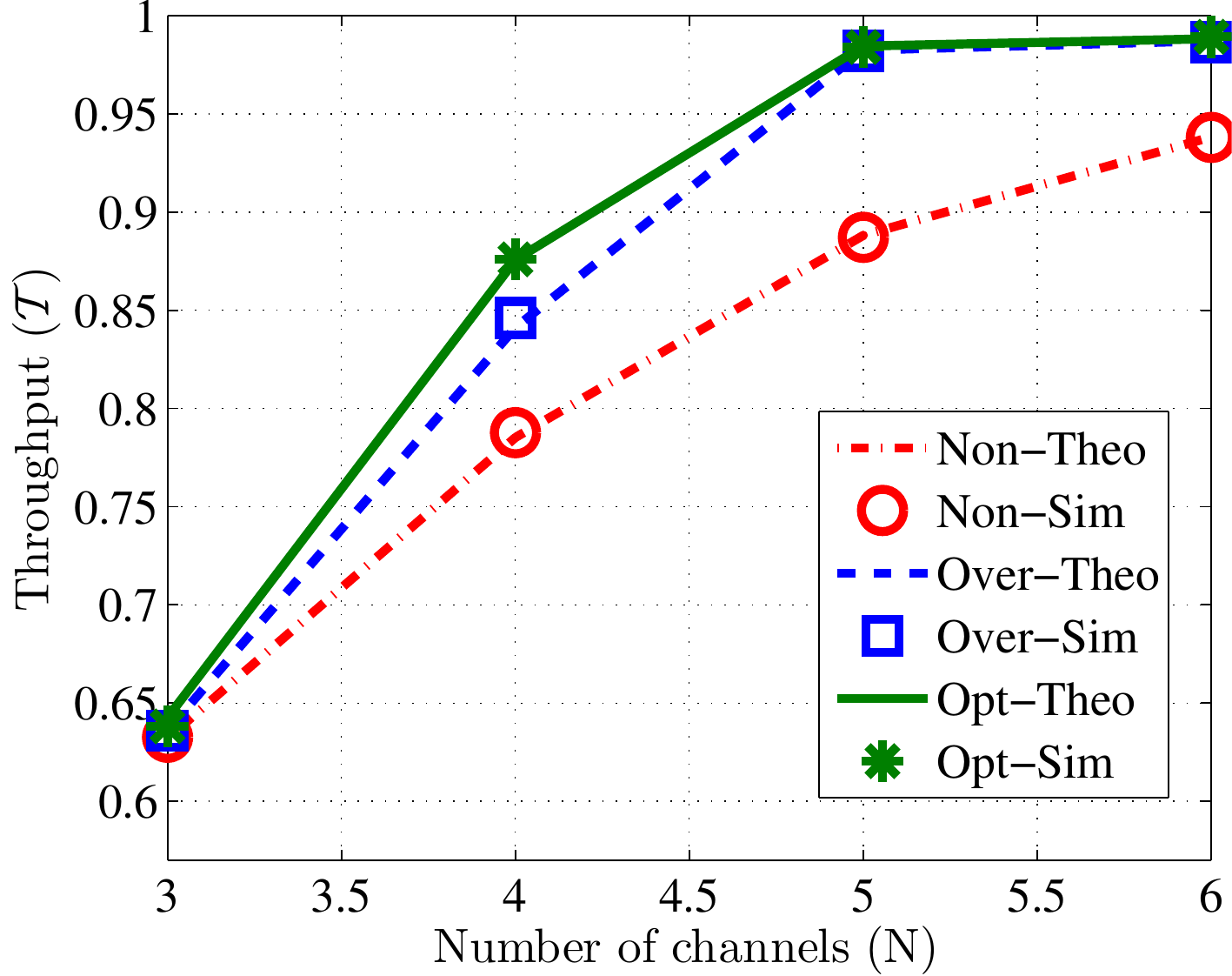}
\caption{Minimum throughput versus the number of channels under max-min fairness (for $M$ = 3, Theo: Theory, Sim: Simulation, Over: Overlapping, Non: Non-overlapping, Opt: Optimal assignment).}
\label{M_3_fairness_optimal_compare}
\end{figure}

\begin{figure}[!t]
\centering
\includegraphics[width=90mm]{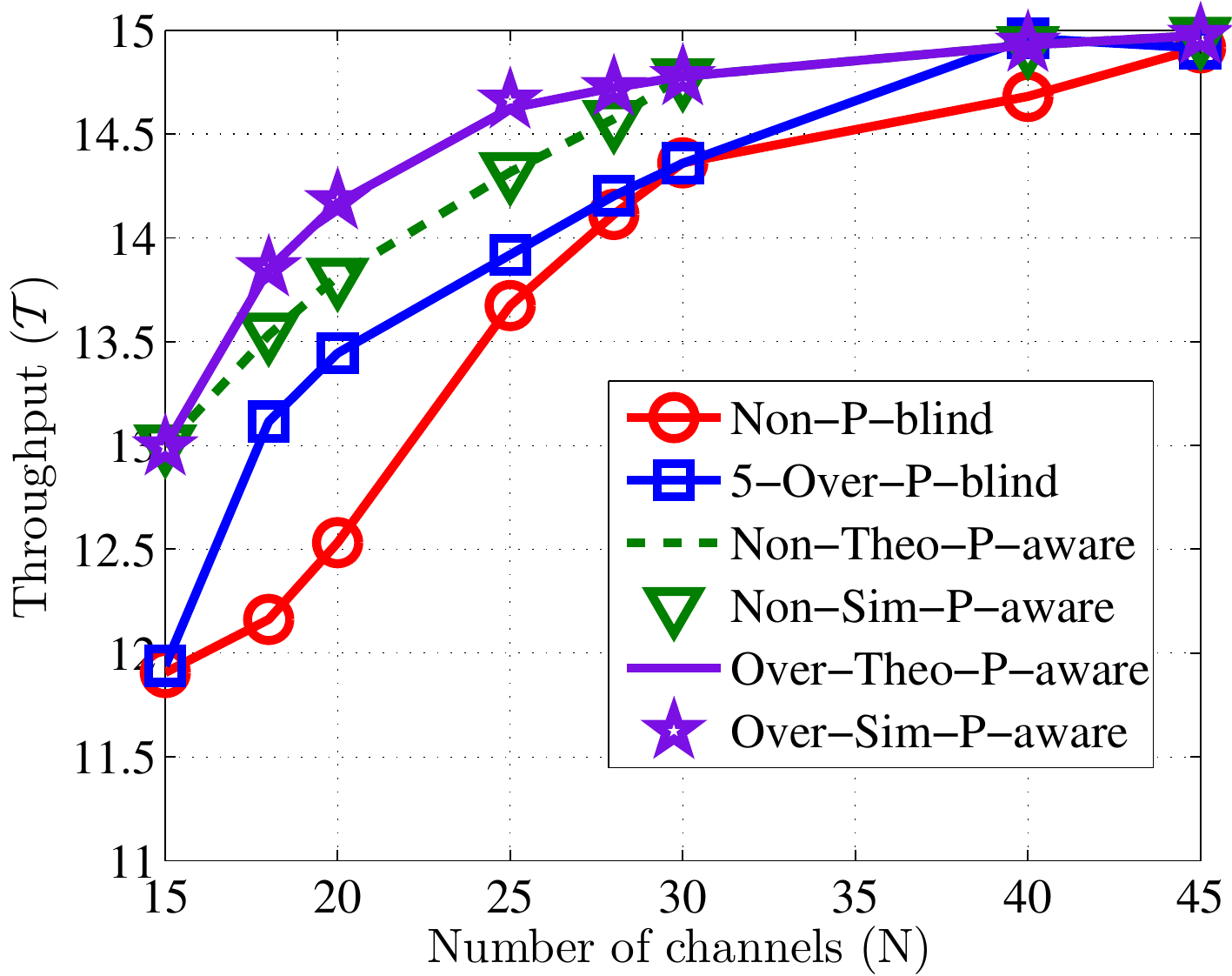}
\caption{Total throughput versus the number of channels under throughput maximization design (for $M$ = 15, Theo: Theory, Sim: Simulation, Over: Overlapping, Non: Non-overlapping, 5-Over: 5-user sharing Overlapping )}
 \label{het_T_M10_Aware_Blind1}
\end{figure}

\begin{figure}[!t]
\centering
\includegraphics[width=90mm]{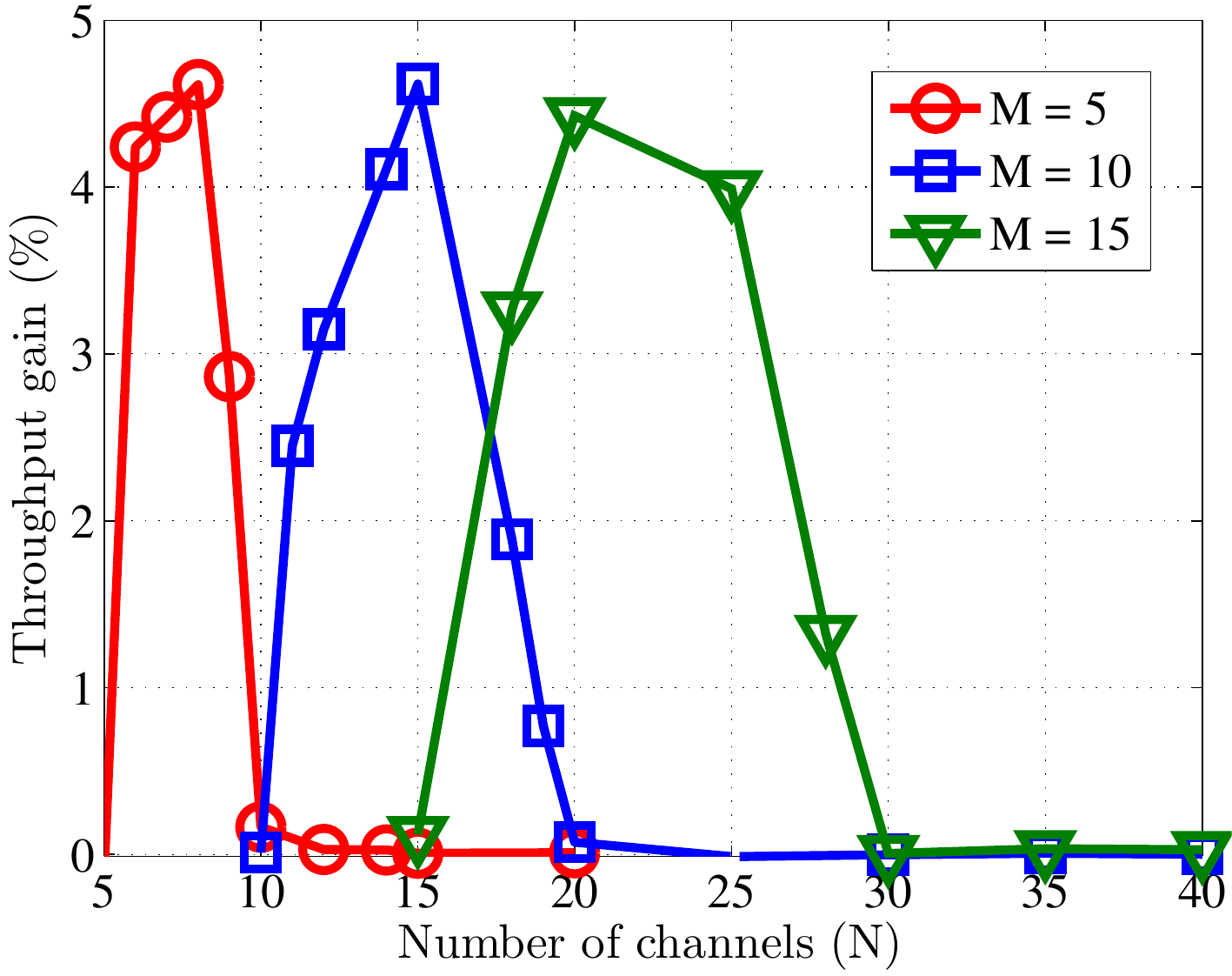}
\caption{Throughput gain between Alg.~\ref{mainalg_ove_Chapter4} and Alg.~\ref{mainalg_non_Chapter4} versus the number of channels}
\label{het_G_M_351015}
\end{figure}

\begin{figure}[!t]
\centering
\includegraphics[width=90mm]{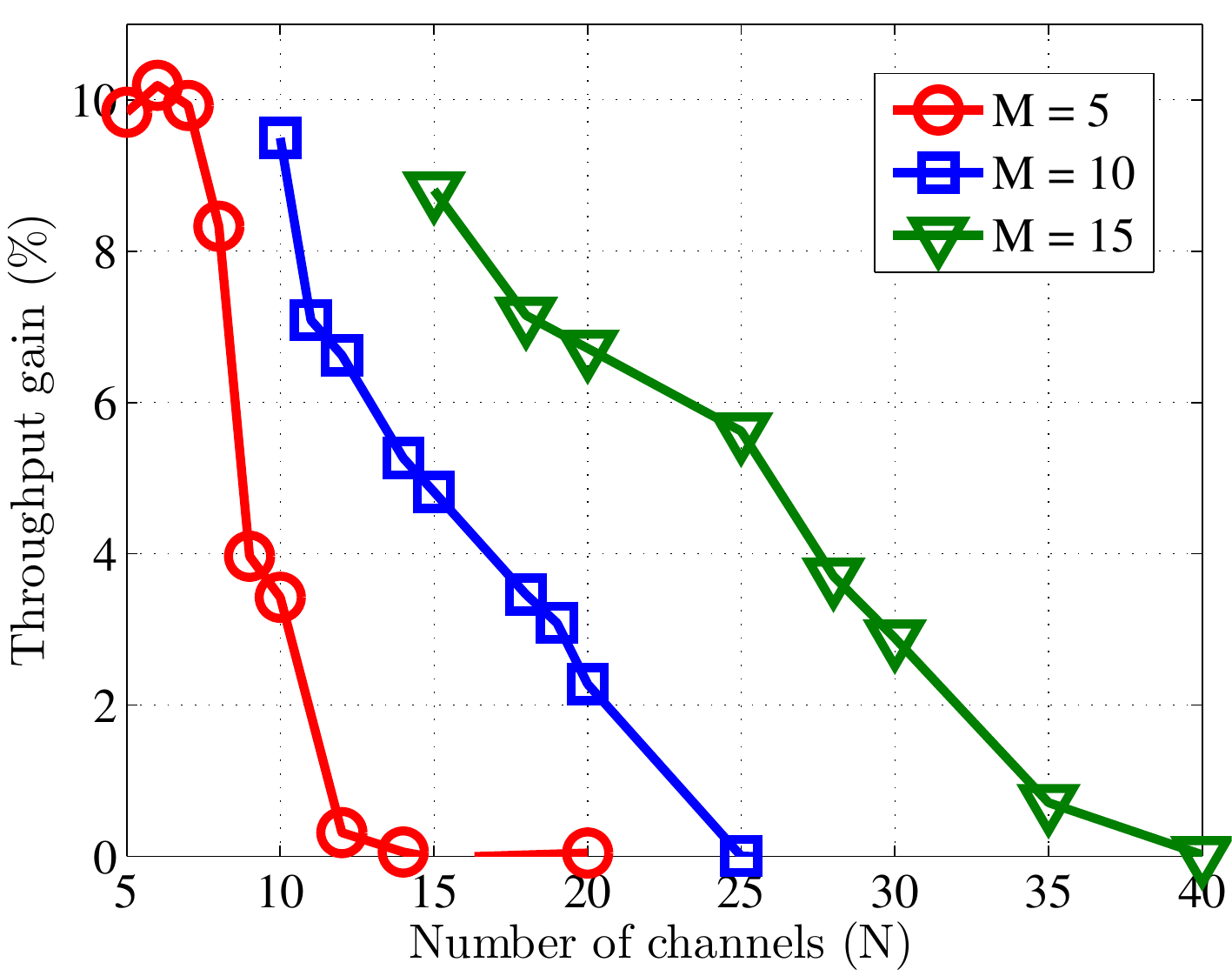}
\caption{Throughput gain between Alg.~\ref{mainalg_ove_Chapter4} and P-blind 5-user sharing versus the number of channels }
\label{Paware_Pblind_M_51015_G}
\end{figure}

\begin{figure}[!t]
\centering
\includegraphics[width=90mm]{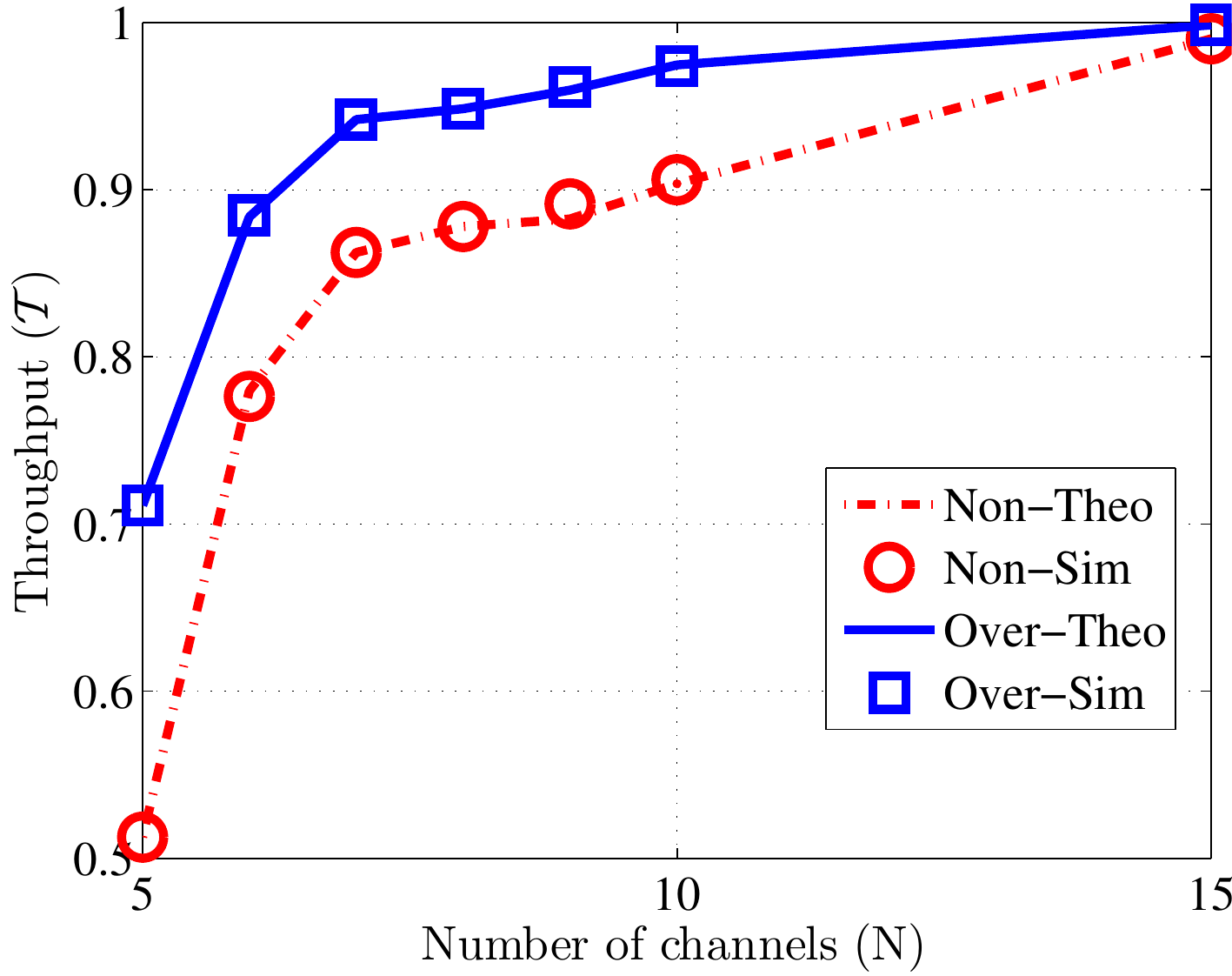}
\caption{Minimum throughput versus the number of channels under max-min fairness (for $M$ = 5, Theo: Theory, Sim: Simulation, Over: Overlapping, Non: Non-overlapping).}
\label{M5_fairness}
\end{figure}

\begin{figure}[!t]
\centering
\includegraphics[width=90mm]{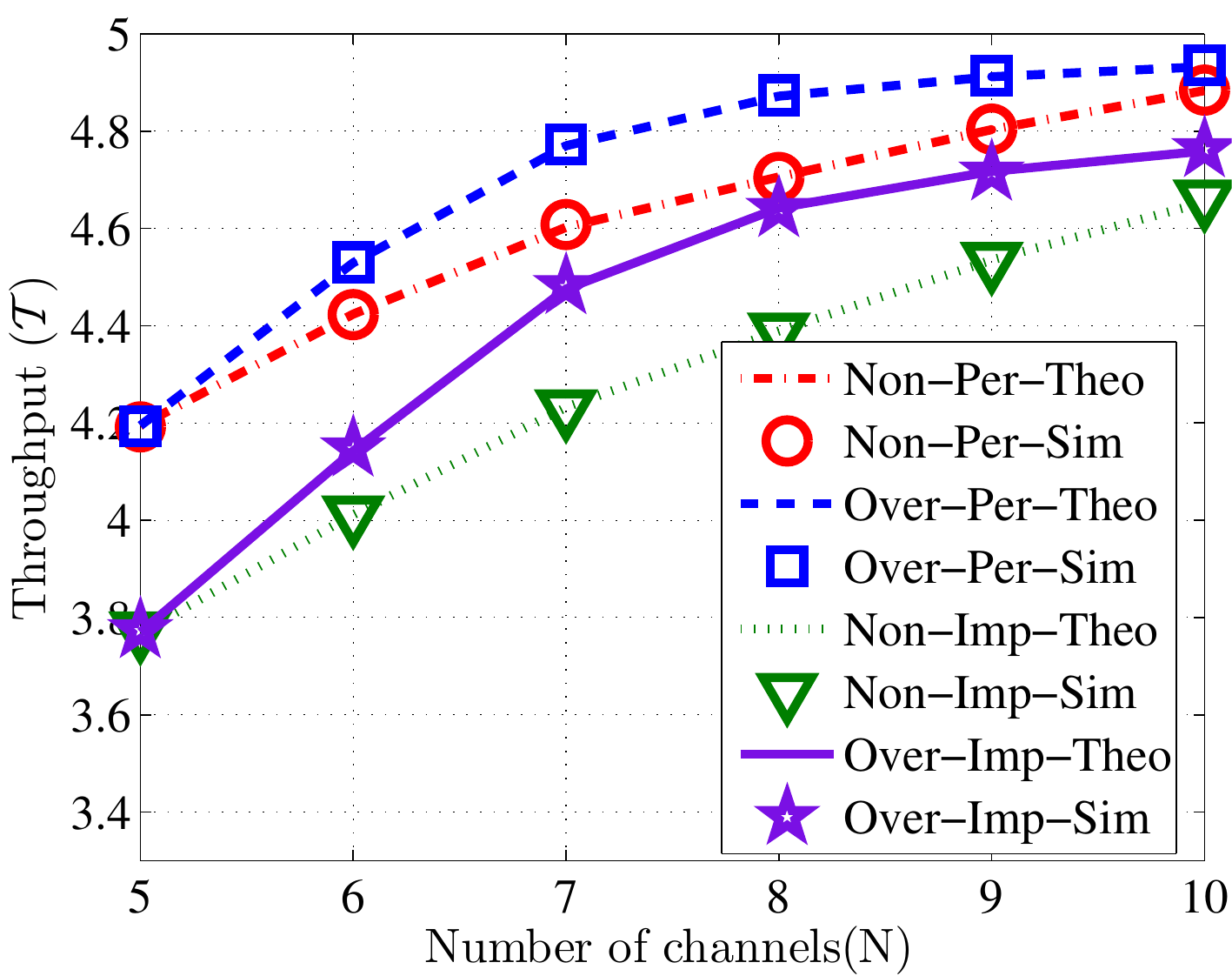}
\caption{Total throughput versus the number of channels under throughput maximization design (for $M = 5, \mathcal{P}_f^{ij} \in \left[0.1, 0.15\right], \mathcal{P}_d^{ij} = 0.9$, Theo: Theory, Sim: Simulation, Over: Overlapping, Non: Non-overlapping, Per: Perfect sensing, Imp: Imperfect sensing).}
\label{M5_Pf_10_tau50}
\end{figure}

We present numerical results to illustrate the throughput performance of the proposed channel assignment algorithms. 
To obtain the results, the probabilities $p_{i,j}$ are randomly realized in the interval [0.7, 0.9] unless stated otherwise. 
We choose the length of control packets as follows: RTS including PHY header 288 bits, CTS including PHY header 240 bits,
which correspond to $t_{\sf RTS}$ = 48$\mu s$, $t_{\sf CTS}$ = 40$\mu s$ for transmission rate of 6Mbps, which is
the basic rate of 802.11a/g standards \cite{wlan}. Other parameters are chosen as follows: cycle time 
$T_{\sf cycle} = 3 ms$; $\theta = 20$$\mu s$, $t_{\sf SIFS}$ = 28$\mu s$,
target collision probability $\epsilon_P$ = 0.03; $t_{\sf SEN}$ and  $t_{\sf SYN}$ are assumed to be negligible so they are ignored.
Note that these values of $\theta$ and $t_{\sf SIFS}$ are typical (interest readers can refer to Tables I and II
in the well-cited reference \cite{bian00} for related information). The value of cycle time $T_{\sf cycle}$ is relatively small given the fact
that practical cognitive systems such as those operating on the TV bands standardized in the 802.22 standard requires 
the spectrum evacuation time of a few seconds \cite{ko10}. We will present the total throughput under throughput maximization
design and the minimum throughput under max-min fairness design in all numerical results. All throughput curves are obtained by averaging over
30 random realizations of  $p_{i,j}$.

\subsection{ MAC Protocol Configuration}

We first investigate interactions between MAC protocol parameters and the achievable throughput performance.
In particular, we plot the average probability of the first collision, which is derived in Section \ref{ConWinCal_Chapter4} versus 
contention window in Fig.~\ref{het_P_M15_Aware_Blind} when Alg.~\ref{mainalg_ove_Chapter4} is used for channel assignment. This figure shows that 
the collision probability first increases then decreases with $N$. This can be
interpreted as follows. When $N$ is relatively small, Alg.~\ref{mainalg_ove_Chapter4} tends to allow
more overlapping channel assignments for increasing number of channels. 
However, more overlapping channel assignments increase the contention level because
more users may want to exploit same channels, which results in larger collision probability.
As $N$ is sufficiently large, a few overlapping channel assignments is needed to achieve
the maximum throughput. Therefore, collision probability decreases with $N$.

We now consider the impact of target collision probability $\epsilon_P$ on the total network throughput,
which is derived in Section \ref{tputana_Chapter4}. Recall that in this analysis collision probability
is not taken into account, which is shown to have negligible errors in Proposition 2. Specifically, we plot
the total network throughput versus $\epsilon_P$ for $M$ = 10 and different values of $N$ in Fig.~\ref{T_Pc_M10_N121518}.
This figure shows that the total throughput slightly increases with $\epsilon_P$. However, the increase is quite marginal
as $\epsilon_P \geq 0.03$. In fact, the required contention window $W$ given in (\ref{Window_Chapter4}) decreases with increasing
$\epsilon_P$ (as can be observed from Fig.~\ref{het_P_M15_Aware_Blind}), which leads to decreasing MAC protocol
overhead $\delta(W)$ as confirmed by (\ref{overhead_Chapter4}) and therefore the increase in the total network throughput.
Moreover, the total throughput may degrade with increasing $\epsilon_P$ because of the increasing number of collisions.
Therefore, we will choose $\epsilon_P=0.03$ to present the following results, which would be reasonable to balance between
 throughput gain due to moderate MAC protocol overhead and throughput loss due to contention collision.

\subsection{Comparisons of Proposed Algorithms versus Optimal Algorithms}

We demonstrate the efficacy of the proposed algorithms by comparing their throughput performances with those obtained
by the optimal brute-force search algorithms for small values of $M$ and $N$. Numerical results are presented for both 
throughput-maximization and max-min fair objectives. In Figs.~ \ref{M_2_Opt_Comparison} and \ref{M_3_Opt_Comparison}, 
we compare the throughputs of the proposed and optimal
algorithms for $M = 2$ and $M = 3$ under the throughput-maximization objective. These figures confirm that Alg.~\ref{mainalg_ove_Chapter4}
achieves throughput very close to that attained by the optimal solution for both values of $M$.

In Figs.~ \ref{M_2_fairness_optimal_compare}, and \ref{M_3_fairness_optimal_compare}, we plot the throughputs achieved by
our proposed algorithm and the optimal algorithm for $M = 2$ and $M = 3$ under the max-min fair objective.
Again, Alg.~\ref{mainalgFO1_Chapter4} achieves throughput very close to the optimal throughput under this design.
In addition, analytical results match simulation results very well
and non-overlapping channel assignment algorithms achieve noticeably lower throughputs than those attained by their overlapping counterparts
if the number of channels is small. It can also be observed that the average throughput per user under the throughput maximization design
is higher than the minimum throughput attained under max-min fair design. This is quite expected since
the max-min fairness trades throughput for fairness.

\subsection{Throughput Performance of Proposed Algorithms}

We illustrate the total throughput $\mathcal{T}$ versus the number of channels obtained by both Alg.~\ref{mainalg_non_Chapter4} and Alg.~\ref{mainalg_ove_Chapter4} where each point is obtained by averaging the throughput over 30 different realizations of $p_{i,j}$ in Fig.~\ref{het_T_M10_Aware_Blind1}. Throughput curves 
due to Alg.~\ref{mainalg_non_Chapter4} and Alg.~\ref{mainalg_ove_Chapter4} are indicated as ``P-ware''  in this figure. In addition, for the comparison purposes, we also show the throughput performance achieved by ``P-blind'' algorithms, which simply allocate channels to users in a round-robin manner without exploiting the heterogeneity
of $p_{i,j}$ (i.e., multiuser diversity gain). For P-blind algorithms, we show the performance of both non-overlapping and overlapping channel assignment algorithms. Here, the overlapping P-blind algorithm allows at most five users to share one particular channel. We have observed through
numerical studies that allowing more users sharing one channel cannot achieve better throughput performance because of the excessive MAC protocol overhead.

As shown in Fig.~\ref{het_T_M10_Aware_Blind1}, the analytical and simulation results achieved by  both proposed algorithms match each other very well.
This validates the accuracy of our throughput analytical model developed in Section \ref{tputana_Chapter4}. 
It also indicates  that the total throughput reaches the maximum value, which is equal to $M$ = 15 as the 
number of channels becomes sufficiently large for both Alg.~\ref{mainalg_non_Chapter4} and Alg.~\ref{mainalg_ove_Chapter4}. This confirms the result stated in Proposition 1. In addition, Alg.~\ref{mainalg_ove_Chapter4} achieves significantly larger throughput than Alg.~\ref{mainalg_non_Chapter4} for low or moderate values of $N$. This performance gain comes from the multiuser diversity gain, which arises due to the spatial dependence of white spaces. For large $N$ (i.e., more than twice the number of users $M$), the negative impact
of MAC protocol overhead prevents Alg.~\ref{mainalg_ove_Chapter4} from performing overlapped channel assignments. Therefore, both Alg.~\ref{mainalg_non_Chapter4} and Alg.~\ref{mainalg_ove_Chapter4} achieve similar throughput performance.

Fig.~\ref{het_T_M10_Aware_Blind1} also indicates that both proposed algorithms outperform the round-robin channel assignment counterparts. In particular, Alg.~\ref{mainalg_non_Chapter4} improves the total throughput significantly compared to the round-robin algorithm under non-overlapping channel assignments. For the overlapping channel assignment schemes, we show the throughput performance of the round-robin assignment algorithms when 5 users are allowed to share one channel (denoted as 5-user sharing in the figure). Although this achieves larger throughput for the round-robin algorithm, it still performs worse compared to the proposed algorithms. 
Moreover, we demonstrate the throughput gain due to Alg.~\ref{mainalg_ove_Chapter4} compared to Alg.~\ref{mainalg_non_Chapter4} for different values
of $N$ and $M$ in Fig.~\ref{het_G_M_351015}. This figure shows that performance gains up to $5\%$ can be achieved 
when the number of channels is small or moderate. In addition, Fig.~\ref{Paware_Pblind_M_51015_G} presents the throughput gain due to Alg.~\ref{mainalg_ove_Chapter4} versus the P-blind algorithm with 5-user sharing. It can be observed that a significant throughput gain of up to 10\% can be achieved for these investigated scenarios.

Fig.~ \ref{M5_fairness} illustrates the throughput of Alg.~\ref{mainalgFN_Chapter4} and Alg.~\ref{mainalgFO1_Chapter4} where $p_{ij}$ are chosen in the range of $\left[0.5,0.9\right]$.
It can be observed that the overlapping channel algorithm also improves the minimum throughput performance compared to the non-overlapping
counterpart significantly. 
Finally, we plot the throughputs achieved by Alg.~\ref{mainalg_non_Chapter4} and Alg.~\ref{mainalg_ove_Chapter4} under perfect
and imperfect spectrum sensing for $M = 5$ in Fig.~\ref{M5_Pf_10_tau50} where the detection probabilities are set as $\mathcal{P}_d^{ij} = 0.9$
while false alarm probabilities are randomly realized as $\mathcal{P}_f^{ij} \in \left[0.1, 0.15\right]$. This figure shows that sensing errors can significantly
degrade the throughput performance of SUs. In addition, the presented results validate the throughput analytical model described
in Section~\ref{Imper_sens_Chapter4}.

\vspace{0.2cm}
\section{Conclusion}
\label{conclusion_Chap4} 

We have investigated the channel assignment problem for cognitive radio networks with hardware-constrained SUs. 
We have first presented the optimal brute-force search algorithm and analyzed its complexity. 
Then, we have developed the following two channel assignment algorithms for throughput maximization: 1) the non-overlapping overlapping channel assignment algorithm and 2) the overlapping channel assignment algorithm. 
In addition, we have developed an analytical model to analyze the saturation throughput that was achieved by the overlapping channel assignment algorithm. 
We have also presented several potential extensions, including the design of max-min fair channel assignment algorithms and throughput analysis, considering imperfect spectrum sensing.
We have validated our results through numerical studies and demonstrated significant throughput gains of the overlapping channel assignment algorithm compared with its non-overlapping and round-robin channel assignment counterparts in different network settings.

%\newpage

%\renewcommand{\baselinestretch}{1.0}

% ---------------------------------------------------------------------------
%: ----------------------- end of thesis sub-document ------------------------
% ---------------------------------------------------------------------------

% this file is called up by thesis.tex
% content in this file will be fed into the main document

%: ----------------------- name of chapter  -------------------------
\chapter{Joint Cooperative Spectrum Sensing and MAC Protocol Design for Multi-channel Cognitive Radio Networks} % top level followed by section, subsection
\zlabel{Chapter5}

%: ----------------------- paths to graphics ------------------------

% change according to folder and file names
\ifpdf
    \graphicspath{{5/figures/PNG/}{5/figures/PDF/}{5/figures/}}
\else
    \graphicspath{{5/figures/EPS/}{5/figures/}}
\fi
%\usepackage{hyperref}

%: ----------------------- contents from here ------------------------

The content of this chapter was published in EURASIP Journal on Wireless Communications and Networking in the following paper:

L.~T.~ Tan, and L.~B.~ Le, ``Joint Cooperative Spectrum Sensing and MAC Protocol Design for Multi-channel Cognitive Radio Networks,'' {\em EURASIP Journal on Wireless Communications and Networking}, 2014 (101), June 2014.

\section{Abstract}
In this paper, we propose a semi-distributed cooperative spectrum sensing (SDCSS) and channel access framework for multi-channel cognitive radio networks (CRNs). 
In particular, we consider a SDCSS scheme where secondary users (SUs) perform sensing and exchange sensing outcomes with each other to locate spectrum holes. 
In addition, we devise the $p$-persistent CSMA-based cognitive MAC protocol integrating the SDCSS
to enable efficient spectrum sharing among SUs. We then perform throughput analysis and develop an algorithm to determine 
 the spectrum sensing and access parameters to maximize the throughput for a given allocation of channel sensing sets. Moreover, we consider the spectrum sensing 
set optimization problem for SUs to maximize the overall system throughput. We present both exhaustive search and low-complexity greedy algorithms
to determine the sensing sets for SUs and analyze their complexity. We also show how our design and analysis can be extended to consider
reporting errors. Finally, extensive numerical results are presented to demonstrate the significant performance gain of our optimized design framework
with respect to non-optimized designs as well as the impacts of different protocol parameters on the throughput performance.

\section{Introduction}

It has been well recognized that cognitive radio is one of the most important technologies that would enable us to meet 
exponentially growing spectrum demand via fundamentally improving the utilization of our precious spectral resources \cite{Zhao07}.
Development of efficient spectrum sensing and access algorithms for cognitive radios are among the key research issues for
successful deployment of this promising technology. 
There is indeed a growing literature on MAC protocol design and analysis for CRNs \cite{Liang08, Cor09, Le11, Le12, Kim08, Su08, Su07, Nan07, Le208, Konda08, Do05} %\cite{Liang08}-\cite{Do05} 
(see \cite{Cor09} for a survey of recent works in this topic). In \cite{Liang08}, it was shown that a significant throughput 
gain can be achieved by optimizing the sensing time under the single-SU setting.
Another related effort along this line was conducted in \cite{Kim08} where sensing-period optimization and optimal channel-sequencing 
algorithms were proposed to efficiently discover spectrum holes and to minimize the exploration delay.

In \cite{Su08}, a control-channel based MAC protocol was proposed for SUs to exploit white spaces in the cognitive ad hoc network. 
In particular, the authors of this paper developed both random and negotiation-based spectrum sensing schemes and performed throughput analysis 
for both saturation and non-saturation scenarios. There exists several other synchronous cognitive MAC protocols, which rely on a control channel 
for spectrum negotiation and access \cite{Su07, Nan07, Le208, Konda08, Do05}. %\cite{Su07}-\cite{Do05}.
In \cite{Le11} and \cite{Le12}, we designed, analyzed, and optimized a window-based MAC protocol to achieve efficient tradeoff between sensing time and contention overhead. 
However, these works considered the conventional single-user-energy-detection-based spectrum sensing scheme, which would only work well if the signal to noise ratio (SNR) 
is sufficiently high. In addition, the MAC protocol in these works was the standard window-based CSMA MAC protocol, which is known to be outperformed by the p-persistent CSMA 
MAC protocol \cite{Cali00}.

Optimal sensing and access design for CRNs were designed by using optimal stopping
theory in \cite{jia08}. In \cite{Sala10}, a multi-channel MAC protocol was proposed considering the distance
among users so that white spaces can be efficiently exploited while satisfactorily protecting primary users (PUs).
Different power and spectrum allocation algorithms were devised to maximize the secondary network throughput in  \cite{taosu10, wang11, zhang113}. %\cite{taosu10}-\cite{zhang113}. 
Optimization of spectrum sensing and access in which either cellular or TV bands can be
employed was performed in \cite{choi11}. These existing works either assumed perfect spectrum sensing or did not consider the
cooperative spectrum sensing in their design and analysis. 

Cooperative spectrum sensing has been proposed to improve the sensing performance where several SUs collaborate with each other to identify 
spectrum holes \cite{Gan07, Gane07, Quan08, Peh09, Cui11, Wei11, Wei09, Seu10, Chaud12}. % \cite{Gan07}-\cite{Seu10}  and \cite{Chaud12}. 
In a typical cooperative sensing scheme, each SU performs sensing independently and then sends its sensing result to a central controller (e.g., an access point (AP)). 
Here, various aggregation rules can be employed to combine these sensing results at the central controller to decide whether or not a particular spectrum band is available for secondary access. 
In \cite{Chaud12}, the authors studied the performance of hard decisions and soft decisions at a fusion center. 
They also investigated the impact of reporting channel errors on the cooperative sensing performance.
Recently, the authors of \cite{Lee13} proposed a novel cooperative spectrum sensing scheme using hard decision combining considering feedback errors.

%In \cite{Peh09}-\cite{Wei09}, 
In \cite{Peh09, Cui11, Wei11, Wei09}, optimization of cooperative sensing under the a-out-of-b rule was studied. 
In \cite{Wei11}, the game-theoretic based method was proposed for cooperative spectrum sensing.
In \cite{Seu10}, the authors investigated the multi-channel scenario where the AP collects statistics from SUs to decide whether it should stop at the current time slot.  
In \cite{Male11, Male13}, two different optimization problems for cooperative sensing were studied.
The first one focuses on throughput maximization where the objective is the probability of false alarm.
The second one attempts to perform interference management where the objective is the probability of detection.
These existing works focused on designing and optimizing parameters for the cooperative spectrum sensing algorithm; however,
they did not consider spectrum access issues. Furthermore, either the single channel setting or homogeneous network scenario (i.e., SUs experience
the same channel condition and spectrum statistics for different channels) was assumed in these works. 

In \cite{zhang11} and \cite{Park11}, the authors conducted design and analysis for cooperative spectrum sensing and MAC protocol design for cognitive radios where parallel spectrum sensing on different channels was assumed to be performed by multiple spectrum sensors at each SU. 
In CRNs with parallel-sensing, there is no need to optimize spectrum sensing sets for SUs. 
These works again considered the homogeneous network and each SU simply 
senses all channels. To the best of our knowledge, existing cooperative spectrum sensing schemes rely
on a central controller to aggregate sensing results for white space detection (i.e., centralized design). In addition, homogeneous
 environments and parallel sensing have been commonly assumed in the literature, which would not be very realistic.

In this work, we consider a general SDCSS and access framework under the heterogeneous environment where statistics of wireless channels, and spectrum holes can be arbitrary and there is no central controller to collect sensing results and make spectrum status decisions. 
In addition, we assume that each SU is equipped with only one spectrum sensor so that SUs have to sense channels sequentially. 
This assumption would be applied to real-world hardware-constrained cognitive radios. 
The considered SDCSS scheme requires SUs to perform sensing on their assigned sets of channels and
then exchange spectrum sensing results with other SUs, which can be subject to errors. 
After the sensing and reporting phases, SUs employ the $p$-persistent CSMA MAC protocol \cite{Cali00} to access one available channel.
In this MAC protocol, parameter $p$ denotes the access probability to the chosen channel if the carrier sensing indicates an available
 channel (i.e., no other SUs transmit on the chosen channel).
It is of interest to determine the access parameter $p $ that can mitigate the collisions and hence enhance the system throughput \cite{Cali00}.
Also, optimization of the spectrum sensing set for each SU (i.e., the set of channels sensed by the SU)  is very critical to achieve good system throughput.
Moreover, analysis and optimization of the joint spectrum sensing and access design become
much more challenging in the heterogeneous environment, which, however, can significantly improve the system performance.
Our current paper aims to resolve these challenges whose contributions can be summarized as follows:
 
\begin{itemize}

\item We propose the distributed $p$-persistent CSMA protocol incorporating SDCSS for multi-channel CRNs. 
Then we analyze the saturation throughput and optimize the spectrum sensing time and access parameters to achieve maximum throughput 
for a given allocation of channel sensing sets. This analysis and optimization are performed in the general heterogeneous scenario assuming that spectrum sensing sets 
for SUs have been predetermined.

\item We study the channel sensing set optimization (i.e., channel assignment) for throughput maximization and devise both
exhaustive search and low-complexity greedy algorithms to solve the underlying NP-hard optimization problem.
Specifically, an efficient solution for the considered problem would only allocate a subset of ``good'' SUs to sense each channel 
so that accurate sensing can be achieved with minimal sensing time.
We also analyze the complexity of the brute-force search and the greedy algorithms.

\item We extend the design and analysis to consider reporting errors as SUs exchange their spectrum sensing results.
In particular, we describe cooperative spectrum sensing model, derive the saturation throughput
considering reporting errors. Moreover, we discuss how the proposed algorithms to optimize the sensing/access parameters and sensing sets
can be adapted to consider reporting errors. Again, all the analysis is performed for the heterogeneous environment.

\item We present numerical results to illustrate the impacts of different parameters on the secondary throughput performance and demonstrate the 
significant throughput gain due to the optimization of different parameters in the proposed framework.

\end{itemize}

The remaining of this paper is organized as follows. Section ~\ref{SystemModel_Chap5} describes system and sensing models. 
MAC protocol design, throughput analysis, and optimization are performed in Section ~\ref{CPCSMA_Chap5} assuming no reporting errors. 
Section ~\ref{Exten} provides further extension for the analysis and optimization considering reporting errors.
Section ~\ref{Results_Chap5} presents numerical results followed by concluding remarks in Section ~\ref{conclusion_Chap5}.
The summary of key variables in the paper is given in Table~\ref{table_Chap5}.
 
%\vspace{10pt}
\section{System Model and Spectrum Sensing Design}
\label{SystemModel_Chap5}

In this section, we describe the system model and spectrum sensing design for the multi-channel CRNs.
Specifically, sensing performances in terms of detection and false alarm probabilities are presented.

\begin{table*} 
\centering
\caption{Summary of Key Variables}
\label{table_Chap5}
\scriptsize
\begin{tabular}{|c||c|c||c||c||c||c||c||c||c||c||c|}
\hline 
\multicolumn{2}{|c|}{\textbf{Variable}} & \multicolumn{10}{c|}{\textbf{Description}}\tabularnewline
\hline 
\multicolumn{12}{|c|}{\textbf{Key variables for no-reporting-error scenario}}\tabularnewline
\hline 
\multicolumn{2}{|c|}{$\mathcal{P}_j\left(\mathcal{H}_0\right)$ ($\mathcal{P}_j\left(\mathcal{H}_1\right)$)} & \multicolumn{10}{l|}{probability that channel $j$ is available (or not available)}\tabularnewline
\hline 
\multicolumn{2}{|c|}{$\mathcal{P}_d^{ij}$($\mathcal{P}_f^{ij}$)} & \multicolumn{10}{l|}{probability of detection (false alarm) experienced by SU $i$ for
channel $j$}\tabularnewline
\hline 
\multicolumn{2}{|c|}{$\mathcal{P}_d^j$ ($\mathcal{P}_f^j$)} & \multicolumn{10}{l|}{probability of detection (false alarm) for channel $j$ under SDCSS}\tabularnewline
\hline 
\multicolumn{2}{|c|}{$\varepsilon^{ij}$, $\gamma^{ij}$} & \multicolumn{10}{l|}{detection threshold, signal-to-noise ratio of the PU's signal}\tabularnewline
\hline 
\multicolumn{2}{|c|}{$\tau^{ij}$, $\tau$} & \multicolumn{10}{l|}{sensing time at SU $i$ on channel $j$, total sensing time}\tabularnewline
\hline 
\multicolumn{2}{|c|}{$N_0$, $f_s$} & \multicolumn{10}{l|}{noise power, sampling frequency}\tabularnewline
\hline 
\multicolumn{2}{|c|}{$a_j$, $b_j$} & \multicolumn{10}{l|}{parameters of $a$-out-of-$b$ rule for channel $j$}\tabularnewline
\hline 
\multicolumn{2}{|c|}{$N$, $M$} & \multicolumn{10}{l|}{total number of SUs, total number of channels}\tabularnewline
\hline 
\multicolumn{2}{|c|}{$\mathcal{S}_j^U$, $\mathcal{S}^U$} & \multicolumn{10}{l|}{set of SUs that sense channel $j$, set of all $N$ SUs}\tabularnewline
\hline 
\multicolumn{2}{|c|}{$\mathcal{S}_i$, $\mathcal{S}$} & \multicolumn{10}{l|}{set of assigned channels for SU $i$, set of all $M$ channels}\tabularnewline
\hline 
\multicolumn{2}{|c|}{$\Phi_l^k$} & \multicolumn{10}{l|}{particular set $k$ of $l$ SUs}\tabularnewline
\hline 
\multicolumn{2}{|c|}{$\Psi_{k_0}^{l_0}$} & \multicolumn{10}{l|}{set $l_0$ of $k_0$ actually available channels }\tabularnewline
\hline 
\multicolumn{2}{|c|}{$\Theta_{k_1}^{l_1}$, $\Omega_{k_2}^{l_2}$} & \multicolumn{10}{l|}{set $l_1$ of $k_1$
available channels (which are indicated by sensing outcomes), }\tabularnewline
\multicolumn{2}{|c|}{} & \multicolumn{10}{l|}{set $l_2$ of $k_2$ misdetected channels (which are indicated by sensing outcomes)}\tabularnewline
\hline 
\multicolumn{2}{|c|}{$\mathcal{NT}$} & \multicolumn{10}{l|}{normalized throughput per one channel}\tabularnewline
\hline 
\multicolumn{2}{|c|}{$\mathcal{T}_p^{ne}$, $\mathcal{T}_{j_2}^{ne}$} & \multicolumn{10}{l|}{conditional throughput: for one particular realization of sensing
outcomes corresponding to 2 sets $\Theta_{k_1}^{l_1}$ and $\Omega_{k_2}^{l_2}$,}\tabularnewline
\multicolumn{2}{|c|}{} & \multicolumn{10}{l|}{for a particular channel $j_2$}\tabularnewline
\hline 
\multicolumn{2}{|c|}{$n_j$, $k_e$} & \multicolumn{10}{l|}{number of SUs who select channel $j$ to access, $k_e=\mid\Theta_{k_1}^{l_1}\bigcup\Omega_{k_2}^{l_2}\mid$}\tabularnewline
\hline 
\multicolumn{2}{|c|}{$T$, $T_R$} & \multicolumn{10}{l|}{cycle time, total reporting time}\tabularnewline
\hline 
\multicolumn{2}{|c|}{$T_S$, $\overline{T}_S$} & \multicolumn{10}{l|}{time for transmission of packet, time for successful RTS/CTS transmission}\tabularnewline
\hline 
\multicolumn{2}{|c|}{$T_I^{i,j}$ ($\overline{T}_I^j$)} & \multicolumn{10}{l|}{$i$-th duration between 2 consecutive RTS/CTS transmission on channel $j$ (its average value)}\tabularnewline
\hline 
\multicolumn{2}{|c|}{$T_C$, $\overline{T}_{cont}^j$} & \multicolumn{10}{l|}{duration of collision, average contention time on channel $j$}\tabularnewline
\hline 
\multicolumn{2}{|c|}{$PD$} & \multicolumn{10}{l|}{propagation delay}\tabularnewline
\hline 
\multicolumn{2}{|c|}{$PS$, $ACK$} & \multicolumn{10}{l|}{lengths of packet and acknowledgment, respectively}\tabularnewline
\hline 
\multicolumn{2}{|c|}{$SIFS$, $DIFS$} & \multicolumn{10}{l|}{lengths of short time interframe space and distributed interframe space, respectively}\tabularnewline
\hline 
\multicolumn{2}{|c|}{$RTS$, $CTS$} & \multicolumn{10}{l|}{lengths of request-to-send and clear-to-send, respectively}\tabularnewline
\hline 
\multicolumn{2}{|c|}{$p$, $\mathcal{P}_C^j$} & \multicolumn{10}{l|}{transmission probability, probability of a generic slot corresponding to collision}\tabularnewline
\hline 
\multicolumn{2}{|c|}{$\mathcal{P}_S^j$, $\mathcal{P}_I^j$} & \multicolumn{10}{l|}{probabilities of a generic slot corresponding to successful transmission, idle slot}\tabularnewline
\hline 
\multicolumn{2}{|c|}{$N_c^j$($\overline{N}_c^j$)} & \multicolumn{10}{l|}{number of collisions before the first successful RTS/CTS exchange
(its average value)}\tabularnewline
\hline 
\multicolumn{2}{|c|}{$f_X^{N_c}$, $f_X^I$} & \multicolumn{10}{l|}{pmfs of $N_c^j$, $T_I^{i,j}$}\tabularnewline
\hline 
\multicolumn{12}{|c|}{\textbf{Key variables as considering reporting errors}}\tabularnewline
\hline 
\multicolumn{2}{|c|}{$\mathcal{P}_e^{i_1i_2}$} & \multicolumn{10}{l|}{probability of reporting errors between SUs $i_1$ and $i_2$}\tabularnewline
\hline 
\multicolumn{2}{|c|}{$\mathcal{P}_d^{i_1i_2j}$ ($\mathcal{P}_f^{i_1i_2j}$)} & \multicolumn{10}{l|}{probabilities of detection (false alarm) experienced by SU $i_1$ on channel $j$ with the sensing result received from SU $i_2$}\tabularnewline
\hline 
\multicolumn{2}{|c|}{$\Theta_{k_1,j_3}^{l_1}$} & \multicolumn{10}{l|}{$l_1$-th set of $k_1$ SUs whose sensing outcomes indicate that channel $j_3$ is vacant}\tabularnewline
\hline 
\multicolumn{2}{|c|}{$\Omega_{k_2,j_4}^{l_2}$} & \multicolumn{10}{l|}{$l_2$-th set of $k_2$ SUs whose sensing outcomes indicate that channel $j_4$ is vacant due to misdetection}\tabularnewline
\hline 
\multicolumn{2}{|c|}{$\Phi_{k_3,j_3}^{l_3}$} & \multicolumn{10}{l|}{$l_3$-th set of $k_3$ SUs in $\Theta_{k_1,j_3}^{l_1}$ who correctly report their sensing information on channel $j_3$ to SU $i_4$}\tabularnewline
\hline 
\multicolumn{2}{|c|}{$\Lambda_{k_4,j_3}^{l_4}$} & \multicolumn{10}{l|}{$l_4$-th set of $k_4$ SUs in $S_{j_3}^U\setminus\Theta_{k_1,j_3}^{l_1}$ who incorrectly report their sensing information on channel $j_3$ to SU $i_4$}\tabularnewline
\hline 
\multicolumn{2}{|c|}{$\Xi_{k_5,j_4}^{l_5}$} & \multicolumn{10}{l|}{$l_5$-th set of $k_5$ SUs in $\Omega_{k_2,j_4}^{l_2}$ who correctly report their sensing information on channel $j_4$ to SU $i_9$}\tabularnewline
\hline 
\multicolumn{2}{|c|}{$\Gamma_{k_6,j_4}^{l_6}$} & \multicolumn{10}{l|}{$l_6$-th set of $k_6$ SUs in $\mathcal{S}_{j_4}^U\setminus\Omega_{k_2,j_4}^{l_2}$ who incorrectly report their sensing information on channel $j_4$ to SU $i_9$}\tabularnewline
\hline 
\multicolumn{2}{|c|}{$\mathcal{S}_{1,i}^a$, $\mathcal{S}_{2,i}^a$} & \multicolumn{10}{l|}{sets of actually available channels and available due to sensing and/or
reporting errors, respectively }\tabularnewline
\hline 
\multicolumn{2}{|c|}{$\mathcal{\hat S}_1^a$, $\mathcal{\hat S}_2^a$} & \multicolumn{10}{l|}{$\mathcal{\hat S}_1^a=\bigcup_{i\in \mathcal{S}^U}\mathcal{S}_{1,i}^a$,
$\mathcal{\hat S}_2^a=\bigcup_{i\in \mathcal{S}^U}\mathcal{S}_{2,i}^a$}\tabularnewline
\hline 
\multicolumn{2}{|c|}{$\mathcal{S}_i^a$, $\mathcal{\hat S}^a$} & \multicolumn{10}{l|}{$\mathcal{S}_i^a=\mathcal{S}_{1,i}^a\bigcup \mathcal{S}_{2,i}^a$, $\mathcal{\hat S}^a=\mathcal{\hat S}_1^a\bigcup \mathcal{\hat S}_2^a$}\tabularnewline
\hline 
\multicolumn{2}{|c|}{$k_e^i$, $k_{max}$} & \multicolumn{10}{l|}{$k_e^i=\mid \mathcal{S}_i^a \mid$, $k_{max}=\mid \mathcal{\hat S}^a\mid$}\tabularnewline
\hline 
\multicolumn{2}{|c|}{$\Psi_j^a$, $\Psi^a$ } & \multicolumn{10}{l|}{set of SUs whose SDCSS outcomes indicate that channel $j$ is available,
$\Psi^a=\bigcup_{j\in \mathcal{\hat S}^a}\Psi_j^a$}\tabularnewline
\hline 
\multicolumn{2}{|c|}{$N_j$, $N_{max}$} & \multicolumn{10}{l|}{$N_j=\mid\Psi_j^a\mid$, $N_{max}=\mid\Psi^a\mid$}\tabularnewline
\hline 
\multicolumn{2}{|c|}{$\mathcal{T}_p^{re}$, $\mathcal{T}_{j_2}^{re}$} & \multicolumn{10}{l|}{conditional throughput for one particular realization of sensing
outcomes and for a particular channel $j_2$, respectively}\tabularnewline
\hline
\end{tabular}
\end{table*}

\subsection{System Model}
\label{System}

 %Fig. 1
\begin{figure}[!t] 
\centering
\includegraphics[width=90mm]{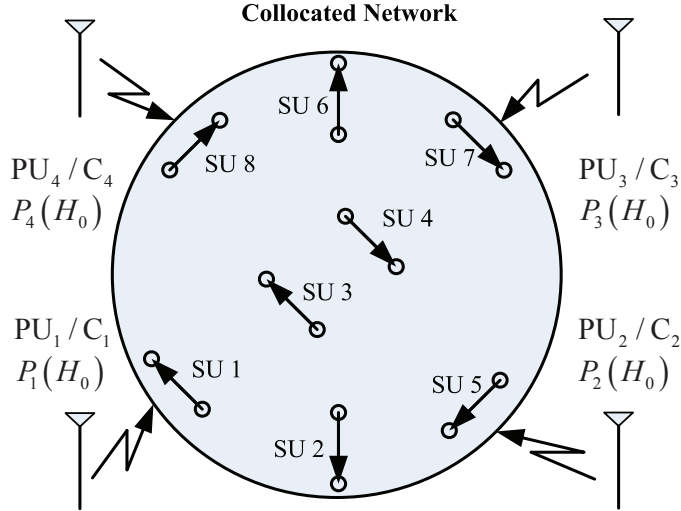}
\caption{Considered network and spectrum sharing model (PU: primary user, SU: secondary user, and $C_i $ is the channel $i $ corresponding to $\text{PU}_i $)}
\label{Fig1}
\end{figure}

We consider a network setting where $N$ pairs of SUs opportunistically exploit white spaces
 in $M$ channels for data transmission. For simplicity, we refer to pair $i$ of SUs simply as SU $i$. 
We assume that each SU can exploit only one available channel for transmission (i.e., SUs are equipped with 
narrow-band radios). We will design a synchronized MAC protocol integrating SDCSS for channel access. We assume 
that each channel is either in the idle or busy state for each predetermined periodic interval, which is referred to as
a cycle in this paper.

We further assume that each pair of SUs can overhear transmissions from other pairs of SUs (i.e., collocated networks). 
There are $M$ PUs each of which may or may not use one corresponding channel for its data transmission
in each cycle. In addition, it is assumed that transmission from any pair of SUs on a particular channel will affect the primary receiver
which receives data on that channel. The network setting under investigation is shown in Fig.~\ref{Fig1} where $C_i$
denotes channel $i$ that belongs to PU $i$.
 
\subsection{Semi-Distributed Cooperative Spectrum Sensing}
\label{Ss}

We assume that each SU $i$ is assigned a set of channels $\mathcal{S}_i $ where
it senses all channels in this assigned set at beginning of each cycle in a sequential manner (i.e., sense one-by-one).
Optimization of such channel assignment will be considered in the next section.
Upon completing the channel sensing, each SU $i$ exchanges the sensing results (i.e., idle/busy status
of all channels in $\mathcal{S}_i $) with other SUs for further processing. 
Here, the channel status of each channel can be represented by one bit (e.g., 1 for idle and 0 for busy status).
Upon collecting sensing results, each SU will decide idle/busy status for all channels. Then, 
SUs are assumed to employ a distributed MAC protocol to perform access resolution so that only
the winning SUs on each channel are allowed to transmit data. The detailed MAC protocol design will be presented later.

Let $\mathcal{H}_0$ and $\mathcal{H}_1$ denote the events that a particular PU is idle and active on its corresponding channel in any cycle, respectively. 
In addition, let $\mathcal{P}_j \left( \mathcal{H}_0 \right)$ and  $\mathcal{P}_j \left( \mathcal{H}_1 \right) = 1 - 
\mathcal{P}_j \left( \mathcal{H}_0 \right)$ be the probabilities that channel $j$ is available and not available for secondary access, respectively.
We assume that SUs employ an energy detection sensing scheme and let $f_s$ be the sampling frequency used in the
sensing period for all SUs. There are two important performance measures,
which are used to quantify the sensing performance, namely detection and false alarm probabilities. In particular, a
detection event occurs when a SU successfully senses a busy channel and false alarm
represents the situation when a spectrum sensor returns a busy status for an idle channel (i.e., the transmission opportunity
is overlooked). 
 
Assume that transmission signals from PUs are complex-valued PSK signals while the noise at the SUs is independent and identically distributed circularly 
symmetric complex Gaussian $\mathcal{CN}\left( {0,{N_0}} \right)$ \cite{Liang08}. Then, the
detection and false alarm probabilities experienced by SU $i$ for the channel $j$ can be calculated as \cite{Liang08}
\beqn
\label{eq1}
\mathcal{P}_d^{ij}\left( \varepsilon ^{ij} ,\tau^{ij}  \right) = \mathcal{Q}\left( \left( \frac{\varepsilon ^{ij} }{N_0} - \gamma ^{ij}  - 1 \right)\sqrt {\frac{\tau^{ij} f_s}{2\gamma ^{ij}  + 1}}  \right), 
\eeqn
\beqn
 \mathcal{P}_f^{ij}\left( \varepsilon ^{ij} ,\tau^{ij}  \right) = \mathcal{Q}\left( \left( \frac{\varepsilon ^{ij} }{N_0} - 1 \right)\sqrt {\tau^{ij} f_s}  \right) \hspace{1.5cm} \nonumber \\ 
 = \mathcal{Q}\left( \sqrt {2\gamma ^{ij}  + 1} \mathcal{Q}^{ - 1}\left( \mathcal{P}_d^{ij}\left(  \varepsilon ^{ij} ,\tau^{ij}   \right) \right)+\sqrt {\tau^{ij} f_s} \gamma ^{ij}  \right),  \label{eq2}
\eeqn
where $i \in \left[ {1,N} \right]$ is the SU index, $j \in \left[ {1,M} \right]$ is the channel index, ${\varepsilon ^{ij}} $ is the detection threshold for 
the energy detector, ${\gamma ^{ij}} $ is the signal-to-noise ratio (SNR) of the PU's signal at the SU, $f_s$ is the sampling frequency, $N_0$ is the noise power, 
$\tau^{ij}$ is the sensing time of SU $i$ on channel $j$, and $\mathcal{Q}\left( . \right)$ is defined as 
$\mathcal{Q}\left( x \right) = \left( {1/\sqrt {2\pi } } \right)\int_x^\infty  {\exp \left( { - {t^2}/2} \right)dt}$. 

We assume that a general cooperative sensing scheme, namely $a$-out-of-$b$ rule, is employed by each SU to determine the idle/busy status
of each channel based on reported sensing results from other SUs. Under this scheme, an SU will declare that a channel is busy
if $a$ or more messages out of $b$ sensing messages report that the underlying channel is busy. 
The a-out-of-b rule covers different rules including OR, AND and majority rules as special cases. In particular, $a=1$ corresponds to the OR rule; 
if $a=b$ then it is the AND rule; and the majority rule has $a=\left\lceil b/2\right\rceil$. 

\begin{figure*}%[!t]
\centering
\includegraphics[width=60mm]{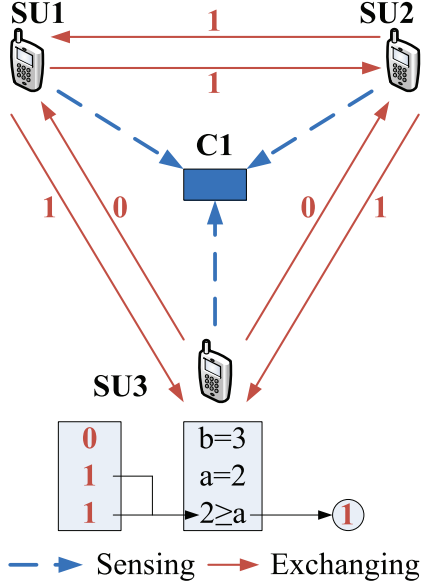}
\caption{Example for SDCSS on 1 channel.}
\label{DCSS_eg}
\end{figure*}

To illustrate the operations of the $a$-out-of-$b$ rule, let us consider a simple example shown in Fig.~\ref{DCSS_eg}.
Here, we assume that 3 SUs collaborate to sense channel one with $a = 2 $ and $b = 3 $.
After sensing channel one, all SUs exchange their sensing outcomes. 
SU3 receives the reporting results comprising two ``1'' and one ``0'' where ``1'' means that the channel is busy and ``0''  means 
channel is idle. Because the total number of ``1s'' is two which is larger than or equal to $a=2$, SU3 outputs the ``1'' in the final sensing result,
namely the channel is busy.

Let us consider a particular channel $j$. Let $\mathcal{S}_j^U$ denote the set of SUs that sense channel $j$, $b_j=\left|\mathcal{S}_j^U\right|$
be the number of SUs sensing channel $j$, and $a_j $ be the number of messages indicating that the underlying channel is busy. 
Then, the final decision on the spectrum status of channel $j$ under the $a$-out-of-$b$ rule has detection and false alarm probabilities that can be written as \cite{Wei11}
\beqn
\mathcal{P}_u^j\left( {\vec \varepsilon ^j}, {\vec \tau^j} , a_j  \right) = \sum_{l=a_j}^{b_j} \sum_{k=1}^{C_{b_j}^l} \prod_{i_1 \in \Phi^k_l} \mathcal{P}_u^{i_1j} \prod_{i_2 \in \mathcal{S}_j^{U} \backslash \Phi^k_l} \mathcal{\bar P}_u^{i_2j} \label{eq1_css_1},
\eeqn
where $u$ represents $d$ or $f$ as we calculate the probability of detection $\mathcal{P}_d^j$ or false alarm $\mathcal{P}_f^j$, respectively; $\mathcal{\bar P}$ is defined as
$\mathcal{\bar P} = 1-\mathcal{P}$; $\Phi^k_l$ in (\ref{eq1_css_1}) denotes a particular set with $l$ SUs whose sensing outcomes suggest that channel $j$ is busy given that 
this channel is indeed busy and idle as $u$ represents $d$ and $f$, respectively. Here, we generate all possible combinations of $\Phi^k_l$ where there are indeed
$C_{b_j}^l $ combinations. Also, ${\vec \varepsilon ^j} = \left\{\varepsilon ^{ij} \right\}$, ${\vec \tau^j} = \left\{\tau^{ij}\right\}$, $i \in \mathcal{S}_j^U$ represent the
 set of detection thresholds and sensing times, respectively. 
For brevity, $\mathcal{P}_d^j\left({\vec \varepsilon ^j}, {\vec \tau^j}, a_j  \right) $ and  $\mathcal{P}_f^j\left( {\vec \varepsilon ^j}, {\vec \tau^j}, a_j  \right)$ are sometimes written as $\mathcal{P}_d^j$ and $\mathcal{P}_f^j$ in the following.

Each SU exchanges the sensing results on its assigned channels with other SUs over a control channel, which is assumed to be always available (e.g., it is owned by the secondary network). 
To avoid collisions among these  message exchanges, we assume that there are $N$ reporting time slots for $N$ SUs each of which
has length equal to $t_r$. Hence, the total time for exchanging sensing results among SUs is $Nt_r$. 
Note that the set of channels assigned to SU $i$ for sensing, namely $\mathcal{S}_i $, is a subset of all channels and these sets can be different for different SUs. 
An example of channel assignment (i.e., channel sensing sets) is presented in Table \ref{table}. In this table, SU 4 is not assigned any channel. Hence, this 
SU must rely on the sensing results of other SUs to determine the spectrum status. 

\begin{table} 
\centering
\caption{Channel Assignment Example for SUs (x denotes an assignment)}
\label{table}
\begin{tabular}{|c|c|c|c|c|c|c|}
\cline{3-7} 
\multicolumn{2}{c|}{} & \multicolumn{5}{c|}{\textbf{Channel}}\tabularnewline
\cline{3-7} 
\multicolumn{2}{c|}{} & 1 & 2 & 3 & 4 & 5\tabularnewline
\hline 
 & 1 & x &  & x & x & \tabularnewline
\cline{2-7} 
 & 2 & x & x &  &  & \tabularnewline
\cline{2-7} 
\textbf{SU} & 3 &  & x & x & x & \tabularnewline
\cline{2-7} 
 & 4 &  &  &  &  & \tabularnewline
\cline{2-7} 
 & 5 & x &  &  &  & x\tabularnewline
\hline
\end{tabular}
\end{table}

\noindent
\textit{Remark 1:} In practice, the idle/busy status of primary system on a particular channel can be arbitrary and would not be synchronized with 
the operations of the SUs (i.e., the idle/busy status of any channel can change in the middle of a cycle). 
Hence, to strictly protect the PUs, SUs should continuously scan the spectrum of interest and evacuate from an exploited channel as soon as
the PU changes from an idle to a busy state. However, this continuous spectrum monitoring would be very costly to implement 
since each SU should be equipped with two half-duplex transceivers to perform spectrum sensing and access at the same time.
A more efficient protection method for PUs is to perform periodic spectrum sensing where SUs perform spectrum sensing at the beginning of each fixed-length
 interval and exploits available frequency bands for data transmission during the remaining time of the interval.
In this paper, we assume that the idle/busy status of each channel remains the same in each cycle, which enables us to analyze the system throughput. 
In general, imposing this assumption would not sacrifice the accuracy of our throughput analysis if PUs maintain their idle/busy status for a 
sufficiently long time. This is actually the case for many practical scenarios such as in the TV bands, as reported by several recent studies \cite{Stevenson09}. 
In addition, our MAC protocol that is developed under this assumption would result in very few collisions with PUs because the cycle time is quite small compared 
to the typical intervals over which the active/idle statuses of PUs change.

\vspace{10pt}
\section{Performance Analysis and Optimization for Cognitive MAC Protocol}
\label{CPCSMA_Chap5}

We present the cognitive MAC protocol design, performance analysis, and optimization for the multi-channel CRNs in this section.

\subsection{Cognitive MAC Protocol Design}
\label{MACDesign}

\begin{figure*}%[!t]
\centering
\includegraphics[width=120mm]{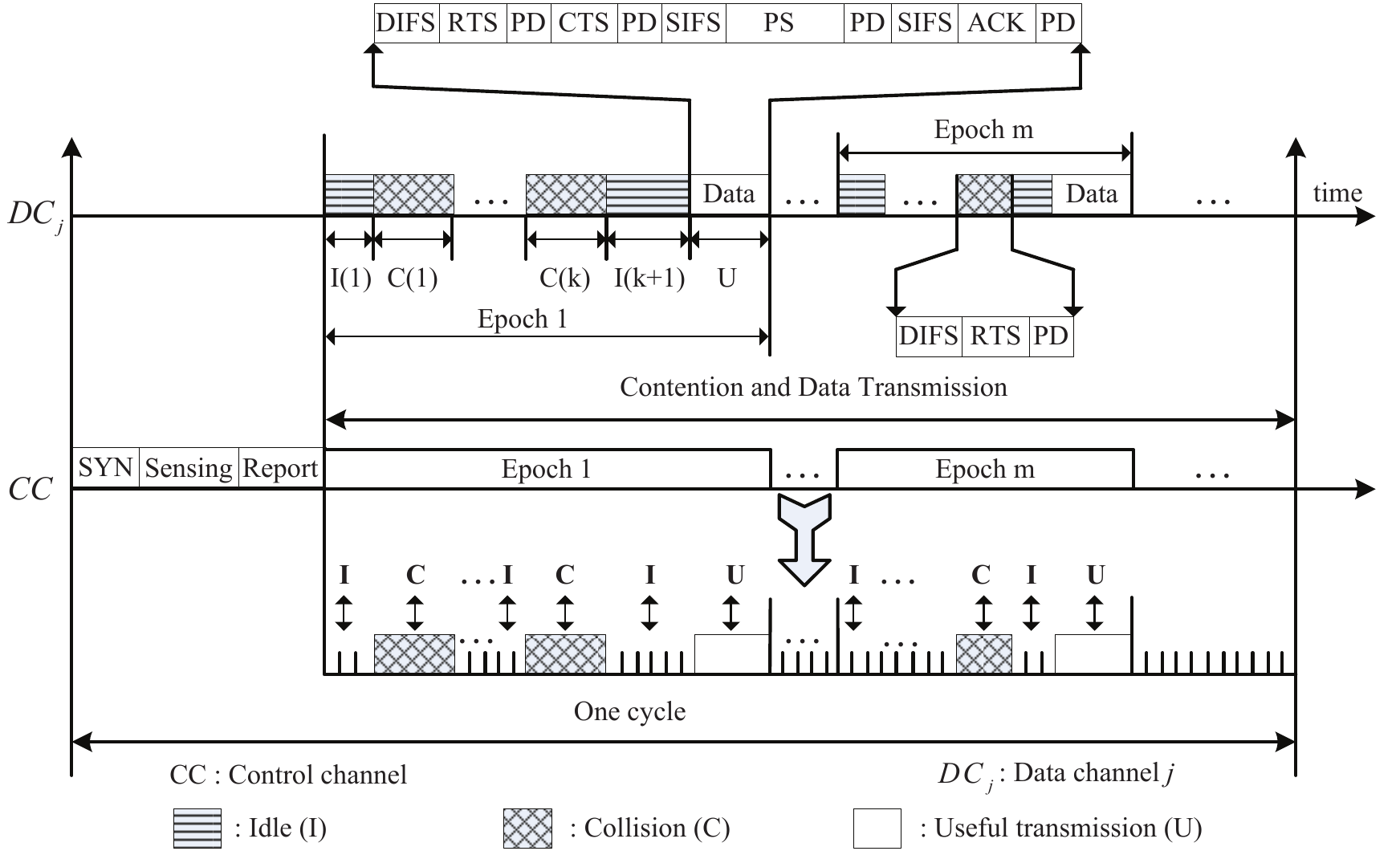}
\caption{Timing diagram of cognitive $p$-persistent CSMA protocol for one specific channel $j$.}
\label{Fig3_1}
\end{figure*}

We assume that time is divided into fixed-size cycles and it is assumed that SUs can perfectly synchronize with each other (i.e., there is no synchronization error) \cite{Konda08}.
We propose a synchronized multi-channel MAC protocol for dynamic spectrum sharing as follows.
The MAC protocol has four phases in each cycle as illustrated in Fig.~\ref{Fig3_1}.
The beacon signal is sent on the control channel to achieve synchronization in the first phase \cite{Konda08} which is presented in the simple manner as follows.
At the beginning of this phase, each SU senses the beacon signal from the volunteered synchronized SU which is the first SU sending the beacon. 
If an SU does not receive any beacon, it selects itself as the volunteered SU and sends out the beacon for synchronization.
In the second phase, namely the sensing phase of length $\tau$, all SUs simultaneously perform spectrum sensing on their assigned channels. 
Here, we have $\tau = \max_{i} \tau^i$, where $\tau^i = \sum_{j \in \mathcal{S}_i} \tau^{ij}$ is total sensing time of SU $i$, $\tau^{ij}$ is the sensing time of SU $i$ on channel $j$, and $\mathcal{S}_i$ is the set of channels assigned for SU $i$. 
We assume that one separate channel is assigned as a control channel which is used to exchange sensing results for reporting as well as broadcast a beacon signal for synchronization.
This control channel is assumed to be always available (e.g., it is owned by the secondary network).
In the third phase, all SUs exchange their sensing results with each other via the control channel.
Based on these received sensing results, each SU employs SDCSS techniques to decide the channel status of all channels and hence has a set of available channels.
Then each SU transmitter will choose one available channel randomly (which is used for contention and data transmission) and inform it to the corresponding SU receiver via the control channel.

In the fourth phase, SUs will participate in contention and data transmission on their chosen channels. 
We assume that the length of each cycle is sufficiently large so that SUs can transmit several packets during this data contention and transmission phase. 
In particular, we employ the $p$-persistent CSMA principle \cite{Cali00} to devise our cognitive MAC protocol.
In this protocol, each SU attempts to transmit on the chosen channel with a probability of $p$ if it senses an available channel (i.e., no other SUs transmit data on its chosen channel). 
In case the SU decides not to transmit (with probability of $1-p$), it will sense the channel and attempt to transmit again in the next slot with probability $p$.
If there is a collision, the SU will wait until the channel is available and attempt to transmit with probability $p$ as before.

The standard 4-way handshake with RTS/CTS (request-to-send/clear-to-send) \cite{bian00} will be employed to reserve a channel for data transmission. 
So the SU choosing to transmit on each available channel  exchanges RTS/CTS messages before transmitting its actual data packet. 
An acknowledgment (ACK) from the receiver is transmitted to the transmitter for successful reception of any packet.
The detailed timing diagram of this MAC protocol is presented in Fig.~\ref{Fig3_1}. 

\noindent
\textit{Remark 2:} For simplicity, we consider the fixed control channel in our design.
However, extensions to consider dynamic control channel selections to avoid the congestion can be adopted in our proposed framework.
More information on these designs can be found in \cite{mo08}.

\vspace{10pt}
\label{TputPCSMA}
\subsection{Saturation Throughput Analysis}

In this section, we analyze the saturation throughput of the proposed cognitive $p$-persistent CSMA protocol
assuming that there are no reporting errors in exchanging the spectrum sensing results among SUs.
Because there are no reporting errors, all SUs acquire the same sensing results for each channel, which implies that they make the same final sensing decisions
since the same a-out-b aggregation rule is employed for each channel. In the analysis, transmission time is counted in terms of 
contention time slot, which is assumed to be $v$ seconds. Each data packet is assumed to be of fixed size of $PS$ time slots. Detailed timing diagram 
of the $p$-persistent CSMA MAC protocol is illustrated in Fig.~\ref{Fig3_1}.

Any particular channel alternates between idle and busy periods from the viewpoint of the secondary system  where each busy period
 corresponds to either a collision or a successful transmission. We use the term ``epoch'' to refer to the interval between two consecutive successful transmissions. 
This means an epoch starts with an idle period followed by some alternating collision periods and idle periods before ending with a successful transmission period. 
Note that an idle period corresponds to the interval between two consecutive packet transmissions (collisions or successful transmissions).

Recall that each SU chooses one available channel randomly for contention and transmission according to the final cooperative sensing outcome.
We assume that upon choosing a channel, an SU keeps contending and accessing this channel until the end of the current cycle.
In the case of missed detection (i.e., the PU is using the underlying channel but the sensing outcome suggests that the channel is available), there will be 
collisions between SUs and the PU. Therefore,  RTS and CTS exchanges will not be successful in this case even though SUs cannot 
differentiate whether they collide with other SUs or the PU. 
Note that channel accesses of SUs due to missed detections do not contribute to the secondary system throughput.

To calculate the throughput for the secondary network, we have to consider all scenarios of idle/busy statuses of all channels and possible
mis-detection and false alarm events for each particular scenario. Specifically,
the normalized throughput per one channel achieved by our proposed MAC protocol, $\mathcal{NT} \left( \left\{\tau^{ij}\right\}, \left\{a_j\right\}, p, \left\{\mathcal{S}_i\right\} \right) $ can be written as 
\beqn
\mathcal{NT} = \sum_{k_0=1}^M \sum_{l_0=1}^{C_M^{k_0}} \prod_{j_1 \in \Psi_{k_0}^{l_0}} \mathcal{P}_{j_1} \left(\mathcal{H}_0\right) \prod_{j_2 \in \mathcal{S} \backslash \Psi_{k_0}^{l_0}} \mathcal{P}_{j_2} \left(\mathcal{H}_1\right) \times \label{NTSC_11}\\
\sum_{k_1=1}^{k_0} \sum_{l_1=1}^{C_{k_0}^{k_1}} \prod_{j_3 \in \Theta_{k_1}^{l_1}} \mathcal{\bar P}_f^{j_3} \prod_{j_4 \in \Psi_{k_0}^{l_0} \backslash \Theta_{k_1}^{l_1}} \mathcal{P}_f^{j_4}  \times \label{NTSC_21}\\
\sum_{k_2 = 0}^{M-k_0} \sum_{l_2=1}^{C_{M-k_0}^{k_2}} \prod_{j_5 \in \Omega_{k_2}^{l_2}} \mathcal{\bar P}_d^{j_5} \prod_{j_6 \in \mathcal{S} \backslash \Psi_{k_0}^{l_0} \backslash \Omega_{k_2}^{l_2}} \mathcal{P}_d^{j_6} \times  \label{NTSC_31}\\
\mathcal{T}_p^{\sf ne} \left(\tau,\left\{a_j\right\},p  \right). \label{NTSC_41}
\eeqn
The quantity (\ref{NTSC_11}) represents the probability that there are $k_0$ available channels, which may or may not  be correctly determined by the SDCSS. 
Here, $\Psi_{k_0}^{l_0} $ denotes a particular set of $k_0 $ available channels out of $M$ channels whose index is $l_0$. 
In addition, the quantity (\ref{NTSC_21}) describes the probability that the SDCSS indicates $k_1 $ available channels whereas the remaining available channels are 
overlooked due to sensing errors where $\Theta_{k_1}^{l_1} $ denotes the $l_1 $-th set with $k_1 $ available channels. 
For the quantity in (\ref{NTSC_31}), $k_2$ represents the number of channels that are not available but the sensing outcomes indicate that they are available 
(i.e., due to misdetection) where $\Omega_{k_2}^{l_2}$ denotes the $l_2$-th set with $k_2$ mis-detected channels. 
The quantity in (\ref{NTSC_31}) describes the probability that the sensing outcomes due to SUs incorrectly indicates $k_2$ available channels. Finally,
$\mathcal{T}_p^{\sf ne} \left(\tau,\left\{a_j\right\},p  \right)$ in (\ref{NTSC_41}) denotes the conditional throughput for a particular realization of sensing
outcomes corresponding to two sets $\Theta_{k_1}^{l_1} $ and $\Omega_{k_2}^{l_2}$. 

Therefore, we have to derive the conditional throughput $\mathcal{T}_p^{\sf ne} \left(\tau,\left\{a_j\right\},p \right)$ to complete the throughput analysis, which
is pursued in the following. Since each SU randomly chooses one available channel according to the SDCSS for contention and access, the number of SUs
actually choosing a particular available channel is a random number. In addition, the SDCSS suggests that channels in $\Theta_{k_1}^{l_1} \cup \Omega_{k_2}^{l_2}$
are available for secondary access but only channels in $\Theta_{k_1}^{l_1}$ are indeed available and can contribute to the secondary throughput 
 (channels in $\Omega_{k_2}^{l_2}$ are misdetected by SUs). Let $\left\{n_{j}\right\} = \left\{n_1, n_2, \ldots, n_{k_e}\right\}$
be the vector describing how SUs choose channels for access where  $k_e = \left|\Theta_{k_1}^{l_1} \cup \Omega_{k_2}^{l_2}\right|$ and $n_{j}$
denotes the number of SUs choosing channel $j$ for access. Therefore, the conditional throughput  
$\mathcal{T}_p^{\sf ne} \left(\tau,\left\{a_j\right\},p \right)$ can be calculated as follows:
\beqn
\label{T_p_cal}
\mathcal{T}_p^{\sf ne}  \left(\tau,\left\{a_j\right\},p \right) = \sum_{\left\{n_{j}\right\}: \: \sum _{j \in \Theta_{k_1}^{l_1} \cup \Omega_{k_2}^{l_2}} n_{j}=N} 
\mathcal{P}\left(\left\{n_{j}\right\}\right) \times \label{T_p_cal_11} \\
\sum_{j_2 \in \Theta_{k_1}^{l_1}} \frac{1}{M}  \mathcal{T}_{j_2}^{\sf ne} \left(\tau,\left\{a_{j_2}\right\},p \left|n=n_{j_2}\right.\right)  
\mathcal{I} \left(n_{j_2}>0\right) \label{T_p_cal_22}, \label{thputpnv}
\eeqn
where $\mathcal{P}\left(\left\{n_{j}\right\}\right)$ in (\ref{T_p_cal_11}) represents the probability that the channel access vector $\left\{n_{j}\right\}$ is realized (each channel $j$ where $j \in \Theta_{k_1}^{l_1} \cup \Omega_{k_2}^{l_2}$ is selected by $n_{j} $ SUs). 
The sum in (\ref{thputpnv}) describes the normalized throughput per channel due to a particular realization of the access vector $\left\{n_{j}\right\}$. 
Therefore, it is equal to the total throughput achieved by all available channels (in the set $\Theta_{k_1}^{l_1}$) divided by the total number of channels $M$.
Here, $\mathcal{T}_{j_2}^{\sf ne} \left(\tau,\left\{a_{j_2}\right\},p \left|n=n_{j_2}\right.\right) $ denotes the conditional throughput achieved by a particular channel $j_2$ when there are $n_{j_2}$ contending on this channel and $\mathcal{I} \left(n_{j_2}>0\right)$ represents the indicator function, which is equal to zero if $n_{j_2}=0$ (i.e., no SU chooses channel $j_2$) and equal to one, otherwise. 
Note that the access of channels in the set $\Omega_{k_2}^{l_2}$ due to missed detection does not contribute to the system throughput, which explains why we do not include these channels in the sum in (\ref{thputpnv}).

Therefore,  we need to drive $\mathcal{P}\left(\left\{n_{j}\right\}\right)$ and $\mathcal{T}_{j_2}^{\sf ne} \left(\tau,\left\{a_{j_2}\right\},p \left|n=n_{j_2}\right.\right)$ to determine the normalized throughput. Note that the sensing outcome due to the SDCSS is the same for all SUs and each SU chooses one channel in the set of
 $k_e = \left|\Theta_{k_1}^{l_1} \cup \Omega_{k_2}^{l_2}\right|$ channels randomly. Therefore, the probability $\mathcal{P}\left(\left\{n_{j}\right\}\right)$  can be
 calculated as follows:
\beqn
\mathcal{P}\left(\left\{n_{j}\right\}\right) &=& \left( {\begin{array}{*{20}{c}}
   N  \\
   {\left\{n_{j}\right\}}  \\
\end{array}} \right) \left(\frac{1}{k_e}\right)^{\sum _{j \in \Theta_{k_1}^{l_1} \cup \Omega_{k_2}^{l_2}} n_{j}} \\
&=& \left( {\begin{array}{*{20}{c}}
   N  \\
   {\left\{n_{j}\right\}}  \\
\end{array}} \right) \left(\frac{1}{k_e}\right)^N,
\eeqn
where $\left(\!\!\!\! {\begin{array}{*{20}{c}}
   N  \\
   {\left\{n_{j}\right\}}  \\
\end{array}}\!\!\!\!\right)$ is the multinomial coefficient which is defined as $ \left(\!\!\!\! {\begin{array}{*{20}{c}}
   N  \\
   {\left\{n_{j}\right\}}  \\
\end{array}} \!\!\!\!\right) = \left(\!\!\!\! {\begin{array}{*{20}{c}}
   N  \\
   {n_1, n_2, \ldots, n_k}  \\
\end{array}} \!\!\!\!\right) = \frac{N!}{n_1! n_2! \ldots n_k!}$.

The calculation of the conditional throughput $\mathcal{T}_{j_2}^{\sf ne} \left(\tau,\left\{a_{j_2}\right\},p \left|n=n_{j_2}\right.\right)$ must account for 
the overhead due to spectrum sensing and exchanges of sensing results among SUs. Let us define $T_R = Nt_r $ where $t_r$ is the report time from each SU to 
all the other SUs; $\tau = \max_{i} \tau^i$ is the total the sensing time; ${\bar T}^{j_2}_{\sf cont}$ is the average total  time due to contention, collisions, 
and RTS/CTS exchanges before a successful packet transmission; $T_S$ is the total time for transmissions of data packet, ACK control packet, and overhead between these data
and ACK packets. Then, the conditional throughput $\mathcal{T}_{j_2}^{\sf ne} \left(\tau,\left\{a_{j_2}\right\},p \left|n=n_{j_2}\right.\right)$ can be written as 
\beqn
\label{con_T}
\mathcal{T}_{j_2}^{\sf ne} \left(\tau,\left\{a_{j_2}\right\},p \left|n=n_{j_2}\right.\right) = \left\lfloor \frac{T-\tau-T_R}{{\bar T}^{j_2}_{\sf cont} + T_S}\right\rfloor \frac{T_S}{T},
\eeqn
where $\left\lfloor  .  \right\rfloor $ denotes the floor function and recall that $T$ is the duration of a cycle. Note
that $\left\lfloor \frac{T - \tau - T_R}{{\bar T}^{j_2}_{\sf cont} + T_S} \right\rfloor$ denotes the average number of successfully transmitted packets
in one particular cycle excluding the sensing and reporting phases. Here, we omit the length of the synchronization phase, which is assumed to be negligible.

To calculate ${\bar T}^{j_2}_{\sf cont} $, we define some further parameters as follows.
Let denote $T_C$ as the duration of the collision; 
${\bar T}_S$ is the required time for successful RTS/CTS transmission. These quantities can be calculated under the 4-way handshake mechanism as \cite{Cali00}
\beqn
\label{TCTSTI}
\left\{ {\begin{array}{*{20}{c}}
   T_S = PS + 2SIFS + 2PD + ACK   \hfill\\
   {\bar T}_S = DIFS + RTS + CTS + 2PD \hfill  \\
   T_C = RTS + DIFS + PD \hfill  \\
\end{array}} \right.,
\eeqn
where $PS$ is the packet size, $ACK$ is the length of an ACK packet, $SIFS$ is the length of a short interframe space, $DIFS$ is the length of a distributed interframe space, 
$PD$ is the propagation delay where $PD$ is usually very small compared to the slot size $v$. 

Let $T_I^{i,j_2}$ be the $i$-th idle duration between two consecutive RTS/CTS transmissions
 (they can be collisions or successes) on a particular channel $j_2$. Then, $T_I^{i,j_2}$ can be calculated based on its probability mass function (pmf),
 which is derived in the following. Recall that all quantities are defined in terms of number of time slots. Now, suppose there are  $n_{j_2}$ SUs
choosing channel $j_2$, let $\mathcal{P}_S^{j_2}$, $\mathcal{P}_C^{j_2}$ and $\mathcal{P}_I^{j_2}$ be the probabilities of a generic slot corresponding to a successful transmission, a collision and an idle slot, respectively. These quantities are calculated as follows
\beqn
\mathcal{P}_S^{j_2} = n_{j_2}p\left(1-p\right)^{n_{j_2}-1} \\
\mathcal{P}_I^{j_2} = \left(1-p\right)^{n_{j_2}} \\
\mathcal{P}_C^{j_2} = 1-\mathcal{P}_S^{j_2}-\mathcal{P}_C^{j_2},
\eeqn
where $p$ is the transmission probability of an SU in a generic slot. 
Note that ${\bar T}^{j_2}_{\sf cont}$ is a random variable (RV) consisting of several intervals corresponding to idle periods, collisions, and one successful RTS/CTS transmission. 
Hence this quantity for channel $j_2$ can be written as 
\beqn
\label{T_cont}
{\bar T}^{j_2}_{\sf cont} = \sum_{i=1}^{N_c^{j_2}} \left(T_C+ T_I^{i,{j_2}}\right) + T_I^{N_c^{j_2}+1,{j_2}} + {\bar T}_S,
\eeqn
where $N_c^{j_2}$ is the number of collisions before the first successful RTS/CTS exchange. 
Hence it is a geometric RV with parameter $1-\mathcal{P}_C^{j_2}/\mathcal{\bar P}_I^{j_2}$ (where $\mathcal{\bar P}_I^{j_2} = 1 - \mathcal{P}_I^{j_2}$). 
Its pmf can be expressed as
\beqn
\label{N_c_cal_Chap5}
 f_{X}^{N_c} \left(x\right) = \left(\frac{\mathcal{P}_C^{j_2}}{\mathcal{\bar P}_I^{j_2}}\right)^{x} \left(1-\frac{\mathcal{P}_C^{j_2}}{\mathcal{\bar P}_I^{j_2}}\right), \: x = 0, 1, 2, \ldots
\eeqn
Also, $T_I^{i,j_2}$ represents the number of consecutive idle slots, which is also a geometric RV with parameter $1-\mathcal{P}_I^{j_2}$ with the following pmf
\beqn
\label{T_I_cal_Chap5}
f_{X}^{I} \left(x\right) = \left(\mathcal{P}_I^{j_2}\right)^{x} \left(1-\mathcal{P}_I^{j_2}\right), \: x = 0, 1, 2, \ldots
\eeqn
Therefore, ${\bar T}^{j_2}_{\sf cont}$  can be written as follows \cite{Cali00}:
\beqn
{\bar T}^{j_2}_{\sf cont}  = {\bar N}_c^{j_2}T_C + {\bar T}_I^{j_2} \left({\bar N}_c^{j_2}+1\right) + {\bar T}_S \label{T_contgeo},
\eeqn
where ${\bar T}_I^{j_2}$ and ${\bar N}_c^{j_2}$ can be calculated as
\beqn
{\bar T}_I^{j_2} &=& \frac{\left(1-p\right)^{n_{j_2}}}{1-\left(1-p\right)^{n_{j_2}}} \\
{\bar N}_c^{j_2} &=& \frac{1-\left(1-p\right)^{n_{j_2}}}{n_{j_2}p\left(1-p\right)^{n_{j_2}-1}}-1. 
\eeqn
These expressions are obtained by using the  pmfs of the corresponding RVs given in (\ref{N_c_cal_Chap5}) and (\ref{T_I_cal_Chap5}), respectively \cite{Cali00}.

%\vspace{10pt}
\subsection{Semi-Distributed Cooperative Spectrum Sensing and $p$--persistent CSMA Access Optimization}
\label{OpCSPCSMA}

We determine optimal sensing and access parameters to maximize the normalized throughput for our proposed SDCSS and $p$-persistent CSMA protocol. 
Here, we assume that the sensing sets $\mathcal{S}_j^U$ for different channels $j$ have been given. Optimization of these sensing sets
is considered in the next section. Note that the optimization performed in this paper is different from those in \cite{Le11}, \cite{Le12} because the MAC protocols
and sensing algorithms in the current and previous works are different. The normalized 
throughput optimization problem can be presented as 
\beqn
 \mathop {\max} \limits_{ \left\{\tau^{ij}\right\}, \left\{a_j\right\} , p} \quad \mathcal{NT}_p \left( \left\{\tau^{ij}\right\}, \left\{a_j\right\}, p, \left\{\mathcal{S}_i\right\}  \right)  \hspace{1.4cm} \label{probp1}\\ 
 \mbox{s.t.}\,\,\,\, \mathcal{P}_d^j \left( {\vec \varepsilon ^j}, {\vec \tau^j}, a_j  \right) \geq \mathcal{\widehat{P}}_d ^j, \: j \in \left[1, M\right] \hspace{1cm} \label{probp2}\\
 \quad \quad 0 < \tau^{ij}  \le {T},  \: 0 \leq p \leq 1, \hspace{1cm} \label{probp3}
\eeqn
where $\mathcal{P}_d^j$ is the detection probability for channel $j$; $\mathcal{\widehat{P}}_d ^j$ denotes the target detection probability;
$\vec \varepsilon ^j$ and $\vec \tau^j$ represent the vectors of detection thresholds and sensing times on channel $j$, respectively;
$a_j$ describes the parameter of the $a_j$-out-of-$b_j$ aggregation rule for SDCSS on channel $j$ with $b_j = |\mathcal{S}_j^U|$
where recall that $\mathcal{S}_j^U$ is the set of SUs sensing channel $j$. The optimization variables for this problem are
sensing times $\tau^{ij}$ and parameters $a_j$ of the sensing aggregation rule, and transmission probability $p$ of the MAC protocol.

It was shown in \cite{Liang08} that the constraints on detection probability should be met with equality at optimality
under the energy detection scheme and single-user scenario. This is quite intuitive since lower detection probability
implies smaller sensing time, which leads to higher throughput. This is still the case for our considered multi-user
scenario as can be verified by the conditional throughput formula (\ref{con_T}). Therefore, we can set
$\mathcal{P}_d^j\left( {\vec \varepsilon ^j}, {\vec \tau^j} , a_j  \right) = \mathcal{\widehat{P}}_d ^j$ to solve
the optimization problem (\ref{probp1})-(\ref{probp3}).

However, $\mathcal{P}_d^j \left( {\vec \varepsilon ^j}, {\vec \tau^j} , a_j  \right)$ is a function of $\mathcal{P}_d^{ij}$
for all SUs $i \in \mathcal{S}_j^U$ since we employ the SDCSS scheme in this paper. Therefore, to simplify the optimization 
we set $\mathcal{P}_d^{ij}=\mathcal{{P}}_d^{j*}$ for all SUs $i \in \mathcal{S}_j^U$ (i.e., all SUs are required to achieve the same detection probability
for each assigned channel). Then, we can calculate $\mathcal{{P}}_d^{j*}$ by using (\ref{eq1_css_1}) for a given value of $\mathcal{\widehat{P}}_d ^j$. In addition, 
we can determine $\mathcal{P}_f^{ij}$ with the obtained value of $\mathcal{{P}}_d^{j*}$ by using (\ref{eq2}), which is the function of sensing time $\tau^{ij}$.

Even after these steps, the optimization problem (\ref{probp1})-(\ref{probp3}) is still very difficult to solve. In fact, it is
the mixed integer non-linear problem since the optimization variables $a_j$ take integer values while other variables take real values.
Moreover, even the corresponding optimization problem achieved by relaxing $a_j$ to real variables is a difficult and non-convex problem to
solve since the throughput in the objective function (\ref{probp1}) given in (\ref{NTSC_41}) is a complicated and non-linear function of optimization variables.

Given this observation, we have devised Alg. \ref{mainalg_pg} to determine the solution for this optimization problem based on the coordinate-descent searching techniques. 
The idea is that at one time we fix all variables while searching for the optimal value of the single variable. This operation is performed sequentially for 
all variables until convergence is achieved. Since the normalized throughput given in (\ref{NTSC_41}) is quite insensitive with respect to $p$, we attempt to determine the 
optimized values for $\left(\left\{{\bar \tau}^{ij}\right\}, \left\{{\bar a}_j\right\} \right)$ first for different values of $p$ (steps 3--11 in Alg.~\ref{mainalg_pg}) before searching
the optimized value of $p$ in the outer loop (step 12 in Alg.~\ref{mainalg_pg}). This algorithm converges to the fixed point solution since we improve
 the objective value over iterations (steps 4--9). 
This optimization problem is non-convex in general. However, we can obtain its optimal solution easily by using the bisection search
technique since the throughput function is quite smooth \cite{Fan10}.
For some specific cases such as in homogeneous systems \cite{Le11, Peh09, Wei09}, the underlying optimization problem is convex, which can be solved efficiently by
using standard convex optimization algorithms. 

\begin{algorithm}[h]%\leesize
\algsetup{linenosize=\scriptsize}
  \scriptsize  %%\small \footnotesize \scriptsize  %\tiny
\caption{\textsc{Optimization of Sensing and Access Parameters }}
\label{mainalg_pg}
%\algsetup{indent=1.5em}
\begin{algorithmic}[1]

\STATE Assume we have the sets of all SU $i$, $\left\{\mathcal{S}_i\right\}$. Initialize $\tau^{ij}$, $j \in \mathcal{S}_i$, the sets of $\left\{a_j\right\}$ for all channel $j$ and $p$.

\STATE For each chosen $p \in \left[0, 1\right]$, find ${\bar \tau}^{ij}$ and $\left\{{\bar a}_j\right\}$ as follows:

\FOR {each possible set $\left\{a_j\right\}$}

\REPEAT

\FOR  {$i = 1$ \text{to} $N$}

\STATE Fix all $\tau^{i_1j}$, $i_1 \neq i$.

\STATE Find the optimal ${\bar \tau}^{ij}$ as ${\bar \tau}^{ij} = \mathop {\argmax} \limits_{0 < \tau^{ij} \leq T} \mathcal{NT}_p\left( \left\{\tau^{ij}\right\}, \left\{a_j\right\}, p\right)$.

\ENDFOR

\UNTIL {convergence}

\ENDFOR

\STATE The best $\left(\left\{{\bar \tau}^{ij}\right\}, \left\{{\bar a}_j\right\}\right)$ is determined for each value of $p$ as $\left(\left\{{\bar \tau}^{ij}\right\}, \left\{{\bar a}_j\right\}\right) = \mathop {\argmax} \limits_{\left\{{ a}_j\right\}, \left\{{\bar \tau}^{ij} \right\}} \mathcal{NT} \left({\bar \tau}^{ij}, \left\{a_j\right\}, p\right)$.

\STATE The final solution $\left( \left\{{\bar \tau}^{ij}\right\}, \left\{{\bar a}_j\right\}, {\bar p}  \right)$ is determined as $\left( \left\{{\bar \tau}^{ij}\right\}, \left\{{\bar a}_j\right\}, {\bar p}  \right) = \mathop {\argmax} \limits_{ \left\{{\bar \tau}^{ij}\right\}, \left\{{\bar a}_j\right\}, p } \mathcal{NT} \left(\left\{{\bar \tau}^{ij}\right\}, \left\{{\bar a}_j\right\}, p\right)$.

\end{algorithmic}
\end{algorithm}

\subsection{Optimization of Channel Sensing Sets}
\label{GACAPp}

For the CRNs considered in the current work, the network throughput strongly depends on the availability of different channels, the spectrum sensing time, and the
sensing quality. Specifically, long sensing time $\tau$ reduces the communications time on the available channels in each cycle of length $T$, which, therefore,
decreases the network throughput. In addition, poor spectrum sensing performance can also degrade the network throughput since SUs can either overlook
available channels (due to false alarm) or access busy channels (due to missed detection). Thus, the total throughput of SUs can be enhanced by optimizing the
access parameter $p$ and sensing design, namely optimizing the assignments of channels to SUs (i.e., optimizing the sensing sets for SUs)
and the corresponding sensing times. 

Recall that we have assumed the channel sensing sets for SUs are fixed to optimize the sensing and access parameters in the previous section.
In this section, we attempt to determine an efficient channel assignment solution (i.e., channel sensing sets) by solving the following problem
\beqn
\label{eq11a}
\max \limits_{ \left\{\mathcal{S}_i\right\}, \left\{a_j\right\}} \mathcal{NT} \left( \left\{{\bar \tau}^{ij}\right\}, \left\{a_j\right\}, {\bar p},  
\left\{\mathcal{S}_i\right\} \right).
\eeqn
Note that the optimal values of $a_j$ can only be determined if we have fixed the channel sensing set $\mathcal{S}_j^U$ for each channel $j$. This is because
we aim to optimize the $a_j$-out-of-$b_j$ aggregation rule of the SDCSS scheme for each channel $j$ where $b_j = |\mathcal{S}_j^U|$. Since $a_j$ takes integer values 
and optimization of channel sensing sets $\mathcal{S}_j^U$ also involves integer variables where we have to determine the set of SUs $\mathcal{S}_j^U$
assigned to sense each channel $j$. Therefore, the optimization problem (\ref{eq11a}) is the non-linear integer program, which is NP-hard \cite{Lee12}.
In the following, we present both brute-force search algorithm and low-complexity greedy algorithm to solve this problem.

\subsubsection{Brute-force Search Algorithm}
\label{BFSA}

Due to the non-linear and combinatorial structure of the formulated channel assignment problem, it would be impossible to explicitly determine the optimal closed form solution
for problem (\ref{eq11a}). However, we can employ the brute-force search (i.e., the exhaustive search) to determine the best channel assignment. 
Specifically, we can enumerate all possible channel assignment solutions. Then, for each channel assignment solution (i.e., sets $\mathcal{S}_j^U$ for all
channels $j$), we employ Alg.~\ref{mainalg_pg} to determine the best spectrum sensing and accessing parameters  $\left\{\tau^{ij}\right\}, \left\{a_j\right\}, p$ and
calculate the corresponding total throughput by using the throughput analytical model in \ref{TputPCSMA}. The channel assignment achieving the
maximum throughput together with its best spectrum sensing and accessing parameters provides the best solution for the optimization problem (\ref{eq11a}).

\subsubsection{Low-Complexity Greedy Algorithm}
\label{PGA}

We propose another low-complexity and greedy algorithm to find the solution for this problem, which is described in  Alg.~\ref{ChanAAp}. 
In this algorithm, we perform the initial channel assignment in step 1, which works as follows. We first temporarily assign all channels for each SU. 
Then, we run Alg. \ref{mainalg_pg} to find the optimal sensing times for this temporary assignment, i.e., to determine $\left\{{\bar \tau}^{ij}\right\}$, which
is used to assign one SU to each channel so that the total sensing time is minimized. In particular, the initial channel assignments are set according to the solution
of the optimization problem (\ref{probp10})-(\ref{probp11}) presented in the following. 
\beqn
 \mathop {\min} \limits_{ \left\{x_{ij}\right\} } \quad \sum_{i,j} \overline{\tau}^{ij} x_{ij} \hspace{1cm} \label{probp10}\\ 
 \mbox{s.t.} \,\,\,\, \sum_{i} x_{ij} = 1, \: j \in \left[1, M\right]. \hspace{0cm} \label{probp11}
\eeqn
where $x_{ij}$ are binary variables representing the channel assignments where
$x_{ij}=1$ if channel $j$ is allocated for SU $i$ (i.e., $j \in \mathcal{S}_i$) and $x_{ij}=0$, otherwise.
We employ the well-known Hungarian algorithm \cite{Kuhn55} to solve this problem. 
Then, we perform further channel assignments in steps 2-18 of Alg.~\ref{ChanAAp}.
Specifically, to determine one channel assignment in each iteration, we temporarily
assign one channel to the sensing set $\mathcal{S}_i$ of each SU $i$ and calculate
the increase of throughput for such channel assignment $\Delta T_{ij}$ with the optimized channel and access parameters
obtained by using Alg.~\ref{mainalg_pg} (step 6). We then
search for the best channel assignment $\left({\bar i}, {\bar j}\right) = \mathop {\argmax} \limits_{i, j \in \mathcal{S} \backslash \mathcal{S}_i} 
\Delta T_{ij} $ and actually perform the corresponding channel assignment if $\Delta T_{\bar i \bar j} >\delta$ (steps 7--10).

In Alg.~\ref{ChanAAp}, $\delta>0$ is a small number which is used in the stopping condition for this algorithm (step 11). 
In particular, if the increase of the normalized throughput due to the new channel assignment is negligible in any iteration
 (i.e., the increase of throughput is less than $\delta$) then the algorithm terminates. Therefore, we can choose
$\delta$ to efficiently balance the achievable throughput performance with the algorithm running time.  
In the numerical studies, we will choose $\delta $ equal to $10^{-3} \times {\mathcal NT}_c $.

The convergence of Alg.~\ref{ChanAAp} can be explained as follows. Over the course of this algorithm, we attempt
to increase the throughput by performing additional channel assignments. It can be observed that
we can increase the throughput by allowing i) SUs to achieve better sensing performance or ii) SUs to reduce their sensing times.
However, these two goals could not be achieved concurrently due to the following reason. 
If SUs wish to improve the sensing performance via cooperative spectrum sensing, we should assign more channels to each of them.
However, SUs would spend longer time sensing the assigned channels with the larger sensing sets, which would ultimately  decrease the throughput.
Therefore, there would exist a point when we cannot improve the throughput by performing further channel assignments, which implies that Alg.~\ref{ChanAAp} must converge.

There is a key difference in the current work and \cite{Le12} regarding the sensing sets of SUs.
Specifically,  the sets of assigned channels are used for spectrum sensing and access in \cite{Le12}.
However, the sets of assigned channels are used for spectrum sensing only in the current work. In addition,
the sets of available channels for possible access at SUs are determined based on the reporting results, which
may suffer from communications errors. 
We will investigate the impact of reporting errors on the throughput performance in Section ~\ref{Exten}.

\begin{algorithm}[h]%\leesize
\algsetup{linenosize=\scriptsize}
  \scriptsize  %%\small \footnotesize \scriptsize  %\tiny
\caption{\textsc{Greedy Algorithm}}
\label{ChanAAp}
%\algsetup{indent=1.5em}
\begin{algorithmic}[1]

\STATE  Initial channel assignment is obtained as follows:

\begin{itemize}

\item Temporarily perform following channel assignments $\widetilde{\mathcal{S}}_i = \mathcal{S}$, $i \in \left[1, N\right]$. Then, run Alg. \ref{mainalg_pg} to obtain
 optimal sensing and access parameters $\left( \left\{{\bar \tau}^{ij}\right\},  \left\{{\bar a}_j\right\}, {\bar p} \right)$. 

\item Employ Hungarian algorithm \cite{Kuhn55} to allocate each channel to exactly one SU to minimize the total cost
where the cost of assigning channel $j$ to SU $i$ is ${\bar \tau}^{ij}$ (i.e., to solve the optimization problem (\ref{probp10})-(\ref{probp11})).

\item The result of this Hungarian algorithm is used to build the initial channel assignment sets $\left\{\mathcal{S}_i\right\}$
for different SU $i$. 
\end{itemize}

\STATE Set $\text{continue} = 1$. %, $k=1$.

\WHILE {\text{continue} = 1}

\STATE Optimize sensing and access parameters for current channel assignment solution $\left\{\mathcal{S}_i \right\}$ by using Alg. \ref{mainalg_pg}.

\STATE Calculate the normalized throughput $\mathcal{NT}_{\sf c} = \mathcal{NT} \left( \left\{{\bar \tau}^{ij}\right\}, \left\{{\bar a}_j\right\}, {\bar p}, 
\left\{\mathcal{S}_i \right\} \right)$ for the optimized sensing and access parameters.

\STATE Each SU $i$ calculates the increase of throughput if it is assigned one further potential channel $j$ as $\Delta T_{ij} = \mathcal{NT} \left( \left\{{\bar \tau}^{ij}\right\}, \left\{{\bar a}_j\right\}, {\bar p}, \left\{\widetilde{\mathcal{S}}_i \right\}\right) - \mathcal{NT}_{\sf c}$ where $\widetilde{\mathcal{S}}_i = \mathcal{S}_i \cup j$,
 $\widetilde{\mathcal{S}}_l = \mathcal{S}_l, \: l \neq i$, and 
$\left\{{\bar \tau}^{ij}\right\}, \left\{{\bar a}_j\right\}, {\bar p}$ are determined
by using Alg. \ref{mainalg_pg} for the temporary assignment sets  $\left\{\widetilde{\mathcal{S}}_i \right\}$.

\STATE Find the ``best'' assignment $\left({\bar i}, {\bar j}\right)$ as $\left({\bar i}, {\bar j}\right) = \mathop {\argmax} \limits_{i, j \in \mathcal{S} \backslash \mathcal{S}_i} \Delta T_{ij} $.

\IF {$\Delta T_{\bar i \bar j} >\delta$}

\STATE Assign channel $\bar j$ to SU $\bar i$: ${\mathcal{S}}_{\overline{i}} = \mathcal{S}_{\overline{i}} \cup \overline{j}$.

\ELSE

\STATE Set $\text{continue} =0$.

\ENDIF

\ENDWHILE

\IF {$\text{continue} =1$}

\STATE Return to step 2.

\ELSE 

\STATE Terminate the algorithm.

\ENDIF

\end{algorithmic}
\end{algorithm}

\subsection{Complexity Analysis}
\label{ComAnap}

In this section, we analyze the complexity of the proposed brute-force search and low-complexity greedy algorithms.

\subsubsection{Brute-force Search Algorithm}
\label{ComAna1p}

To determine the complexity of the brute-force search algorithm, we need to calculate the 
number of possible channel assignments. Since each channel can be either allocated or not allocated
to any SU, the number of channel assignments is $2^{MN}$. Therefore, the complexity of the brute-force search algorithm is $\mathcal{O}\left(2^{MN}\right)$. 
Note that to obtain the best channel assignment solution, we must run Alg. \ref{mainalg_pg} to find the best sensing and access parameters for each potential
channel assignment, calculate the throughput achieved by such optimized configuration, and compare all the throughput values to determine the best solution.

\subsubsection{Low-complexity Greedy Algorithm}
\label{ComAna2p}

In step 1, we run Hungarian algorithm to perform the first channel assignment for each SU $i$. 
The complexity of this operation can be upper-bounded by $\mathcal{O}\left(M^2N\right)$ (see \cite{Kuhn55} for more details). 
In each iteration in the assignment loop (i.e., steps 2-18), each SU $i$ needs to calculate the increases of throughput for different potential channel assignments. 
Then, we select the assignment resulting in maximum increase of throughput.
Hence, the complexity involved in these tasks is upper-bounded by $MN$ since there are at most $M$ channels to assign for each of $N$ SUs. 
Also, the number of assignments to perform is upper bounded by $MN$ (i.e., iterations of the main loop). 
Therefore, the complexity of the assignment loop is upper-bounded by $M^2N^2$. 
Therefore, the total worst-case complexity of Alg. \ref{ChanAAp} is $\mathcal{O}\left(M^2N+M^2N^2\right) = \mathcal{O}\left(M^2N^2\right)$, which is much lower than that of the brute-force search algorithm.
As a result, Table ~\ref{table1} in Section ~\ref{Results_Chap5} demonstrates that our proposed greedy algorithms achieve the throughput performance very close to that achieved by the brute-force search algorithms albeit they require much lower computational complexity.

\subsection{Practical Implementation Issues}
\label{Prac_Imp}

In our design, the spectrum sensing and access operation is distributed, however, channel assignment is performed in centralized manner.
In fact, one SU is pre-assigned as a cluster head, which conducts channel assignment for SUs (i.e., determine channel sensing
sets for SUs).
For fairness, we can assign the SU as the cluster head in the round-robin manner. 
To perform channel assignment, the cluster head is responsible for estimating $\mathcal{P}_j\left(\mathcal{H}_0\right)$.
Upon determining the channel sensing sets for all SUs, the cluster head will forward the results to the SUs.
Then based on these pre-determined sensing sets, SUs will perform spectrum sensing and run the underlying MAC protocol to access the channel distributively in each cycle.
It is worth to emphasize that the sensing sets for SUs are only determined once the probabilities $\mathcal{P}_j\left(\mathcal{H}_0\right)$ change, which would be quite infrequent in practice (e.g., in the time scale of hours or even days). 
Therefore, the estimation cost for $\mathcal{P}_j\left(\mathcal{H}_0\right)$ and all involved communication overhead due to sensing set optimization operations would be acceptable.

\vspace{10pt}
\section{Consideration of Reporting Errors}
\label{Exten}

In this section, we consider the impact of reporting errors on the performance of the proposed joint SDCSS and access design.
Note that each SU relies on the channel sensing results received from other SUs in $\mathcal{S}_j^U$ to determine the sensing
outcome for each channel $j$. If there are reporting errors then different SUs may receive
different channel sensing results, which lead to different final channel sensing decisions. The throughput analysis, therefore,
must account for all possible error patterns that can occur in reporting channel sensing results.  
We will present the cooperative sensing model and throughput analysis considering reporting errors in the following.

\subsection{Cooperative Sensing with Reporting Errors}
\label{CS_w_RE}

In the proposed SDCSS scheme, each SU $i_1 $ collects sensing results for each channel $j $ from all SUs $i_2 \in \mathcal{S}_j^U $ who are assigned to sense channel $j$.
In this section, we consider the case where there can be errors in reporting the channel sensing results among SUs. We assume that the channel sensing
result for each channel transmitted by one SU to other SUs is represented by a single bit whose 1/0 values indicates that the underlying channel is available and busy, respectively.
In general, the error probability of the reporting message between SUs $i_1$ and $i_2$ depends on the employed modulation scheme and the signal to noise ratio (SNR) of the
communication channel between the two SUs. We denote the bit error probability of transmitting the reporting bit from SU  $i_2$ to SU $i_1$ as $\mathcal{P}_e^{i_1i_2} $. 
In addition, we assume that the error processes of different reporting bits for different SUs are independent.
Then, the probability of detection and probability of false alarm experienced by SU $i_1$ on channel $j$ with the sensing result received from SU $i_2$ can be 
written as 
\beqn
\mathcal{P}_{u,e}^{i_1i_2j} = \left\{ \begin{array}{*{20}{c}}
   \mathcal{P}_u^{i_2j} \left(1-\mathcal{P}_e^{i_1i_2}\right) + \left(1-\mathcal{P}_u^{i_2j}\right) \mathcal{P}_e^{i_1i_2} & {\mbox{if } i_1 \neq i_2}  \\
   \mathcal{P}_u^{i_2j}  & {\mbox{if } i_1 = i_2}  \\
\end{array} \right.  \label{eq1_dcss_1}
\eeqn
where $u \equiv d$ and $u \equiv f$ represents probabilities of detection and false alarm, respectively. 
Note that we have $\mathcal{P}_e^{i_1i_2} = 0$ if $i_1=i_2=i$ since there is no sensing result exchange involved in this case.
As SU $i$ employs the $a_j$-out-of-$b_j$ aggregation rule for channel $j$, the probabilities of detection and false alarm for SU $i$ on channel $j$ 
can be calculated as
\beqn
\mathcal{\tilde{P}}_u^{ij}\left( {\vec \varepsilon ^j}, {\vec \tau^j} , a_j  \right) = \sum_{l=a_j}^{b_j} \sum_{k=1}^{C_{b_j}^l} \prod_{i_1 \in \Phi^l_k} \mathcal{P}_{u,e}^{ii_1j} \prod_{i_2 \in \mathcal{S}_j^{U} \backslash \Phi^l_k} \mathcal{\bar P}_{u,e}^{ii_2j}. \label{Pu_rep_1}
\eeqn
Again, $u \equiv d$ and $u \equiv f$ represent the corresponding probabilities of detection or  false alarm, respectively. 
Recall that $\mathcal{S}_j^U $ represents the set of SUs who are assigned to sense channel $j$; thus, we have $b_j = |\mathcal{S}_j^U|$
and $1 \leq a_j \leq b_j = \left| \mathcal{S}_j^U \right|$.    
For brevity, $\mathcal{\tilde P}_u^{ij}\left({\vec \varepsilon ^j}, {\vec \tau^j}, a_j  \right) $ is written as $\mathcal{\tilde P}_u^{ij}$ in the following.

\subsection{Throughput Analysis Considering Reporting Errors}
\label{TputanaCRE}

\begin{figure}[!t]%[!t]
\centering
\includegraphics[width=90mm]{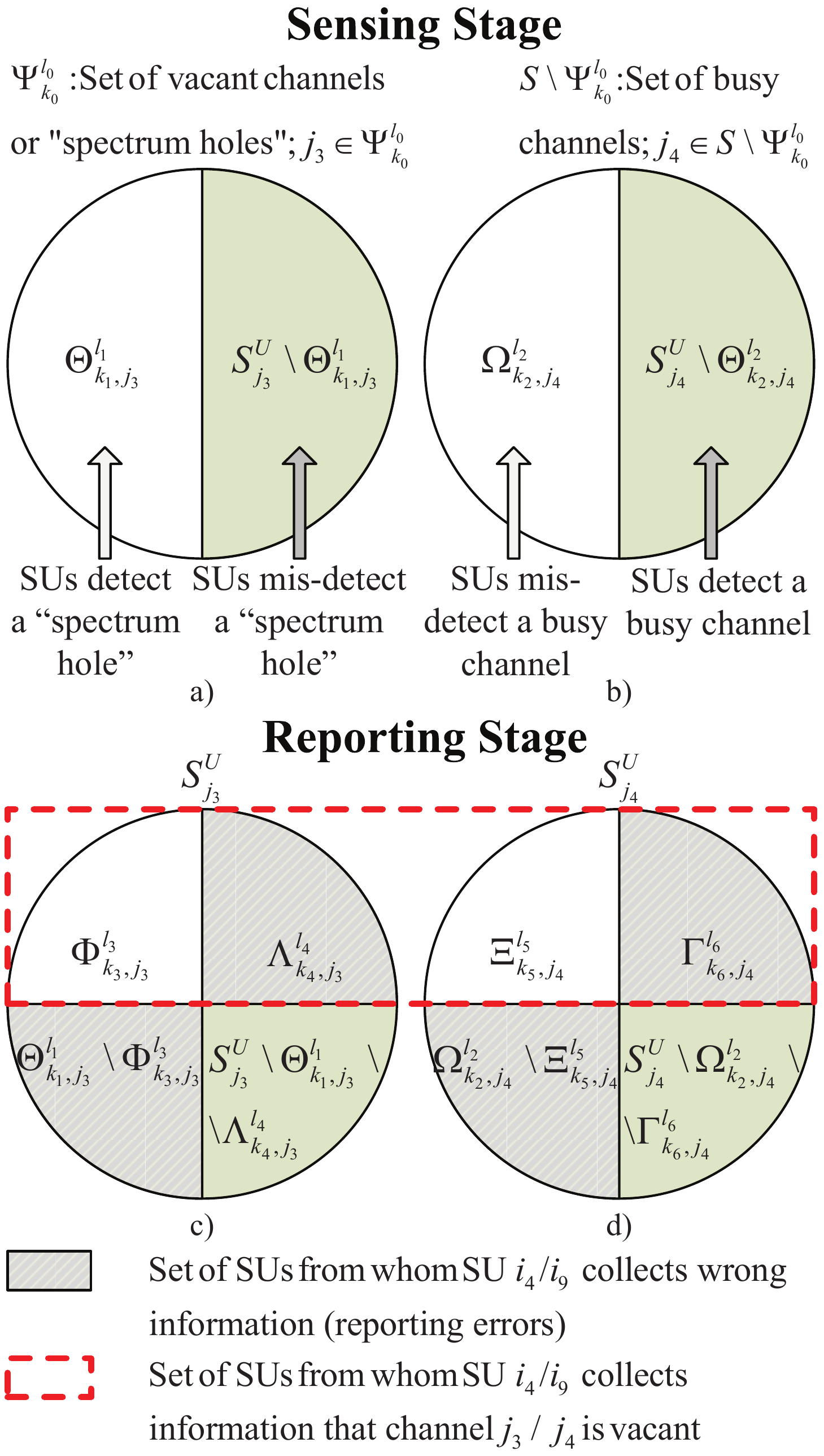}
\caption{Illustration of different sets in one combined scenario.}
\label{Fig0}
\end{figure}

In order to analyze the saturation throughput for the case there are reporting errors, we have to
consider all possible scenarios due to the idle/busy status of all channels, sensing outcomes given by different SUs, and error/success 
events in the sensing result exchange processes. For one such combined scenario we have to derive the total conditional
throughput due to all available channels. Illustration of different involved sets for one combined scenario of following analysis is presented in Fig.~\ref{Fig0}.
In particular, the normalized throughput considering reporting errors can be expressed as follows:
\beqn
\mathcal{NT} = \sum_{k_0=1}^M \sum_{l_0=1}^{C_M^{k_0}} \prod_{j_1 \in \Psi_{k_0}^{l_0}} \mathcal{P}_{j_1} \left(\mathcal{H}_0\right) \prod_{j_2 \in \mathcal{S} \backslash \Psi_{k_0}^{l_0}} \mathcal{P}_{j_2} \left(\mathcal{H}_1\right) \times \label{NTSC_1}\\
\prod_{j_3 \in \Psi_{k_0}^{l_0}} \sum_{k_1=0}^{ |\mathcal{S}_{j_3}^U| } \sum_{l_1=1}^{C_{|\mathcal{S}_{j_3}^U|}^{k_1}} \prod_{i_0 \in \Theta_{k_1,j_3}^{l_1}} 
\mathcal{\bar P}_f^{i_0,j_3} 
\prod_{i_1 \in  \mathcal{S}_{j_3}^U \backslash \Theta_{k_1,j_3}^{l_1}} \mathcal{P}_f^{i_1,j_3} \times \label{NTSC_2} \\
\prod_{j_4 \in \mathcal{S} \backslash \Psi_{k_0}^{l_0}} \sum_{k_2 = 0}^{|\mathcal{S}_{j_4}^U|} \sum_{l_2=1}^{C_{|\mathcal{S}_{j_4}^U|}^{k_2}} \prod_{i_2 \in \Omega_{k_2,j_4}^{l_2}} \mathcal{\bar P}_d^{i_2,j_4} \prod_{i_3 \in \mathcal{S}_{j_4}^U \backslash \Omega_{k_2,j_4}^{l_2}} \mathcal{P}_d^{i_3,j_4} \times  \label{NTSC_3}
\eeqn
\beqn
\prod_{i_4 \in  \mathcal{S}^U }  \sum_{k_3=0}^{k_1 } \sum_{l_3=1}^{C_{k_1 }^{k_3}} \prod_{i_5 \in \Phi_{k_3,j_3}^{l_3}}  \mathcal{\bar P}_e^{i_4,i_5} \prod_{i_6 \in \Theta_{k_1,j_3}^{l_1} \backslash \Phi_{k_3,j_3}^{l_3}}  \mathcal{P}_e^{i_4,i_6} \times \label{NTSC_40} \\
\sum_{k_4=0 }^{|\mathcal{S}_{j_3}^U| - k_1 } \sum_{l_4=1}^{C_{|\mathcal{S}_{j_3}^U| - k_1 }^{k_4 }} \prod_{i_7 \in \Lambda_{k_4,j_3}^{l_4 }}  \mathcal{P}_e^{i_4,i_7} \prod_{i_8 \in \mathcal{S}_{j_3}^U \backslash \Theta_{k_1,j_3}^{l_1} \backslash \Lambda_{k_4,j_3}^{l_4}}  \mathcal{\bar P}_e^{i_4,i_8} \times \label{NTSC_4} \\
\prod_{i_9 \in  \mathcal{S}^U }  \sum_{k_5=0}^{k_2 } \sum_{l_5=1}^{C_{k_2 }^{k_5}} \prod_{i_{10} \in \Xi_{k_5,j_4}^{l_5}}  \mathcal{\bar P}_e^{i_9,i_{10}} \prod_{i_{11} \in \Omega_{k_2,j_4}^{l_2} \backslash \Xi_{k_5,j_4}^{l_5}}  \mathcal{P}_e^{i_9,i_{11}} \times  \label{NTSC_50} \\
\sum_{k_6=0 }^{|\mathcal{S}_{j_4}^U| - k_2 } \sum_{l_6=1}^{C_{|\mathcal{S}_{j_4}^U| - k_2 }^{k_6 }} \prod_{i_{12} \in \Gamma_{k_6,j_4}^{l_6 }} \!\!\!\! \mathcal{P}_e^{i_9,i_{12}} \!\!\!\!\!\!\!\! \prod_{i_{13} \in \mathcal{S}_{j_4}^U \backslash \Omega_{k_2,j_4 }^{l_2 } \backslash \Gamma_{k_6,j_4}^{l_6 }} \!\!\!\!\!\!\!\! \mathcal{\bar P}_e^{i_9,i_{13}} \times \label{NTSC_5} \\
\mathcal{T}_p^{\sf re} \left(\tau,\left\{a_j\right\},p  \right), \label{NTSC_6}
\eeqn
where $\mathcal{T}_p^{\sf re} \left(\tau,\left\{a_j\right\},p  \right)$ denotes the conditional throughput for one
combined scenario discussed above. 
In (\ref{NTSC_1}), we generate all possible sets where $k_0 $ channels are available for secondary access (i.e., they are not used by PUs)
while the remaining channels are busy. There are $C_{M}^{k_0}$ such sets and $\Psi_{k_0}^{l_0} $ represents one particular set of available channels. 
The first product term in (\ref{NTSC_1}) denotes the probability that all channels in $\Psi_{k_0}^{l_0} $ are available while the second product term describes
 the probability that the remaining channels are busy. %Note that $\bar{\Psi}_{k_0}^{l_0} = \mathcal{S} \backslash \Psi_{k_0}^{l_0}$. 

Then, for one particular channel $j_3 \in \Psi_{k_0}^{l_0}$, we generate all possible sets with $k_1 $ SUs in $\mathcal{S}_{j_3}^U $ ($\mathcal{S}_{j_3}^U $ is the set of SUs 
who are assigned to sense channel $j_3 $) whose sensing results indicate that channel $j_3 $ is available in (\ref{NTSC_2}).
There are $C_{\left|\mathcal{S}_{j_3}^U\right|}^{k_1 } $ sets and $\Theta_{k_1,j_3}^{l_1} $ denotes one such typical set.
Again, the first product term in (\ref{NTSC_2}) is the probability that the sensing outcomes of all SUs in $\Theta_{k_1,j_3}^{l_1} $ indicate that channel $j_3 $ is available; and 
the second term is the probability that the sensing outcomes of all SUs in the remaining set $\mathcal{S}_{j_3}^U \backslash \Theta_{k_1,j_3}^{l_1} $ indicate 
that channel $j_3 $ is not available.

In (\ref{NTSC_3}), for one specific channel $j_4 \in \mathcal{S} \backslash \Psi_{k_0}^{l_0}$, we generate all possible sets with $k_2 $ SUs in 
$\mathcal{S}_{j_4}^U $ whose sensing outcomes indicate that channel $j_4 $ is available due to missed detection.
There are $C_{\left|\mathcal{S}_{j_4}^U\right|}^{k_2} $ such sets and $\Omega_{k_2,j_4}^{l_2} $ is a typical one.
Similarly, the first product term in (\ref{NTSC_3}) is the probability that the sensing outcomes of all SUs in $\Omega_{k_2,j_4}^{l_2} $ indicate 
that channel $j_4 $ is available; and the second term is the probability that the sensing outcomes of all SUs in the remaining set $\mathcal{S}_{j_4}^U 
\backslash \Omega_{k_2,j_4}^{l_2} $ indicate that channel $j_4 $ is not available.

Recall that for any specific channel $j$, each SU in $\mathcal{S}^U $ (the set of all SUs) receives sensing results from 
a group of SUs who are assigned to sense the channel $j$. In (\ref{NTSC_40}), we consider all possible error events due to message exchanges from 
SUs in $\Theta_{k_1,j_3}^{l_1} $.
The first group denoted as $\Phi_{k_3,j_3}^{l_3}$ includes SUs in $\Theta_{k_1,j_3}^{l_1} $ has its sensing results received at SU  $i_4 \in \mathcal{S}^U $ indicating that 
channel $j_3 $ available (no reporting error) while the second group of SUs $  \Theta_{k_1,j_3}^{l_1} \backslash  \Phi_{k_3,j_3}^{l_3}$ has the sensing results 
received at SU  $i_4 \in \mathcal{S}^U $ suggesting that channel $j_3 $ is not available due to reporting errors. For each of these two groups, we generate all possible
 sets of SUs of different sizes and capture the corresponding
probabilities. In particular, we generate all sets with $k_3 $ SUs $i_5 \in \Phi_{k_3,j_3}^{l_3}$ where SU $i_4 $ collects correct sensing information from SUs $i_5 $ 
(i.e., there is no error on the channel between $i_4 $ and $i_5 $). Similar expression is presented for the second group in which we generate all sets of 
$k_4$ SUs $i_6  \in \Theta_{k_1,j_3}^{l_1} \backslash  \Phi_{k_3,j_3}^{l_3} $ where SU $i_4 $ collects wrong sensing information from each SU $i_6 $ (i.e., there is an error on 
the channel between $i_4 $ and $i_6 $). Similarly, we present the possible error events due to exchanges of sensing results from the set of SUs $\mathcal{S}_{j_3}^U \backslash \Theta_{k_1,j_3}^{l_1} $ in (\ref{NTSC_4}).

In (\ref{NTSC_50}) and (\ref{NTSC_5}), we consider all possible error events due to sensing result exchanges
for channel $j_4 \in \mathcal{S} \backslash \Psi_{k_0}^{l_0}$. Here, each SU in $\mathcal{S}^U $ collects sensing result information from two sets of SUs in $\Omega_{k_2,j_4}^{l_2} $
and $\mathcal{S}_{j_4}^U \backslash \Omega_{k_2,j_4}^{l_2} $, respectively. 
The first set includes SUs in $\Omega_{k_2,j_4}^{l_2} $ whose sensing results indicate that channel $j_4 $ available due to missed detection, while the second set includes 
SUs in $\mathcal{S}_{j_4}^U \backslash \Omega_{k_2,j_4}^{l_2} $ whose sensing results indicate that channel $j_4 $ is not available. 
Possible outcomes for the message exchanges due to the first set $\Omega_{k_2,j_4}^{l_2} $ are captured in (\ref{NTSC_50}) where we present the outcomes
for two groups of this first set. For group one, we generate all sets with $k_5 $ SUs $i_{10} \in \Xi_{k_5,j_4}^{l_5}$ where SU $i_9 $ collects correct sensing information from SUs $i_{10} $ (i.e., there is no error on the channel between $i_9 $ and $i_{10} $). 
For group two, we consider the remaining sets of SUs in $\Omega_{k_2,j_4}^{l_2}  \backslash  \Xi_{k_5,j_4}^{l_5} $ where SU $i_9 $ receives
wrong sensing information from each SU $i_{11} $ (i.e., there is an error on the channel between $i_9 $ and $i_{11} $).
Similar partitioning of the set $\mathcal{S}_{j_4}^U \backslash \Omega_{k_2,j_4}^{l_2} $ into two groups $\Gamma_{k_6,j_4}^{l_6 }$ and $\mathcal{S}_{j_4}^U \backslash \Omega_{k_2,j_4}^{l_2} \backslash \Gamma_{k_6,j_4}^{l_6 }$ with the corresponding message reporting error patterns is captured in (\ref{NTSC_5}).

For each combined scenario whose probability is presented above, each SU $i$ has collected sensing result information for each
channel, which is the sensing results obtained by itself or received from other SUs.
Then, each SU $i$ determines the idle/busy status of each channel $j$ by applying the $a_j$-out-of-$b_j$ rule on the collected sensing information. 
Let $\mathcal{S}^a_i$ be set of channels, whose status is ``available'' as being suggested by the $a_j$-out-of-$b_j$ rule at SU $i$. 
According to our design MAC protocol, SU $i$ will randomly select one channel in the set $\mathcal{S}^a_i$ to perform contention and transmit its data.
In order to obtain the conditional throughput $\mathcal{T}_p^{\sf re} \left(\tau,\left\{a_j\right\},p  \right)$ for one particular
combined scenario, we have to reveal the contention operation on each actually available channel, which is presented in the following.

Let  $\mathcal{S}^a_i =  \mathcal{S}^a_{1,i} \cup \mathcal{S}^a_{2,i} $ where channels in $\mathcal{S}^a_{1,i}$ are actually available and
channels in $\mathcal{S}^a_{2,i}$ are not available but the SDCSS policy suggests the opposite due to sensing and/or reporting errors.
Moreover, let $\mathcal{\hat S}^a_1 = \bigcup_{i \in \mathcal{S}^U} \mathcal{S}^a_{1,i} $ be the set of actually available channels, which are detected by
 all SUs by using the  SDCSS policy. Similarly, we define $\mathcal{\hat S}^a_2 = \bigcup_{i \in \mathcal{S}^U} \mathcal{S}^a_{2,i} $ 
as the set of channels indicated as available by some SUs due to errors. Let $k_e^i = \left|\mathcal{S}^a_i\right|$ be the number of available channels at SU $i $;
then SU $i$ chooses one channel in  $\mathcal{S}^a_i$ to transmit data with probability  $1/k_e^i$.
In addition, let $\mathcal{\hat S}^a = \mathcal{\hat S}^a_1 \cup \mathcal{\hat S}^a_2 $ be set of all ``available'' channels each of which is determined as being available by at 
least one SU and let $k_{\sf{max}} = \left|\mathcal{\hat S}^a\right|$ be the size of this set.

%To calculate the throughput for each channel $j$, let $\Psi_j^a $ be the set of SUs whose SDCSS outcomes indicate that channel $j $ is available
%and let us define $\Psi^a = \bigcup_{j \in \mathcal{\hat S}^a} \Psi_j^a $. In addition, let us define $N_j = \left|\Psi_j^a \right|$ and $N_{\text{max}} = \left|\Psi^a \right|$, 
%which are the sizes of these sets, respectively. Moreover, we assume that channels in $\mathcal{\hat S}^a$ are indexed by $1, 2, \ldots, k_{\text{max}}$.
%Similar to the throughput analysis without reporting errors, we consider all possible sets $\left\{ n_{j} \right\} = \left\{n_1, n_2, \ldots, n_{k_{\text{max}}} \right\}$
%where $n_j$ is the number of SUs choosing channel $j$ to transmit data. Then, we can calculate the conditional throughput as follows:

To calculate the throughput for each channel $j$, let $\Psi_j^a $ be the set of SUs whose SDCSS outcomes indicate that channel $j $ is available
and let $\Psi^a = \bigcup_{j \in \hat{\mathcal S}^a} \Psi_j^a $ be the set of SUs whose SDCSS outcomes indicate that at least one channel in the assigned spectrum sensing set is available. 
In addition, let us define $N_j = \left|\Psi_j^a \right|$ and $N_{\sf{max}} = \left|\Psi^a \right|$, which describe the sizes of these sets, respectively.
It is noted that $N_{\sf{max}} \leq N $ due to the following reason. 
In any specific combination that is generated in Eqs. (\ref{NTSC_1})--(\ref{NTSC_5}), there can be some SUs, denoted as $\left\{i\right\}$, whose sensing outcomes 
indicate that all channels in the assigned spectrum sensing sets are not available (i.e., not available for access).
Therefore, we have $\Psi^a = {\mathcal S}^U \backslash \left\{i\right\}$, which implies $N_{\sf{max}} \leq N $ where $N = \left|{\mathcal S}^U \right| $.
Moreover, we assume that channels in $\hat{\mathcal S}^a$ are indexed by $1, 2, \ldots, k_{\sf{max}}$.
Similar to the throughput analysis without reporting errors, we consider all possible sets $\left\{ n_{j} \right\} = \left\{n_1, n_2, \ldots, n_{k_{\sf{max}}} \right\}$
where $n_j$ is the number of SUs choosing channel $j$ for access. Then, we can calculate the conditional throughput as follows:
\beqn
\mathcal{T}_p^{\sf re} \left(\tau,\left\{a_j\right\},p \right)  = \sum_{\left\{n_{j_1}\right\}: \sum _{j_1 \in \mathcal{\hat S}^a } n_{j_1}=N_{\text{max}}} 
\mathcal{P}\left(\left\{N_{j_1},n_{j_1}\right\}\right) \times \hspace{0.0cm} \label{T_p_cal_re1} \\
\sum_{j_2 \in \mathcal{\hat S}_1^a} \frac{1}{M} \mathcal{T}_{j_2}^{\sf re} \left(\tau,\left\{a_{j_2}\right\},p \left|n=n_{j_2}\right.\right)   \mathcal{I} \left(n_{j_2}>0\right). \label{T_p_cal_re2} 
\eeqn
Here $\mathcal{P}\left(\left\{N_{j_1},n_{j_1}\right\}\right)$ is the probability that each channel $j_1 $ ($j_1 \in \mathcal{\hat S}^a $) is selected by $n_{j_1} $ SUs
for $j_1=1, 2, \ldots, k_{\text{max}}$. This probability can be calculated as
\beqn
\label{P_Nj1_nj1}
\mathcal{P}\left(\left\{N_{j_1},n_{j_1}\right\}\right) = \left( {\begin{array}{*{20}{c}}
   {\left\{N_{j_1}\right\}}  \\
   {\left\{n_{j_1}\right\}}  \\
\end{array}} \right) \prod_{i \in \Psi^a}\left(\frac{1}{k_e^i}\right),
\eeqn
where  $\left( {\begin{array}{*{20}{c}}
   {\left\{N_{j_1}\right\}}  \\
   {\left\{n_{j_1}\right\}}  \\
\end{array}} \right) $
describes the number of ways to realize the access vector $\left\{ n_{j} \right\}$ for $k_{\sf{max}}$ channels, which can be obtained by using the enumeration technique as follows. 
For a particular way that the specific set of $n_1 $ SUs ${\mathcal S}_{1}^{n_1}$ choose channel one (there are $C^{n_1}_{N_1}$ such ways), we can express the set of remaining SUs 
that can choose channel two as $\Psi^a_{(2)} = \Psi_2^a \backslash ({\mathcal S}_{1}^{n_1} \cap \Psi_2^a) $. 
We then consider all possible ways that $n_2 $ SUs in the set $\Psi^a_{(2)} $ choose channel two and we denote this set of SUs as ${\mathcal S}_{2}^{n_2} $ (there are $C^{n_2}_{\widetilde{N}_2}$ 
such ways where $\widetilde{N}_2 = |\Psi^a_{(2)}|$). Similarly, we can express the set of SUs that can choose channel three as $\Psi^a_{(3)} = \Psi^a_3 \backslash ((\cup_{i=1}^2 {\mathcal S}_{i}^{n_i}) \cap \Psi^a_3) $ and consider all possible ways that $n_3$ SUs in the set $\Psi^a_{(3)}$ can choose channel three, and so on. 
This process is continued until $n_{k_{\sf{max}}}$ SUs choose channel $k_{\sf{max}}$. Therefore, the number of ways to realize the access vector $\left\{ n_{j} \right\}$ can be 
determined by counting all possible cases in the enumeration process.

%where  $\left( {\begin{array}{*{20}{c}}
   %{\left\{N_{j_1}\right\}}  \\
   %{\left\{n_{j_1}\right\}}  \\
%\end{array}} \right) $
%describes the number of ways to realize the access vector $\left\{ n_{j} \right\}$ for $k_{\text{max}}$ channels,
%which can be obtained by using the enumeration technique as follows. For any particular way under which the specific set of $n_1 $ SUs $\mathcal{S}_{1}^{n_1}$ choose channel one
 %(there are $C^{n_1}_{N_1}$ such ways), we determine the set of remaining SUs that can choose channel two as $\Psi^a_{(2)} = \Psi^a \backslash \mathcal{S}_{1}^{n_1}$. We then realize all particular ways
%where $n_2$ SUs in the set $\Psi^a_{(2)}$ choose channel two and we denote this set of SUs as $\mathcal{S}_{2}^{n_2}$  (there are $C^{n_2}_{\widetilde{N}_2}$ such ways where $\widetilde{N}_2 = |\Psi^a_{(2)}|$). We then obtain the set of SUs that can choose channel three as $\Psi^a_{(3)} = \Psi^a_{(2)} \backslash \mathcal{S}_{2}^{n_2}$
%and realize all particular ways under which $n_3$ SUs in the set $\Psi^a_{(3)}$ choose channel three, and so on. This process is continued until $n_{k_{\text{max}}}$ SUs choose
%channel $k_{\text{max}}$ in the last step. The number of ways to realize $\left\{ n_{j} \right\}$ can be determined by counting all possible realizations in the enumeration process.

The product term in (\ref{P_Nj1_nj1}) is due to the fact that each SU $i$ chooses one available with probability $1/k_e^i$. 
The conditional throughput $\mathcal{T}_{j_2}^{\sf re} \left(\tau, \left\{a_{j_2}\right\},p\left| n=n_{j_2}\right. \right) $ is calculated by using the same expression (\ref{con_T})
given in Section \ref{CPCSMA_Chap5}. In addition, only actually available channel $j_2 \in \mathcal{\hat S}_1^a$ can contribute the total throughput, which explains the throughput 
sum in (\ref{T_p_cal_re2}).

\subsection{Design Optimization with Reporting Errors}

The optimization of channel sensing/access parameters as well as channel sensing sets can be conducted in the same manner with that
in Section \ref{CPCSMA_Chap5}. However, we have to utilize the new throughput analytical model presented in Section \ref{TputanaCRE} in this case. Specifically, Algs. \ref{mainalg_pg} and \ref{ChanAAp} can 
still be used to determine the optimized sensing/access parameters and channel sensing sets, respectively. Nonetheless,
we need to use the new channel sensing model capturing reporting errors in Section \ref{CS_w_RE} in these algorithms. In particular,
 from the equality constraint on the detection probability, i.e., $\mathcal{P}_d^j\left( {\vec \varepsilon ^j}, {\vec \tau^j} , a_j  \right) = \mathcal{\widehat{P}}_d ^j$,
we have to use (\ref{eq1_dcss_1}) and (\ref{Pu_rep_1}) to determine ${P}_d^{ij}$ (and the corresponding ${P}_f^{ij}$)
assuming that ${P}_d^{ij}$ are all the same for all pairs $\left\{i,j\right\}$ as what we have done in Section \ref{CPCSMA_Chap5}.

\vspace{10pt}
\section{Numerical Results}
\label{Results_Chap5}

To obtain numerical results in this section, the key parameters for the proposed MAC protocol are chosen as follows:
cycle time is $T = 100 ms$; the slot size is $v=20{\mu} s$, which is the same as in IEEE 802.11p standard; packet size is $PS = 450$ slots (i.e., $450v$); 
propagation delay $PD = 1 {\mu} s$; $SIFS = 2$ slots; $DIFS = 10 $ slots; $ACK = 20$ slots; $CTS = 20$ slots; $RTS = 20$ slots;
sampling frequency for spectrum sensing is $f_s = 6 MHz$;  and $t_r = 80 {\mu} s$.  %bandwidth of PUs' QPSK signals is $6 MHz$;
The results presented in all figures except Fig. \ref{Fig9} correspond to the case where there is no reporting error.

\begin{table*} % [!t]
\centering
\caption{Throughput vs probability of vacant channel (MxN=4x4)}
\label{table1}
\footnotesize
\begin{tabular}{|c|c|c|c|c|c|c|c|c|c|c|c|}
\cline{3-12} 
\multicolumn{2}{c|}{} & \multicolumn{10}{c|}{$\mathcal{P}_j\left(\mathcal{H}_0\right)$}\tabularnewline
\cline{3-12} 
\multicolumn{2}{c|}{} & 0.1 & 0.2 & 0.3 & 0.4 & 0.5 & 0.6 & 0.7 & 0.8 & 0.9 & 1\tabularnewline
\hline 
 & Greedy &0.0816  &  0.1524   & 0.2316   & 0.2982   & 0.3612   & 0.4142  &  0.4662   & 0.5058   & 0.5461  &  0.5742 \tabularnewline
\cline{2-12} 
$\mathcal{NT}$ & Optimal & 0.0817  &  0.1589  &  0.2321  &  0.3007  &  0.3613  &  0.4183  &  0.4681  &  0.5087  &  0.5488  &  0.5796 \tabularnewline
\cline{2-12} 
 & Gap (\%) & 0.12    &    4.09   & 0.22  &  0.83   &     0.03  &  0.98    &     0.40  &  0.57  &  0.49    &   0.93 \tabularnewline
\hline
\end{tabular}
\end{table*}

To investigate the efficacy of our proposed low-complexity channel assignment algorithm (Alg. \ref{ChanAAp}), we compare the throughput performance achieved by the optimal brute-force search and greedy channel assignment algorithm in Table ~\ref{table1}. 
In particular, we show normalized throughput $\mathcal{NT}$ versus probabilities $\mathcal{P}_j\left(\mathcal{H}_0\right)$ for these two algorithms and the relative gap between them. 
Here, the probabilities $\mathcal{P}_j\left(\mathcal{H}_0\right)$ for different channels $j$  are chosen to be the same and we choose $M = 4$ channels and $N = 4$ SUs. 
To describe the SNR of different SUs and channels, we use  $\left\{ i,j \right\}$ to denote a combination of channel $j$ and SU $i$ who senses this channel. 
The SNR setting for different combinations of SUs and channels $\left\{i, j\right\}$ is performed for two groups of SUs as $\gamma_1^{ij} = -15 dB$: channel 1: $\left\{1,1\right\}, \left\{2,1\right\}, \left\{3,1\right\} $; channel 2: $\left\{2,2\right\}, \left\{4,2\right\} $; channel 3: $\left\{1,3\right\},\left\{4,3\right\}$; and channel 4: $\left\{1,4\right\}, \left\{3,4\right\}$. 
The remaining combinations correspond to the SNR value $\gamma_2^{ij} = -20 dB$ for group two. 
The results in this table confirms that the throughput gaps between our greedy algorithm and the brute-force optimal search algorithm are quite small, which are less that 1\% for all except the case two presented in this table. 
These results confirm that our proposed greedy algorithm works well for small systems (i.e., small M and N).
In the following, we investigate the performance of our proposed algorithms for larger systems.

\begin{figure*}%[!t]
\centering
\includegraphics[width=90mm]{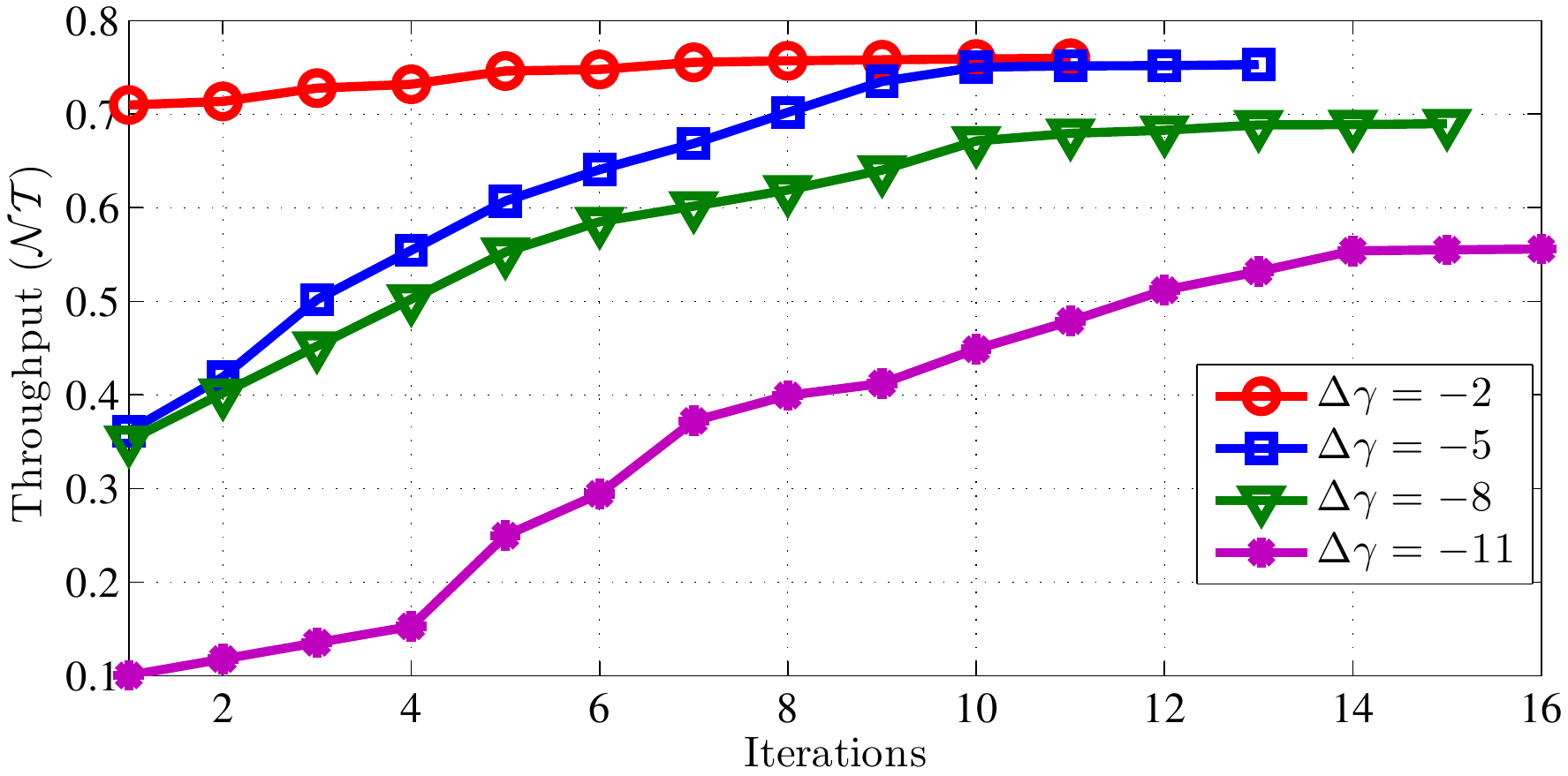}
\caption{Convergence illustration for Alg.~\ref{ChanAAp}.}
\label{Convergence_NT_Iter}
\end{figure*}

\begin{figure}[!t]
\centering
\includegraphics[width=90mm]{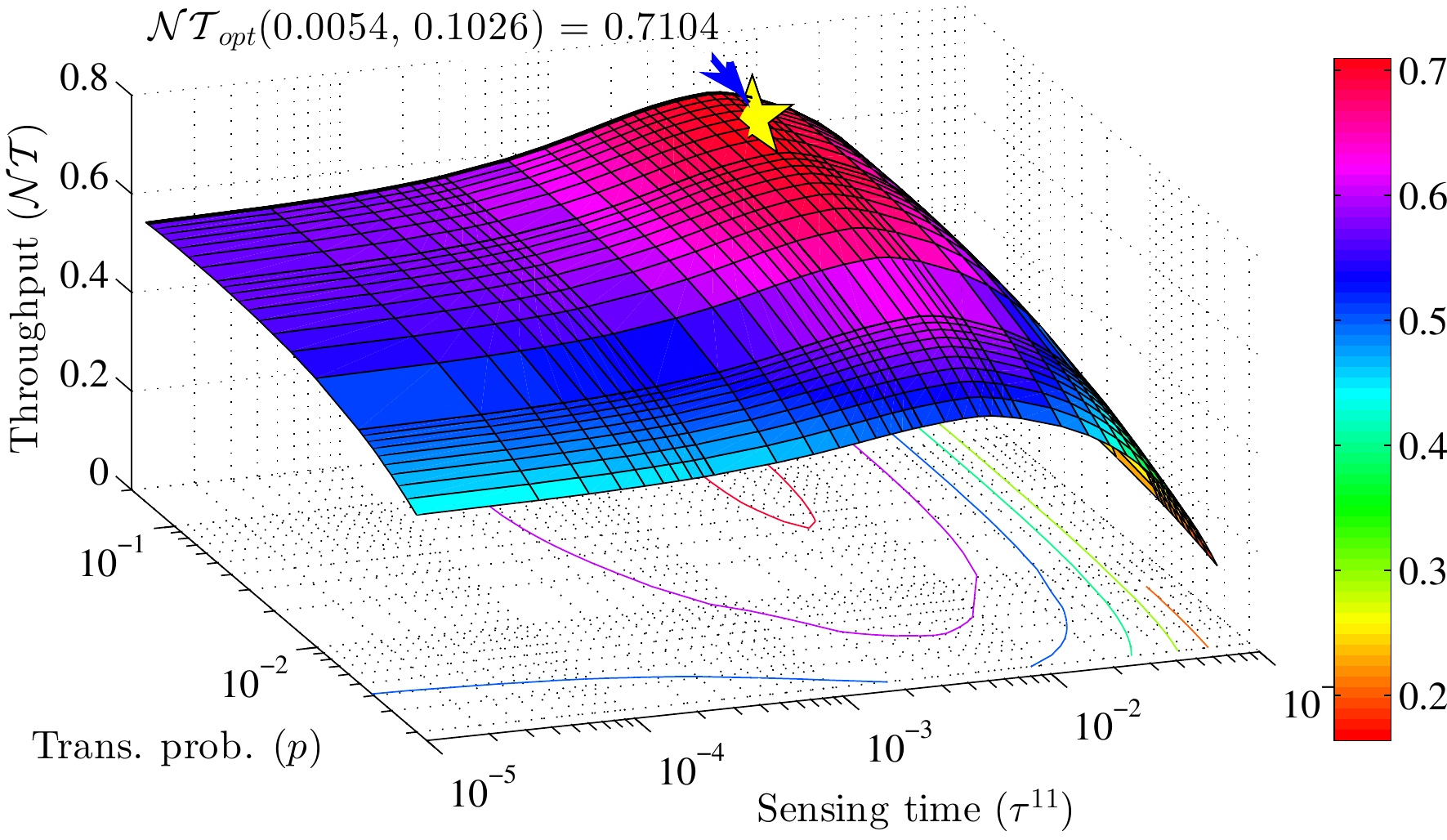}
\caption{Normalized throughput versus transmission probability $p$ and sensing time $\tau ^{11}$ for $\Delta \gamma = -7$, $N=10$ and $M=4$.}
\label{Fig4}
\end{figure}

\begin{figure}[!t]
\centering
\includegraphics[width=90mm]{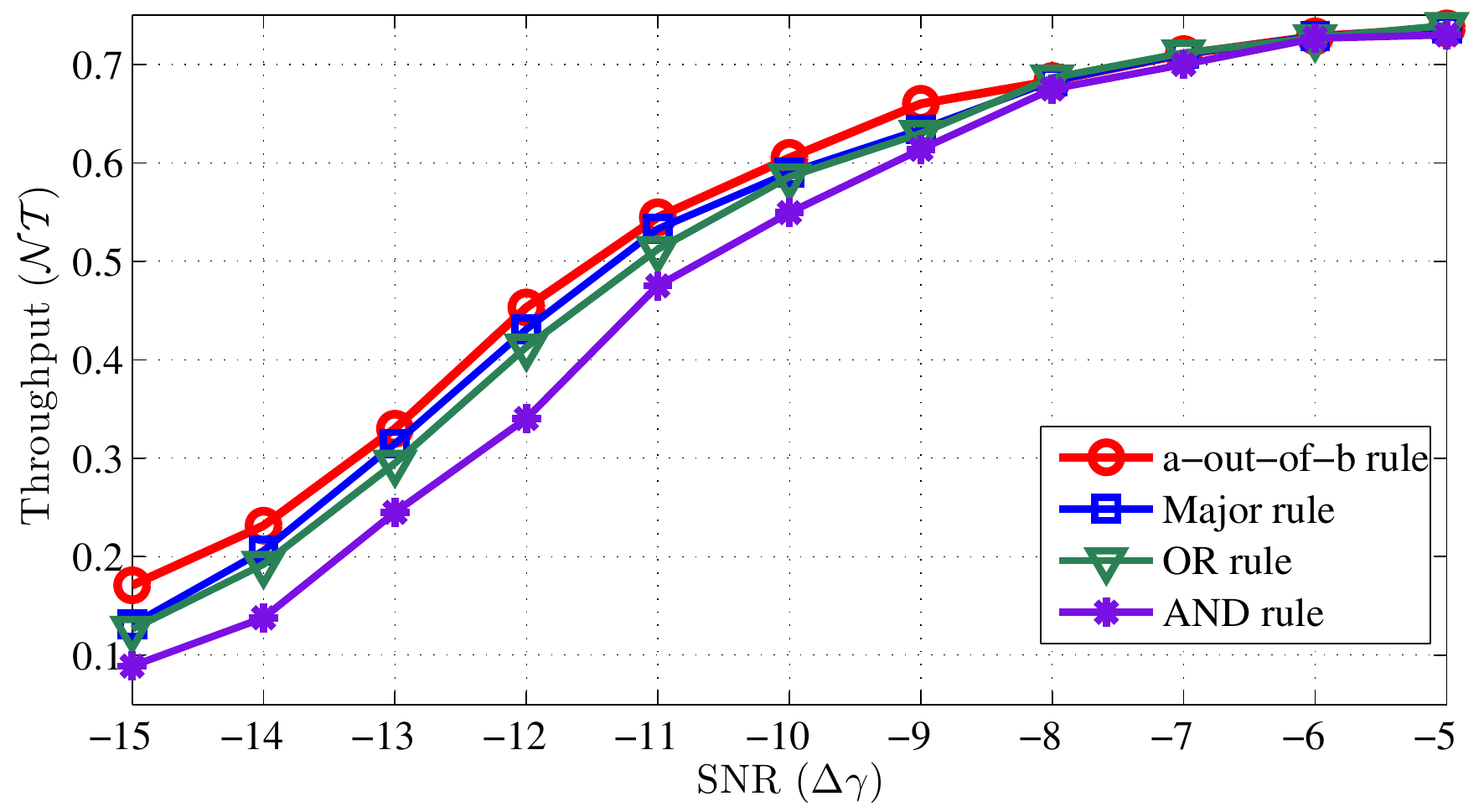}
\caption{Normalized throughput versus SNR shift $\Delta \gamma$ for $N=10$ and $M=4$ under 4 aggregation rules.}
\label{Fig5}
\end{figure}

\begin{figure}[!t]
\centering
\includegraphics[width=90mm]{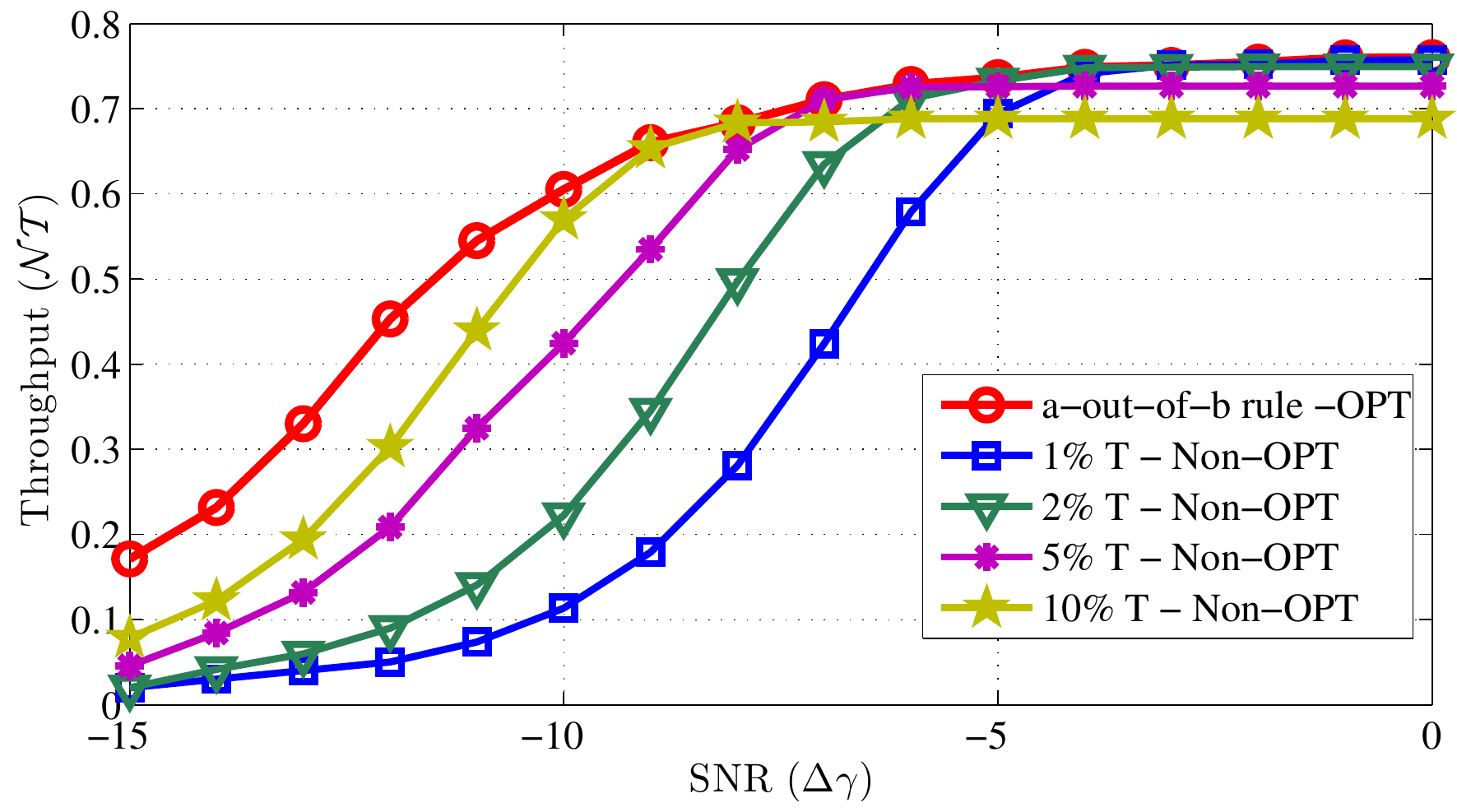}
\caption{Normalized throughput versus SNR shift $\Delta \gamma$ for $N=10$ and $M=4$ for optimized and non-optimized scenarios.}
\label{Fig6}
\end{figure}

\begin{figure}[!t]
\centering
\includegraphics[width=90mm]{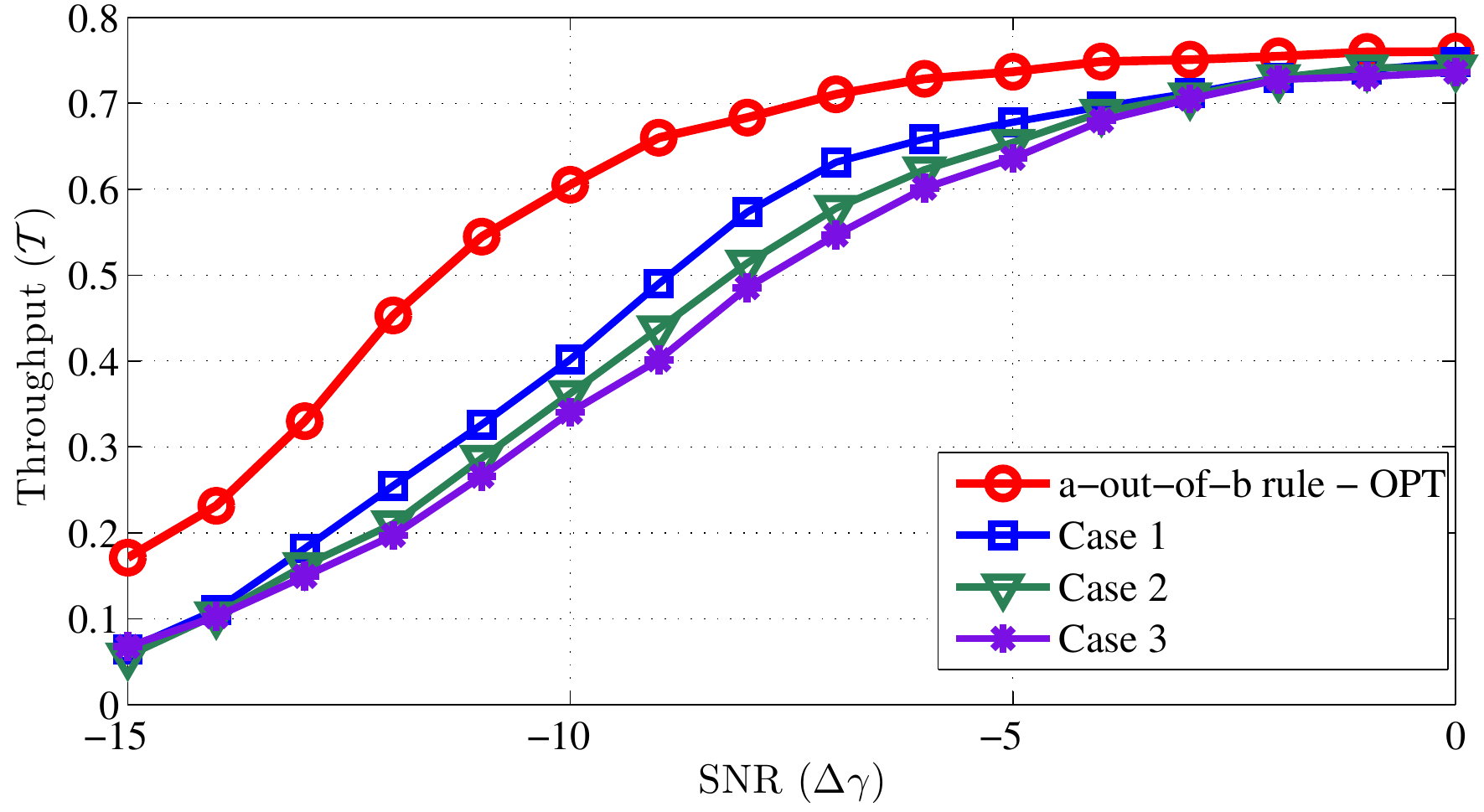}
\caption{Normalized throughput versus SNR shift $\Delta \gamma$ for $N=10$ and $M=4$ for optimized and RR channel assignments.}
\label{Fig7}
\end{figure}

To investigate the performance of our proposed algorithm for a typical system, we consider the network setting with $N=10$ and $M=4$. We divide SUs into 2 groups where
 the received SNRs at SUs due to the transmission from PU $i$ is equal
 to $\gamma^{ij}_{1,0} = -15 dB$ and $\gamma^{ij}_{2,0} = -10 dB$ (or their shifted values described later) for the two groups, respectively. 
Again, to describe the SNR of different SUs and channels, we use  $\left\{ i,j \right\}$ to denote a combination of channel $j$ and SU $i$ who senses this channel. 
The combinations of the first group corresponding to $\gamma_{1,0}^{ij} = -10 dB$ are chosen as follows: channel 1: $\left\{1,1\right\}, \left\{2,1\right\}, \left\{3,1\right\} $; channel 2: $\left\{2,2\right\}, \left\{4,2\right\}, \left\{5,2\right\}$; channel 3: $\left\{4,3\right\}, \left\{6,3\right\}, \left\{7,3\right\}$; and channel 4: $\left\{1,4\right\}, \left\{3,4\right\}, \left\{6,4\right\}, \left\{8,4\right\}, \left\{9,4\right\}, \left\{10,4\right\}$. The remaining combinations belong to the second group with
the SNR equal to $\gamma_{2,0}^{ij} = -15 dB$. To obtain results for different values of SNRs, we consider different shifted sets of SNRs  where $\gamma_{1}^{ij} $ and 
$\gamma_{2}^{ij}$ are shifted by 
$\Delta \gamma$ around their initial values $\gamma_{1,0}^{ij} = -15 dB$ and $\gamma_{2,0}^{ij} = -10 dB$ as $\gamma_{1}^{ij} = \gamma_{1,0}^{ij} + \Delta \gamma $ and $\gamma_{2}^{ij} = \gamma_{2,0}^{ij} + \Delta \gamma $. For example, as $\Delta \gamma = -10$, the resulting SNR values are $\gamma_1^{ij} = -25 dB$ and $\gamma_2^{ij} = -20 dB$. 
These parameter settings are used to obtain the results presented in Figs.~\ref{Convergence_NT_Iter},~\ref{Fig4},~\ref{Fig5},~\ref{Fig6}, and \ref{Fig7} in the following.

Fig.~\ref{Convergence_NT_Iter} illustrates the convergence of Alg. 2 where
we show the normalized throughput ${\mathcal NT}_p$ versus the iterations for $\Delta \gamma = -2, -5, -8 $ and $-11 dB$.
For simplicity, we choose $\delta $ equals $10^{-3} \times {\mathcal NT}_c $ in Alg. 2. 
This figure confirms that Alg.~\ref{ChanAAp} converges after about 11, 13, 15 and 16 iterations  for $\Delta \gamma = -2,-5, -8,$ and $-11 dB$, respectively.
In addition, the normalized throughput increases over the iterations as expected.

Fig.~\ref{Fig4} presents normalized throughput $\mathcal{NT}_p$ versus transmission probability $p$ and sensing time $\tau^{11}$ for 
the SNR shift equal to $\Delta \gamma = -7$ where the sensing times for other pairs of SUs and channels are optimized as in Alg. \ref{mainalg_pg}. 
This figure shows that channel sensing and access parameters can strongly impact the throughput of the secondary network,
which indicates the need to optimize them. This figure shows that the optimal values of  $p$ and  $\tau^{11}$ are around
 $\left({\bar \tau}^{11}, {\bar p}\right)= \left(0.0054 s, 0.1026\right)$ to achieve the maximum normalized throughput of $\mathcal{NT}_p = 0.7104 $. 
It can be observed that normalized throughput $\mathcal{NT}_p$ is less sensitive to transmission probability $p$ 
while it varies more significantly as the sensing time $\tau^{11}$ deviates from the optimal value.
In fact, there can be multiple available channels which each SU can choose from. Therefore, the contention level on each
available channel would not be very intense for most values of $p$. This explains why the throughput is not
very sensitive to the access parameter $p$.

% Fig. 5

In Fig.~\ref{Fig5}, we compare the normalized throughput of the secondary network as each SU employs four different aggregation
 rules, namely AND, OR, majority, and the optimal a-out-of-b rules. The
four throughput curves in this figure represent the optimized normalized throughput values achieved by using Algs. \ref{mainalg_pg} and \ref{ChanAAp}. 
For the OR, AND, majority rules, we do not need to find optimized $a_j$ parameters for different channels $j$ in Alg. \ref{mainalg_pg}. 
Alternatively, $a_j = 1 $, $a_j = b_j $ and $a_j = \left\lceil b/2 \right\rceil $ correspond to the OR, AND and majority rules, respectively.
It can be seen that the optimal a-out-of-b rule achieves the highest throughput among the considered rules. 
Moreover, the performance gaps between the optimal a-out-of-b rule and other rule tends to be larger for smaller SNR values.
% Fig. 6

In Fig.~\ref{Fig6}, we compare the throughput performance as the sensing times are  optimized by using Alg. \ref{mainalg_pg} and they are fixed at different fractions of the cycle time in Alg. \ref{mainalg_pg}.  
For fair comparison, the optimized a-out-of-b rules are used in both schemes with optimized and non-optimized sensing times.
For the non-optimized scheme, we employ Alg. \ref{ChanAAp} for channel assignment; however, we do not optimize the sensing times in Alg. \ref{mainalg_pg}. 
Alternatively, $\tau^{ij}$ is chosen from the following values: $1\% T$, $2\%T$, $5\%T$ and $10\%T$ where $T$ is the cycle time. 
Furthermore, for this non-optimized scheme, we still find an optimized value of ${\bar a}_j$ for each channel $j$ (corresponding to the sensing phase) and the optimal value of ${\bar p} $ (corresponding to the access phase) in Alg. \ref{mainalg_pg}. 
This figure confirms that the optimized design achieves the largest throughput.
Also, small sensing times can achieve good throughput performance at the high-SNR regime but result in poor performance if the SNR values are low.
In contrast, too large sensing times (e.g., equal $10\%T $) may become inefficient if the SNR values are sufficiently large. 
These observations again illustrate the importance of optimizing the channel sensing and access parameters.

% Fig. 7

\begin{table}[!t]
\centering
\caption{Round-robin Channel Assignment (x denotes an assignment)}
\label{table_round}
\begin{tabular}{|c|c|c|c|c|c|c|c|c|c|c|c|c|c|}
\cline{3-14} 
\multicolumn{2}{c|}{} & \multicolumn{12}{c|}{\textbf{Channel}}\tabularnewline
\cline{3-14} 
\multicolumn{1}{c}{} &  & \multicolumn{4}{c|}{\textbf{Case 1}} & \multicolumn{4}{c|}{\textbf{Case 2}} & \multicolumn{4}{c|}{\textbf{Case 3}}\tabularnewline
\cline{3-14} 
\multicolumn{2}{c|}{} & 1 & 2 & 3 & 4 & 1 & 2 & 3 & 4 & 1 & 2 & 3 & 4\tabularnewline
\hline 
 & 1 & x &  &  &  & x & x &  &  & x & x & x & \tabularnewline
\cline{2-14} 
 & 2 &  & x &  &  &  & x & x &  &  & x & x & x\tabularnewline
\cline{2-14} 
 & 3 &  &  & x &  &  &  & x & x &  &  & x & x\tabularnewline
\cline{2-14} 
 & 4 &  &  &  & x &  &  &  & x &  &  &  & x\tabularnewline
\cline{2-14} 
\textbf{SU} & 5 & x &  &  &  & x & x &  &  & x & x & x & \tabularnewline
\cline{2-14} 
 & 6 &  & x &  &  &  & x & x &  &  & x & x & x\tabularnewline
\cline{2-14} 
 & 7 &  &  & x &  &  &  & x & x &  &  & x & x\tabularnewline
\cline{2-14} 
 & 8 &  &  &  & x &  &  &  & x &  &  &  & x\tabularnewline
\cline{2-14} 
 & 9 & x &  &  &  & x & x &  &  & x & x & x & \tabularnewline
\cline{2-14} 
 & 10 &  & x &  &  &  & x & x &  &  & x & x & x\tabularnewline
\hline
\end{tabular}
\end{table}

We compare the normalized throughput under our optimized design and the round-robin (RR) channel assignment strategies in Fig.~\ref{Fig7}. 
For RR channel assignment schemes, we first allocate channels for SUs as described in Table~\ref{table_round} (i.e., we consider three different RR channel assignments). 
In the considered round-robin channel assignment schemes, we assign at most 1, 2 and 3 channels for each SU corresponding to cases 1, 2 and 3 as shown in Table~\ref{table_round}.
In particular, we sequentially assign channels with increasing indices for the next SUs until exhausting (we then repeat this procedure for the following SU).
Then, we only employ Alg. \ref{mainalg_pg} to optimize the sensing and access parameters for these RR channel assignments. 
Fig.~\ref{Fig7} shows that the optimized design achieves much higher throughput than those due to RR channel assignments. 
These results confirm that channel assignments for cognitive radios play a very important role in maximizing the spectrum utilization for CRNs.
In particular, if it would be sufficient to achieve good sensing and throughput performance if we assign a small number of nearby SUs to sense any particular channel instead of requiring all SUs to sense the channel. 
This is because ``bad SUs'' may not contribute to improve the sensing performance but result in more sensing overhead, which ultimately decreases the throughput of 
the secondary network.

% Fig. 8
\begin{figure}[!t]
\centering
\includegraphics[width=90mm]{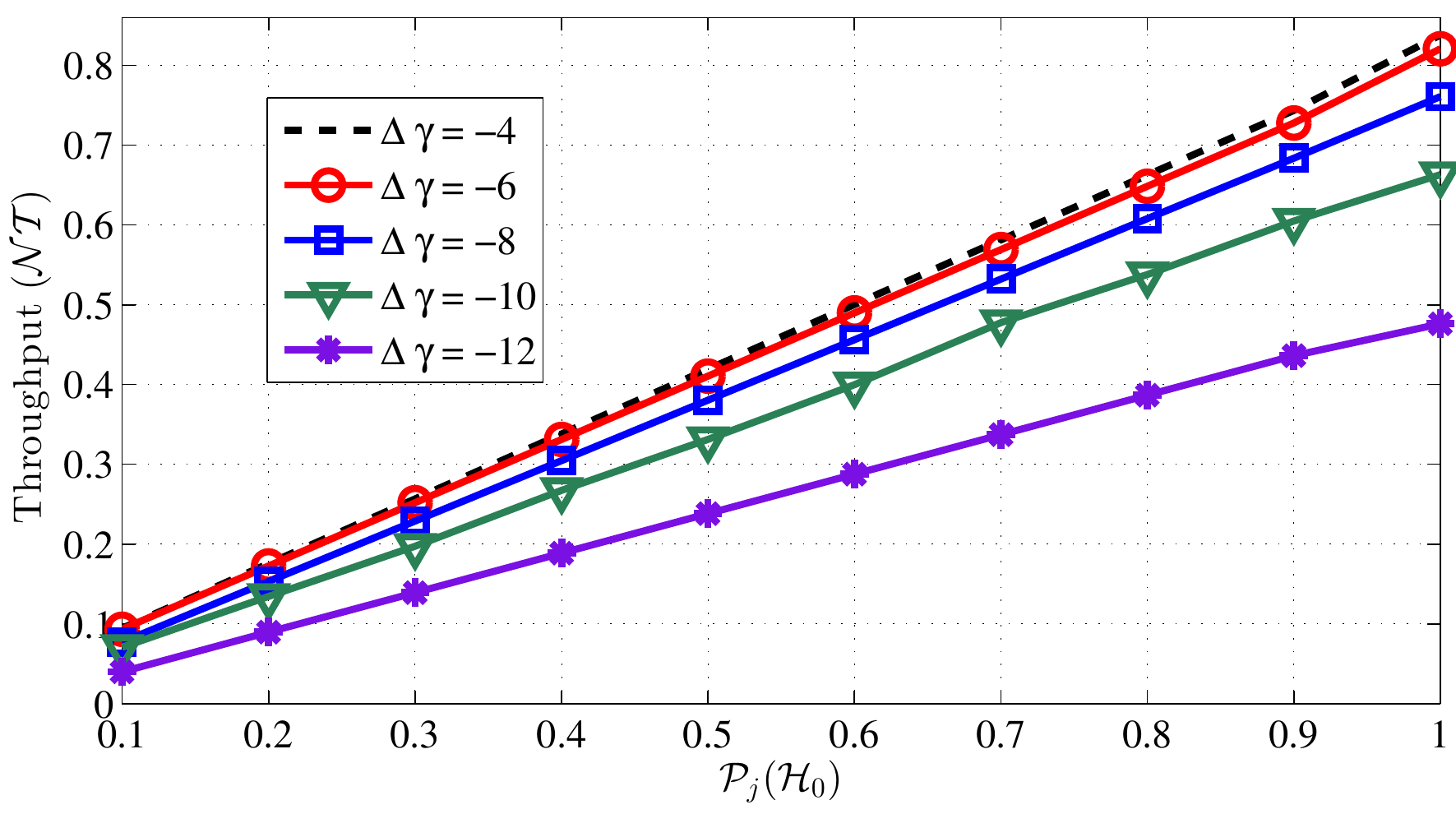}
\caption{Normalized throughput versus probability of having vacant channel $\mathcal{P}_j \left(\mathcal{H}_0\right)$ for $N=10$ and $M=4$ for optimized 
channel assignments and a-out-of-b aggregation rule.}
\label{Fig8}
\end{figure}

In Fig.~\ref{Fig8}, we consider the impact of PUs' activities on throughput performance of the secondary network. In particular, we vary the probabilities of having 
idle channels for secondary spectrum access ($\mathcal{P}_j \left(\mathcal{H}_0\right)$) in the range of $\left[0.1,1\right]$. For larger values of $\mathcal{P}_j \left(\mathcal{H}_0\right)$, there are more opportunities for SUs to find spectrum holes to transmit data, which results in higher throughput and vice versa. Moreover, this figure
shows that the normalized throughput increases almost linearly with $\mathcal{P}_j \left(\mathcal{H}_0\right)$. Also as the $\Delta \gamma$ increases (i.e., higher SNR), 
the throughput performance  can be improved significantly. However,  the improvement becomes negligible if the SNR values are sufficiently large (for $\Delta \gamma$ in $\left[-6,-4\right]$). This is because for large SNR values, the required sensing time is sufficiently small, therefore, further increase of SNR does not reduce the sensing time much to improve 
the normalized throughput.

% Fig. 9
\begin{figure}[!t]
\centering
\includegraphics[width=90mm]{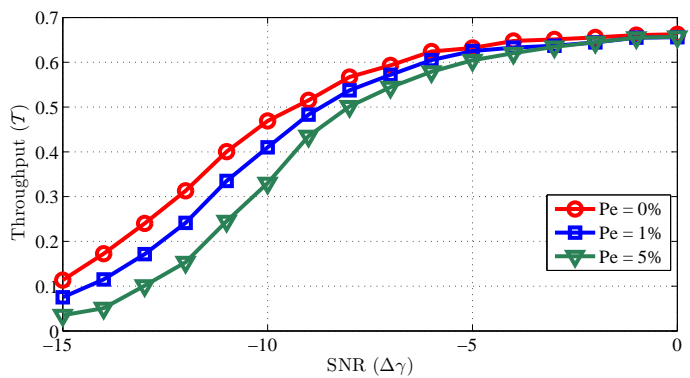}
\caption{Normalized throughput versus SNR shift $\Delta \gamma$ for $N=4$ and $M=3$ for optimized channel assignments and a-out-of-b aggregation rules.}
\label{Fig9}
\end{figure}

Finally, we study the impact of reporting errors on the throughput performance by using the extended throughput analytical model in Section ~\ref{Exten}. The network setting under
investigation has $N=4 $ SUs and $M=3 $ channels. Again, we use notation $\left\{ i,j \right\}$ to represent a combination of channel $j$ and SU $i$. 
The combinations with $\gamma_{10}^{ij} = -10 dB$ are chosen as follows: channel 1: $\left\{1,1\right\}, \left\{2,1\right\}, \left\{3,1\right\} $; channel 2: $\left\{2,2\right\}, \left\{4,2\right\}$; channel 3: $\left\{1,3\right\}, \left\{4,3\right\}$. The remaining combinations correspond to $\gamma_{20}^{ij} = -15 dB$. 
We assume that the reporting errors between every pair of 2 SUs are the same, which is denoted as $P_e$.
In Fig.~\ref{Fig9}, we show the achieved throughput as $P_e = 0\% $, $P_e = 1\% $ and $P_e = 5\% $ under optimized design. 
We can see that when $P_e$ increases, the normalized throughput decreases quite significantly if the SNR is sufficiently low.
However, in the high-SNR regime, the throughput performance is less sensitive to the reporting errors.

\vspace{0.2cm}
\section{Conclusion}
\label{conclusion_Chap5} 

We have proposed a general analytical and optimization framework for SDCSS and access design in multi-channel CRNs. 
In particular, we have proposed the $p$-persistent CSMA MAC protocol integrating the SDCSS mechanism. Then,
we have analyzed the throughput performance of the proposed design and have developed an efficient algorithm to optimize its sensing and
access parameters. Moreover, we have presented both optimal brute-force search and low-complexity algorithms to determine efficient channel sensing sets 
and have analyzed their complexity. We have also extended the framework to consider reporting errors in exchanging sensing results among SUs.
Finally, we have evaluated the impacts of different parameters on the throughput performance of the proposed design
and illustrated the significant performance gap between the optimized and non-optimized designs. Specifically, it has been
confirmed that optimized sensing and access parameters as well as channel assignments can achieve considerably better throughput
performance than that due to the non-optimized design. 
In the future, we will extend SDCSS and MAC protocol design for the multihop CRNs.

% ---------------------------------------------------------------------------
%: ----------------------- end of thesis sub-document ------------------------
% ---------------------------------------------------------------------------

% this file is called up by thesis.tex
% content in this file will be fed into the main document

%: ----------------------- name of chapter  -------------------------
%\chapter{Asynchronous Full--Duplex MAC Protocol for Cognitive Radio Networks} % top level followed by section, subsection
\chapter{Design and Optimal Configuration of Full--Duplex MAC Protocol for Cognitive Radio Networks Considering Self--Interference} % top level followed by section, subsection
\zlabel{Chapter6}

%: ----------------------- paths to graphics ------------------------

% change according to folder and file names
\ifpdf
    \graphicspath{{6/figures/PNG/}{6/figures/PDF/}{6/figures/}}
\else
    \graphicspath{{6/figures/EPS/}{6/figures/}}
\fi
%\usepackage{hyperref}

%: ----------------------- contents from here ------------------------

The content of this chapter was submitted in IEEE Access in the following paper:

L.~T.~ Tan, and L.~B.~ Le, ``Design and Optimal Configuration of Full--Duplex MAC Protocol for Cognitive Radio Networks
Considering Self--Interference,'' in {\em IEEE Access}, 2015 (Under review).

%L.~T.~ Tan, and L.~B.~ Le, ``Distributed MAC Protocol Design for Full-Duplex Cognitive Radio Networks,'' in 2015 IEEE Global Communications Conference (IEEE GLOBECOM 2015), San Diego, CA, USA, December, 2015.

\section{Abstract}

In this paper, we propose an adaptive Medium Access Control (MAC) protocol for full-duplex (FD) cognitive radio networks in which
FD secondary users (SUs) perform channel contention followed by concurrent spectrum sensing and transmission,
and transmission only with maximum power in two different stages (called the FD sensing and transmission stages, respectively)
in each contention and access cycle. The proposed FD cognitive MAC (FDC-MAC) protocol does not require synchronization among SUs
and it efficiently utilizes the spectrum and mitigates the self-interference in the FD transceiver.
We then develop a mathematical model to analyze the throughput performance of the FDC-MAC protocol where 
both half-duplex (HD) transmission (HDTx) and FD transmission (FDTx) modes are considered.
Then, we study the FDC-MAC configuration optimization through adaptively
controlling the spectrum sensing duration and transmit power level of the FD sensing stage
where we prove that there exists optimal sensing time and transmit power to achieve the maximum throughput
and we develop an algorithm to configure the proposed FDC-MAC protocol. Extensive numerical results are presented 
to illustrate the characteristic of the optimal FDC-MAC configuration and the impacts of protocol
parameters and the self-interference cancellation quality on the throughput performance.
Moreover, we demonstrate the significant throughput gains of the FDC-MAC protocol 
with respect to existing half-duplex MAC (HD MAC) and single-stage FD MAC protocols.
%
%In this paper, we propose a novel Medium Access Control (MAC) protocol for full-duplex cognitive radio networks where
%full-duplex (FD) secondary users (SUs) can perform spectrum sensing and access spectrum holes simultaneously.
%The proposed asynchronous FD cognitive MAC (FDC-MAC) protocol does not require synchronization among SUs
%and it can efficiently utilize the spectrum and mitigate the self-interference in the FD transceiver through adaptively
%controlling the spectrum sensing duration and transmit power level. 
%We then develop a mathematical model to analyze the throughput performance of the proposed FDC-MAC protocol. 
%Both half-duplex (HD) transmission (HDTx) and FD transmission (FDTx) modes are considered in the design and analysis.
%Then, we study the parameter configuration optimization problem for the proposed FDC-MAC protocol
%where we prove that there exists an optimal sensing time to achieve the maximum throughput. 
%Moreover, we develop an algorithm to configure the MAC protocol so that efficient self-interference management and sensing overhead control can be achieved.
%Finally, extensive numerical results are presented to gain further insights and to evaluate the performance of our MAC design. In particular,
 %we demonstrate that some existing cognitive MAC protocols correspond to special cases of our proposed FDC-MAC design under specific optimized configuration conditions.

\section{Introduction}

Engineering MAC protocols for efficient sharing of white spaces is an important research topic in cognitive radio networks (CRNs).
One critical requirement for the cognitive MAC design is that transmissions on the licensed frequency band from primary users (PUs) should  be satisfactorily protected from the SUs' spectrum
 access. Therefore, a cognitive MAC protocol for the secondary network must realize both the spectrum sensing and access functions so that timely detection of the PUs' communications and effective 
spectrum sharing among SUs can be achieved. Most existing research works on cognitive MAC protocols have focused on the design and analysis of HD MAC 
(e.g., see \cite{Cor09, Yu09} and the references therein).

Due to the HD constraint, SUs typically employ a two-stage sensing/access procedure where they perform spectrum sensing in the first stage before accessing
available spectrum for data transmission in the second stage \cite{Liang08, Le11, Le12, Zhao07, Kim08, wang08}. %\cite{Liang08}--\cite{wang08}. 
This constraint also requires SUs be synchronized during
the spectrum sensing stage, which could be difficult to achieve in practice.   
In fact, spectrum sensing enables SUs to detect white spaces that are not occupied by PUs \cite{Yu09, Liang08, Le11, Le12, Haykin09, Axell12}; %\cite{Yu09}--\cite{Le12}, \cite{Haykin09, Axell12}; %\cite{Yu09, Liang08, Le11, Le12, Haykin09, Axell12}; %\cite{Yu09}--\cite{Le12};
therefore, imperfect spectrum sensing can reduce the spectrum utilization due to  failure in detecting white spaces and potentially result in collisions 
with active PUs. Consequently, sophisticated design and parameter configuration of cognitive MAC protocols must be conducted to
achieve good performance while appropriately protecting PUs \cite{Cor09, Konda08, Le11, Le12, Zhao07, Kim08, wang08}. %\cite{Cor09}, \cite{Le11}--\cite{wang08}, \cite{Konda08}.  
As a result, traditional MAC protocols \cite{Thorpe14, bian00, Cali00, Chhaya97, Akyildiz99} %\cite{Thorpe14}--\cite{Akyildiz99} %\cite{Thorpe14, Bian00, Cali00, Chhaya97, Akyildiz99} %\cite{Thorpe14}--\cite{Akyildiz99} 
adapted to the CRN may not provide satisfactory performance.

In general, HD-MAC protocols may not exploit white spaces very efficiently since significant sensing time may be required, which would otherwise be utilized for data transmission. 
Moreover, SUs may not timely detect the PUs' activity during their transmissions, which can cause severe interference to active PUs.
Thanks to recent advances on FD technologies, a FD radio can transmit and receive data simultaneously on the same frequency band \cite{Duarte12, Everett14, Sabharwal14, Korpi14, Choi10, Jain11}. %\cite{Duarte12}--\cite{Jain11}. %\cite{Duarte12, Everett14, Sabharwal14, Korpi14, Choi10, Jain11}. 
%  \cite{Duarte12}--\cite{Choi10}. 
One of the most critical issues of wireless FD communication is the presence of self-interference, which is caused by power leakage from the transmitter to the receiver of a FD transceiver.
The self-interference may indeed lead to serious communication performance degradation of FD wireless systems.
Despite recent advances on self-interference cancellation (SIC) techniques \cite{Everett14, Sabharwal14, Korpi14} %\cite{Everett14}--\cite{Korpi14} %\cite{Everett14, Sabharwal14, Korpi14}  %\cite{Everett14}--\cite{Korpi14} 
(e.g., propagation SIC, analog-circuit SIC, and digital baseband SIC), 
self-interference still exists due to various reasons such as the limitation of hardware and channel estimation errors.

\subsection{Related Works}

There are some recent works that propose to exploit the FD communications for MAC-level channel access in multi-user wireless networks
\cite{Jain11, Duarte14, Goyal13, Ramirez13, Choi15}. %\cite{Jain11}--\cite{Choi15}. %\cite{Jain11, Duarte14, Goyal13, Ramirez13, Choi15}. % \cite{Jain11}, \cite{Duarte14}--\cite{Choi15}. 
%Specifically, the wireless system model composes multiple nodes and a single access point (AP) where all the nodes including an AP operates in a FD mode.
In \cite{Jain11}, the authors develop a centralized MAC protocol to support asymmetric data traffic where network nodes may transmit data 
packets of different lengths, and they propose to mitigate the hidden node problem by employing a busy tone.
To overcome this hidden node problem, Duarte et al. propose to adapt the standard 802.11 MAC protocol with the RTS/CTS handshake in \cite{Duarte14}.
Moreover, Goyal et al. in \cite{Goyal13} extend this study to consider interference between two nodes due to their concurrent transmissions.
Different from conventional wireless networks, designing MAC protocols in CRNs is more challenging because the spectrum sensing function must be
efficiently integrated into the MAC protocol.
In addition, the self-interference must be carefully addressed in the simultaneous spectrum sensing and access to mitigate its negative
impacts on the sensing and throughput performance.

The FD technology has been employed for more efficient spectrum access design in cognitive radio networks
 \cite{Cheng14, tan2015distributed, Kim12, Kim15} %\cite{Cheng14}--\cite{Kim15}  
where SUs can perform sensing and transmission simultaneously. % \cite{Afifi14, Cheng14, report, Kim12, Kim15} %\cite{Afifi14}--\cite{Kim15}   
%In general, the self-interference due to simultaneous sensing and transmission may lead to degradation of the SUs' spectrum sensing performance. 
In \cite{Cheng14}, a FD MAC protocol is developed which allows simultaneous spectrum access of the SU and PU networks 
where both PUs and SUs are assumed to employ the $p$-persistent MAC protocol for channel contention resolution and access.
This design is, therefore, not applicable to the hierarchical spectrum access in the CRNs where PUs should have higher spectrum access priority 
compared to SUs. 

In our previous work \cite{tan2015distributed}, we propose the FD MAC protocol by using the standard backoff mechanism as
in the 802.11 MAC protocol where we employ concurrent FD sensing and access during data transmission as well as frame fragmentation.
Moreover, engineering of a cognitive FD relaying network is 
considered in \cite{Kim12, Kim15}, where various resource allocation algorithms to improve the outage probability are proposed.
In addition, the authors in \cite{Ramirez13} develop the joint routing and distributed resource allocation for FD wireless networks.
In \cite{Choi15}, Choi et al. study the distributed power allocation for a hybrid FD/HD system where all network nodes operate in the HD mode but 
the access point (AP) communicates by using the FD mode.
In practice, it would be desirable to design an adaptable MAC protocol, which can be configured to operate in an
optimal fashion depending on specific channel and network conditions. This design will be pursued in our current work.

\subsection{Our Contributions}

In this paper, we make a further bold step in designing, analyzing, and optimizing an adaptive
 FDC--MAC protocol for CRNs, where the self-interference and imperfect spectrum
sensing are explicitly considered. In particular, the contributions of this paper can be summarized 
as follows.

\begin{enumerate}

\item We propose a novel FDC--MAC protocol that can efficiently exploit the FD transceiver for 
spectrum spectrum sensing and access of the white space without requiring synchronization among SUs. 
In this protocol, after the $p$-persistent based channel contention phase, the winning SU enters
the data phase consisting of two stages, i.e., concurrent sensing and transmission
in the first stage (called FD sensing stage) and transmission only in the second stage (called transmission stage).
The developed FDC--MAC protocol, therefore, enables the optimized configuration of transmit
power level and sensing time during the FD sensing stage to mitigate the self-interference and appropriately protect the active PU.
After the FD sensing stage, the SU can transmit with the maximum power to achieve the highest throughput.  

\item We develop a mathematical model for throughput performance analysis of the proposed FDC-MAC protocol 
considering the imperfect sensing, self-interference effects, and the dynamic status changes of the PU.
In addition, both one-way and two-way transmission scenarios, which are called  HD transmission
(HDTx) and FD transmission (FDTx) modes, respectively, are considered in the analysis. Since the PU can change its idle/active 
status during the FD sensing and transmission stages, different potential status-change scenarios are studied in the analytical model.

\item We study the optimal configuration of FDC-MAC protocol parameters including the
 SU's sensing duration and transmit power to maximize the achievable throughput under
both FDTx and HDTx modes. We prove that there exists an optimal sensing time to
achieve the maximum throughput for a given transmit power value during the FD sensing
stage under both FDTx and HDTx modes. Therefore, optimal protocol
parameters can be determined through standard numerical search methods.

\item
Extensive numerical results are presented to illustrate the impacts of different 
protocol parameters on the throughput performance and the optimal configurations of the proposed FDC-MAC protocol.
Moreover, we show the significant throughput enhancement of the proposed FDC-MAC protocol compared to existing cognitive
MAC protocols, namely the HD MAC protocol and a single-stage FD MAC protocol with concurrent sensing and access.
Specifically, our FDC-MAC protocol achieves higher throughput with the increasing maximum power while
the throughput of the single-stage FD MAC protocols decreases with the maximum power in the high power regime
due to the self-interference. Moreover, the proposed FDC-MAC protocol leads to significant higher throughput
than that due to the HD MAC protocol.

\end{enumerate}

The remaining of this paper is organized as follows. Section ~\ref{SystemModel} describes the system and PU models. FDC--MAC protocol design, and
throughput analysis for the proposed FDC--MAC protocol are performed in Section ~\ref{MACNonFrag}.
Then, Section ~\ref{FDC_MAC_Configuration} studies the optimal configuration of the proposed FDC--MAC protocol to achieve the
maximum secondary throughput.
Section ~\ref{Results} demonstrates numerical results followed by concluding remarks in Section ~\ref{conclusion}.

\section{System and PU Activity Models}
\label{SystemModel}

\subsection{System Model}
\label{System}

We consider a cognitive radio network where $n_0$ pairs of SUs opportunistically exploit white spaces on 
a frequency band for communications. We assume that each SU is equipped with a FD transceiver, which can perform sensing and transmission simultaneously.
However, the sensing performance of each SU is impacted by the self-interference from its transmitter since the transmitted power is leaked into
the received signal. We denote $I(P)$ as the average self-interference power, which is modeled as
 $I(P) = \zeta \left(P\right)^{\xi}$ \cite{Duarte12} where $P$ is the SU's transmit power, $\zeta$ and $\xi$ ($0 \leq \xi \leq 1$) are 
predetermined coefficients which represent the quality of self-interference cancellation (QSIC).
In this work, we design a asynchronous cognitive MAC protocol where no synchronization is required among SUs and between SUs and PUs.
We assume that different pairs of SUs can overhear transmissions from the others (i.e., a collocated network is assumed). 
In the following, we refer to pair $i$ of SUs as SU $i$ for brevity.

%
%We consider a cognitive radio network where $n_0$ pairs of SUs opportunistically exploit white spaces on 
%one frequency band for communications. We assume that each SU is equipped with a FD transceiver, which can perform sensing and transmission simultaneously.
%However, the sensing performance of SUs is impacted by self-interference from its transmission since the transmitted power is leaked into
%the received signal. We denote $I(P)$ as the average self-interference power, which is modeled as
 %$I(P) = \zeta \left(P\right)^{\xi}$ \cite{Duarte12} where $P$ is the SU's transmit power, $\zeta$ and $\xi$ ($0 \leq \xi \leq 1$) are 
%predetermined coefficients which represent the quality of self-interference cancellation (QSIC).
%In this work, we design a asynchronous cognitive MAC protocol where no synchronization is required among SUs and between SUs and PUs.
%We assume that different pairs of SUs can overhear transmissions from the others (i.e., a collocated network is assumed). 
%In the following, we refer to pair $i$ of SUs as SU $i$ for brevity.

\subsection{Primary User Activity}
\label{PUAM}

We assume that the PU's idle/active status follows two independent random processes. We say that the channel is available and busy 
for SUs' access if the PU is in the idle and active (or busy) states, respectively. Let $\mathcal{H}_0$ and $\mathcal{H}_1$ denote the events 
that the PU is idle and active, respectively. To protect the PU, we assume that SUs must stop their transmissions and evacuate from the busy channel 
within the maximum delay of $T_{\sf eva}$, which is referred to as channel evacuation time. 

Let $\tau_{\sf ac}$ and $\tau_{\sf id}$ denote the random variables which represent the durations of active and idle channel states, respectively.
We denote probability density functions (pdf) of $\tau_{\sf ac}$ and $\tau_{\sf id}$ as $f_{\tau_{\sf ac}}\left(t\right)$ and $f_{\tau_{\sf id}}\left(t\right)$, respectively. 
While most results in this paper can be applied to general pdfs $f_{\tau_{\sf ac}}\left(t\right)$ and $f_{\tau_{\sf id}}\left(t\right)$, we mostly consider the
exponential pdf in the analysis.
In addition, let  $\mathcal{P}\left(\mathcal{H}_0\right) = \frac{{\bar \tau}_{\sf id}}{{\bar \tau}_{\sf id}+{\bar \tau}_{\sf ac}}$ and  $\mathcal{P}\left( \mathcal{H}_1 \right) = 
1 - \mathcal{P}\left(\mathcal{H}_0\right)$ present the probabilities that the channel is available and busy, respectively
where ${\bar \tau}_{\sf id}$ and ${\bar \tau}_{\sf ac}$ denote the average values of $\tau_{\sf ac}$ and $\tau_{\sf id}$, respectively.
We assume that the probabilities that $\tau_{\sf ac}$ and $\tau_{\sf id}$ are smaller than $T_{\sf eva}$ are sufficiently small (i.e., the PU
changes its status slowly) so that we can ignore events with multiple idle/active status changes in one channel evacuation interval $T_{\sf eva}$.

\section{Full-Duplex Cognitive MAC Protocol}
\label{MACNonFrag}

In this section, we describe the proposed FDC-MAC protocol and conduct its throughput
analysis considering imperfect sensing, self-interference of the FD transceiver, and
dynamic status change of the PUs.

%In this section, we describe the proposed FDC-MAC protocol and conduct its throughput
%analysis considering imperfect sensing and self-interference of the FD transceiver.

\subsection{FDC-MAC Protocol Design}

%Fig. 1
\begin{figure*}[!t]
\centering
\includegraphics[width=150mm]{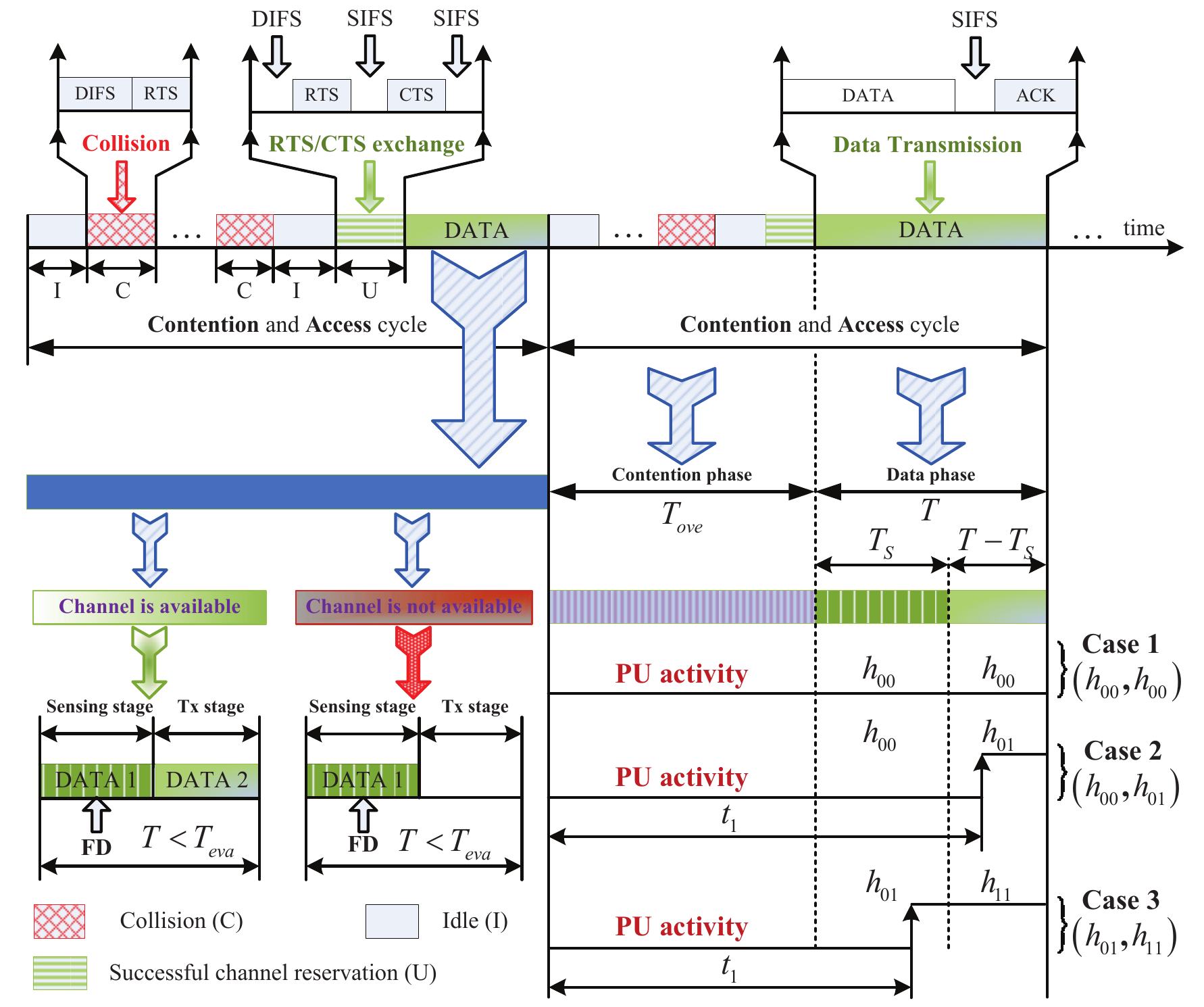}
\caption{Timing diagram of the proposed full-duplex cognitive MAC protocol.}
\label{Sentime_FDMAC_SF}
\end{figure*}

The proposed FDC-MAC protocol integrates three important elements of cognitive MAC protocol,
namely contention resolution, spectrum sensing, and access functions.
Specifically, SUs employ the $p$-persistent CSMA principle \cite{Cali00} for contention resolution
where each SU with data to transmit attempts to capture an available channel with a probability $p$ after the channel is sensed to be idle during the standard DIFS interval  (DCF Interframe Space). 
If a particular SU decides not to transmit (with probability of $1-p$), it will sense the channel and attempt to transmit again in the next slot of length $\sigma$ 
with probability $p$.
%If there is a collision, the SU will wait until the channel is available and attempt to transmit with probability $p$ as before.
To complete the reservation, the four-way handshake with Request-to-Send/Clear-to-Send (RTS/CST) exchanges \cite{bian00} is employed to reserve 
the available channel for transmission.
Specifically, the secondary transmitter sends RTS to the secondary receiver and waits until it successfully receives the CTS from the secondary receiver. 
All other SUs, which hear the RTS and CTS exchange from the winning SU, defer to access the channel for a duration equal to the data transmission time, $T$.
Then, an acknowledgment (ACK) from the SU's receiver is transmitted to its corresponding transmitter to notify the successful reception of a packet.
Furthermore, the standard small interval, namely SIFS (Short Interframe Space), is used before the transmissions of CTS, ACK, and data frame as in the
standard 802.11 MAC protocol \cite{bian00}.

In our design, the data phase after the channel contention phase comprises two stages where the SU performs
concurrent sensing and transmission in the first stage with duration $T_S$ and transmission only in the second stage with duration $T-T_S$.
Here, the SU exploits the FD capability of its transceiver to realize concurrent sensing and transmission 
the first stage (called FD sensing stage)  where the sensing outcome at the end of this stage (i.e., an idle or active channel status) determines its 
further actions as follows.
Specifically, if the sensing outcome indicates an available channel then the SU transmits data in the second stage; otherwise, it remains silent for the remaining time of the 
data phase with duration $T-T_S$.

We assume that the duration of the SU's data phase $T$ is smaller than the channel evacuation time $T_{\sf eva}$ so timely evacuation from the busy channel can be 
realized with reliable FD spectrum sensing. Therefore, our design allows to protect the PU with evacuation delay at most $T$ if
the MAC carrier sensing during the contention phase
and the FD spectrum sensing in the data phase are perfect. Furthermore, we assume that the SU transmits at power levels $P_{\sf sen} \leq P_{\sf max}$ and 
$P_{\sf dat} = P_{\sf max}$ during
the FD sensing and transmission stages, respectively where $P_{\sf max}$ denotes the maximum power and
 the transmit power $P_{\sf sen}$ in the FD sensing stage will be optimized to effectively mitigate the 
self-interference and achieve good sensing-throughput tradeoff. The timing diagram of the proposed FDC--MAC protocol is illustrated 
in Fig.~\ref{Sentime_FDMAC_SF}.

We allow two possible operation modes in the transmission stage. 
The first is the HD transmission mode (HDTx mode) where there is only one direction of data transmission from the SU transmitter to the SU receiver.
In this mode, there is no self-interference in the transmission stage. The second is the FD transmission mode (FDTx mode) where two-way communications
between the pair of SUs are assumed (i.e., there are two data flows between the two SU nodes in opposite directions).
In this mode, the achieved throughput can be potentially enhanced (at most doubling the throughput of the HDTx mode) but self-interference must be 
taken into account in throughput quantification.

Our proposed FDC--MAC protocol design indeed enables flexible and adaptive configuration, which can efficiently exploit the capability of the FD transceiver.  
Specifically, if the duration of the FD sensing stage is set equal to the duration of the whole data phase (i.e., $T_S = T$), then the SU performs 
concurrent sensing and transmission for the 
whole data phase as in our previous design \cite{tan2015distributed}. This configuration may degrade the achievable throughput since the transmit power during the 
FD sensing stage is typically set smaller  $P_{\sf max}$  to mitigate the self-interference and achieve the required sensing performance. We will refer
the corresponding MAC protocol with $T_S = T$ as one-stage FD MAC in the sequel.

Moreover, if we set the SU transmit power $P_{\sf sen}$ in the sensing stage equal to zero, i.e., $P_{\sf sen} = 0$, 
then we achieve the traditional two-stage HD cognitive MAC protocol where sensing and transmission are performed sequentially in two
 different stages \cite{Le11, Le12}. Moreover, the proposed FDC--MAC protocol is more flexible than existing designs \cite{tan2015distributed}, \cite{Le11, Le12} 
since different existing designs can be achieved through suitable configuration of the protocol parameters of our FDC--MAC protocol. 
It will be demonstrated that the proposed FDC--MAC protocol achieves significant better throughput than that of the existing cognitive MAC protocols.
In the following, we present the throughput analysis based on which the protocol configuration optimization can be performed.

\subsection{Throughput Analysis}
\label{Tput_Ana_MACNonFrag}

We now conduct the saturation throughput analysis for the secondary network where all SUs are assumed to always have data to transmit.
The resulting throughput can be served as an upper bound for the throughput in the non-saturated scenario \cite{bian00}.
This analysis is performed by studying one specific contention and access cycle (CA cycle) with the contention phase and data phase
 as shown in Fig.~\ref{Sentime_FDMAC_SF}.
Without loss of generality, we will consider the normalized throughput achieved per one unit  of system bandwidth (in bits/s/Hz). 
Specifically, the normalized throughput of the FDC--MAC protocol can be expressed as 
\beqn
\label{NT_NonFrag}
\mathcal{NT} = \frac{\mathcal{B}}{T_{\sf ove} + T},
\eeqn
where $T_{\sf ove}$ represents the time overhead required for one successful channel reservation (i.e., successful RTS/CTS exchanges), 
$\mathcal{B}$ denotes the amount of data (bits) transmitted in one CA cycle per one unit of system bandwidth, which is expressed in bits/Hz.
To complete the throughput analysis, we derive the quantities $T_{\sf ove}$ and $\mathcal{B}$ in the remaining of this subsection.

\subsubsection{Derivation of $T_{\sf ove}$}

The average time overhead for one successful channel reservation can be calculated as
\beqn \label{tover}
T_{\sf ove} = {\overline T}_{\sf cont} + 2SIFS + 2PD + ACK,
\eeqn
where $ACK$ is the length of an ACK message, $SIFS$ is the length of a short interframe space, and $PD$ is the propagation delay
where $PD$ is usually small compared to the slot size $\sigma$, and ${\overline T}_{\sf cont}$ denotes the average time overhead
due to idle periods, collisions, and successful transmissions of RTS/CTS messages in one CA cycle. For
better presentation of the paper, the derivation of ${\overline T}_{\sf cont}$ is given in Appendix \ref{AppenA0}.

\subsubsection{Derivation of $\mathcal{B}$}

To calculate $\mathcal{B}$, we consider all possible cases that capture the activities of SUs and status changes of the PU in the FDC-MAC data phase of duration $T$.
Because the PU's activity is not synchronized with the SU's transmission, the PU can change its idle/active status any time.
We assume that there can be at most one transition between the idle and active states of the PU during one data phase interval. 
This is consistent with the assumption on the slow status changes of the PU as described in Section~\ref{PUAM} since $T < T_{\sf eva}$. 
Furthermore, we assume that the carrier sensing of the FDC-MAC protocol is perfect; therefore, the PU is idle at the beginning of the FDC-MAC data phase.
Note that the PU may change its status during the SU's FD sensing or transmission stage, which requires us to consider different possible events in the data phase. 

We use $h_{ij}$ ($i, j \in \left\{ 0, 1 \right\}$) to represent events capturing status changes of the PU in the FD sensing stage and transmission stage
where $i$ = 0 and $i$ = 1 represent the idle and active states of the PU, respectively.
For example, if the PU is idle during the FD sensing stage and becomes active during the transmission stage, then we represent
this event as $\left(h_{00}, h_{01}\right)$ where sub-events $h_{00}$ and $h_{01}$ represent the status changes in the FD sensing
and transmission stages, respectively. Moreover, if the PU changes from the idle to the active state during
the FD sensing stage and remains active in the remaining of the data phase, then we represent this event as $\left(h_{01}, h_{11}\right)$

It can be verified that we must consider the following three cases with the corresponding status changes of the PU during the FDC-MAC data phase to analyze $\mathcal{B}$.
\begin{itemize}
\item \textbf{Case 1}: The PU is idle for the whole FDC-MAC data phase  (i.e., there is no PU's signal in both FD sensing and transmission stages) and we denote this
event as $\left(h_{00}, h_{00}\right)$. The average number of bits (in bits/Hz) transmitted during the data phase in this case is denoted as $\mathcal{B}_1$.

\item \textbf{Case 2}: The PU is idle during the FD sensing stage but the PU changes from the idle to the active status in the transmission stage. 
We denote the event corresponding to this case as $\left(h_{00}, h_{01}\right)$ where $h_{00}$ and $h_{01}$ capture the sub-events in the FD sensing and transmission stages, respectively.
The average number of bits (in bits/Hz) transmitted during the data phase in this case is represented by $\mathcal{B}_2$.

\item \textbf{Case 3}: The PU is first idle then becomes active during the FD sensing stage and it remains active during the whole transmission stage.
Similarly we denote this event as $\left(h_{01}, h_{11}\right)$ and the average number of bits (in bits/Hz) transmitted during the data phase in this case is denoted as $\mathcal{B}_3$.
\end{itemize}

Then, we can calculate $\mathcal{B}$ as follows:
\beqn
\mathcal{B} = \mathcal{B}_1 + \mathcal{B}_2 + \mathcal{B}_3.
\eeqn
To complete the analysis, we will need to derive $\mathcal{B}_1$, $\mathcal{B}_2$, and $\mathcal{B}_3$, which are given in Appendix \ref{AppenB1}.

\section{FDC--MAC Protocol Configuration for Throughput Maximization}
\label{FDC_MAC_Configuration}

In this section, we  study the optimal configuration of the proposed FDC--MAC protocol to achieve the maximum 
 throughput while satisfactorily protecting the PU. 

%In this section, we  study the optimal configuration of the proposed FDC--MAC protocol to achieve the maximum secondary
 %throughput while satisfactorily protecting the PU. 

\subsection{Problem Formulation}
\label{TputOpt}

Let $\mathcal{NT}(T_S, p, P_{\sf sen})$ denote the normalized secondary throughput, which is the function of the sensing time $T_S$, transmission probability
 $p$, and the SU's transmit power $P_{\sf sen}$ in the FD sensing stage. 
In the following, we assume a fixed frame length $T$, which is set smaller the required evacuation time $T_{\sf eva}$ to achieve timely evacuation from a busy channel for
the SUs. We are interested in determining suitable configuration for $p$, $T_S$ and $P_{\sf sen}$ to maximize the secondary throughput, $\mathcal{NT}(T_S, p, P_{\sf sen})$.
In general, the optimal  transmission probability $p$ should balance between reducing collisions among SUs and limiting the protocol overhead.
However, the achieved throughput is less sensitive to the transmission probability  $p$ as will be demonstrated later via the numerical study.
Therefore, we will seek to optimize the throughput over $P_{\sf sen}$ and $T_S$ for a reasonable and fixed value of $p$. 

 For brevity, we express the throughput as a function of $P_{\sf sen}$ and $T_S$ only, i.e., $\mathcal{NT}(T_S, P_{\sf sen})$.
Suppose that the PU requires that the average detection probability is at least $\overline{\mathcal{P}}_{d}$.
Then, the throughput maximization problem can be stated as follows:
%\vspace{0.05cm}
%\noindent
%\textbf{Problem 1:} 
%\vspace{0.0cm}
\begin{equation}
\label{eq3a}% eq3
\begin{array}{l}
 {\mathop {\max }\limits_{T_S,p, P_{\sf sen}}} \quad {\mathcal{NT}} \left(T_S, P_{\sf sen}\right)  \\ 
 \mbox{s.t.}\,\,\,\, \hat{\mathcal{P}}_{d}\left(\varepsilon,T_S\right) \geq \mathcal{\overline P}_{d},  \\
 \quad \quad 0 \leq P_{\sf sen} \leq P_{\sf max}, \quad 0 \leq T_S \leq T,\\
 \end{array}\!\!
\end{equation}
where $P_{\sf max}$ is the maximum power for SUs, and $T_S$ is upper bounded by $T$.
In fact, the first constraint on $\hat{\mathcal{P}}_{d}\left(\varepsilon,T_S\right)$ implies that the spectrum sensing should be sufficiently reliable to protect the PU which can
be achieved  with sufficiently large sensing time $T_S$. 
Moreover, the SU's transmit
 power $P_{\sf sen}$ must be appropriately set to achieve good tradeoff between the network throughput and self-interference mitigation. 

\subsection{Parameter Configuration for FDC--MAC Protocol}

To gain insights into the parameter configuration of the FDC--MAC protocol, we first study the optimization with respect to the sensing time $T_S$ for a given $P_{\sf sen}$.
For any value of $T_S$, we would need to set the sensing detection threshold $\varepsilon$ %and SU's transmit power $P_{\sf sen}$ 
so that  the detection probability constraint
is met with equality, i.e., $\mathcal{\hat P}_d \left(\varepsilon,T_S\right) = \mathcal{\overline P}_d$ as in \cite{Liang08, Le11}. 
Since the detection probability is smaller in \textbf{Case 3} (i.e., the PU changes from the idle to 
active status during the FD sensing stage of duration $T_S$) compared to that in \textbf{Case 1} and \textbf{Case 2} (i.e.,
the PU remains idle during the FD sensing stage) considered in the previous section, we only need to consider \textbf{Case 3}
to maintain the detection probability constraint. The average probability of detection for the FD sensing in \textbf{Case 3} can be expressed as
\beqn
\label{P_average}
\mathcal{\hat P}_d = \!  \int_{0}^{T_S}  \mathcal{P}_d^{01}(t) \! f_{\tau_{\sf id}}\!\!\left(t\left|0 \leq t \leq T_S\right.\right) dt,
\eeqn
where $t$ denotes the duration from the beginning of the FD sensing stage to the instant when the PU changes to the active state, and
$f_{\tau_{\sf id}}\left(t\left|\mathcal{A}\right.\right)$ is the pdf of $\tau_{\sf id}$ conditioned on event $\mathcal{A}$
capturing the condition $0 \leq t \leq T_S$, which is given as
\beqn
\label{cond_pdf_tau_id}
f_{\tau_{\sf id}}\left(t\left|\mathcal{A}\right.\right) = \frac{f_{\tau_{\sf id}}\left(t\right)}{\Pr\left\{\mathcal{A}\right\}} = \frac{\frac{1}{{\bar \tau}_{\sf id}} \exp(-\frac{t}{{\bar \tau}_{\sf id}})}{1-\exp(-\frac{T_S}{{\bar \tau}_{\sf id}})}.
\eeqn
Note that $\mathcal{P}_d^{01}(t)$ is derived in Appendix \ref{CAL_P_F_P_D} and $f_{\tau_{\sf id}} \left(t\right)$ is given in (\ref{pdf_tau_ac_id}). 

We consider the following single-variable optimization problem for a given $P_{\sf sen}$:
%\vspace{0.0cm}
%\noindent
%\textbf{Problem 2:} \\
\begin{equation}
\label{NT_NonFrag_OPT_TS}
{\mathop {\max }_{0< T_S \leq T} } \quad {\mathcal{NT}} \! \left(T_S,  {P}_{\sf sen} \right). %\left|_{P_{\sf sen} = {\overline P}_{\sf sen}} \right.}  
\end{equation}

We characterize the properties of function $\mathcal{NT}(T_S, P_{\sf sen})$ with respect to $T_S$ for a given  $P_{\sf sen}$ in the following theorem
whose proof is provided in Appendix~\ref{Prosp1}. For simplicity,
the throughput function is written as  $\mathcal{NT}(T_S)$.

\vspace{0.2cm}
\noindent
\textbf{Theorem 1:} The objective function ${{\mathcal{NT}} \!\!\left( { T_S } \right)}$ of (\ref{NT_NonFrag_OPT_TS}) satisfies
the following properties
\begin{enumerate}
\item 
$\mathop {\lim }\limits_{T_S  \to 0} \frac{\partial {\mathcal{NT}}}{\partial T_S } =  + \infty$,

\item 
\begin{enumerate}
%\item \textbf{Case 1}: HDTX mode or FDTX mode with 
\item For HDTx mode with $\forall P_{\sf sen}$ and FDTx mode with $P_{\sf sen} < \overline{P}_{\sf sen}$, we have $\mathop {\lim }\limits_{T_S  \to T}   \frac{\partial {\mathcal{NT}}}{\partial T_S }  < 0$,

\item For FDTx mode with $P_{\sf sen} > \overline{P}_{\sf sen}$, we have $\mathop {\lim }\limits_{T_S  \to T}   \frac{\partial {\mathcal{NT}}}{\partial T_S }  > 0$,
\end{enumerate}

\item $\frac{\partial^2 {\mathcal{NT}}}{\partial T_S^2} < 0$, $\forall T_S$,

\item The objective function ${{\mathcal{NT}} \!\!\left( T_S \right)}$ is bounded from above,
\end{enumerate}
where $\overline{P}_{\sf sen} =  N_0 \left[\left(1+\frac{P_{\sf dat}}{N_0+ \zeta P_{\sf dat}^\xi}\right)^2-1\right]$ is the
critical value of ${P}_{\sf sen}$ such that 
 $\mathop {\lim } \limits_{T_S  \to T} \frac{\partial \mathcal{NT}}{\partial T_S} = 0$.
%\beqn
%\label{P_sen_threshold}
%\overline{P}_{\sf sen} =  N_0 \left[\left(1+\frac{P_{\sf dat}}{N_0+\zeta P_{\sf dat}^\xi}\right)^2-1\right].
%\eeqn

%\begin{proof} The proof is provided in Appendix~\ref{Prosp1}. \end{proof}

We would like to discuss the properties stated in Theorem 1. 
For the HDTx mode with $\forall P_{\sf sen}$ and FDTx mode with low $P_{\sf sen}$, then properties 1, 2a, and 4 imply that there must be at least 
one $T_S$ in  $\left[0,T\right]$ that maximizes $\mathcal{NT}\left( T_S \right)$. The third property implies that  this maximum is indeed unique. 
Moreover, for the FDTx with high $P_{\sf sen}$, then properties 1, 2b, 3 and 4 imply that ${{\mathcal{NT}} \!\!\left( { T_S } \right)}$ increases in $\left[0,T\right]$.
Hence, the throughput ${{\mathcal{NT}} \!\!\left( { T_S } \right)}$ achieves its maximum with sensing time $T_S = T$.
We propose an algorithm to determine optimal $\left(T_S, P_{\sf sen}\right)$, which is summarized in Algorithm~\ref{OPT_Throughput_GMAC}.
Here, we can employ the bisection scheme and other numerical methods to determine the optimal value $T_S$ for a given $P_{\sf sen}$.

\begin{algorithm}[h]%\leesize
%\scriptsize
\caption{\textsc{FDC-MAC Configuration Algorithm}}
\label{OPT_Throughput_GMAC}
%\algsetup{indent=1.5em}
\begin{algorithmic}[1]

\FOR {each considered value of $P_{\sf sen} \in [0,P_{\sf max}]$}

\STATE Find optimal $T_S$ for problem (\ref{NT_NonFrag_OPT_TS}) using the bisection method as ${\overline T}_S \left(P_{\sf sen}\right) = \mathop {\argmax} \limits_{0 \leq T_S \leq T} \mathcal{NT} \left(T, P_{\sf sen}\right)$.

\ENDFOR

\STATE The final solution $\left(T_S^*, P_{\sf sen}^*\right)$ is determined as $\left(T_S^*, P_{\sf sen}^*\right) = \mathop {\argmax} \limits_{P_{\sf sen}, {\overline T}_S \left(P_{\sf sen}\right)} \mathcal{NT} \left(T_S\left(P_{\sf sen}\right), P_{\sf sen}\right)$.

\end{algorithmic}
\end{algorithm}

\section{Numerical Results}
\label{Results}

For numerical studies, we set the key parameters for the FDC--MAC protocol as follows: mini-slot duration is $\sigma = 20 {\mu} s$; $PD = 1 {\mu} s$; $SIFS = 2\sigma$ ${\mu} s$; $DIFS = 10\sigma$ ${\mu} s$; $ACK = 20\sigma$ ${\mu} s$; $CTS = 20\sigma$ ${\mu} s$; $RTS = 20\sigma$ ${\mu} s$. Other parameters are chosen as follows unless stated otherwise: the
sampling frequency $f_s = 6$ MHz; bandwidth of PU's signal $6$ MHz; $\mathcal{\overline P}_d = 0.8$; $T = 15$ ms; $p = 0.0022$; the SNR of the PU 
signal at each SU $\gamma_P = \frac{P_p}{N_0} = -20$ dB; varying self-interference parameters $\zeta$ and $\xi$. Without loss of generality, the noise power is
normalized to one; hence, the SU transmit power $P_{\sf sen}$ becomes $P_{\sf sen} = SNR_s$; and we set $P_{\sf max} = 15$dB.

% Fig. 2
\begin{figure}[!t]
\centering
\includegraphics[width=80mm]{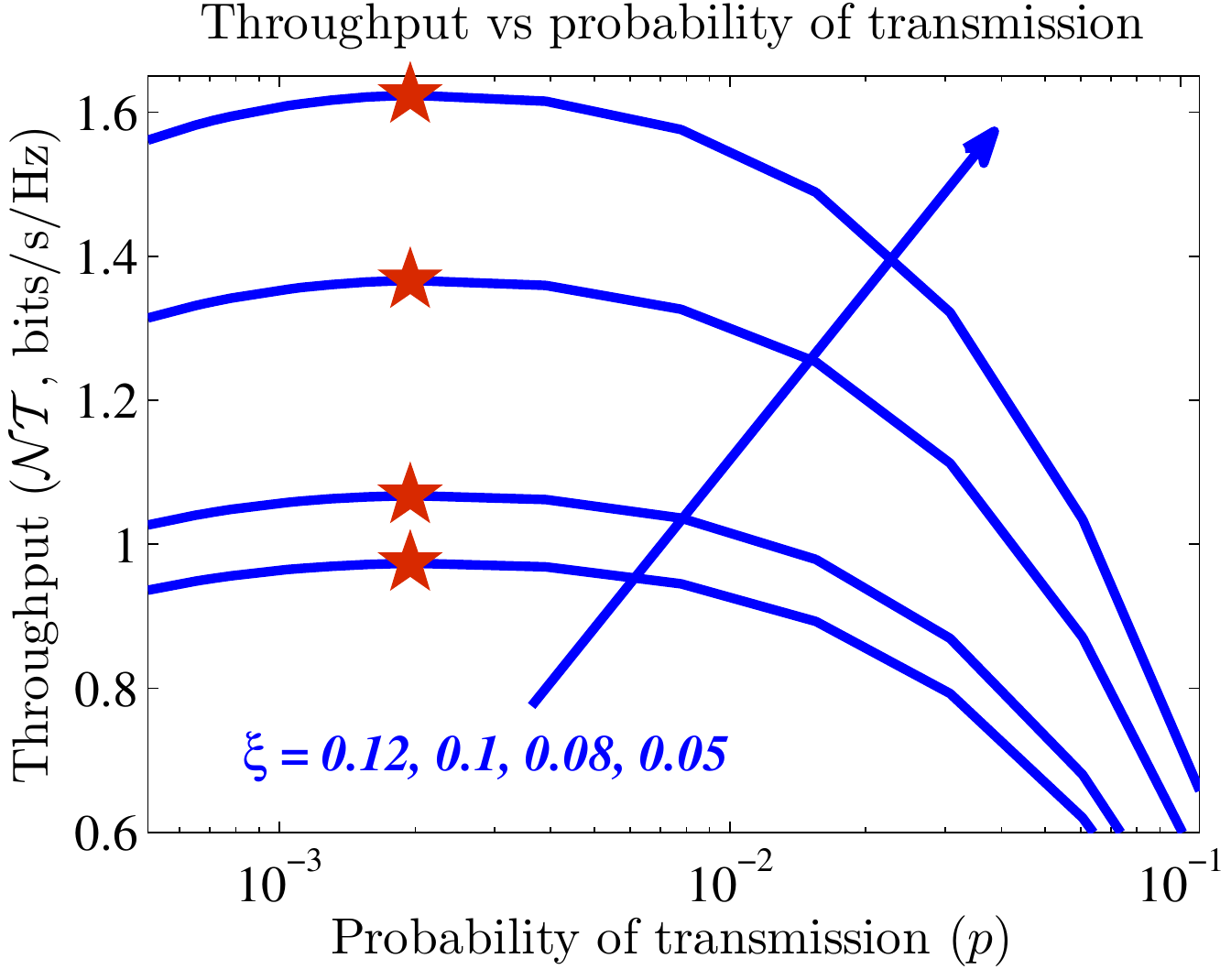}
\caption{Normalized throughput versus transmission probability $p$ for $T = 18$ ms, ${\bar \tau}_{\sf id} = 1000$ ms, ${\bar \tau}_{\sf ac} = 100$ ms, and varying $\xi$.}
\label{T_VS_p_nofrag_xi}
\end{figure}

% Fig. 3
\begin{figure}[!t]
\centering
\includegraphics[width=80mm]{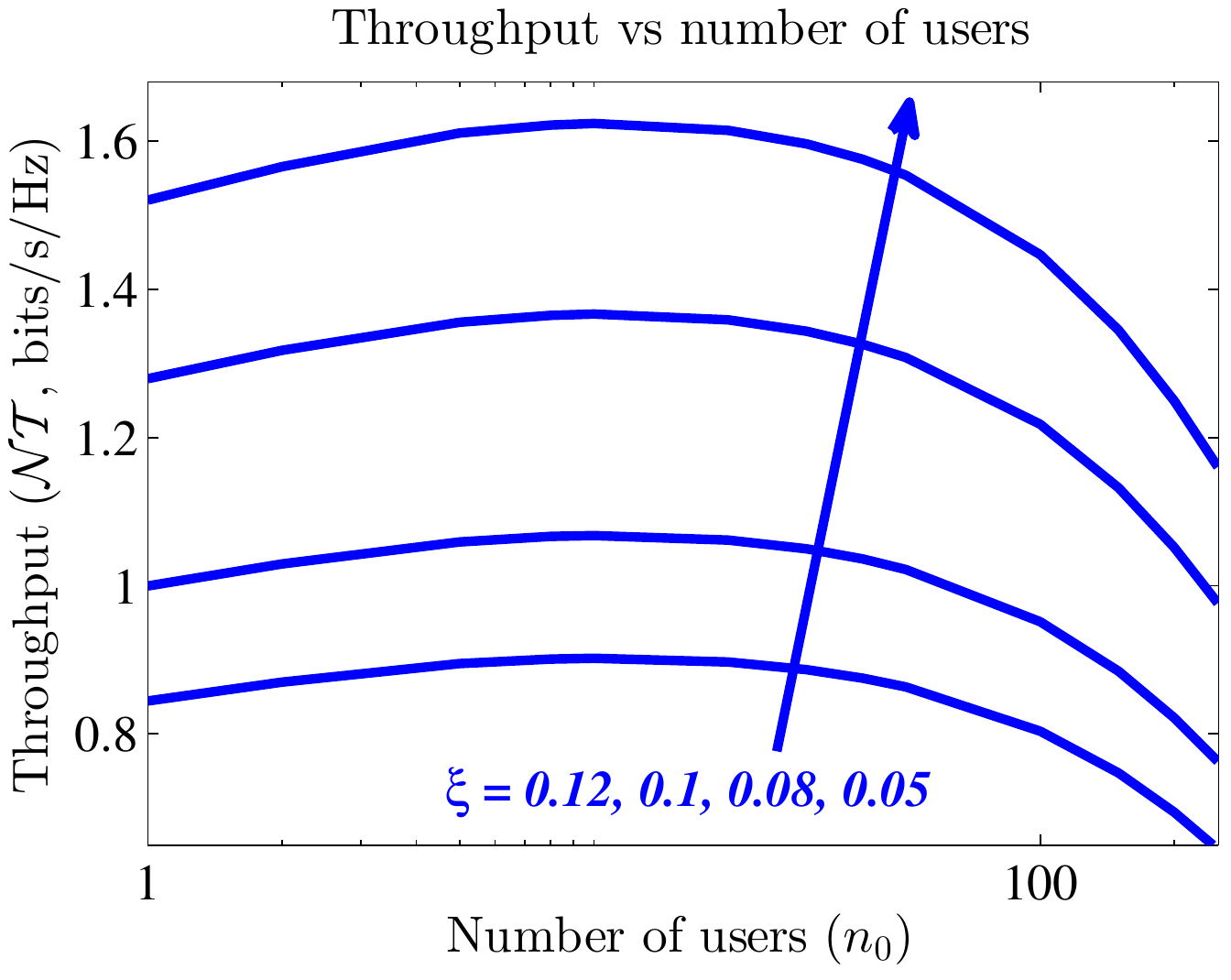}
\caption{Normalized throughput versus the number of SUs $n_0$ for $T = 18$ ms, $p = 0.0022$, ${\bar \tau}_{\sf id} = 1000$ ms, ${\bar \tau}_{\sf ac} = 100$ ms, and varying $\xi$.}
\label{T_VS_n0_xi_nofrag}
\end{figure}

We first study the impacts of self-interference parameters on the throughput performance with the following parameter setting: $\left({\bar \tau}_{\sf id}, {\bar \tau}_{\sf ac}\right) = 
\left(1000, 100\right)$ ms, $P_{\sf max} = 25$ dB, $T_{\sf eva} = 40$ ms, $\zeta = 0.4$, $\xi$ is varied in $\xi = \left\{0.12, 0.1, 0.08, 0.05\right\}$, and $P_{\sf dat} = P_{\sf max}$.
Recall that the self-interference depends on the transmit power $P$ as  $I(P) = \zeta \left(P\right)^{\xi}$ where $P= P_{\sf sen}$ and $P = P_{\sf dat}$
in the FD sensing and transmission stages, respectively. 
Fig.~\ref{T_VS_p_nofrag_xi} illustrates the variations of the throughput versus the transmission probability  $p$.
It can be observed that when $\xi$ decreases (i.e., the self-interference is smaller), the achieved throughput increases.
This is because SUs can transmit with higher power while still maintaining the sensing constraint during the FD sensing stage, which leads to throughput improvement. 
The optimal $P_{\sf sen}$ corresponding to these values of $\xi$ are $P_{\sf sen} = SNR_s = \left\{25.00, 18.01, 14.23, 11.28\right\}$ dB and 
the optimal probability of transmission is $p^* = 0.0022$ as indicated by a star symbol.
Therefore, to obtain all other results in this section, we set $p^* = 0.0022$.

Fig.~\ref{T_VS_n0_xi_nofrag} illustrates the throughput performance versus number of SUs $n_0$ when we keep the same parameter settings
as those for Fig.~\ref{T_VS_p_nofrag_xi} and $p^* = 0.0022$.
Again, when $\xi$ decreases (i.e., the self-interference becomes smaller), the achieved throughput increases. 
In this figure, the optimal $SNR_s$ achieving the maximum throughput corresponding to the considered values of $\xi$ 
are $P_{\sf sen} = SNR_s = \left\{25.00, 18.01, 14.23, 11.28\right\}$ dB, respectively.

% Fig. 4
\begin{figure}[!t]
\centering
\includegraphics[width=80mm]{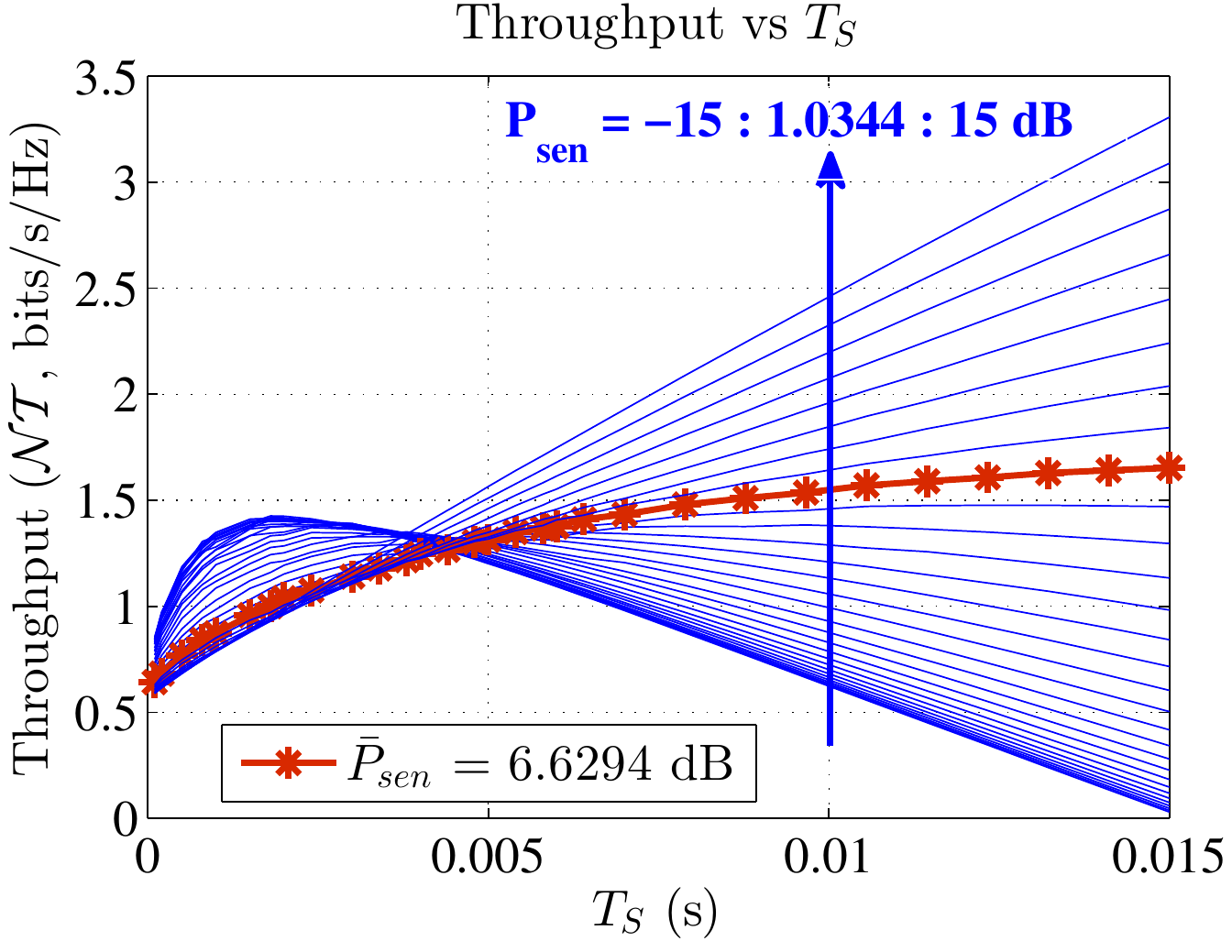}
\caption{Normalized throughput versus SU transmit power $P_{\sf sen}$ and sensing time $T_S$ for $p = 0.0022$, ${\bar \tau}_{\sf id} = 500$ ms, ${\bar \tau}_{\sf ac} = 50$ ms, $n_0 = 40$, 
$\xi =1$, $\zeta = 0.7$ and FDTx with $P_{\sf dat} = 15$ dB.}
\label{T_vs_Tsen_FDTX_Largezetaxi}
\end{figure}

We now verify the results stated in Theorem 1 for the FDTx mode. Specifically,
Fig.~\ref{T_vs_Tsen_FDTX_Largezetaxi} shows the throughput performance for the scenario where the QSIC is very low with large $\xi$ and $\zeta$
where we set the network parameters as follows: $p = 0.0022$, ${\bar \tau}_{\sf id} = 500$ ms, ${\bar \tau}_{\sf ac} = 50$ ms, $n_0 = 40$, $\xi =1$, $\zeta = 0.7$, and $P_{\sf dat} = 15$ dB.
Moreover, we can obtain $\overline{P}_{\sf sen}$ as in (\ref{P_sen_threshold}) in Appendix~\ref{Prosp1}, that is $\overline{P}_{\sf sen} = 6.6294$ dB.
In this figure, the curve indicated by asterisks, which corresponds to $P_{\sf sen} = \overline{P}_{\sf sen}$, shows the monotonic increase of the throughput with sensing
time $T_S$ and other curves corresponding to $P_{\sf sen} > \overline{P}_{\sf sen}$ have the same characteristic.
In contrast, all remaining curves (corresponding to $P_{\sf sen} < \overline{P}_{\sf sen}$) first increase to the maximum values and then decrease as we increase $T_S$.

% Fig. 5
\begin{figure}[!t]
\centering
\includegraphics[width=80mm]{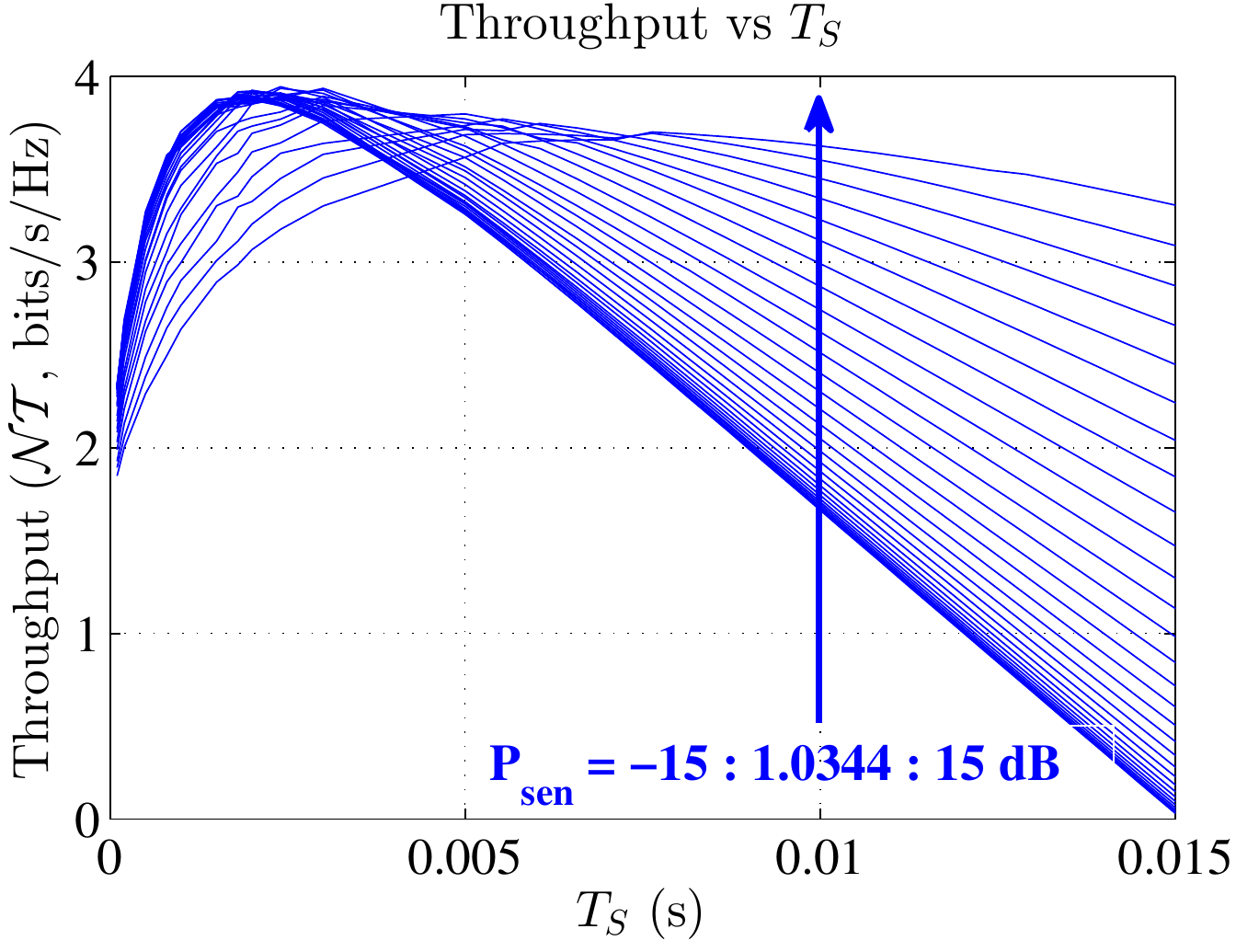}
\caption{Normalized throughput versus SU transmit power $P_{\sf sen}$ and sensing time $T_S$ for $p = 0.0022$, ${\bar \tau}_{\sf id} = 500$ ms, ${\bar \tau}_{\sf ac} = 50$ ms, $n_0 = 40$, $\xi =1$, $\zeta = 0.08$ and FDTx with $P_{\sf dat} = 15$ dB.}
\label{T_vs_Tsen_FDTX_Smallzetaxi}
\end{figure}

Fig.~\ref{T_vs_Tsen_FDTX_Smallzetaxi} illustrates the throughput performance for the very high QSIC with small $\xi$ and $\zeta$
where we set the network parameters as follows: $p = 0.0022$, ${\bar \tau}_{\sf id} = 500$ ms, ${\bar \tau}_{\sf ac} = 50$ ms, $n_0 = 40$, $\xi =1$, $\zeta = 0.08$, and $P_{\sf dat} = 15$ dB.
Moreover, we can obtain $\overline{P}_{\sf sen}$ as in (\ref{P_sen_threshold}) in Appendix~\ref{Prosp1} that is $\overline{P}_{\sf sen} = 19.9201$ dB.
We have $P_{\sf sen} < P_{\sf max} = 15dB < \overline{P}_{\sf sen}$ in this scenario; hence, all the curves first increases to the maximum throughput and then decreases with
the increasing $T_S$. Therefore, we have correctly validated the properties stated in Theorem 1.

% Fig. 6
\begin{figure}[!t]
\centering
\includegraphics[width=80mm]{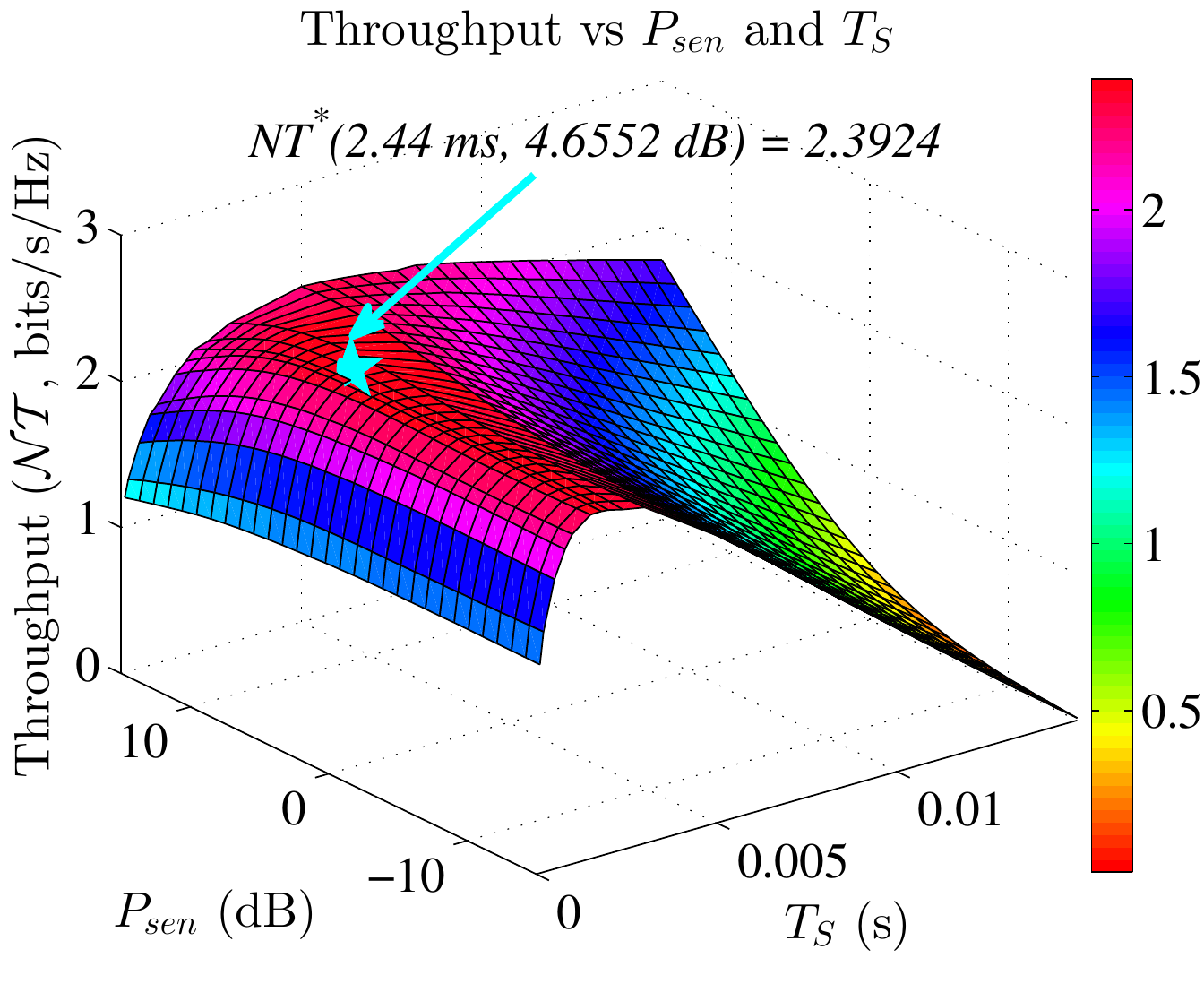}
\caption{Normalized throughput versus SU transmit power $P_{\sf sen}$ and sensing time $T_S$ for $p = 0.0022$, ${\bar \tau}_{\sf id} = 150$ ms, ${\bar \tau}_{\sf ac} = 50$ ms, 
$n_0 = 40$, $\xi =0.95$, $\zeta = 0.08$ and FDTx with $P_{\sf dat} = 15$ dB.}
\label{P2_15dB_tau1505015ms}
\end{figure}

Now we investigate the throughput performance versus SU transmit power $P_{\sf sen}$ and sensing time $T_S$ for the case of high QSIC with $\xi =0.95$ and $\zeta = 0.08$.
Fig.~\ref{P2_15dB_tau1505015ms} shows the throughput versus the SU transmit power $P_{\sf sen}$ and sensing time $T_S$ for the FDTx mode with $P_{\sf dat} = 15$ dB, $p = 0.0022$, 
${\bar \tau}_{\sf id} = 150$ ms, ${\bar \tau}_{\sf ac} = 50$ ms, and $n_0 = 40$. 
It can be observed that there exists an optimal configuration of the SU transmit power $P_{\sf sen}^* = 4.6552$ dB and sensing time $T_S^* = 2.44$ ms to achieve the 
maximum throughput $\mathcal{NT}\left(T_S^*, P_{\sf sen}^*\right) = 2.3924$, which is indicated by a star symbol.
These results confirm that SUs must set appropriate sensing time and transmit power for the FDC--MAC protocol to achieve the maximize throughput,
which cannot be achieved by setting $T_s = T$ as proposed in existing designs such as in \cite{tan2015distributed}. %\cite{Afifi14} and \cite{report}.

% Fig. 7
%\begin{figure}[!t]
%\centering
%\includegraphics[width=80mm]{P2_15dB_tau1505015ms_Mediumzetaxi}
%\caption{Normalized throughput versus SU transmit power $P_{\sf sen}$ and sensing time $T_S$ for $p = 0.0022$, ${\bar \tau}_{\sf id} = 150$ ms, ${\bar \tau}_{\sf ac} = 50$ ms, $n_0 = 40$, 
%$\xi =0.95$, $\zeta = 0.2$ and FDTx with $P_{\sf dat} = 15$ dB.}
%\label{P2_15dB_tau1505015ms_Mediumzetaxi}
%\end{figure}
%
%Fig.~\ref{P2_15dB_tau1505015ms_Mediumzetaxi} shows the throughput versus SU transmit power $P_{\sf sen}$ and sensing time $T_S$ for the FDTx mode under
%the medium QSIC with $\xi =0.95$, and $\zeta = 0.2$ where other parameters are set as follows:
%$P_{\sf dat} = 15$ dB, $p = 0.0022$, ${\bar \tau}_{\sf id} = 150$ ms, ${\bar \tau}_{\sf ac} = 50$ ms, and $n_0 = 40$. 
%Again, there exists an optimal configuration of SU transmit power $P_{\sf sen}^* = -9.8276$ dB and sensing time $T_S^* = 1.8$ ms to achieve the maximum throughput $\mathcal{NT}\left(T_S^*, P_{\sf sen}^*\right) = 1.7871$, which is indicated by a star symbol.
%Moreover, we can see that the FDC--MAC tends to choose an optimal configuration close to the HD sensing design in this case (i.e., the SU transmitter would not transmit data during the spectrum sensing stage) since $P_{\sf sen}^*$ is closer to zero comparing to the optimal design shown in Fig.~\ref{P2_15dB_tau1505015ms}.

% Fig. 8
\begin{figure}[!t]
\centering
\includegraphics[width=80mm]{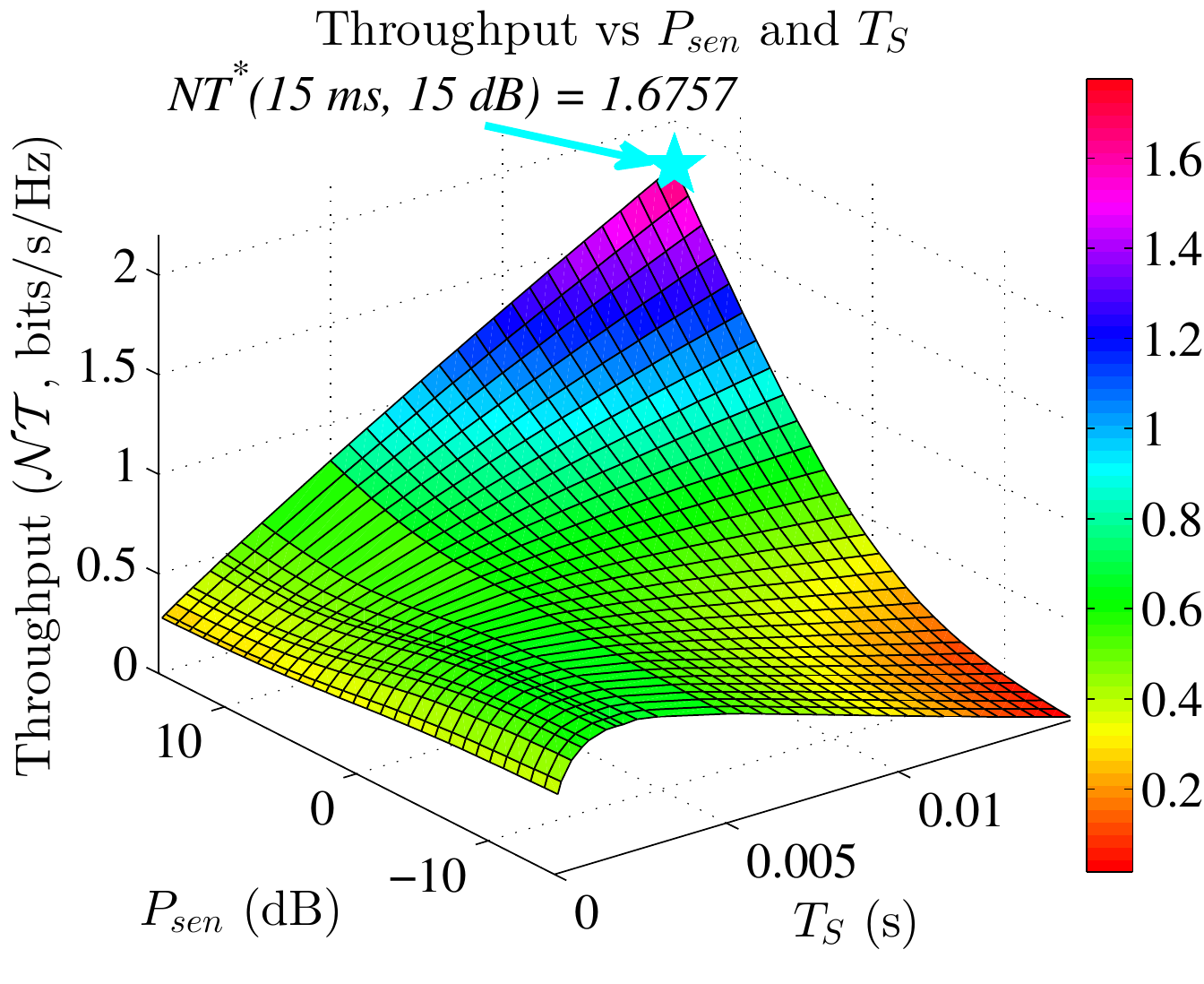}
\caption{Normalized throughput versus SU transmit power $P_{\sf sen}$ and sensing time $T_S$ for $p = 0.0022$, ${\bar \tau}_{\sf id} = 150$ ms, ${\bar \tau}_{\sf ac} = 50$ ms, $n_0 = 40$, $\xi =0.95$, $\zeta = 0.8$ and FDTx with $P_{\sf dat} = 15$ dB.}
\label{P2_15dB_tau1505015ms_Largezetaxi}
\end{figure}

In Fig.~\ref{P2_15dB_tau1505015ms_Largezetaxi}, we present the throughput versus the SU transmit power $P_{\sf sen}$ and sensing time $T_S$ 
for the low QSIC scenario where $p = 0.0022$, ${\bar \tau}_{\sf id} = 150$ ms, ${\bar \tau}_{\sf ac} = 50$ ms, $P_{\sf max} = 15$ dB, $n_0 = 40$, $\xi =0.95$, and $\zeta = 0.8$. 
The optimal configuration of SU transmit power $P_{\sf sen}^* = 15$ dB and sensing time $T_S^* = 15$ ms to achieve the maximum throughput $\mathcal{NT}\left(T_S^*, P_{\sf sen}^*\right) = 1.6757$
 is again indicated by a star symbol. Under this optimal configuration, the FD sensing is performed during the whole data phase (i.e., there is no transmission stage). 
In fact, to achieve the maximum throughput, the SU must provide the satisfactory sensing performance and attempt to achieve high transmission rate.
Therefore, if the QSIC is low, the data rate achieved during the transmission stage can be lower than that in the FD sensing stage because of
the very strong self-interference in the transmission stage. Therefore, setting longer FD sensing time enables to achieve more satisfactory sensing
performance and higher transmission rate, which explains that the optimal configuration should set $T_S^* = T$ for the low QSIC scenario.
This protocol configuration corresponds to existing design in \cite{tan2015distributed}, which is a special case of the proposed FDC--MAC protocol. % \cite{Afifi14} and \cite{report}.

% Fig. 9
\begin{figure}[!t]
\centering
\includegraphics[width=80mm]{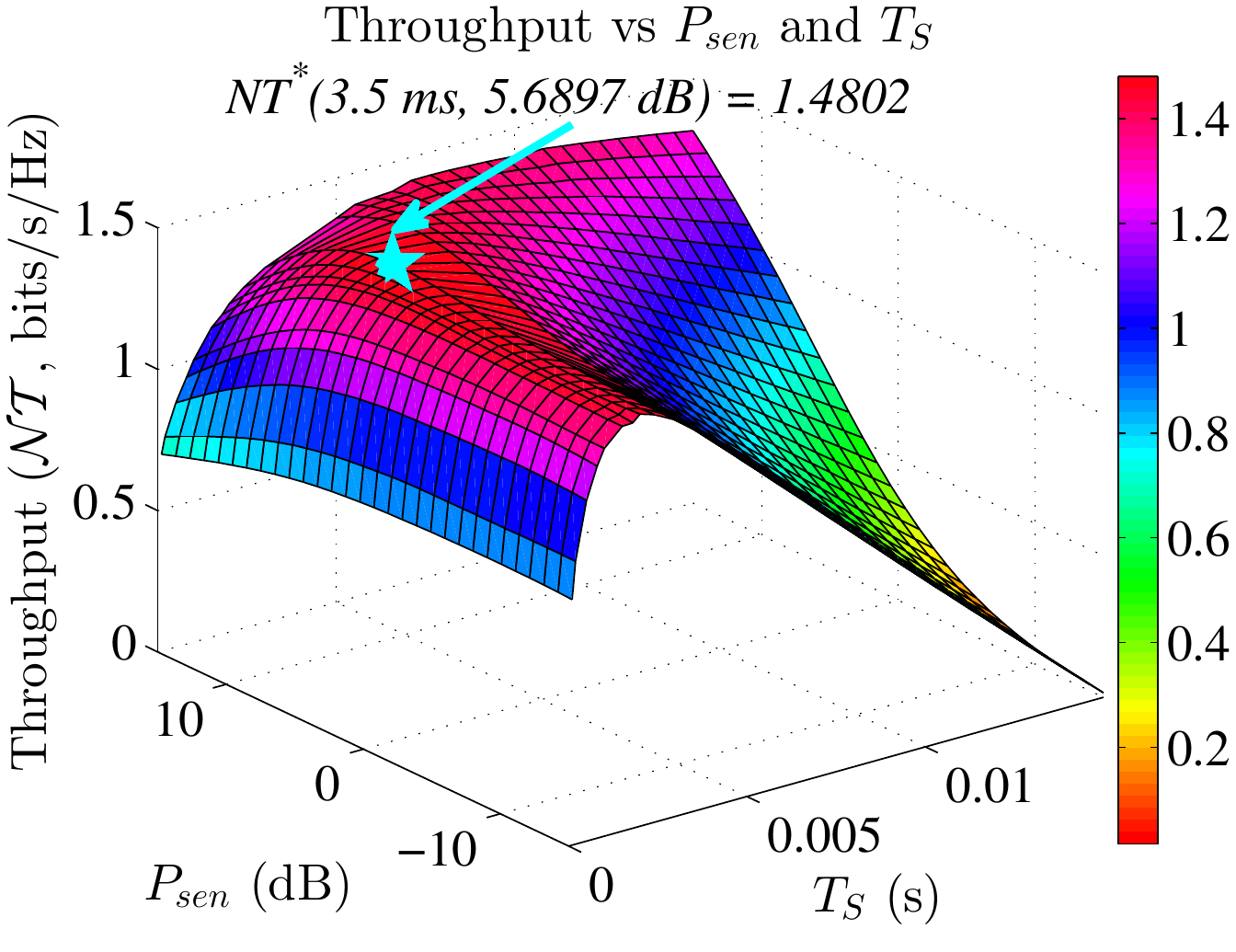}
\caption{Normalized throughput versus SU transmit power $P_{\sf sen}$ and sensing time $T_S$ for $p = 0.0022$, ${\bar \tau}_{\sf id} = 150$ ms, ${\bar \tau}_{\sf ac} = 50$ ms, $n_0 = 40$, $\xi =0.95$, $\zeta = 0.08$ and HDTx.}
\label{P2_15dB_tau1505015ms_HDTX}
\end{figure}

We now investigate the throughput performance with respect to the SU transmit power $P_{\sf sen}$ and sensing time $T_S$ for the HDTx mode.
Fig.~\ref{P2_15dB_tau1505015ms_HDTX} illustrates the throughput performance for the high QSIC scenario with $\xi =0.95$ and $\zeta = 0.08$.
It can be observed that there exists an optimal configuration of SU transmit power $P_{\sf sen}^* = 5.6897$ dB and sensing time $T_S^* = 3.5$ ms to achieve the maximum throughput $\mathcal{NT}\left(T_S^*, P_{\sf sen}^*\right) = 1.4802$, which is indicated by a star symbol.
The maximum achieved throughput of the HDTx mode is lower than that in the FDTx mode presented in Fig.~\ref{P2_15dB_tau1505015ms}.
This is because with high QSIC, the FDTx mode can transmit more data than the HDTx mode in the transmission stage.

% Fig. 10
\begin{figure}[!t]
\centering
\includegraphics[width=80mm]{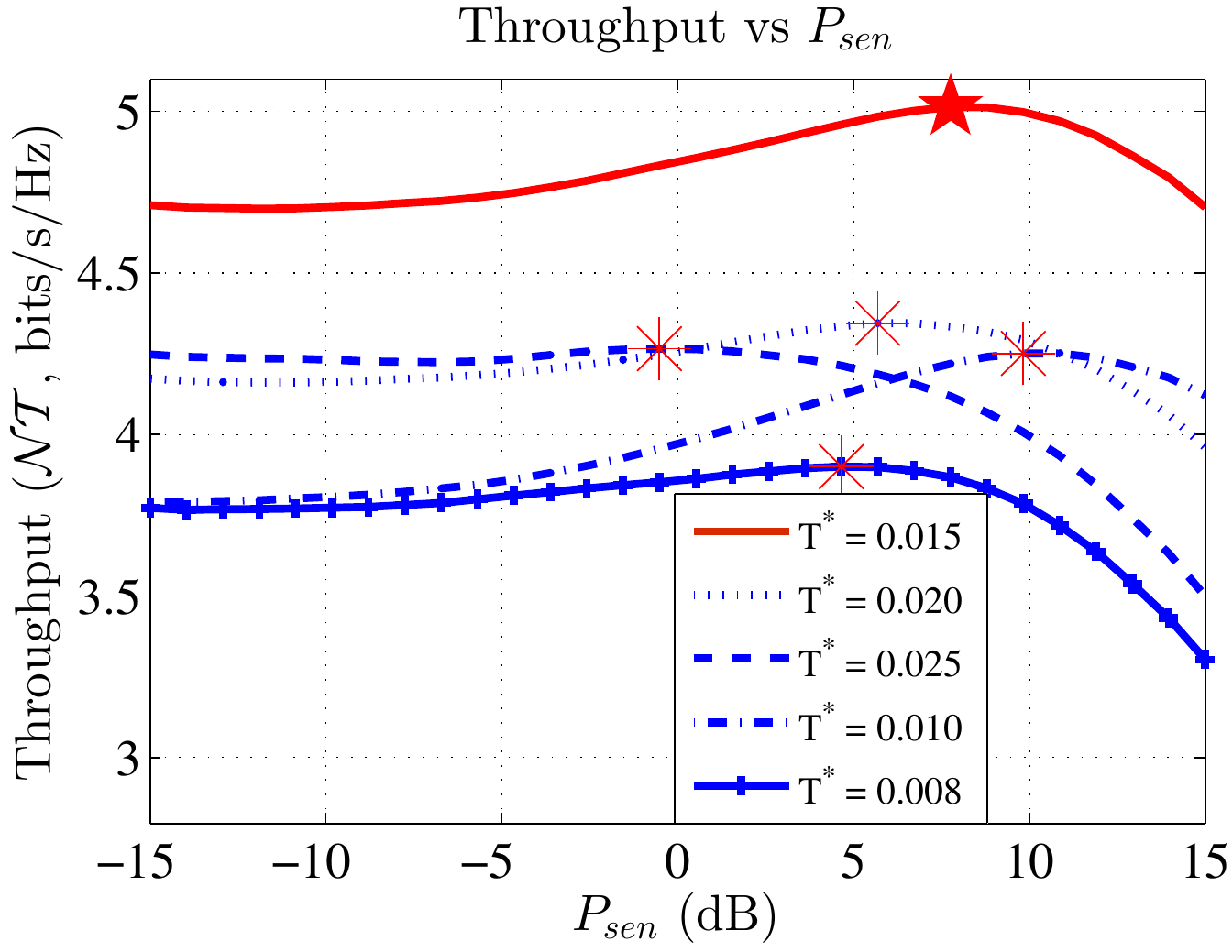}
\caption{Normalized throughput versus SU transmit power $P_{\sf sen}$ for $T_S = 2.2$ ms, $p = 0.0022$, ${\bar \tau}_{\sf id} = 1000$ ms, ${\bar \tau}_{\sf ac} = 50$ ms, $n_0 = 40$, 
$\xi =0.95$, $\zeta = 0.08$, varying $T$, and FDTx with $P_{\sf dat} = 15$ dB.}
\label{tauall_Tsens}
\end{figure}

In Fig.~\ref{tauall_Tsens}, we show the throughput versus the SU transmit power $P_{\sf sen}$ for $T_S = 2.2$ ms,  $p = 0.0022$, ${\bar \tau}_{\sf id} = 1000$ ms, ${\bar \tau}_{\sf ac} = 50$ ms,
 $n_0 = 40$, $\xi =0.95$, $\zeta = 0.08$ and various values of $T$ (i.e., the data phase duration) for the FDTx mode with $P_{\sf dat} = 15$ dB. 
For each value of $T$, there exists the optimal SU transmit power $P_{\sf sen}^*$ which is indicated by an asterisk.
It can be observed that as $T$ increases from 8 ms to 25 ms, the achieved maximum throughput first increases then decreases with $T$.
Also in the case with $T^* = 15$ ms, the SU achieves the largest throughput which is indicated by a star symbol.
Furthermore, the achieved throughput significantly decreases when the pair of $\left(T,P_{\sf sen}\right)$ deviates from the optimal values, $\left(T^*,P_{\sf sen}^*\right)$.

% Fig. 11
\begin{figure}[!t]
\centering
\includegraphics[width=80mm]{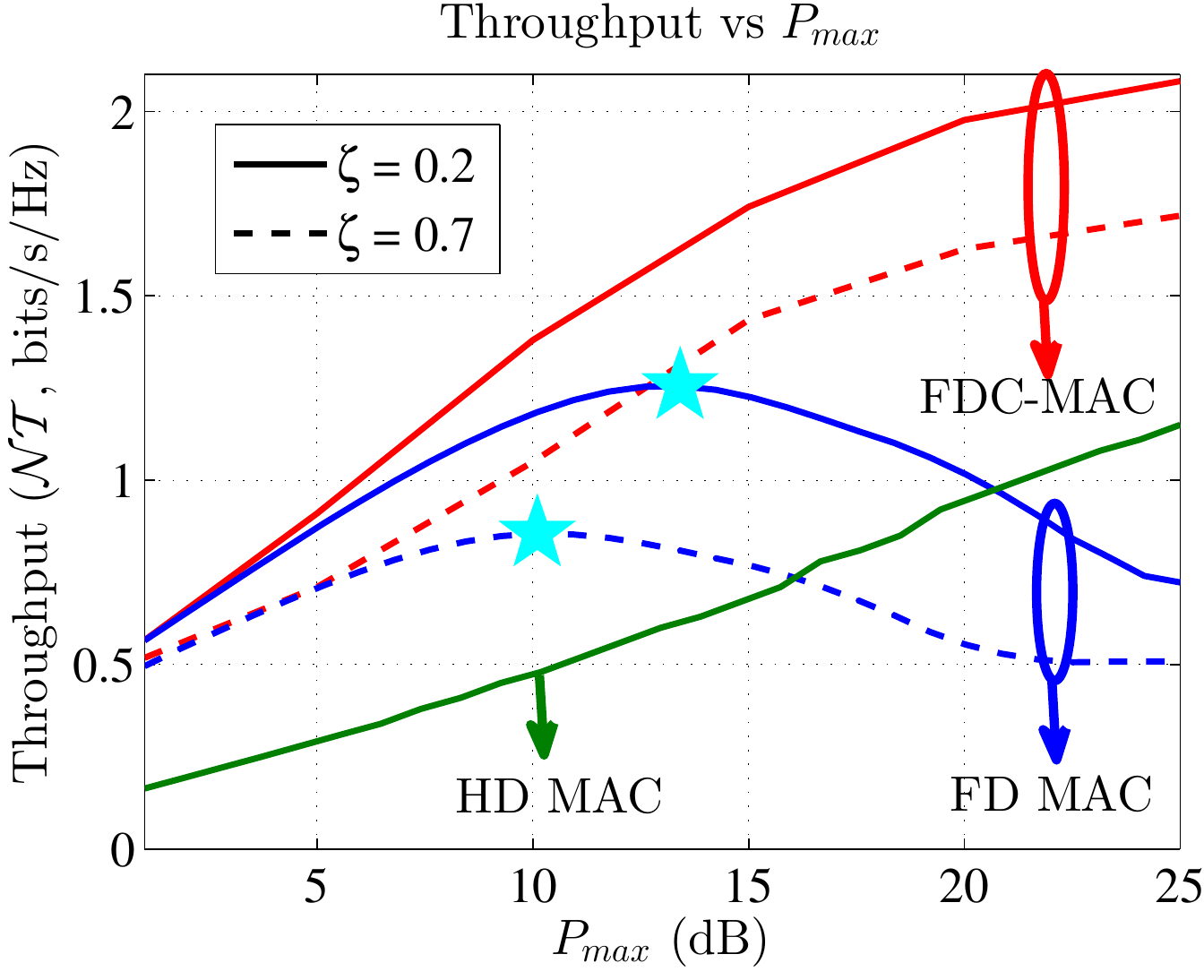}
\caption{Normalized throughput versus $P_{\sf max}$ for ${\bar \tau}_{\sf id} = 150$ ms, ${\bar \tau}_{\sf ac} = 75$ ms, $n_0 = 40$, $\xi =0.85$, $n_0 = 40$, 
$\zeta = \left\{0.2, 0.7\right\}$, and FDTx with $P_{\sf dat} = P_{\sf max}$ dB.}
\label{T_vs_Pmax_tauidac_150_75}
\end{figure}

Finally, we compare the throughput of our proposed FDC-MAC protocol,  the single-stage FD MAC protocol
where FD sensing is performed during the whole data phase \cite{tan2015distributed} and the HD MAC protocol without
exploiting concurrent sensing and transmission during the sensing interval in Fig.~\ref{T_vs_Pmax_tauidac_150_75}.
For brevity, the single-stage FD MAC protocol is refereed to as FD MAC in this figure.
The parameter settings are as follows: ${\bar \tau}_{\sf id} = 150$ ms, ${\bar \tau}_{\sf ac} = 75$ ms, $n_0 = 40$, $\xi =0.85$, $n_0 = 40$, 
$\zeta = \left\{0.2, 0.7\right\}$, and FDTx with $P_{\sf dat} = P_{\sf max}$ dB.
For fair comparison, we first obtain the optimal configuration of the single-stage FD MAC protocol, i.e., then we use $\left(T^*, p^*\right)$ for the HD MAC protocol and 
FDC-MAC protocol.
For the single-stage FD MAC protocol, the transmit power  is set to $P_{\sf max}$ because there is only a single stage where the SU performs 
sensing and transmission simultaneously during the data phase.
In addition, the HD MAC protocol will also transmit with the maximum transmit power $P_{\sf max}$ to achieve the highest throughput.
For both studied cases of $\zeta= \left\{0.2, 0.7\right\}$, our proposed FDC-MAC protocol significantly outperforms the other two protocols.
Moreover, the single-stage FD MAC protocol \cite{tan2015distributed} with power allocation outperforms the HD MAC protocol at the corresponding optimal power level
required by the single-stage FD MAC protocol. However, both single-stage FDC-MAC and HD MAC protocols achieve increasing throughput with higher $P_{\sf max}$ while
the single-stage FD MAC protocol has the throughput first increased then decreased as $P_{\sf max}$ increases. This demonstrates
that the self-interference has the very negative impact on the throughput performance of the single-stage FD MAC protocol, which 
is efficiently mitigated by our proposed FDC-MAC protocol.

\section{Conclusion}
\label{conclusion} 

In this paper, we have proposed the FDC--MAC protocol for cognitive radio networks, analyzed its throughput
performance, and studied its optimal parameter configuration. The design and analysis have taken into
account the FD communication capability and the self-interference of the FD transceiver.
We have shown that there exists an optimal FD sensing time to achieve the maximum throughput.
In addition, we have presented extensive numerical results to demonstrate the impacts
of self-interference and protocol parameters on the throughput performance. In particular, we have shown that
the FDC--MAC protocol achieves significantly higher throughput the HD MAC protocol, which confirms
that the FDC--MAC protocol can efficiently exploit the FD communication capability. Moreover,
the FDC--MAC protocol results in higher throughput with the increasing maximum power budget while
the throughput of the single-stage FD MAC can decrease in the high power regime. This result validates
the importance of adopting the two-stage procedure in the data phase and the optimization of sensing time
and transmit power during the FD sensing stage to mitigate the negative self-interference impact.

%In this paper, we have proposed the FDC--MAC protocol for cognitive radio networks, analyzed its throughput
%performance, and studied its optimal parameter configuration. The design and analysis have taken into
%account the FD communication capability and the self-interference of the FD transceiver.
%We have shown that there exists an optimal FD sensing time to achieve the maximum throughput.
%In addition, we have presented extensive numerical results to demonstrate the impacts
%of self-interference and protocol parameters on the throughput performance. In particular, we have shown that
%certain existing designs (e.g., parallel sensing and transmission during the whole data phase) may result in 
%good performance in only some specific scenarios, which are not optimal in the general setting.  

%\appendices
\section{Appendices}

\subsection{Derivation of ${\overline T}_{\sf cont}$}
\label{AppenA0}

To calculate ${\overline T}_{\sf cont}$, we define some further parameters as follows.
Denote $T_{\sf coll}$ as the duration of the collision and
$T_{\sf succ}$ as the required time for successful RTS/CTS transmission. These quantities can be calculated as follows \cite{Cali00}:
\beqn
\label{TCTSTI}
\left\{ {\begin{array}{*{20}{c}}
   T_{\sf succ} = DIFS + RTS + SIFS + CTS + 2PD \hfill  \\
   T_{\sf coll} = DIFS + RTS + PD, \hfill  \\
\end{array}} \right.
\eeqn
where $DIFS$ is the length of a DCF (distributed coordination function) interframe space, $RTS$ and $CTS$ denote
the lengths of the RTS and CTS messages, respectively.

As being shown in Fig.~\ref{Sentime_FDMAC_SF}, there can be several idle periods and collisions before one successful channel reservation.
Let $T_{\sf idle}^i$ denote the $i$-th idle duration between two consecutive RTS/CTS exchanges, which can be collisions or successful exchanges. 
Then, $T_{\sf idle}^i$ can be calculated based on its probability mass function (pmf),  which is derived as follows. 
In the following, all relevant quantities are defined in terms of the number of time slots.
With $n_0$ SUs joining the contention resolution, let $\mathcal{P}_{\sf succ}$, $\mathcal{P}_{\sf coll}$ and $\mathcal{P}_{\sf idle}$ denote
 the probabilities that a particular generic slot corresponds to a successful transmission, a collision, and an idle slot, respectively. These 
probabilities can be calculated as follows:
\beqn
\mathcal{P}_{\sf succ} &=& n_0p\left(1-p\right)^{n_0-1} \\
\mathcal{P}_{\sf idle} &=&  \left(1-p\right)^{n_0} \hspace{1cm} \\
\mathcal{P}_{\sf coll} &=&  1-\mathcal{P}_{\sf succ}-\mathcal{P}_{\sf idle},
\eeqn
where $p$ is the transmission probability of an SU in a generic slot. 
In general, the interval ${ T}_{\sf cont}$, whose average value is  ${\overline T}_{\sf cont}$ given in (\ref{tover}), consists of several intervals 
corresponding to idle periods, collisions, and one successful RTS/CTS transmission. 
Hence, this quantity can be expressed as 
\beqn
\label{T_cont}
{T}_{\sf cont} = \sum_{i=1}^{N_{\sf coll}} \left(T_{\sf coll}+ T_{\sf idle}^i\right) + T_{\sf idle}^{N_{\sf coll}+1} + T_{\sf succ},
\eeqn
where $N_{\sf coll}$ is the number of collisions before the successful RTS/CTS exchange and 
$N_{\sf coll}$ is a geometric random variable (RV) with parameter $1-\mathcal{P}_{\sf coll}/\mathcal{\overline P}_{\sf idle}$ where $\mathcal{\overline P}_{\sf idle} = 1 - \mathcal{P}_{\sf idle}$. 
Therefore, its pmf can be expressed as
\beqn
\label{N_c_cal}
 f_{X}^{N_{\sf coll}} \left(x\right) = \left(\frac{\mathcal{P}_{\sf coll}}{\mathcal{\overline P}_{\sf idle}}\right)^{x} \left(1-\frac{\mathcal{P}_{\sf coll}}{\mathcal{\overline P}_{\sf idle}}\right), \: x = 0, 1, 2, \ldots
\eeqn
Also, $T_{\sf idle}$ represents the number of consecutive idle slots, which is also a geometric RV with parameter $1-\mathcal{P}_{\sf idle}$ with the following pmf
\beqn
\label{T_I_cal}
f_{X}^{T_{\sf idle}} \left(x\right) = \left(\mathcal{P}_{\sf idle}\right)^{x} \left(1-\mathcal{P}_{\sf idle}\right), \: x = 0, 1, 2, \ldots
\eeqn
Therefore, ${\overline T}_{\sf cont}$ (the average value of ${T}_{\sf cont}$) can be written as follows \cite{Cali00}:
\beqn
{\overline T}_{\sf cont}  = {\overline N}_{\sf coll}T_{\sf coll} + {\overline T}_{\sf idle} \left({\overline N}_{\sf coll}+1\right) + T_{\sf succ} \label{T_contgeo},
\eeqn
where ${\overline T}_{\sf idle}$ and ${\overline N}_{\sf coll}$ can be calculated as
\beqn
{\overline T}_{\sf idle} &=& \frac{\left(1-p\right)^{n_0}}{1-\left(1-p\right)^{n_0}} \\
{\overline N}_{\sf coll} &=& \frac{1-\left(1-p\right)^{n_0}}{n_0p\left(1-p\right)^{n_0-1}}-1. 
\eeqn
These expressions are obtained by using the  pmfs of the corresponding RVs given in (\ref{N_c_cal}) and (\ref{T_I_cal}), respectively \cite{Cali00}.

\subsection{Derivations of $\mathcal{B}_1$, $\mathcal{B}_2$, $\mathcal{B}_3$}
\label{AppenB1}

We will employ a pair of parameters $\left(\theta, \varphi\right)$ to represent the HDTX and FDTX modes where ($\left(\theta, \varphi\right) = \left(0, 1\right)$) for
HDTx mode and  ($\left(\theta, \varphi\right) = \left(1, 2\right)$) for the FDTx mode. Moreover,
since the transmit powers in the FD sensing and transmission stages are different, which are equal to $P_{\sf sen}$ and $P_{\sf dat}$, respectively,
we define different SNRs and SINRs in these two stages as follows: $\gamma_{S1} = \frac{P_{\sf sen}}{N_0}$ and $\gamma_{S2} = \frac{P_{\sf sen}}{N_0+P_p}$
are the SNR and SINR achieved by the SU in the FD sensing stage with and without the presence of the PU, respectively; 
$\gamma_{D1} = \frac{P_{\sf dat}}{N_0+\theta I}$ and $\gamma_{D2} = \frac{P_{\sf dat}}{N_0+P_p+\theta I}$ for $I = \zeta P_{\sf dat}^\xi$ are the SNR and SINR
achieved by the SU in the transmission stage with and without the presence of the PU, respectively. It can be seen that we have accounted for
the self-interference for the FDTx mode during the transmission stage in $\gamma_{D1}$ by noting that $\theta=1$ in this case. The parameter $\varphi$
for the HDTx and FDTx modes will be employed to capture the throughput for one-way and two-way transmissions in these modes, respectively.

The derivations of $\mathcal{B}_1$, $\mathcal{B}_2$, and $\mathcal{B}_3$ require us to consider different possible
sensing outcomes in the FD sensing stage. In particular, we need to determine the detection probability $\mathcal{P}_d^{ij}$, which is the probability
of correctly detecting the PU given the PU is active, and the false alarm probability $\mathcal{P}_f^{ij}$, which is the
probability of the erroneous sensing of an idle channel, for each event $h_{ij}$ capturing the state changes of the PU. 
In the following analysis, we assume the exponential distribution for ${\tau}_{\sf ac}$ and ${\tau}_{\sf id}$ where ${\bar \tau}_{\sf ac}$ and ${\bar \tau}_{\sf id}$ denote the
corresponding average values of these active and idle intervals.
Specifically, let $f_{\tau_{\sf x}}\left(t\right)$ denote the pdf of $\tau_{\sf x}$ (${\sf x}$ represents ${\sf ac}$ or ${\sf id}$ in the pdf of $\tau_{\sf ac}$ or $\tau_{\sf id}$, respectively) 
then
\beqn
\label{pdf_tau_ac_id}
f_{\tau_{\sf x}}\left(t\right) =  \frac{1}{{\bar \tau}_{\sf x}} \exp(-\frac{t}{{\bar \tau}_{\sf x}}).
\eeqn
Similarly, we employ $T_S^{ij}$ and $T_D^{ij}$ to denote the number of bits transmitted on one unit of system bandwidth during the FD sensing and transmission stages under
the PU's state-changing event $h_{ij}$, respectively. 
%Note that $T_D^{ij}$, and $T_S^{ij}$ ($i,j \in \left\{0,1\right\}$) are calculated according to Appendix \ref{TPUT_FRAGMEN}, while $\mathcal{P}_f^{00}$ and $\mathcal{P}_d^{01}$ are derived from Appendix \ref{CAL_P_F_P_D}.

We can now  calculate $\mathcal{B}_1$ as follows:
%\beqn
%\mathcal{B}_1 = \mathcal{P}\left(\mathcal{H}_0\right) \int_{t=T_{\sf ove}+T}^\infty T_1^{00} f_{\tau_{\sf id}}(t) dt = \mathcal{P}\left(\mathcal{H}_0\right) T_1^{00} \exp\left(-\frac{T_{\sf ove}+T}{\bar{\tau}_{\sf id}}\right)
%\eeqn
\beqn
\mathcal{B}_1 = \mathcal{P}\left(\mathcal{H}_0\right) \int_{t=T_{\sf ove}+T}^\infty T_1^{00} f_{\tau_{\sf id}}(t) dt %\hspace{1cm} \nonumber\\
= \mathcal{P}\left(\mathcal{H}_0\right) T_1^{00} \exp\left(-\frac{T_{\sf ove}+T}{\bar{\tau}_{\sf id}}\right), \hspace{0.85cm}
\eeqn
where $\mathcal{P}\left(\mathcal{H}_0\right)$ denotes the probability of the idle state of the PU, and
  $\mathcal{P}_f^{00}$ is the false alarm probability for 
event $h_{00}$ given in Appendix~\ref{CAL_P_F_P_D}. Moreover, $T_1^{00} = \mathcal{P}_f^{00} T_S^{00} + (1-\mathcal{P}_f^{00})(T_S^{00}+T_D^{00})$, $T_S^{00} = T_S \log_2 \left(1+\gamma_{S1}\right)$, 
$T_D^{00} = \varphi \left(T-T_S\right) \log_2 \left(1+\gamma_{D1}\right)$ where $T_S^{00}$ and $T_D^{00}$ denote the number of bits transmitted (over one Hz of system bandwidth) in the FD sensing
and transmission stages of the data phase, respectively. 
After some manipulations, we achieve
%\beqn 
%\mathcal{B}_1 = \mathcal{K}_e  \exp \left(\frac{T}{\Delta\tau}\right) \left[T_S \log_2 \left(1+\gamma_{S1}\right) +  \varphi \left(1 - \mathcal{P}_f^{00}\right) \left(T-T_S\right) \log_2 \left(1+\gamma_{D1}\right)\right],
%\eeqn
%\beqn 
%\mathcal{B}_1 = \mathcal{K}_e  \exp \left(\frac{T}{\Delta\tau}\right) \left[T_S \log_2 \left(1+\gamma_{S1}\right) + \right. \hspace{0.4cm} \nonumber\\
%\left.  \varphi \left(1 - \mathcal{P}_f^{00}\right) \left(T-T_S\right) \log_2 \left(1+\gamma_{D1}\right)\right], 
%\eeqn
\beqn 
\mathcal{B}_1 = \mathcal{K}_e  \exp \left(\frac{T}{\Delta\tau}\right) \left[T_S \log_2 \left(1+\gamma_{S1}\right) +   \varphi \left(1 - \mathcal{P}_f^{00}\right) \left(T-T_S\right) \log_2 \left(1+\gamma_{D1}\right)\right], 
\eeqn
where $\mathcal{K}_e = \mathcal{P} \left(\mathcal{H}_0\right) \exp \left(-\left(\frac{T_{\sf ove}}{{\bar \tau}_{\sf id}}+\frac{T}{{\bar \tau}_{\sf ac}}\right)\right)$ and $\frac{1}{\Delta\tau} = \frac{1}{\bar{\tau}_{\sf ac}} - \frac{1}{\bar{\tau}_{\sf id}}$.

Moreover, we can calculate $\mathcal{B}_2$ as
%\beqn \label{eqT2}
%\mathcal{B}_2 = \mathcal{P}\left(\mathcal{H}_0\right) \int_{t_1 =T_{\sf ove}+T_S}^{T_{\sf ove}+T} \int_{t_2 =T_{\sf ove}+T-t_1}^\infty T_2^{01}(t_1) f_{\tau_{\sf id}}(t_1)f_{\tau_{\sf ac}}(t_2) dt_1 dt_2,
%\eeqn
\beqn \label{eqT2}
\mathcal{B}_2 = \mathcal{P}\left(\mathcal{H}_0\right) \int_{t_1 =T_{\sf ove}+T_S}^{T_{\sf ove}+T} \int_{t_2 =T_{\sf ove}+T-t_1}^\infty %\hspace{2.4cm} \nonumber\\
T_2^{01}(t_1) f_{\tau_{\sf id}}(t_1)f_{\tau_{\sf ac}}(t_2) dt_1 dt_2,
\eeqn
where $T_2^{01}(t_1) = \mathcal{P}_f^{00} T_S^{00} + (1-\mathcal{P}_f^{00})(T_S^{00} +T_D^{01}\left(\bar{t}_1\right))$,
 %$T_D^{01}\left({t}_1\right) = \varphi \left[\bar{t}_1 \log_2 \left(1+\gamma_{D1}\right) + \left(T-T_S-\bar{t}_1\right) \log_2 \left(1+\gamma_{D2}\right)\right]$, 
 $T_D^{01}\left({t}_1\right) = \varphi \left(T-T_S-\bar{t}_1\right) \log_2 \left(1+\gamma_{D2}\right) + \varphi \bar{t}_1 \log_2 \left(1+\gamma_{D1}\right)$, 
and $\bar{t}_1 = t_1 - \left(T_{\sf ove}+T_S\right)$. In this expression, $t_1$ denotes the interval from the beginning of the CA cycle to the instant when the PU changes to the active state from
an idle state. Again, $T_S^{00}$ and $T_D^{01}$ denote the amount of data transmitted in the FD sensing and transmission stages for this case, respectively.
 After some manipulations, we achieve
%\beqn
%\mathcal{B}_2 = \mathcal{K}_e \frac{\Delta\tau}{{\bar \tau}_{\sf id}} \left\{\left(\exp\left(\frac{T}{\Delta\tau}\right)-\exp\left(\frac{T_S}{\Delta\tau}\right)\right) \left[T_S \log_2 \left(1\!+\!\gamma_{S1}\right) \!- \! \varphi \Delta\tau \left(1\!-\!\mathcal{P}_f^{00}\right) \log_2 \left(\frac{1\!+\!\gamma_{D1}}{1\!+\!\gamma_{D2}}\right)\right] \right. \nonumber \\
%\left. + \varphi \left(T - T_S\right) \left(1\!-\!\mathcal{P}_f^{00}\right) \left[\!\exp\left(\!\frac{T}{\Delta\tau}\!\right) \log_2 \!\left(\!1\!+\!\gamma_{D1}\!\right) \!-\! \exp\left(\!\frac{T_S}{\Delta\tau}\!\right) \log_2 \left(1\!+\!\gamma_{D2}\right)\! \right]\! \right\}.
%\eeqn
%\beqn
%\mathcal{B}_2 = \mathcal{K}_e \frac{\Delta\tau}{{\bar \tau}_{\sf id}} \left\{\left(\exp\left(\frac{T}{\Delta\tau}\right)-\exp\left(\frac{T_S}{\Delta\tau}\right)\right) \times \right. \nonumber \hspace{1.12cm}\\
 %\left[T_S \log_2 \left(1\!+\!\gamma_{S1}\right) \!- \! \varphi \Delta\tau \left(1\!-\!\mathcal{P}_f^{00}\right) \log_2 \left(\frac{1\!+\!\gamma_{D1}}{1\!+\!\gamma_{D2}}\right)\right] \hspace{0.34cm} \nonumber \\
%+ \varphi \left(T - T_S\right) \left(1\!-\!\mathcal{P}_f^{00}\right) \times \hspace{4.2cm} \nonumber\\
%\left. \left[\!\exp\left(\!\frac{T}{\Delta\tau}\!\right) \log_2 \!\left(\!1\!+\!\gamma_{D1}\!\right) \!-\! \exp\left(\!\frac{T_S}{\Delta\tau}\!\right) \log_2 \left(1\!+\!\gamma_{D2}\right)\! \right]\! \right\}.
%\eeqn
\beqn
\mathcal{B}_2 = \mathcal{K}_e \frac{\Delta\tau}{{\bar \tau}_{\sf id}} \left\{\left(\exp\left(\frac{T}{\Delta\tau}\right)-\exp\left(\frac{T_S}{\Delta\tau}\right)\right) 
 \left[T_S \log_2 \left(1\!+\!\gamma_{S1}\right) \!- \! \varphi \Delta\tau \left(1\!-\!\mathcal{P}_f^{00}\right) \log_2 \left(\frac{1\!+\!\gamma_{D1}}{1\!+\!\gamma_{D2}}\right)\right] \right.  \nonumber \\
\left. + \varphi \left(T - T_S\right) \left(1\!-\!\mathcal{P}_f^{00}\right) 
 \left[\!\exp\left(\!\frac{T}{\Delta\tau}\!\right) \log_2 \!\left(\!1\!+\!\gamma_{D1}\!\right) \!-\! \exp\left(\!\frac{T_S}{\Delta\tau}\!\right) \log_2 \left(1\!+\!\gamma_{D2}\right)\! \right]\! \right\}. \hspace{2.35cm}
\eeqn

Finally, we can express $\mathcal{B}_3$ as follows:
%\beqn
%\mathcal{B}_3 = \mathcal{P}\left(\mathcal{H}_0\right) \!\!\! \int_{t_1 =T_{\sf ove}}^{T_{\sf ove}+T_S} \!\!\!\int_{t_2 =T_{\sf ove}\!+\!T-t_1}^\infty \!\!\! \left[\mathcal{P}_d^{01} \left(\bar{t}_1\right) T_S^{01} \left(\bar{t}_1\right) \!+\! (1\!-\!\mathcal{P}_d^{01} \left(\bar{t}_1\right))(T_S^{01} \left(\bar{t}_1\right) \!+ \! T_D^{11})\right]    f_{\tau_{\sf id}}(t_1)f_{\tau_{\sf ac}}(t_2) dt_1 dt_2
%\eeqn
\beqn
\mathcal{B}_3 \!=\! \mathcal{P}\!\left(\mathcal{H}_0\right) \!\!\!\! \int_{t_1 =T_{\sf ove}}^{T_{\sf ove}\!+\!T_S} \!\!\!\!\!\int_{t_2 =T_{\sf ove}\!+\!T\!-\!t_1}^\infty %\nonumber \hspace{2.4cm}\\
\!\!\!\!\!\!\!\!\!\left[\!\mathcal{P}_d^{01} \!\left(\bar{t}_1\right) \!T_S^{01} \!\left(\bar{t}_1\right) \!\!+\!\! (1\!\!-\!\!\mathcal{P}_d^{01} \!\left(\bar{t}_1\right))(T_S^{01} \!\left(\bar{t}_1\right) \!\!+\! \! T_D^{11})\!\right] \!%\nonumber \\
f_{\tau_{\sf id}}(t_1)f_{\tau_{\sf ac}}(t_2) dt_1 dt_2
\eeqn
where $T_S^{01} \left(\bar{t}_1\right) = \bar{t}_1 \log_2 \left(1+\gamma_{S1}\right) + \left(T_S-\bar{t}_1\right) \log_2 \left(1+\gamma_{S2}\right)$, $T_D^{11} = 
 \varphi \left(T-T_S\right) \log_2 \left(1+\gamma_{D2}\right)$, $\bar{t}_1 = t_1 - T_{\sf ove}$, and $t_1$ is the same as in (\ref{eqT2}). Here, $T_S^{01}$ and $T_D^{11}$ denote the amount of data delivered
in the FD sensing and transmission stages for the underlying case, respectively. After some manipulations, we attain
%\beqn
%\mathcal{B}_3 = \mathcal{K}_e \int_{t =0}^{T_S} \left[T_S^{01} \left(t\right) + T_D^{11} -\mathcal{P}_d^{01} \left(t\right) T_D^{11}\right]  f_{\tau_{\sf id}}(t) \exp\left(\frac{t}{\bar{\tau}_{\sf ac}}\right) dt = \mathcal{B}_{31} + \mathcal{B}_{32},
%\eeqn
\beqn
\mathcal{B}_3 = \mathcal{K}_e \int_{t =0}^{T_S} \left[T_S^{01} \left(t\right) + T_D^{11} -\mathcal{P}_d^{01} \left(t\right) T_D^{11}\right] %\nonumber \\
f_{\tau_{\sf id}}(t) \exp\left(\frac{t}{\bar{\tau}_{\sf ac}}\right) dt = \mathcal{B}_{31} + \mathcal{B}_{32},
\eeqn
where
%\beqn
%\mathcal{B}_{31} = \mathcal{K}_e \int_{t =0}^{T_S} \left[T_S^{01} \left(t\right) + T_D^{11}\right] f_{\tau_{\sf id}}(t) \exp \left(\frac{t}{\bar{\tau}_{\sf ac}}\right) dt \hspace{8.6cm} \nonumber \\
%= \mathcal{K}_e \frac{\Delta\tau}{\bar{\tau}_{\sf id}} \!\left\{\!\Delta\tau \!\left[\!\left(\!\frac{T_S}{\Delta\tau} \!-\! 1\!\right) \exp\! \left(\!\frac{T_S}{\Delta\tau}\!\right) \!+\! 1\right] \!\log_2 \!\left(\!\frac{1\!+\!\gamma_{S1}}{1\!+\!\gamma_{S2}}\!\right)  \!+\! \left[\!\exp \!\left(\!\frac{T_S}{\Delta\tau}\!\right) \!-\! 1\!\right] \! \left[T_D^{11} \!+\! T_S \!\log_2 \!\left(1 \!+\! \gamma_{S2}\right) \right] \! \right\}
%\eeqn
\beqn
\mathcal{B}_{31} = \mathcal{K}_e \int_{t =0}^{T_S} \left[T_S^{01} \left(t\right) + T_D^{11}\right] f_{\tau_{\sf id}}(t) \exp \left(\frac{t}{\bar{\tau}_{\sf ac}}\right) dt \nonumber \hspace{1.5cm}\\
= \mathcal{K}_e \frac{\Delta\tau}{\bar{\tau}_{\sf id}} \!\left\{\!\Delta\tau \!\left[\!\left(\!\frac{T_S}{\Delta\tau} \!-\! 1\!\right) \exp\! \left(\!\frac{T_S}{\Delta\tau}\!\right) \!+\! 1\right] \log_2 \!\left(\!\frac{1\!+\!\gamma_{S1}}{1\!+\!\gamma_{S2}}\!\right) \right. \nonumber \\
\left. + \left[\exp \left(\frac{T_S}{\Delta\tau}\right) - 1\right] \left[T_D^{11} + T_S \log_2 \left(1+\gamma_{S2}\right) \right]\right\}, \hspace{0.6cm}
\eeqn
and
\beqn \label{T32bar}
\mathcal{B}_{32} = -\mathcal{K}_e T_D^{11} \bar{T}_{32},
\eeqn
where $\bar{T}_{32} = \int_{t =0}^{T_S} \mathcal{P}_d^{01} \left(t\right) f_{\tau_{\sf id}}(t) \exp \left(\frac{t}{\bar{\tau}_{\sf ac}}\right) dt$.

\subsection{False Alarm and Detection Probabilities}
\label{CAL_P_F_P_D}

We derive the detection and false alarm probabilities for FD sensing and two PU's state-changing events $h_{00}$ and $h_{01}$ in this appendix.
Assume that the transmitted signals from the PU and SU are circularly symmetric complex Gaussian (CSCG) signals while the noise at the secondary
receiver is independently and identically distributed CSCG $\mathcal{CN}\left( {0,{N_0}} \right)$ \cite{Liang08}. 
Under FD sensing, the  false alarm probability for event $h_{00}$ can be derived using the similar method as in \cite{Liang08},
which is given as
\beqn
\mathcal{P}_f^{00} = \mathcal{Q} \left[\left(\frac{\epsilon}{N_0+I(P_{\sf sen})}-1\right)\sqrt{f_sT_S}\right],
\eeqn
where $\mathcal{Q} \left(x\right) = \int_x^{+\infty} \exp \left(-t^2/2\right) dt$; $f_s$, $N_0$, $\epsilon$, $I(P_{\sf sen})$ are the sampling frequency, the noise power, the detection
threshold and the self-interference, respectively; $T_S$ is the FD sensing duration. 

%$\mathcal{P}_f^{00} = \mathcal{Q} \left[\left(\frac{\epsilon}{N_0+I}-1\right)\sqrt{f_sT_S}\right]$,

The  detection probability for event $h_{01}$ is given as
\beqn
\mathcal{P}_d^{01} \!\! =  \mathcal{Q} \left( \!\!\frac{\left(\!\! \frac{\epsilon}{N_0+I(P_{\sf sen})}- \frac{T_S-t}{T_S}\gamma_{PS}-1\right) 
 \!\sqrt{f_sT_S}}{\sqrt{\frac{T_S-t}{T_S}\left(\gamma_{PS}+1\right)^2+\frac{t}{T_S}}} \!\! \right), 
\eeqn
where $t$ is the interval from the beginning of the data phase to the instant when the PU changes its state,
 $\gamma_{PS} = \frac{P_p}{N_0+I(P_{\sf sen})}$ is the signal-to-interference-plus-noise ratio (SINR) of the PU's signal at the SU. 

\subsection{Proof of Proposition 1}
\label{Prosp1}

The first derivative of $\mathcal{NT}$ can be written as follows:
\beqn
\frac{\partial \mathcal{NT}}{\partial T_S} = \frac{1}{T_{\sf ove}+T} \sum_{i=1}^3 \frac{\partial \mathcal{B}_i}{\partial T_S}.
\eeqn
We derive the first derivative of $\mathcal{B}_i$ ($i = 1, 2, 3$) in the following. Toward this end, we will employ the approximation of $\exp \left(x\right) \approx 1 + x$, $x = \frac{T_x}{\tau_x}$, $T_x \in \left\{T, T_S, T-T_S\right\}$, $\tau_x \in \left\{{\bar \tau}_{\sf id}, {\bar \tau}_{\sf ac}, \Delta \tau\right\}$ where recall that $\frac{1}{\Delta\tau} = \frac{1}{\bar{\tau}_{\sf ac}} - \frac{1}{\bar{\tau}_{\sf id}}$. This approximation holds under the assumption that $T_x << \tau_x$ since we can omit all higher-power terms $x^n$ for 
$n > 1$ from the Maclaurin series expansion
of function $\exp \left(x\right)$. Using this approximation, we can express the first derivative of $\mathcal{B}_1$ as
%\beqn
%\frac{\partial \mathcal{B}_1}{\partial T_S} = \mathcal{K}_e \exp \left(\frac{T}{\Delta \tau}\right) \left\{\log_2 \left(1+\gamma_{S1}\right)  - \varphi\left[\left(T-T_S\right)\frac{\partial \mathcal{P}_f^{00}}{\partial T_S} + \left(1 - \mathcal{P}_f^{00}\right)\right]\log_2 \left(1+\gamma_{D1}\right) \right\},
%\eeqn
\beqn
\frac{\partial \mathcal{B}_1}{\partial T_S} = \mathcal{K}_e \exp \left(\frac{T}{\Delta \tau}\right) \left\{\log_2 \left(1\!+\!\gamma_{S1}\right) %\right. \hspace{2.5cm} \nonumber\\
\!- \!\varphi\left[\left(T-T_S\right)\frac{\partial \mathcal{P}_f^{00}}{\partial T_S} \!+\! \left(1 \!- \!\mathcal{P}_f^{00}\right)\right]\log_2 \left(1\!+\!\gamma_{D1}\right) \right\}
\eeqn
where $\frac{\partial \mathcal{P}_f^{00}}{\partial T_S}$ is the first derivative of $\mathcal{P}_f^{00}$ whose derivation is given in Appendix \ref{APPROX_PF}.

Moreover, the first derivative of $\mathcal{B}_2$ can be written as
\beqn
\frac{\partial \mathcal{B}_2}{\partial T_S} &=& \mathcal{K}_e \frac{\Delta \tau}{{\bar \tau}_{\sf id}} 
\left\{\!\left[\exp \left(\frac{T}{\Delta \tau}\right) - \left(1+\frac{T}{\Delta \tau}\right) \exp \left(\frac{T_S}{\Delta \tau}\right)\right] \log_2 \left(1+\gamma_{S1}\right) \right. \hspace{3cm} \nonumber \\ % \hspace{5cm}\\
&& - \varphi\frac{\partial \mathcal{P}_f^{00}}{\partial T_S} \!\left[\Delta \tau \!\left(\!\exp\! \left(\!\frac{T_S}{\Delta \tau}\!\right)\! - \!\exp \!\left(\!\frac{T}{\Delta \tau}\!\right)\!\right) \!\log_2\! \left(\!\frac{1\!+\!\gamma_{D1}}{1\!+\!\gamma_{D2}}\!\right) \right. \nonumber \\  % \hspace{5.1cm}\\
&& \left. + \left(T\!\!-\!\!T_S\right) \!\!\left(\!\!\exp \!\left(\!\!\frac{T}{\Delta \tau}\!\!\right) \!\log_2 \!\left(1\!+\!\gamma_{D1}\right) \!-\! \exp \!\left(\!\!\frac{T_S}{\Delta \tau}\!\!\right) \!\log_2 \!\left(1\!+\!\gamma_{D2}\right) \!\!\right) \!\!\right] \nonumber \\  %\hspace{2.5cm}\\
&& + \varphi \left(1 - \mathcal{P}_f^{00}\right) \left[-\frac{T-T_S}{\Delta \tau} \exp \left(\frac{T_S}{\Delta \tau}\right) \log_2 \left(1+\gamma_{D2}\right) + \exp \left(\frac{T_S}{\Delta \tau}\right) \log_2 \left(\frac{1+\gamma_{D1}}{1+\gamma_{D2}}\right) \right. \nonumber \\  % \\
&& \left.\left. - \left(\exp \!\left(\!\frac{T}{\Delta \tau}\!\right) \log_2 \!\left(1\!+\!\gamma_{D1}\right) \!-\! \exp \left(\!\frac{T_S}{\Delta \tau}\!\right) \log_2 \!\left(\!1\!+\!\gamma_{D2}\!\right)\right) \right] \right\}. %\hspace{1.9cm}
\eeqn

Finally, the first derivative of $\mathcal{B}_3$ can be written as
\beqn
\frac{\partial \mathcal{B}_3}{\partial T_S} = \frac{\partial \mathcal{B}_{31}}{\partial T_S} + \frac{\partial \mathcal{B}_{32}}{\partial T_S},
\eeqn
where
%\beqn
%\frac{\partial \mathcal{B}_{31}}{\partial T_S} =\mathcal{K}_e \frac{\Delta \tau}{{\bar \tau}_{\sf id}} \left\{\Delta \tau \left[1+\left(\frac{T_S}{\Delta \tau}-1\right) \exp \left(\frac{T_S}{\Delta \tau}\right)\right] \log_2 \left(\frac{1+\gamma_{S1}}{1+\gamma_{S2}}\right) \right. \hspace{2.0cm} \nonumber \\
%\left.+\left(\exp \left(\frac{T_S}{\Delta \tau}\right)-1\right) \left[T_S \log_2 \left(1+\gamma_{S2}\right) +\varphi\left(T-T_S\right) \log_2 \left(1+\gamma_{D2}\right)\right]\right\}.
%\eeqn
\beqn
\frac{\partial \mathcal{B}_{31}}{\partial T_S} = \mathcal{K}_e \frac{\Delta \tau}{{\bar \tau}_{\sf id}} \left\{\Delta \tau \left[1+\left(\frac{T_S}{\Delta \tau}-1\right) \exp \left(\frac{T_S}{\Delta \tau}\right)\right] \log_2 \left(\frac{1+\gamma_{S1}}{1+\gamma_{S2}}\right) \right. \hspace{2cm}\nonumber\\
\left.+\left(\exp \left(\frac{T_S}{\Delta \tau}\right)-1\right) \left[T_S \log_2 \left(1+\gamma_{S2}\right) +\varphi\left(T-T_S\right) \log_2 \left(1+\gamma_{D2}\right)\right]\right\}. 
\eeqn
To obtain the derivative for $\mathcal{B}_{32}$, we note that $1 \leq \exp \left(\frac{t}{\bar{\tau}_{\sf ac}}\right) \leq \exp \left(\frac{T_S}{\bar{\tau}_{\sf ac}}\right)$ for $\forall t \in \left[0, T_S\right]$. Moreover,
from the results in (\ref{P_average}) and (\ref{cond_pdf_tau_id}) and using the definition of $\bar{T}_{32}$ in (\ref{T32bar}), we have $\mathcal{\overline P}_d \left(1-\exp\left(\frac{-T_S}{{\bar \tau}_{\sf id}}\right)\right) \leq \bar{T}_{32} \leq \mathcal{\overline P}_d \left(1-\exp\left(\frac{-T_S}{{\bar \tau}_{\sf id}}\right)\right) \exp \left(\frac{T_S}{\bar{\tau}_{\sf ac}}\right)$.
Using these results, the first derivative of $\mathcal{B}_{32}$ can be expressed as
\beqn
\frac{\partial \mathcal{B}_{32}}{\partial T_S} = - \mathcal{K}_e \mathcal{\overline P}_d \varphi\frac{T-2T_S}{{\bar \tau}_{\sf id}} \log_2 \left(1+\gamma_{D2}\right).
\eeqn

Therefore, we have obtained the first derivative of  $\mathcal{NT}$ and we are ready to prove the first statement of Theorem 1.
Substitute $T_S = 0$ to the derived $\frac{\partial \mathcal{NT}}{\partial T_S}$ and use the approximation $\exp \left(x\right) \approx 1 + x$, 
 we yield the following result after some manipulations
\beqn
\mathop {\lim } \limits_{T_S \to 0} \frac{\partial \mathcal{NT}}{\partial T_S} = - K_0 K_1 \mathop {\lim } \limits_{T_S  \to 0} \frac{\partial \mathcal{P}_f^{00}}{\partial T_S},
\eeqn
where $K_0 = \frac{1}{T_{\sf ove}+T} \mathcal{K}_e$ and
%\beqn
%K_1 = \varphi \left[T \left(1+\frac{T}{\Delta \tau}\right) + \frac{T^2}{{\bar \tau}_{\sf id}}\right] \log_2 \left(1+\gamma_{D1}\right) 
%+ \varphi \frac{T\Delta \tau}{{\bar \tau}_{\sf id}} \log_2 \left(1+\gamma_{D2}\right).
%\eeqn
\beqn
K_1 = \varphi \left[T \left(1+\frac{T}{\Delta \tau}\right) + \frac{T^2}{{\bar \tau}_{\sf id}}\right] \log_2 \left(1+\gamma_{D1}\right) % \hspace{1.2cm} \nonumber\\
+ \varphi \frac{T\Delta \tau}{{\bar \tau}_{\sf id}} \log_2 \left(1+\gamma_{D2}\right).
\eeqn
It can be verified that $K_0 > 0$, $K_1 > 0$ and $\mathop {\lim } \limits_{T_S  \to 0} \frac{\partial \mathcal{P}_f^{00}}{\partial T_S} = - \infty$ by using the derivations in Appendix \ref{APPROX_PF};
 hence, we have $\mathop {\lim } \limits_{T_S  \to 0} \frac{\partial \mathcal{NT}}{\partial T_S} = +\infty >0$. This completes the proof of the first statement of the theorem.

We now present the proof for the second statement of the theorem.
Substitute $T_S = T$ to $\frac{\partial \mathcal{NT}}{\partial T_S}$ and utilize the approximation $\exp \left(x\right) \approx 1 + x$, we yield 
\beqn
\mathop {\lim } \limits_{T_S  \to T} \frac{\partial \mathcal{NT}}{\partial T_S} = \frac{1}{T_{\sf ove}+T} \sum_{i=1}^3 \frac{\partial \mathcal{B}_i}{\partial T_S} (T),
\eeqn
where we have
\beqn
%\frac{\partial \mathcal{B}_1}{\partial T_S} (T) &=& \mathcal{K}_e \left(1+\frac{T}{\Delta \tau}\right) 
%\left[\log_2\left(1+\gamma_{S1}\right) - \varphi \left(1-\mathcal{P}_f^{00} (T)\right) \log_2 \left(1+\!\gamma_{D1}\right)\right] \label{deriT1}  \\
%\eeqn
%\beqn
\frac{\partial \mathcal{B}_1}{\partial T_S} (T) &= &\mathcal{K}_e \left(1+\frac{T}{\Delta \tau}\right) %\times \hspace{3.9cm} \nonumber\\
\left[\log_2\left(1+\gamma_{S1}\right) - \varphi \left(1-\mathcal{P}_f^{00} (T)\right) \log_2 \left(1+\!\gamma_{D1}\right)\right] \label{deriT1}\\
%\eeqn
%\beqn
\frac{\partial \mathcal{B}_2}{\partial T_S} (T) &= &- \mathcal{K}_e \frac{T}{{\bar \tau}_{\sf id}} \log_2 \left(1+\gamma_{S1}\right)  \hspace{3.2cm}\\
%\eeqn
%\beqn
\frac{\partial \mathcal{B}_{31}}{\partial T_S} (T) &=& \mathcal{K}_e \! \frac{T}{{\bar \tau}_{\sf id}} \! \left[\log_2 \!\left(\!1\!+\!\gamma_{S1}\!\right)\left(\!1\!+\!\gamma_{S2}\!\right)\!-\! \varphi \! \log_2 \! \left(\!1\!+\!\gamma_{D2}\!\right)\right]  \\
%\eeqn
%\beqn
\frac{\partial \mathcal{B}_{32}}{\partial T_S} (T) &=& \mathcal{K}_e \varphi \frac{T}{{\bar \tau}_{\sf id}} \mathcal{\overline P}_d \log_2 \left(1+\gamma_{D2}\right).  \hspace{2.6cm} \nonumber
\eeqn

Omit all high-power terms in the expansion of $\exp(x)$ (i.e., $x^n$ with $n>1$) where $x = \frac{T_x}{\tau_x}$, $T_x \in \left\{T, T_S, T-T_S\right\}$, $\tau_x \in \left\{{\bar \tau}_{\sf id}, {\bar \tau}_{\sf ac}, \Delta \tau\right\}$, we yield
\beqn \label{ThpT0}
\mathop {\lim } \limits_{T_S  \to T} \frac{\partial \mathcal{NT}}{\partial T_S} \approx \frac{1}{T_{\sf ove}+T}  \frac{\partial \mathcal{B}_1}{\partial T_S} (T). 
\eeqn

We consider the HDTx and FDTx modes in the following. For the HDTx mode, we have $\varphi = 1$ and $\theta = 0$. 
Then, it can be verified that $\mathop {\lim } \limits_{T_S  \to T} \frac{\partial \mathcal{NT}}{\partial T_S} < 0$ by using the results in (\ref{deriT1}) and (\ref{ThpT0}).
This is because we have $\log_2\left(1+\gamma_{S1}\right) - \left(1-\mathcal{P}_f^{00} (T)\right) \log_2 \left(1+\gamma_{D1}\right) \approx \log_2\left(1+\gamma_{S1}\right) - \log_2 \left(1+\gamma_{D1}\right) < 0$ (since
we have $\mathcal{P}_f^{00} (T) \approx 0$ and $\gamma_{S1} \leq \gamma_{D1}$).

For the FDTx mode, we have $\varphi = 2$, $\theta = 1$, and
also $\gamma_{S1} = \frac{P_{\sf sen}}{N_0}$ and $\gamma_{D1} = \frac{P_{\sf dat}}{N_0+I(P_{\sf dat})} = \frac{P_{\sf dat}}{N_0+ \zeta P_{\sf dat}^\xi}$.
We would like to define a critical value of ${P}_{\sf sen}$ which satisfies $\mathop {\lim } \limits_{T_S  \to T} \frac{\partial \mathcal{NT}}{\partial T_S} = 0$ 
to proceed further. Using the result in (\ref{deriT1}) and (\ref{ThpT0}) as well as the approximation $\mathcal{P}_f^{00} (T) \approx 0$, and
by solving $\mathop {\lim } \limits_{T_S  \to T} \frac{\partial \mathcal{NT}}{\partial T_S} = 0$ we yield
\beqn
\label{P_sen_threshold}
\overline{P}_{\sf sen} =  N_0 \left[\left(1+\frac{P_{\sf dat}}{N_0+\zeta P_{\sf dat}^\xi}\right)^2-1\right].
\eeqn
Using (\ref{deriT1}), it can be verified that if $P_{\sf sen} > \overline{P}_{\sf sen}$ then  $\mathop {\lim } \limits_{T_S  \to T} \frac{\partial \mathcal{NT}}{\partial T_S} > 0$;
otherwise, we have $\mathop {\lim } \limits_{T_S  \to T} \frac{\partial \mathcal{NT}}{\partial T_S} \leq 0$. So we have completed the proof for the second statement
of Theorem 1.

To prove the third statement of the theorem, we derive the second derivative of $\mathcal{NT}$ as
 \beq
\frac{\partial^2 \mathcal{NT}}{\partial T_S^2} = \frac{1}{T_{\sf ove}+T} 
\sum_{i=1}^3 \frac{\partial^2 \mathcal{B}_i}{\partial T_S^2},
\eeq
where we have 
%\beqn \label{2derT1}
%\frac{\partial^2 \mathcal{B}_1}{\partial T_S^2} = - \mathcal{K}_e\varphi \exp \left(\frac{T}{\Delta \tau}\right)  \log_2\left(1+\gamma_{D1}\right) 
%\left[\left(T-T_S\right) \frac{\partial^2 \mathcal{P}_f^{00}}{\partial T_S^2} - 2\frac{\partial \mathcal{P}_f^{00}}{\partial T_S}\right],
%\eeqn
\beqn \label{2derT1}
\frac{\partial^2 \mathcal{B}_1}{\partial T_S^2} = - \mathcal{K}_e\varphi \exp \left(\frac{T}{\Delta \tau}\right)  \log_2\left(1+\gamma_{D1}\right) %\times \hspace{1.5cm} \nonumber\\
\left[\left(T-T_S\right) \frac{\partial^2 \mathcal{P}_f^{00}}{\partial T_S^2} - 2\frac{\partial \mathcal{P}_f^{00}}{\partial T_S}\right],
\eeqn
where $\frac{\partial^2 \mathcal{P}_f^{00}}{\partial T_S^2}$ is the second derivative of $\mathcal{P}_f^{00}$ and according to the derivations in Appendix~\ref{APPROX_PF}, we have 
$\frac{\partial^2 \mathcal{P}_f^{00}}{\partial T_S^2} > 0$, $\frac{\partial \mathcal{P}_f^{00}}{\partial T_S} < 0$, $\forall T_S$.
Therefore, we yield $\frac{\partial^2 \mathcal{B}_1}{\partial T_S^2} < 0$ $\forall T_S$.

Consequently, we have the following upper bound for $\frac{\partial^2 \mathcal{B}_1}{\partial T_S^2}$ by omitting the term $\exp \left(\frac{T}{\Delta \tau}\right) > 1$ in (\ref{2derT1})
\beqn
\frac{\partial^2 \mathcal{B}_1}{\partial T_S^2} \leq \mathcal{K}_e \left[h_1(T_S) + h_2(T_S) \right],
\eeqn
where 
\beqn
h_1(T_S) &=& - \varphi \left(T-T_S\right) \frac{\partial^2 \mathcal{P}_f^{00}}{\partial T_S^2} \log_2\left(1+\gamma_{D1}\right), \nonumber \\
%\eeqn
%\beqn
h_2(T_S) &=& 2 \varphi \frac{\partial \mathcal{P}_f^{00}}{\partial T_S} \log_2\left(1+\gamma_{D1}\right). \nonumber
\eeqn
Moreover, we have
\beqn
\!\!\!\!\!\!\!\!\!\! \frac{\partial^2 \mathcal{B}_2}{\partial T_S^2} \!\!\!\! &=& \!\!\!\! \frac{\mathcal{K}_e \Delta \tau} {{\bar \tau}_{\sf id}}\!\! \left\{\!-\frac{2+\frac{T_S}{\Delta \tau}} {\Delta \tau} \exp \left(\frac{T_S}{\Delta \tau}\right)\log_2 \left(1+\gamma_{S1}\right)  \right. \nonumber  \hspace{5.0cm}\\
&-&\!\!\!\! \varphi\frac{\partial^2 \mathcal{P}_f^{00}}{\partial T_S^2} \left[\!\Delta \tau \left(\!\exp\! \left(\!\frac{T_S}{\Delta \tau}\!\right)\! - \!\exp \left(\!\frac{T}{\Delta \tau}\!\right)\!\right) \!\log_2\! \left(\!\frac{1+\gamma_{D1}}{1+\gamma_{D2}}\!\right) \right. \nonumber \\ % \hspace{4.9cm}\\
&+&\!\!\!\! \left.\left(T\!\!-\!\!T_S\right) \!\!\left(\!\!\exp \!\left(\!\!\frac{T}{\Delta \tau}\!\!\right) \!\log_2 \!\left(1\!+\!\gamma_{D1}\right) \!-\! \exp \!\left(\!\!\frac{T_S}{\Delta \tau}\!\!\right) \!\log_2 \!\left(1\!+\!\gamma_{D2}\right) \!\!\right) \!\!\right] \nonumber \\ % \hspace{2.5cm}\\
&+&\!\!\!\!  2\varphi\frac{\partial \mathcal{P}_f^{00}}{\partial T_S} \left[\left(\!\exp\! \left(\!\frac{T}{\Delta \tau}\!\right)\! - \!\exp \left(\!\frac{T_S}{\Delta \tau}\!\right)\!\right) \!\log_2 \!\left(1\!+\!\gamma_{D1}\right) 
\!+ \! \frac{T\!\!-\!\!T_S}{\Delta \tau} \exp \!\left(\!\!\frac{T_S}{\Delta \tau}\!\!\right) \!\log_2 \!\left(1\!+\!\gamma_{D2}\right)\right] \nonumber \\ %\hspace{0.2cm}\\
&-&\!\!\!\! \left.\varphi\left(\!1 \!-\! \mathcal{P}_f^{00}\right) \frac{T\!\!-\!\!T_S}{\Delta \tau} \exp \!\left(\!\!\frac{T_S}{\Delta \tau}\!\!\right) \log_2 \!\left(1\!+\!\gamma_{D2}\right) \!+\! \varphi\left(1 \!-\! \mathcal{P}_f^{00}\right) \!\frac{1}{\Delta \tau} \!\exp \!\left(\!\!\frac{T_S}{\Delta \tau}\!\!\right) \!\log_2\! \left(\!\frac{1\!+\!\gamma_{D1}}{1\!+\!\gamma_{D2}}\!\right)\!\right\}.
\eeqn
%%\beqn
%%\frac{\partial^2 \mathcal{B}_2}{\partial T_S^2} = \frac{\mathcal{K}_e \Delta \tau} {{\bar \tau}_{\sf id}} \left\{-\frac{2+\frac{T_S}{\Delta \tau}} {\Delta \tau} \exp \left(\frac{T_S}{\Delta \tau}\right)\log_2 \left(1+\gamma_{S1}\right)  \right.\\
%%-\varphi\frac{\partial^2 \mathcal{P}_f^{00}}{\partial T_S^2} \left[\!\Delta \tau \left(\!\exp\! \left(\!\frac{T_S}{\Delta \tau}\!\right)\! - \!\exp \left(\!\frac{T}{\Delta \tau}\!\right)\!\right) \!\log_2\! \left(\!\frac{1+\gamma_{D1}}{1+\gamma_{D2}}\!\right) +\right. \\
%%\left.\left(T\!\!-\!\!T_S\right) \!\!\left(\!\!\exp \!\left(\!\!\frac{T}{\Delta \tau}\!\!\right) \!\log_2 \!\left(1\!+\!\gamma_{D1}\right) \!-\! \exp \!\left(\!\!\frac{T_S}{\Delta \tau}\!\!\right) \!\log_2 \!\left(1\!+\!\gamma_{D2}\right) \!\!\right) \!\!\right] \\
%%+ 2\varphi\frac{\partial \mathcal{P}_f^{00}}{\partial T_S} \left[\left(\!\exp\! \left(\!\frac{T}{\Delta \tau}\!\right)\! - \!\exp \left(\!\frac{T_S}{\Delta \tau}\!\right)\!\right)\log_2 \!\left(1\!+\!\gamma_{D1}\right) \right. \\
%%\left.+\frac{T\!\!-\!\!T_S}{\Delta \tau} \exp \!\left(\!\!\frac{T_S}{\Delta \tau}\!\!\right) \log_2 \!\left(1\!+\!\gamma_{D2}\right)\right] \\
%%-\varphi\left(1 - \mathcal{P}_f^{00}\right) \frac{T\!\!-\!\!T_S}{\Delta \tau} \exp \!\left(\!\!\frac{T_S}{\Delta \tau}\!\!\right) \log_2 \!\left(1\!+\!\gamma_{D2}\right) \\
%%\left.+\varphi\left(1 - \mathcal{P}_f^{00}\right) \frac{1}{\Delta \tau} \exp \!\left(\!\!\frac{T_S}{\Delta \tau}\!\!\right) \log_2\! \left(\!\frac{1+\gamma_{D1}}{1+\gamma_{D2}}\!\right)\right\}
%%\eeqn
Therefore, we can approximate $\frac{\partial^2 \mathcal{B}_2}{\partial T_S^2}$ as follows:
\beqn
\frac{\partial^2 \mathcal{B}_2}{\partial T_S^2} = \mathcal{K}_e \left[h_3(T_S) + h_4(T_S) + h_5(T_S)\right],
\eeqn
where 
\beqn
h_3(T_S) &=& -\frac{\left(2+\frac{T_S}{\Delta \tau}\right)\left(1+\frac{T_S}{\Delta \tau}\right)}{{\bar \tau}_{\sf id}}  \log_2 \left(1+\gamma_{S1}\right), \nonumber \\
%\eeqn
%\beqn
h_4(T_S) &=& -\varphi\left(1 - \mathcal{P}_f^{00}\right) \frac{T\!\!-\!\!T_S}{{\bar \tau}_{\sf id}} \!\left(1+\!\frac{T_S}{\Delta \tau}\!\!\right) \log_2 \!\left(1\!+\!\gamma_{D2}\right) \hspace{4cm} \nonumber \\
&& - \varphi\frac{\partial^2 \mathcal{P}_f^{00}}{\partial T_S^2} \left(\!T\! - \!T_S\!\right) \left[\!\frac{T}{{\bar \tau}_{\sf id}} \!\log_2\! \left(\!1+\!\gamma_{D1}\!\right)\! -\! \frac{T_S}{{\bar \tau}_{\sf id}} \!\log_2\! \left(\!1+\!\gamma_{D2}\!\right) \right] \hspace{1cm} \nonumber \\
&&  + 2\varphi\frac{\partial \mathcal{P}_f^{00}}{\partial T_S} \frac{T-T_S}{{\bar \tau}_{\sf id}} \!\!\left[\log_2\! \left(\!\frac{1\!+\!\gamma_{D1}}{1\!+\!\gamma_{D2}}\!\right) \!+\! \frac{T_S}{\Delta \tau} \log_2 \!\left(\!1\!+\!\gamma_{D2}\!\right) \right], \nonumber \\
%\eeqn
%\beqn
%h_4(T_S) = -\varphi\left(1 - \mathcal{P}_f^{00}\right) \frac{T\!\!-\!\!T_S}{{\bar \tau}_{\sf id}} \!\left(1+\!\frac{T_S}{\Delta \tau}\!\!\right) \log_2 \!\left(1\!+\!\gamma_{D2}\right)\\
%- \varphi\frac{\partial^2 \mathcal{P}_f^{00}}{\partial T_S^2} \left(\!T\! - \!T_S\!\right) \left[\!\frac{T}{{\bar \tau}_{\sf id}} \!\log_2\! \left(\!1+\!\gamma_{D1}\!\right)\! -\! \frac{T_S}{{\bar \tau}_{\sf id}} \!\log_2\! \left(\!1+\!\gamma_{D2}\!\right) \right]\\
%+ 2\varphi\frac{\partial \mathcal{P}_f^{00}}{\partial T_S} \frac{T-T_S}{{\bar \tau}_{\sf id}} \!\!\left[\log_2\! \left(\!\frac{1\!+\!\gamma_{D1}}{1\!+\!\gamma_{D2}}\!\right) \!+\! \frac{T_S}{\Delta \tau} \log_2 \!\left(\!1\!+\!\gamma_{D2}\!\right) \right]
%\eeqn
%\beqn
h_5(T_S) &=& \varphi\left(1 - \mathcal{P}_f^{00}\right) \frac{1}{{\bar \tau}_{\sf id}} \!\left(1+\!\frac{T_S}{\Delta \tau}\!\!\right) \log_2\! \left(\!\frac{1+\gamma_{D1}}{1+\gamma_{D2}}\!\right). \nonumber 
\eeqn
In addition, we have 
%\beqn
%\frac{\partial^2 \mathcal{B}_{31}}{\partial T_S^2} =  \frac{\mathcal{K}_e}{{\bar \tau}_{\sf id}} \exp \!\left(\!\frac{T_S}{\Delta \tau}\!\right) \left\{\!\left(\!1 \!+\! \frac{T}{\Delta \tau}\!\right)  \!\log_2 \!\left(1 \!+\! \gamma_{S1}\right) 
%\!+\! \log_2 \!\left(1 \!+\! \gamma_{S2}\right) \!+\! \varphi\left(\!\frac{T-T_S}{\Delta \tau}\!-\! 2\right) \!\log_2 \left(1 \!+\! \gamma_{D2}\right)\right\}.
%\eeqn
\beqn
\frac{\partial^2 \mathcal{B}_{31}}{\partial T_S^2} =  \frac{\mathcal{K}_e}{{\bar \tau}_{\sf id}} \exp \!\left(\!\!\frac{T_S}{\Delta \tau}\!\!\right) \left\{\left(1 + \frac{T}{\Delta \tau}\right)  \log_2 \left(1+\gamma_{S1}\right) \right. \hspace{0.1cm} \nonumber\\
\left.+ \log_2 \left(1+\gamma_{S2}\right) + \varphi\left(\!\frac{T-T_S}{\Delta \tau}\!-2\right)\log_2 \left(1+\gamma_{D2}\right)\right\}.
\eeqn
We can approximate $\frac{\partial^2 \mathcal{B}_{31}}{\partial T_S^2}$ as follows:
\beqn
\frac{\partial^2 \mathcal{B}_{31}}{\partial T_S^2} = \mathcal{K}_e \left[h_6(T_S) + h_6(T_S)\right],
\eeqn
where
\beqn
h_6(T_S) &=& \frac{1}{{\bar \tau}_{\sf id}} \log_2 \left(1\!+\!\gamma_{S1}\right) \left(1\!+\!\gamma_{S2}\right), \nonumber \\
%\eeqn
%\beqn
h_7(T_S) &=& - \frac{2\varphi}{{\bar \tau}_{\sf id}}\log_2 \left(1\!+\!\gamma_{D2}\right).
\eeqn
Finally, we have
\beqn
\frac{\partial^2 \mathcal{B}_{32}}{\partial T_S^2} = \mathcal{K}_e h_8(T_S),
\eeqn
where 
\beqn
h_8(T_S) = \frac{2 \varphi \mathcal{\bar P}_d}{{\bar \tau}_{\sf id}} \log_2 \left(1\!+\!\gamma_{D2}\right).
\eeqn
The above analysis yields $\frac{\partial^2 \mathcal{NT}}{\partial T_S^2} = \mathcal{K}_e \sum_{i=1}^8 h_i(T_S)$. 
Therefore, to prove that $\frac{\partial^2 \mathcal{NT}}{\partial T_S^2} < 0$, we should prove that $h(T_S) < 0$ since $\mathcal{K}_e >0$ where
\beqn
h(T_S) = \sum_{i=1}^8 h_i(T_S).
\eeqn
It can be verified that $h_1(T_S) < 0$ and $h_4(T_S) < 0$, $\forall T_S$ because $\frac{\partial^2 \mathcal{P}_f^{00}}{\partial T_S^2} > 0$, $\frac{\partial \mathcal{P}_f^{00}}{\partial T_S} < 0$
according to Appendix \ref{APPROX_PF} and $\gamma_{D2} < \gamma_{D1}$. Moreover, we have
\beqn
h_3(T_S) < -\frac{2}{{\bar \tau}_{\sf id}}  \log_2 \left(1+\gamma_{S1}\right),
\eeqn
and because $\gamma_{S1} > \gamma_{S2}$, we have
\beqn
h_3(T_S) < -\frac{1}{{\bar \tau}_{\sf id}}  \log_2 \left(1+\gamma_{S1}\right)\left(1+\gamma_{S2}\right) = -h_6(T_S).
\eeqn
Therefore, we have $h_3(T_S) + h_6(T_S) < 0$.
Furthermore, we can also obtain the following result $h_7(T_S) + h_8(T_S) \leq 0$ because $\mathcal{\bar P}_d \leq 1$.
To complete the proof, we must prove that $h_2 (T_S) + h_5 (T_S) \leq 0$, which is equivalent to
\beqn
\label{EQN_LAST}
-2 {\bar \tau}_{\sf id} \frac{\partial \mathcal{P}_f^{00}}{\partial T_S} \frac{\log_2\left(1+\gamma_{D1}\right)}{\log_2\! \left(\!\frac{1+\gamma_{D1}}{1+\gamma_{D2}}\!\right)}  \geq \left(1 - \mathcal{P}_f^{00}\right)  \!\left(1+\!\frac{T_S}{\Delta \tau}\!\!\right),
\eeqn
where according to Appendix \ref{APPROX_PF}
\beqn
\frac{\partial \mathcal{P}_f^{00}}{\partial T_S} = - \frac{\bar{\gamma}\sqrt{f_s T_S}}{2\sqrt{2\pi}T_S}  \exp \left(-\frac{\left(\bar{\alpha}+\bar{\gamma} \sqrt{f_s T_S}\right)^2}{2} \right),
\eeqn
where $\bar{\alpha} = \left(\bar{\gamma}_1+1\right)\mathcal{Q}^{-1}\left(\overline{\mathcal{P}}_d\right)$. 
It can be verified that (\ref{EQN_LAST}) indeed holds because the LHS of (\ref{EQN_LAST}) is always larger than to 2 while the RHS of (\ref{EQN_LAST}) is always less than 2. 
Hence, we have completed the proof of the third statement of Theorem 1.

Finally,  the  fourth statement in the theorem obviously holds because $\mathcal{B}_i$ ($i$ = 1, 2, 3) are all bounded from above.
Hence, we have completed the proof of Theorem 1.

\subsection{Approximation of $\mathcal{P}_f^{00}$ and Its First and Second Derivatives}
\label{APPROX_PF}

We can approximate $\mathcal{\hat P}_d$ in (\ref{P_average}) as follows:
\beqn
\mathcal{\hat P}_d = \mathcal{Q} \left[\left(\frac{\epsilon}{N_0+I}-\bar{\gamma}-1\right)\frac{\sqrt{f_sT_S}}{\bar{\gamma}_1+1}\right],
\eeqn
where $\bar{\gamma}$ and $\bar{\gamma}_1$ are evaluated by a numerical method.
Hence, $\mathcal{P}_f^{00}$ can be calculated as we set $\mathcal{\hat P}_d = \overline{\mathcal{P}}_d$, which is given as follows:
\beqn
\mathcal{P}_f^{00} = \mathcal{Q} \left(\bar{\alpha}+\bar{\gamma}\sqrt{f_sT_S}\right),
\eeqn
where $\bar{\alpha} = \left(\bar{\gamma}_1+1\right)\mathcal{Q}^{-1}\left(\overline{\mathcal{P}}_d\right)$.

We now derive the first derivative of $\mathcal{P}_f^{00}$ as
\beqn
\frac{\partial \mathcal{P}_f^{00}}{\partial T_S} = - \frac{\bar{\gamma}\sqrt{f_s T_S}}{2\sqrt{2\pi}T_S}  \exp \left(-\frac{\left(\bar{\alpha}+\bar{\gamma} \sqrt{f_s T_S}\right)^2}{2} \right).
\eeqn
It can be seen that $\frac{\partial \mathcal{P}_f^{00}}{\partial T_S} <0$ since $\bar{\gamma} > 0$.
Moreover, the second derivative of $\mathcal{P}_f^{00}$ is
\beqn
\frac{\partial^2 \mathcal{P}_f^{00}}{\partial T_S^2} = \frac{\bar{\gamma}\sqrt{f_s T_S}}{4\sqrt{2\pi}T_S^2}  \left(1+\frac{1}{2}y\bar{\gamma} \sqrt{f_s T_S} \right)\exp \left(-\frac{y^2}{2} \right),
\eeqn
where $y=\bar{\alpha}+\bar{\gamma} \sqrt{f_s T_S}$.

We can prove that $\frac{\partial^2 \mathcal{P}_f^{00}}{\partial T_S^2} > 0$ by considering two different cases as follows. 
For the first case with $\frac{\bar{\alpha}^2}{\bar{\gamma}^2f_s} \leq T_S \leq T$ ($0 \leq \mathcal{P}_f^{00} \leq 0.5$), this statement holds since $y > 0$.
For the second case with $0 \leq T_S \leq \frac{\bar{\alpha}^2}{\bar{\gamma}^2f_s}$ ($0.5 \leq \mathcal{P}_f^{00} \leq 1$), $y \leq 0$, then we have
 $0< y-\bar{\alpha} = \bar{\gamma} \sqrt{f_s T_S} \leq -\bar{\alpha}$ and $0  \leq -y \leq -\bar{\alpha}$.
By applying the Cauchy-Schwarz inequality, we obtain $0 \leq -y (y-\bar{\alpha}) \leq \frac{\bar{\gamma}^2}{4} < 1 <2$; hence $1+\frac{1}{2}y\bar{\gamma} \sqrt{f_s T_S} > 0$. 
This result implies that $\frac{\partial^2 \mathcal{P}_f^{00}}{\partial T_S^2} > 0$.

\chapter{Conclusions and Further Works} % top level followed by section, subsection
\zlabel{Chapter7}

%: ----------------------- paths to graphics ------------------------

% change according to folder and file names
\ifpdf
    \graphicspath{{7/figures/PNG/}{7/figures/PDF/}{7/figures/}}
\else
    \graphicspath{{7/figures/EPS/}{7/figures/}}
\fi

%: ----------------------- contents from here ------------------------

Although cognitive radio technology is an important paradigm shift to solve the spectrum scarcity problem for future wireless networks, 
many challenges remain to be resolved to achieve the benefits offered by this technology.  Our dissertation focuses 
on the design, analysis, and optimization of joint spectrum
sensing and access design for CRNs under different practically relevant network settings. 
The developed techniques enable a CRN to efficiently exploit idle spectrum over time, 
frequency, and space for data transmission.
In this chapter, we summarize our research contributions and discuss some future research directions.

\section{Major Research Contributions}
\label{MRCon_Chap7}

We have developed three different CMAC design frameworks addressing different network scenarios for HD CRNs
as well as an adaptive FDC-MAC framework for FD CRNs. These research outcomes have been resulted in
three journal publications  \cite{Le11, Le12}, \cite{tan2014joint} and its corresponding
conference publications \cite{Tanconf2012, Tanconf2012b, Tanconf2013, Tanconf2015, tan2015distributed}
as well as one journal under submission \cite{tan2014joint}.

In the first contribution, we have proposed the MAC protocols for CRNs with parallel sensing that explicitly take into account 
spectrum-sensing operation and imperfect sensing performance. In addition, we have performed throughput analysis for the proposed MAC protocols 
and determined their optimal configurations for throughput maximization. 
These studies have been conducted for both single- and multiple-channel scenarios subject to protection constraints for primary receivers. 

In the second contribution, we have investigated the MAC protocol design, analysis, and channel assignment issues for CRNs
 with sequential sensing. For the channel assignment, we have presented the optimal brute-force and low-complexity algorithms and analyzed their complexity. 
In particular, we have developed two greedy channel assignment algorithms for throughput maximization, namely non-overlapping and overlapping 
channel assignment algorithms.
In addition, we have proposed an analytical model to quantify the saturation throughput of the overlapping channel assignment algorithm. 
We have also presented several potential extensions including the design of max-min fair channel assignment algorithms and consideration of 
imperfect spectrum sensing. 
 
In the third contribution, we study a general SDCSS and access framework for the heterogeneous cognitive environment where 
channel statistics and spectrum holes on different channels can be arbitrary. Moreover,
no central controller is required to collect sensing results and make spectrum sensing decisions. 
In particular, the design is based on the distributed $p$-persistent CSMA protocol incorporating SDCSS for multi-channel CRNs. 
We have performed saturation throughput analysis and optimization of spectrum sensing time and access parameters to
 achieve the maximum throughput for a given allocation of channel sensing sets. 
Afterward we have studied the channel sensing set optimization (i.e., channel assignment) for throughput maximization and 
investigated both exhaustive search and low-complexity greedy algorithms to solve the underlying optimization problem.
Then we have extended the design and analysis to consider reporting errors during the exchanges of spectrum sensing results.

In the last contribution, we have proposed the FDC--MAC protocol for FD CRNs, analyzed its throughput 
performance, and studied its optimal parameter configuration. 
The design and analysis take into account the FD communication capability and the self-interference of the FD transceiver. 
In particular, SUs employ the standard p-persistent CSMA mechanism for contention resolution then the winning SU performs 
simultaneous sensing and transmission during the sensing stage and transmission only in the transmission stage.
We have also shown that there exists an optimal FD sensing time to achieve the maximum throughput. 
Moreover, we have proposed an algorithm to configure different design parameters including SU's transmit
 power and sensing time to achieve the maximum throughput. 

\section{Further Research Directions}
\label{FRWork_Chap7}

Our research work in this dissertation focuses on the MAC protocol for efficient media sharing and QoS 
provisioning in CRNs. The following research directions are of importance and deserve further investigation.

%\textbf{Multichannel MAC protocol design for FDCRNs}
\subsection{Multi-channel MAC protocol design for FD CRNs}

The MAC protocol design for FD CRNs was proposed for the single-channel scenario in Chapter \zref{Chapter6} where it has been shown
to deliver excellent and flexible performance tradeoffs for FD CRNs \cite{tan2015distributed, tan2015distgen}.
This proposed FDC--MAC protocol design is more general and flexible than existing MAC protocols for FD CRNs.
However, engineering the multi-channel FD CMAC design is more challenging, which will be pursued in the future.
Moreover, we will also consider the channel assignment problem for FD CRNs.

%\textbf{Multihop MAC protocol design for HDCRNs and FDCRNs}
\subsection{CMAC and routing design for multi-hop HD and FD CRNs}

Routing protocol design for HD and FD CRNs presents many interesting open problems to address.
In the  multi-hop communication environments with spectrum heterogeneity,
cross-layer design for CMAC and routing deserves further investigations. In particular, development of suitable coordination and spectrum
sensing schemes to manage interference among concurrent SUs' transmissions, efficiently exploit spectrum holes, and protect PUs
for HD CRNs and FD CRNs is a good direction for further research. 
Finally, study of channel assignment is also a promising research direction in multi-hop CRNs.

%\textbf{Joint MAC protocol and data processing design problem for the smartgrid}
\subsection{Applications of cognitive radio networking techniques for smartgrids}

We plan to investigate the joint cognitive protocol and data processing design for the smartgrid application. 
We are particularly interested in exploiting potential sparsity structure of the smartgrid data so that the data
can be compressed before being transmitted over the smartgrid communication networks \cite{Elda10, Bara10}.
This is quite expected since many types of smartgrid data can be very correlated over both space and time.
Here, existing techniques developed in the compressed sensing field can be applied for processing the smargrid data 
\cite{Tan10, Tan101, Tan2011using}. Cognitive network protocols employed to deliver smartgrid data can 
further degrade the communication performance due to access collisions and the intermittent nature of spectrum holes. 
Therefore, joint design of cognitive protocols and data processing algorithms is important to ensure the 
desirable end-to-end QoS performance.

\section{List of publication}
\label{Pub_Chap7}

\noindent
\textbf{Journals:}

\vspace{0.2cm}
\noindent
[J1] L.~T.~ Tan, and L.~B.~ Le, ``Design and Optimal Configuration of Full--Duplex MAC Protocol for Cognitive Radio Networks
Considering Self-Interference,'' submitted.

\noindent
[J2] L.~T.~ Tan, and L.~B.~ Le, ``Joint Data Compression and MAC Protocol Design for Smartgrids with Renewable Energy,'' submitted.

\noindent
[J3] L.~T.~ Tan, and L.~B.~ Le, ``Joint Cooperative Spectrum Sensing and MAC Protocol Design for Multi-channel Cognitive Radio Networks,'' \emph{EURASIP Journal on Wireless Communications and Networking}, 2014 (101), June 2014.

\noindent
[J4] L.~T.~ Tan, and L.~B.~ Le, ``Channel Assignment with Access Contention Resolution for Cognitive Radio Networks,'' \emph{IEEE Transactions on Vehicular Technology}, vol. 61, no. 6, pp.  2808--2823, July 2012.

\noindent
[J5] L.~T.~ Tan, and L.~B.~ Le, ``Distributed MAC Protocol for Cognitive Radio Networks: Design, Analysis,and Optimization,'' \emph{IEEE Transactions on Vehicular Technology}, vol. 60, no. 8, pp. 3990--4003, Oct. 2011

\vspace{0.3cm}
\noindent
\textbf{Conferences:}

\vspace{0.2cm}
\noindent
[C1] L. T. Tan and L. B. Le, ``Distributed MAC Protocol Design for Full-Duplex Cognitive Radio Networks,'' in \emph{2015 IEEE Global Communications Conference (IEEE GLOBECOM 2015)}, San Diego, CA, USA, Dec. 2015. 

\noindent
[C2] L. T. Tan and L. B. Le, ``Compressed Sensing Based Data Processing and MAC Protocol Design for Smartgrids,'' in \emph{2015 IEEE Wireless Communication and Networking Conference (IEEE WCNC 2015)},  New Orleans, LA USA, pp. 2138 - 2143, March 2015.

\noindent
[C3] L.~T.~ Tan, and L.~B.~ Le, ``General Analytical Framework for Cooperative Sensing and Access Trade-off Optimization,'' in \emph{2013 IEEE Wireless Communication and Networking Conference (IEEE WCNC 2013)}, Shanghai, China, pp. 1697 - 1702, April 2013.

\noindent
[C4] L.~T.~ Tan, and L.~B.~ Le, ``Fair Channel Allocation and Access Design for Cognitive Ad Hoc Networks,'' in \emph{2012 IEEE Global Communications Conference (IEEE GLOBECOM 2012)}, Anaheim, California, USA, pp. 1162 - 1167, Dec. 2012.

\noindent
[C5] L.~T.~ Tan, and L.~B.~ Le, ``Channel Assignment for Throughput Maximization in Cognitive Radio Networks,'' in \emph{2012 IEEE Wireless Communications and Networking Conference (IEEE WCNC 2012)}, Paris, France, pp. 1427-1431, April 2012.

\chapter{Appendix} % top level followed by section, subsection
\zlabel{Appendix}

%: ----------------------- paths to graphics ------------------------

% change according to folder and file names
\ifpdf
    \graphicspath{{8/figures/PNG/}{8/figures/PDF/}{8/figures/}}
\else
    \graphicspath{{8/figures/EPS/}{8/figures/}}
\fi

\section{Channel Assignment for Throughput Maximization in Cognitive Radio Networks}

The content of this appendix was published in \textit{Proc. IEEE WCNC'2012} in the following paper:

L.~T.~ Tan, and L.~B.~ Le, ``Channel Assignment for Throughput Maximization in Cognitive Radio Networks,'' in {\em Proc. IEEE WCNC'2012}, Paris, France, pp. 1427--1431, April 2012.

\newpage

\section{Fair Channel Allocation and Access Design for Cognitive Ad Hoc Networks}

The content of this appendix was published in \textit{Proc. IEEE GLOBECOM'2012} in the following paper:

L.~T.~ Tan, and L.~B.~ Le, ``Fair Channel Allocation and Access Design for Cognitive Ad Hoc Networks,'' in {\em Proc. IEEE GLOBECOM'2012}, Anaheim, California, USA, pp. 1162--1167, December 2012.

\newpage

\section{General Analytical Framework for Cooperative Sensing and Access Trade-off Optimization}

The content of this appendix was published in \textit{Proc. IEEE WCNC'2013} in the following paper:

L.~T.~ Tan, and L.~B.~ Le, ``General Analytical Framework for Cooperative Sensing and Access Trade-off Optimization,'' in {\em Proc. IEEE WCNC'2013}, Shanghai, China, pp. 1697--1702, April 2013.

\newpage

\section{Distributed MAC Protocol Design for Full-Duplex Cognitive Radio Networks}

The content of this appendix was published in \textit{Proc. IEEE GLOBECOM'2015} in the following paper:

L.~T.~ Tan, and L.~B.~ Le, ``Distributed MAC Protocol Design for Full-Duplex Cognitive Radio Networks,'' in {\em Proc. IEEE GLOBECOM'2015}, San Diego, CA, USA, December, 2015.

%\section{Asynchronous Full--Duplex MAC Protocol for Cognitive Radio Networks}

%The content of this chapter was submitted in IEEE Transactions on Cognitive Communications and Networking in the following paper:
%
%L.~T.~ Tan, and L.~B.~ Le, ``Asynchronous Full--Duplex MAC Protocol for Cognitive Radio Networks,'' in {\em IEEE Transactions on Cognitive Communications and Networking}, July 2015 (Under review).

%\include{10_summary/Summary}
%\include{11_summary_francaise/Summary_Francaise}

%\include{9_backmatter/appendices}

%\include{7/discussion}               % discussion of results
%
%\include{8/materials_methods}        % description of lab methods

% --------------------------------------------------------------
%:                  BACK MATTER: appendices, refs,..
% --------------------------------------------------------------

% the back matter: appendix and references close the thesis

%: ----------------------- bibliography ------------------------

% The section below defines how references are listed and formatted
% The default below is 2 columns, small font, complete author names.
% Entries are also linked back to the page number in the text and to external URL if provided in the BibTex file.

% PhDbiblio-url2 = names small caps, title bold & hyperlinked, link to page 
%\begin{multicols}{2} % \begin{multicols}{ # columns}[ header text][ space]
\begin{multicols}{1} % \begin{multicols}{ # columns}[ header text][ space]
%\begin{tiny} % tiny(5) < scriptsize(7) < footnotesize(8) < small (9)
%\begin{footnotesize} % tiny(5) < scriptsize(7) < footnotesize(8) < small (9)
\begin{scriptsize} % tiny(5) < scriptsize(7) < footnotesize(8) < small (9)
\bibliographystyle{Latex/Classes/PhDbiblio-url2} % Title is link if provided
\renewcommand{\bibname}{References} % changes the header; default: Bibliography

\bibliography{9_backmatter/references} % adjust this to fit your BibTex file

%\end{tiny}
%\end{footnotesize}
\end{scriptsize}
\end{multicols}

% --------------------------------------------------------------
% Various bibliography styles exit. Replace above style as desired.

% in-text refs: (1) (1; 2)
% ref list: alphabetical; author(s) in small caps; initials last name; page(s)
%\bibliographystyle{Latex/Classes/PhDbiblio-case} % title forced lower case
%\bibliographystyle{Latex/Classes/PhDbiblio-bold} % title as in bibtex but bold
%\bibliographystyle{Latex/Classes/PhDbiblio-url} % bold + www link if provided

%\bibliographystyle{Latex/Classes/jmb} % calls style file jmb.bst
% in-text refs: author (year) without brackets
% ref list: alphabetical; author(s) in normal font; last name, initials; page(s)

%\bibliographystyle{plainnat} % calls style file plainnat.bst
% in-text refs: author (year) without brackets
% (this works with package natbib)

% --------------------------------------------------------------

% according to Dresden med fac summary has to be at the end
%\include{0_frontmatter/abstract}

%: Declaration of originality
%\include{9_backmatter/declaration}

\end{document}